\documentclass[aps,showpacs,superscriptaddress,
amsmath,amssymb,floatfix,superscriptaddress,twocolumn]{revtex4}
\usepackage{graphicx}
\usepackage{epsfig}
\usepackage{color}
\definecolor{cayenne}{RGB}{128, 0, 0}
\definecolor{green}{RGB}{1, 165, 1}
\definecolor{orange}{RGB}{255,128,0}
\newcommand{\fig}[1]{\parbox{1.5cm}{\epsfig{file=#1.eps,width=1.5cm}}}
\newcommand{\largefig}[1]{\parbox{3.5cm}{\epsfig{file=#1.eps,width=3.5cm}}}
\begin{document}
\title{Curvature corrections to the nonlocal interfacial model for short-ranged forces}
\author{Jos\'e M. Romero-Enrique}
\affiliation{
Departamento de F\'{\i}sica At\'omica, Molecular y
Nuclear, \'Area de F\'{\i}sica Te\'orica, Universidad de Sevilla,
Avenida de Reina Mercedes s/n, 41012 Seville, Spain}
\author{Alessio Squarcini}
\affiliation{
Max-Planck-Institut f\"ur Intelligente Systeme,
Heisenbergstra{\ss}e 3, 70569 Stuttgart, Germany}
\affiliation{
IV. Institut f\"ur Theoretische Physik, Universit\"at Stuttgart,
Pfaffenwaldring 57, 70569 Stuttgart, Germany}
\author{Andrew O. Parry}
\affiliation{
Department of Mathematics, Imperial College London, London SW7 2AZ, United Kingdom}
\author{Paul M. Goldbart}
\altaffiliation{Address from August 2018: Department of Physics, The University of Texas at Austin, Austin, Texas 78712, USA.}
\affiliation{
School of Physics, Georgia Institute of Technology, 837 State Street, Atlanta, Georgia 30332, USA}
\begin{abstract}
{In this paper we revisit the derivation of a nonlocal interfacial  Hamiltonian model for systems with short-ranged intermolecular forces. Starting from a microscopic Landau-Ginzburg-Wilson Hamiltonian with a double parabola 
potential, we reformulate the derivation of the interfacial model using a rigorous boundary integral approach. This is done for three scenarios: a single fluid phase in contact with a nonplanar substrate (i.e., wall); a free interface separating coexisting fluid phases (say, liquid and gas); and finally a liquid-gas interface in contact with a nonplanar confining wall, as is applicable to wetting phenomena. For the first two cases our approaches identifies the correct form of the curvature corrections to the free energy and, for the case of a free interface, it allows us to recast these as an interfacial self-interaction as conjectured previously in the literature. When the interface is in contact with a substrate our approach similarly identifies curvature corrections to the nonlocal binding potential, describing the interaction of the interface and wall, for which we propose a generalized and improved diagrammatic formulation.  
}
\end{abstract}
%\pacs{02.30.Rz,05.20.Jj,68.03.Cd,68.08.Bc,68.35.Md}
\maketitle

\section{Introduction}
While significant progress has been made in the past few decades in understanding the statistical mechanics of inhomogeneous fluids and related interfacial phenomena \cite{Evans,Dietrich,Schick,Forgacs}, from a fundamental perspective many challenges remain for theory. Techniques based on molecular methods such as computer simulations \cite{Lee,Rowley} and density-functional theory \cite{Evans,Evans2} are wide spread, but under some circumstances large length scales emerge which make the use of mesoscopic models, often referred to as effective interfacial Hamiltonian or capillary-wave models, much more convenient and useful \cite{Buff,Weeks,Bedeaux}. These in fact have been pivotal in the development of the fluctuation theory of the thermal-wandering-induced roughness associated with a free or weakly pinned liquid-gas interface \cite{Buff} and also the classification of critical singularities occurring at continuous surface phase transitions such as wetting \cite{Dietrich,Schick,Forgacs,BHL,FH,delfino1} and wedge filling \cite{parry10,parry11,parry12,parry13,romero1,romero2,romero3,romero4,romero5,delfino2}. The search for a link between truly microscopic approaches and these mesoscopic descriptions can be traced back to van der Waals \cite{Rowlinson} and continue to this day, and in the past few years considerable effort has been invested in establishing this connection more rigorously. For example, intrinsic sampling methods use a many-body percolative, approach to identify  the interfacial position from the underlying microscopic molecular configurations, and this has been extensively used in simulations \cite{tarazona1,tarazona2,tarazona3,tarazona4,tarazona5,tarazona6,tarazona7,tarazona8,tarazona9,tarazona10,tarazona11,tarazona12,tarazona13,tarazona14}. A second, related, development has been the attempt to systematically derive an interfacial model for wetting transitions in settings involving short-ranged intermolecular forces from a more microscopic starting point \cite{parry1,parry2,parry3,parry4,parry5,parry6,parry7}. The need for this was originally driven by the significant discrepancy between initial predictions of strong nonuniversality for three-dimensional (3D) critical wetting, based on renormalization-group studies of local, partly phenomenologically justified, interfacial models \cite{BHL,FH}, and the findings of more microscopic Ising model simulation studies, which only reported minor deviations from mean-field-like behavior \cite{binder}.
In attempting to explain this, Fisher and Jin \cite{Jin1,Jin3} set out a very useful systematic basis for the derivation of an interfacial model from an underlying continuum Landau-Ginzburg-Wilson (LGW) Hamiltonian. Their idea was to introduce a constraint that specifies the interfacial configuration [which we denote by $\ell({\mathbf{x}})$] from that of the more microscopic order parameter $m({\mathbf{r}})$. Different options are available, such as the crossing criterion, in which $\ell({\bf{x}})$ is identified as the surface on which the order parameter takes some specified value or, alternatively, integral criteria which
are generalized measures of the local adsorption. Once the interface is defined, the interfacial Hamiltonian $H[\ell]$ is 
identified via the partial trace:
\begin{equation}
e^{-\beta H[\ell]}=\int D me^{-\beta \mathcal{H}_{LGW}[m]} \approx e^{-\beta \mathcal{H}_{LGW}[m_\Xi]},
\end{equation}
where $m_\Xi(\bf{r})$ is the profile that \emph{minimizes} the LGW Hamiltonian $\mathcal{H}_{LGW}[m]$,
subject to the constraint and additional boundary conditions. The Fisher-Jin identification \cite{Jin1,Jin3}, generalized to nonplanar walls, will be the starting point for our entire investigation. Within this scheme, therefore, all that is required is the determination of the constrained profile $m_\Xi(\bf{r})$, which will be a functional of the interfacial configuration (and wall shape). Fisher and Jin obtained this for a planar wall by considering perturbations about the flat interfacial configuration. However, this perturbative approach is inadequate for the purposes at hand because it misidentifies the structure of corrections to the standard local interfacial model. Indeed, this leads to serious problems when carried forward in renormalization-group studies, where it erroneously alters the structure of the well-known global phase diagrams for wetting \cite{Nakanishi}. Later, it was appreciated that the solutions to the constrained mean-field-like equations for $m_{\Xi}(\bf{r})$ could be reformulated using Green's functions \cite{parry1,parry2}. This highlights immediately the {\emph{nonlocal}} nature of the interaction of the interface and substrate, which has a simple diagrammatic representation. Furthermore, it resolves many of the problems associated with the fluctuation theory of critical wetting and in addition yields a description of correlation functions fully consistent with exact sum rules \cite{parry4,parry5}. The 
predictions of this nonlocal model are also consistent with more recent Ising model simulations, which reported deviations
from mean-field
behavior for critical wetting \cite{binder2}. The nonlocal decay of order between a fluctuating interface and particles situated away from it, predicted by this approach, has been seen directly in both Ising model and molecular simulation studies \cite{binder3,tarazona11}.

In this paper we present an alternative and more rigorous derivation of the nonlocal interfacial model to that presented originally \cite{parry1,parry2}, which was still partly physically motivated. Our derivation is based on exact integral representations of the solutions of linear partial differential equations on a closed domain, which are cast as functionals of the solutions at the boundaries. This method, referred to as the boundary element method, has in fact been applied successfully to numerous engineering problems \cite{Brebbia,Katsikadelis}. Here we develop an improved perturbative diagrammatic approach, which is related to the multiple reflection method used in the celebrated analysis
of the wave equation yielding eigenfrequencies in a closed domain \cite{Balian,Balian2,Balian3}. 
When applying this methodology to the evaluation of the interfacial free energy and order-parameter profile of a fluid phase in contact with a structured substrate or of a constrained liquid-gas configuration, we recover, at leading order, the previous nonlocal model but now with curvature corrections. For the case of a constrained liquid-gas interface this leads naturally to an interfacial self-interaction precisely as has been conjectured \cite{PR}. The most detailed application of this method involves the rigorous determination of the binding potential functional for wetting films in contact with a nonplanar  wall. Here, we identify an additional series of diagrams, not present in the original formulation, which arise when the substrate and liquid-gas interface are not parallel. These diagrams are resummed to obtain an alternative version of the nonlocal model which recovers the original version of the nonlocal model in certain limits.

Our paper is arranged as follows. In Sec. II we present the theoretical framework and recall the mathematical tools used in our approach. In Sec. III we apply this to a single phase, which we take to be liquid, in contact with a wall to determine the mean-field excess free-energy functional $F_{wl}[\psi]$, which is a functional of the wall shape $\psi$. In Sec. IV we extend this to a free liquid-gas interface, and determine the interfacial Hamiltonian $H[\ell]$, which is a functional of the (constrained) interfacial configuration $\ell$. Finally, we consider the most involved situation, in which a constrained wetting film is located near to a nonplanar wall, and we determine the binding potential  $W[\ell,\psi]$, which is a functional of both the interface and the wall shapes.

\section{Theoretical framework}
Consider the LGW Hamiltonian defined on a domain 
$\Omega$: 
\begin{eqnarray}
\mathcal{H}_{LGW}[m]=\int_\Omega 
d\mathbf{r} \left\{\frac{1}{2} \left( \boldsymbol\nabla m\right)^2+
\Delta \phi(m)\right\}\nonumber\\
+\int_{\partial \Omega}\ \Phi_s(\mathbf{s}) d\mathbf{s},
\label{HLGW}
\end{eqnarray}
where the shifted potential $\Delta \phi(m)$ corresponds to the excess
contribution, with respect to the bulk, of the free-energy density, and $\Phi_s$ 
is a surface potential defined on the domain boundary $\partial \Omega$. 
Typically, $\Phi_s$ is taken to be a quadratic function 
of $m(\mathbf{s})$, i.e., 
\begin{equation}
\Phi_s(\mathbf{s})=-\frac{g}{2}\left[m(\mathbf{s})+\frac{h_1}{g}\right]^2,
\end{equation}
where $h_1$ and $g$ are a local field and an enhancement parameter, respectively, modeling 
the coupling to the substrate (i.e., wall). Usually, these 
quantities are taken to be equal to their flat-wall counterparts, although 
additional curvature-induced terms may be included phenomenologically. 
Finally, we note that fixed (i.e., Dirichlet) boundary conditions correspond to the limit 
$h_1 \to \infty$ and $g\to -\infty$ with $-h_1/g = m_1$,
where $m_1$ is the fixed value of the order parameter at the wall.

In a zero external field (i.e., $h=0$) and below the bulk critical temperature $T_c$, the shifted potential $\Delta \phi$ has a familiar double-well structure, which we capture via the simple double-parabola (DP) approximation
\begin{equation}
\Delta \phi (m) = \frac{\kappa^2}{2}\left(|m|-m_0\right)^2,
\label{deltaphi}
\end{equation}
where $\kappa$ is the inverse bulk correlation length and $m_0$ 
is the bulk order parameter. In this description, therefore, there 
are two bulk phases having order-parameter values $-m_0$ (which we regard as
the gas phase) and $+m_0$ (which we regard as the liquid phase).  
For a general inhomogeneous situation, we will identify the phase at any point via the
sign of the order parameter. Consequently, we will refer to the phase as gas if $m<0$ and liquid otherwise.
Finally, we adopt a simple crossing criterion of a constrained interfacial configuration, whereby the
interface is defined as the surface on which the order parameter vanishes, 
i.e., $m=0$ \cite{Jin1,Jin3}.

As the constrained profile $m_\Xi$ minimizes the LGW Hamiltonian, it satisfies the mean-field-like Euler-Lagrange equation 
\begin{equation}
\nabla^2 m_\Xi = \kappa^2(m_\Xi-m_b),
\label{pde}
\end{equation}
where $m_b=\pm m_0$, depending on whether the bulk phase is liquid or gas.
This partial differential equation is to be solved subject to the 
natural boundary condition
\begin{equation}
\mathbf{n}\mathbf{\cdot} \boldsymbol\nabla m_\Xi (\mathbf{s})=h_1+gm_\Xi(\mathbf{s}),  
\label{boundary}
\end{equation}
where $\mathbf{n}$ is the outward normal to the boundary of the integration domain $\Omega$. 
If fixed boundary conditions are applied on $\mathbf{s}$, 
we simply set $m_\Xi (\mathbf{s})=m_1$ instead of imposing Eq.~(\ref{boundary}). 

In order to obtain the mean-field free energy, we first consider the situation where there is only one phase in the
integration domain. Multiplying Eq.~(\ref{pde}) by $(m_\Xi-m_b)/2$ and integrating over the domain $\Omega$,
we get
\begin{equation}
\int_\Omega d\mathbf{r} \frac{(m_\Xi-m_b)\nabla^2 m_\Xi}{2} = \int_\Omega \frac{\kappa^2}{2}(m_\Xi-m_b)^2 .
\label{intpde}
\end{equation}
We now use the identity $u\nabla^2 u + \boldsymbol{\nabla}u \mathbf{\cdot} \boldsymbol{\nabla} u = \boldsymbol{\nabla} 
\mathbf{\cdot} (u \boldsymbol{\nabla} u)$ with $u=m_\Xi-m_b$  and apply the divergence theorem to obtain
\begin{eqnarray}
\int_\Omega 
d\mathbf{r} \left\{\frac{1}{2} \left( \boldsymbol\nabla m_\Xi\right)^2+ \frac{\kappa^2}{2}(m_\Xi-m_b)^2\right\}\nonumber\\
=\oint_{\partial \Omega} d\mathbf{s} \frac{m_\Xi(\mathbf{s})-m_b}{2}[ \mathbf{n}\mathbf{\cdot}\boldsymbol\nabla m_\Xi 
(\mathbf{s})] .
\label{mfHLGW0}
\end{eqnarray}
Next we make use of Eq.~(\ref{boundary}) to rewrite the surface contribution to Eq.~(\ref{HLGW}) using
\begin{eqnarray}
\Phi_s(\mathbf{s})&=&-\frac{1}{2}\left(m_\Xi(\mathbf{s})+\frac{h_1}{g}\right)
\left[\mathbf{n}\mathbf{\cdot}\boldsymbol\nabla m_\Xi(\mathbf{s})\right]\nonumber\\
=&-&\left[\frac{m_\Xi(\mathbf{s})-m_b}{2}+\frac{h_1+gm_b}{2g}\right][ \mathbf{n}\mathbf{\cdot}\boldsymbol\nabla m_\Xi 
(\mathbf{s})] .
\end{eqnarray}
Hence, when evaluated at the equilibrium profile $m_\Xi$, the LGW Hamiltonian identifies the free energy as
\begin{eqnarray}
\mathcal{H}_{LGW}[m_\Xi]=\int_\Omega 
d\mathbf{r} \left\{\frac{1}{2} \left( \boldsymbol\nabla m_\Xi\right)^2+ \frac{\kappa^2}{2}(m_\Xi-m_b)^2\right\}\nonumber\\
+\oint_{\partial \Omega}\ \Phi_s(\mathbf{s}) d\mathbf{s},
\label{mfHLGWprev}
\end{eqnarray}
and reduces to
\begin{equation}
\mathcal{H}_{LGW}[m_\Xi] 
=-\frac{h_1+gm_b}{2g}\oint_{\partial \Omega} d\mathbf{s} \ \mathbf{n}\mathbf{\cdot} \boldsymbol\nabla m_\Xi (\mathbf{s}),
\label{mfHLGW}
\end{equation}
a result that will be central to our method.

\begin{figure}
\includegraphics[width=8.6cm]{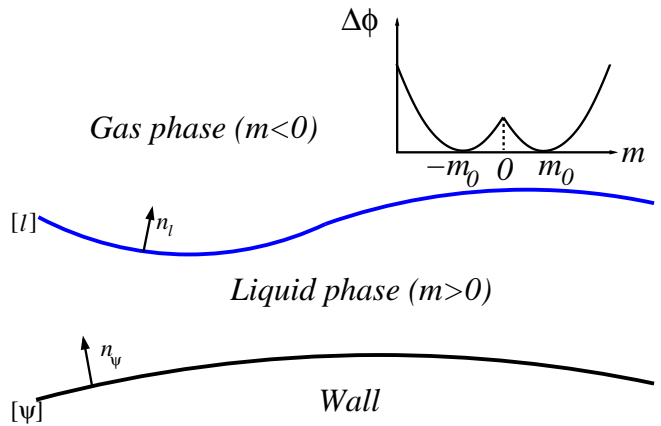}
\caption{Schematic illustration of a nonplanar interfacial configuration (blue line) for a constrained wetting film of liquid at a nonplanar wall (black line). Conventions for the surface normals are shown.
The inset shows the double-parabola approximation for $\Delta \phi (m)$. \label{figure1}}
\end{figure}

In the presence of a wetting layer of a different phase that intrudes between the wall and the bulk (see Fig.~\ref{figure1}) the domain $\Omega$ must be considered to be the union $\Omega = \cup_i \Omega_i$, where each appropriate subdomain $\Omega_i$ has boundaries $\partial \Omega_i$ that lie either on the substrate surface or at the liquid-gas interface.
We define $\partial \Omega_i=\partial \Omega_{1,i} \cup \partial \Omega_{2,i}$, where the boundary
condition (\ref{boundary}) is satisfied in $\partial \Omega_{1,i}$ and $\partial \Omega_{2,i}$ corresponds 
to the gas-liquid interface.
For this case, the generalization of Eq.~(8) identifies the constrained free-energy functional for a given interfacial configuration as
\begin{eqnarray}
&&\mathcal{H}_{LGW}[m_\Xi]=\nonumber\\
&&=\sum_i \Bigg[\int_{\Omega_i} 
d\mathbf{r} \left\{\frac{1}{2} \left( \boldsymbol\nabla m_\Xi\right)^2+ \frac{\kappa^2}{2}[m_\Xi-(m_b)_i]^2\right\}
\nonumber\\
&&\qquad\qquad\qquad\qquad+\int_{\partial \Omega_{1,i}} d\mathbf{s}\ \Phi_s(\mathbf{s})
\Bigg],
\label{mfHLGW2prev}
\end{eqnarray}
which reduces to
\begin{eqnarray}
\mathcal{H}_{LGW}[m_\Xi]
=\sum_i \Bigg[-\frac{h_1+gm_b}{2g}\int_{\partial \Omega_{1,i}} d\mathbf{s} \ \mathbf{n}\mathbf{\cdot} 
\boldsymbol\nabla m_\Xi (\mathbf{s}) 
\nonumber\\
- \frac{(m_b)_i}{2}\int_{\partial \Omega_{2,i}} d\mathbf{s} 
[\mathbf{n}\mathbf{\cdot} \boldsymbol\nabla m_\Xi (\mathbf{s})]\Bigg].
\label{mfHLGW2}
\end{eqnarray}

The mean-field free energy corresponds to the 
interfacial configuration that gives the least free energy. In this sense, the free energy becomes
a functional of the interfacial profile. 

The above results demonstrate that, for the potential, the equilibrium free energy of an interfacial configuration may be determined in terms of the normal derivatives of the order parameter at the substrate and liquid-gas interface (if present). This simplification is not so surprising, given that, for the DP potential, the whole order-parameter profile can be obtained formally in terms of the values at the boundaries. To see this, let us consider  the (rescaled) Green's function that solves
\begin{equation}
{\cal L}K (\mathbf{r},\mathbf{r}_0) \equiv \left(-\nabla^2_{\mathbf{r}} +\kappa^2\right) 
K (\mathbf{r},\mathbf{r}_0) =2\kappa \delta (\mathbf{r}-\mathbf{r}_0),
\label{OZ}
\end{equation} 
and vanishes as $|\mathbf{r}-\mathbf{r}_0|\to \infty$. The subscript in $\nabla^2_{\mathbf{r}}$ denotes
that the nabla operator acts on the argument $\mathbf{r}$ of $K$. Its solution is 
the Ornstein-Zernike correlation function
\begin{equation}
K(\mathbf{r},\mathbf{r}_0)=\frac{\kappa}{2\pi} \frac{\exp(-\kappa|\mathbf{r}-
\mathbf{r}_0|)}{|\mathbf{r}-\mathbf{r}_0|}.
\label{OZ2}
\end{equation}
The second Green's identity for the Hermitian operator ${\cal L}$ states that
\begin{equation}
\int_{\Omega} d\mathbf{r} \left[v{\cal L}u-u{\cal L}v\right]= 
-\int_{\partial \Omega} \left[
v (\mathbf{n}\mathbf{\cdot} \boldsymbol\nabla u)-u(\mathbf{n}\mathbf{\cdot} \boldsymbol\nabla v)\right] d\mathbf{s}
\label{greenidentity}
\end{equation}
for any domain $\Omega$ with boundary $\partial \Omega$, where the outward normal 
is $\mathbf{n}$ and $u$ and $v$ are arbitrary functions. If we choose $u(\mathbf{r})=
K(\mathbf{r},\mathbf{r}_0)/2\kappa$ and $v(\mathbf{r})=m_\Xi(\mathbf{r})-m_b$, and taking into account Eqs. (\ref{pde}) and (\ref{OZ}), then
\begin{eqnarray}
&&[m_\Xi(\mathbf{r})-m_b]\Theta_\Omega(\mathbf{r})=\frac{1}{2\kappa}\int_{\partial \Omega} d\mathbf{s} K(\mathbf{s},\mathbf{r})
[\mathbf{n}\mathbf{\cdot} \boldsymbol\nabla m_\Xi(\mathbf{s})]\nonumber\\&&-\frac{1}{2\kappa}\int_{\partial \Omega} d\mathbf{s}
[m_\Xi(\mathbf{s})-m_b] [\mathbf{n}\mathbf{\cdot} \boldsymbol\nabla_{\mathbf{s}} K(\mathbf{s},\mathbf{r})],
\label{greenidentity2}
\end{eqnarray}
where $\Theta_\Omega(\mathbf{r})$ is the characteristic function of the set $\Omega$, i.e., $\Theta_\Omega
(\mathbf{r})=1$ if $\mathbf{r}\in \Omega$, and $\Theta_\Omega (\mathbf{r})=0$ otherwise (excluding in both 
cases the boundary $\partial \Omega$).
As before, $\partial \Omega_1$ refers to the portion of the boundary where
Eq.~(\ref{boundary}) is satisfied, 
while $\partial \Omega_2$  lies on the appropriate side of
the gas-liquid interface (see Fig.~\ref{figure1}). We can then recast Eq.~(\ref{greenidentity2}) as
\begin{eqnarray}
[m_\Xi(\mathbf{r})&-&m_b]\Theta_\Omega(\mathbf{r})\nonumber\\
&=&\frac{1}{2\kappa}\int_{\partial \Omega_1} d\mathbf{s}
\left(\frac{h_1}{g}+m_b\right) \partial_n K(\mathbf{s},\mathbf{r})\nonumber\\
&+&\frac{m_b}{2\kappa}\int_{\partial \Omega_2} d\mathbf{s} \partial_n K(\mathbf{s},\mathbf{r})
\nonumber\\
&+&\frac{1}{2\kappa}\int_{\partial \Omega_1} d\mathbf{s} \left(K(\mathbf{s},\mathbf{r})-\frac{1}{g}\partial_n 
K(\mathbf{s},\mathbf{r})\right)
\partial_n m_\Xi(\mathbf{s})\nonumber\\&+&\frac{1}{2\kappa}\int_{\partial \Omega_2} d\mathbf{s}
K(\mathbf{s},\mathbf{r}) \partial_n m_\Xi(\mathbf{s}),
\label{greenidentity3}
\end{eqnarray}
where $\partial_n$ denotes the normal derivative $\mathbf{n}\mathbf{\cdot} \boldsymbol\nabla_{\mathbf{s}}$. 

What remains is the determination of the normal derivative $\partial_n m_\Xi$ at each point along the subdomain boundaries. However, Eq.~(\ref{greenidentity3}) itself cannot be used to determine these, and we must use a technique to modify this appropriately. To this end, we first place $\bf{r}$ at a boundary point {\emph{and}} deform the boundary near it by cutting a circular hole of radius $\epsilon$ and  adding a hemispherical cap atop it, so that the point is again inside the sub-domain under consideration. We then evaluate the order parameter at $\bf{r}$ and finally take the 
limit $\epsilon\to 0$.
Assuming the interfaces are smooth, we obtain the integral equation within each domain
\begin{eqnarray}
&&\frac{m_\Xi(\mathbf{s}_0)-m_b}{2}= 
\frac{1}{2\kappa}\int_{\partial \Omega_1} d\mathbf{s}
\left(\frac{h_1}{g}+m_b\right) \partial_n K(\mathbf{s},\mathbf{s}_0)\nonumber\\
&+&\frac{m_b}{2\kappa}\int_{\partial \Omega_2} d\mathbf{s} \partial_n K(\mathbf{s},\mathbf{s}_0)
\nonumber\\
&+&\frac{1}{2\kappa}\int_{\partial \Omega_1} d\mathbf{s} \left(K(\mathbf{s},\mathbf{s}_0)-\frac{1}{g}\partial_n 
K(\mathbf{s},\mathbf{s}_0)\right)
\partial_n m_\Xi(\mathbf{s})\nonumber\\&+&\frac{1}{2\kappa}\int_{\partial \Omega_2} d\mathbf{s}
K(\mathbf{s},\mathbf{s}_0) \partial_n m_\Xi(\mathbf{s}),
\label{greenidentity4}
\end{eqnarray}
where the normal derivative of the Green's function $K$ acts on its first argument and 
integration must be interpreted as its Cauchy principal value.
Consequently, if $\mathbf{s}_0\in \partial \Omega_1$, Eq.~(\ref{greenidentity4}) can be written as
\begin{eqnarray}
&&\frac{1}{2\kappa}\int_{\partial \Omega_1} d\mathbf{s}
\left(\frac{h_1}{g}+m_b\right) \left[\partial_n K(\mathbf{s},\mathbf{s}_0)+\kappa\delta(\mathbf{s}-\mathbf{s}_0)\right]
\nonumber\\
&+&\frac{1}{2\kappa}\int_{\partial \Omega_1} d\mathbf{s} \Bigg(K(\mathbf{s},\mathbf{s}_0)-\frac{1}{g}\partial_n 
K(\mathbf{s},\mathbf{s}_0)\nonumber\\&-&\frac{\kappa}{g}\delta(\mathbf{s}-\mathbf{s}_0)\Bigg)
\partial_n m_\Xi(\mathbf{s})
+\frac{m_b}{2\kappa}\int_{\partial \Omega_2} d\mathbf{s} \partial_n K(\mathbf{s},\mathbf{s}_0)
\nonumber\\&+&\frac{1}{2\kappa}\int_{\partial \Omega_2} d\mathbf{s}
K(\mathbf{s},\mathbf{s}_0) \partial_n m_\Xi(\mathbf{s})=0.
\label{greenidentity5}
\end{eqnarray}
Similarly, if $\mathbf{s}_0\in \partial \Omega_2$, Eq.~(\ref{greenidentity4}) reads
\begin{eqnarray}
&&\frac{1}{2\kappa}\int_{\partial \Omega_1} d\mathbf{s}
\left(\frac{h_1}{g}+m_b\right) \partial_n K(\mathbf{s},\mathbf{s}_0)\nonumber\\
&+&\frac{m_b}{2\kappa}\left(\kappa+\int_{\partial \Omega_2} d\mathbf{s} \partial_n K(\mathbf{s},\mathbf{s}_0)\right)
\nonumber\\
&+&\frac{1}{2\kappa}\int_{\partial \Omega_1} d\mathbf{s} \left(K(\mathbf{s},\mathbf{s}_0)-\frac{1}{g}\partial_n 
K(\mathbf{s},\mathbf{s}_0)\right)
\partial_n m_\Xi(\mathbf{s})\nonumber\\&+&\frac{1}{2\kappa}\int_{\partial \Omega_2} d\mathbf{s}
K(\mathbf{s},\mathbf{s}_0) \partial_n m_\Xi(\mathbf{s})=0.
\label{greenidentity6}
\end{eqnarray}

Under some circumstances, e.g., for certain boundary conditions, Eq. (\ref{greenidentity3}) is not the most 
convenient representation
of the constrained order-parameter profile. 
Another possible representation is the single-layer potential. Let us assume
that the order parameter on the boundary $\partial \Omega$ is known. Now we determine the solutions to the Helmholtz
equation inside and outside $\Omega$ with the same Dirichlet boundary conditions. We can use Eq.~ 
(\ref{greenidentity2}) for these problems, keeping in mind that the normal derivatives are different in each 
problem. Adding these equations, we find the representation, valid everywhere in space,
\begin{equation}
m_\Xi(\mathbf{r})=m_b +\frac{1}{2\kappa}\int_{\partial \Omega} d\mathbf{s} K(\mathbf{s},\mathbf{r}) 
\Psi(\mathbf{s}),
\label{singlelayer}
\end{equation}
where $\Psi(\mathbf{s})=[\partial_n m_\Xi(\mathbf{s})]^{+}-[\partial_n  
m_\Xi(\mathbf{s})]^{-}$, with the plus (minus) sign standing for the interior (exterior) problem to $\Omega$, 
and $\mathbf{n}(\mathbf{s})$ is the outward normal from $\Omega$. The auxiliary
function $\Psi(\mathbf{s})$ can be obtained from the boundary integral equation
\begin{equation}
m_\Xi(\mathbf{s})=m_b+\frac{1}{2\kappa}
\int_{\partial \Omega} d\mathbf{s_0} K(\mathbf{s}_0,\mathbf{s})\Psi(\mathbf{s}_0).
\label{singlelayer2}
\end{equation}
The normal derivatives of the order-parameter profile on the boundary can be related to $\Psi$ as
\begin{equation}
[\partial_n m_\Xi(\mathbf{s})]^{\pm}=\pm \frac{\Psi(\mathbf{s})}{2}+\frac{1}{2\kappa}
\int_{\partial \Omega} d\mathbf{s_0}
\mathbf{n}(\mathbf{s})\mathbf{\cdot}\boldsymbol\nabla_{\mathbf{s}} K(\mathbf{s}_0,\mathbf{s})\Psi(\mathbf{s}_0).
\end{equation}

Alternatively, a double-layer potential representation of the order-parameter profile can be obtained if the normal
derivative of the order parameter on the boundary $\partial \Omega$ is known. We use Eq.~(\ref{greenidentity2}) 
for the solutions to the Helmholtz equation inside and outside $\Omega$ with opposite Neumann boundary 
conditions.
Note that the outward normal for every domain is the inward normal for the other one. By adding
these equations
we again obtain a representation that is valid everywhere in space:
\begin{equation}
\delta m_\Xi(\mathbf{r})=\frac{1}{2\kappa}\int_{\partial \Omega} d\mathbf{s} \mathbf{n}(\mathbf{s})\mathbf{\cdot}
\boldsymbol\nabla_\mathbf{s} K(\mathbf{s},\mathbf{r}) 
\overline{\Psi}(\mathbf{s}),
\label{doublelayer}
\end{equation}
where $\delta m_\Xi(\mathbf{r})\equiv m_\Xi(\mathbf{r})-m_b$ and $m_b=\pm m_0$ is the appropriate 
bulk order parameter in the
region containing $\mathbf{r}$. Here the modified auxiliary function 
$\overline{\Psi}(\mathbf{s})=[\delta m_\Xi(\mathbf{s})]^{-}-
[\delta m_\Xi(\mathbf{s})]^{+}$. The limits of the order parameter on each side of $\partial \Omega$ are
related to $\overline{\Psi}$ as
\begin{equation}
\delta m_\Xi(\mathbf{s}^{\pm})=\mp\frac{\overline{\Psi}(\mathbf{s})}{2}+\frac{1}{2\kappa}
\int_{\partial \Omega} d\mathbf{s_0} \mathbf{n}(\mathbf{s}_0)\mathbf{\cdot}
\boldsymbol\nabla_{\mathbf{s}_0}K(\mathbf{s}_0,\mathbf{s})\overline{\Psi}(\mathbf{s}_0).
\label{doublelayer2}
\end{equation}
On the other hand, $\partial_n m_\Xi(\mathbf{s})\equiv \mathbf{n}(\mathbf{s})\mathbf{\cdot}
\boldsymbol\nabla_\mathbf{s} m_\Xi$ is continuous on $\partial \Omega$:
\begin{equation}
\partial_n m_\Xi(\mathbf{s})=\mathbf{n}(\mathbf{s})\mathbf{\cdot}
\boldsymbol\nabla_\mathbf{s}\left[
\int_{\partial \Omega} d\mathbf{s_0}  \overline{\Psi} (\mathbf{s}_0)
\mathbf{n}(\mathbf{s}_0)\mathbf{\cdot}\boldsymbol\nabla_{\mathbf{s}_0} \frac{K(\mathbf{s}_0,\mathbf{s})}{2\kappa}
\right]
\end{equation}

Finally, we provide some additional relations which will be useful later. On using the Green's identity 
(\ref{greenidentity}) for two Green's functions, it follows that
\begin{eqnarray}
\int_{\partial \Omega} d\mathbf{s} K(\mathbf{s},\mathbf{r}) \mathbf{n}(\mathbf{s})\mathbf{\cdot}\boldsymbol
\nabla_{\mathbf{s}} K(\mathbf{s},\mathbf{r}')\nonumber\\
=\int_{\partial \Omega} d\mathbf{s} K(\mathbf{s},\mathbf{r}') \mathbf{n}(\mathbf{s})\mathbf{\cdot}\boldsymbol
\nabla_{\mathbf{s}} K(\mathbf{s},\mathbf{r}) ,
\label{greenidentity7}
\end{eqnarray}
where $\mathbf{r}$ and $\mathbf{r}'$ are positions inside the domain $\Omega$. If $\mathbf{r}'\to \mathbf{s}'$ on
the boundary $\partial \Omega$ then Eq.~(\ref{greenidentity7}) leads to
\begin{eqnarray}
&&\int_{\partial \Omega} d\mathbf{s} K(\mathbf{s},\mathbf{s}') \mathbf{n}(\mathbf{s})\mathbf{\cdot}\boldsymbol
\nabla_{\mathbf{s}} K(\mathbf{s},\mathbf{r})\nonumber\\
&&=-\kappa K(\mathbf{s}',\mathbf{r})+
\int_{\partial \Omega} d\mathbf{s} K(\mathbf{s},\mathbf{r}) \mathbf{n}(\mathbf{s})\mathbf{\cdot}\boldsymbol
\nabla_{\mathbf{s}} K(\mathbf{s},\mathbf{s}') .
\label{greenidentity8}
\end{eqnarray} 
Finally, if $\mathbf{r}\to \mathbf{s}$ on $\partial \Omega$, then
\begin{eqnarray}
&&\int_{\partial \Omega} d\mathbf{s} K(\mathbf{s}_0,\mathbf{s}) \mathbf{n}(\mathbf{s}_0)\mathbf{\cdot}\boldsymbol
\nabla_{\mathbf{s}_0} K(\mathbf{s}_0,\mathbf{s}')\nonumber\\&&
=
\int_{\partial \Omega} d\mathbf{s}_0 K(\mathbf{s}_0,\mathbf{s}') \mathbf{n}(\mathbf{s})\mathbf{\cdot}\boldsymbol
\nabla_{\mathbf{s}_0} K(\mathbf{s}_0,\mathbf{s}) .
\label{greenidentity9}
\end{eqnarray}

In the following sections we apply this formalism to obtain the interfacial free energies 
relevant to wetting phenomena: (i)~the interfacial free energy of a nonwetting 
bulk phase in contact with a rough substrate, (ii)~the self-interaction 
corresponding to a free liquid-gas interface, and finally (iii)~the binding potential for
a wetting film configuration (see Fig.~\ref{figure1}).

\section{Interfacial free energy of a liquid phase in contact with a nonplanar wall}

The first system that we consider is the simple case of a bulk phase 
in contact with a nonplanar wall when a wetting film is absent. The local height of the wall, above some reference plane (often taken to be the plane $z=0$), is written $\psi({\mathbf{x}})$, where ${\mathbf{x}}=(x,y)$ is the parallel displacement. Without loss of generality, we concentrate on the wall-liquid interface, supposing that the local surface field $h_1$ is positive so that the order parameter has the same (positive) value throughout.    
In this case, the domain $\Omega$ is just 
the set of points for which $z>\psi(\mathbf{x})$. In addition, we suppose that the substrate is chemically 
homogeneous, so 
$h_1$ and $g$ do not vary with position. The equilibrium mean-field configuration $m_\Xi(\bf{r})$ follows from the simple 
minimization of the LGW Hamiltonian, resulting in the Helmholtz equation (\ref{pde}) and the boundary conditions
\begin{eqnarray}
&&\mathbf{n}_\psi\mathbf{\cdot}\boldsymbol \nabla m_\Xi (\mathbf{r})=-h_1-gm_\Xi(\mathbf{r}), \  
\textrm{for}\ \mathbf{r}=(\mathbf{x},\psi(\mathbf{x})); \label{boundary2}\\
&&m_\Xi(z)\to m_b \qquad 
\textrm{for}\quad z\to +\infty,
\label{boundary2b}
\end{eqnarray}
where the bulk magnetization for the liquid phase is $m_b=m_0$. Similar results apply to the wall-gas interface,
for which $h_1$ is negative and $m_b=-m_0$. Here $\mathbf{n}_\psi$ denotes the \emph{inward} normal to the wall.
Since the order parameter does not change sign in $\Omega$, Eq.~(\ref{greenidentity5}) can be recast as
\begin{eqnarray}
\int_\psi d\mathbf{s} \left[K(\mathbf{s},
\mathbf{s}^\prime)
+\frac{1}{g} \partial_{n} K(\mathbf{s},\mathbf{s}^\prime)
-\frac{\kappa}{g}\delta(\mathbf{s}-\mathbf{s}^\prime)
\right] 
q(\mathbf{s})\nonumber\\
=\left(\frac{-h_1}{g}-m_b\right)\left[-\kappa+
\int_\psi d\mathbf{s} 
\partial_{n} K(\mathbf{s},\mathbf{s}^\prime)\right],
\label{inteq1}
\end{eqnarray}
where the integration $\int_\psi$ is over the substrate surface, $\partial_{n}(\mathbf{s})$ is shorthand 
for $\mathbf{n}(\mathbf{s})\mathbf{\cdot}\boldsymbol{\nabla}_{\mathbf{s}}$, $q\equiv \partial_{n} \delta m_\Xi$,  and
$\delta m_\Xi\equiv m_\Xi-m_b$.
Equation~(\ref{inteq1}) can only be solved exactly in a few exceptional circumstances, such as when symmetry arguments can be applied; these include  planar, cylindrical, or spherical substrates, all of which have constant curvature. 
For example, for a planar substrate $q$ is constant over the surface and has the value
\begin{equation}
q=-\frac{\kappa (h_1+gm_b)}{\kappa-g}.
\label{qflat}
\end{equation}
However, the generic solution of Eq.~(\ref{inteq1}) must include the \emph{local} curvature of the substrate,
and it is natural to look for a perturbative solution when the 
deviations from the flat case are small. To this end, let us introduce the principal curvatures 
$k_{1}(\mathbf{s})$ and $k_{2}(\mathbf{s})$ at a point $\mathbf{s}=(\mathbf{x},\psi(\mathbf{x}))$ on the surface. Here 
$R_{i}=1/\kappa_{i}$ are the corresponding radii of curvature and it is convenient to recall that 
$K_{G}=k_{1}k_{2}$ is the Gaussian curvature and $H=(k_{1}+k_{2})/2$ is the mean curvature (or half of 
the total curvature). 
Let us denote by $R$ the minimum of $|R_1|$ and $|R_2|$ so that
$H \sim R^{-1}$. Far from the bulk critical point the bulk correlation length $\kappa^{-1}$  
is microscopically small, so the substrate can be considered flat over several 
correlation lengths provided that $\kappa R \gg 1$.

\subsection{Perturbative approach}
\label{sec_3A}
We now set out our perturbative analysis of Eq.~(\ref{inteq1}). The idea is to expand all elementary building blocks of Eq.~(\ref{inteq1}) [the Ornstein-Zernike (OZ) kernel, its normal derivative $q$, and $d \mathbf{s}$] on the left-hand side and right-hand side of Eq.~(\ref{inteq1}) in powers of the curvature $H$, which can then 
be equated, term by term. 
We suppose that locally the surface is well approximated by a paraboloid in a neighborhood of 
$\mathbf{s}^{\prime}$, where we locate the origin 
of the coordinates. Consider now a point on the substrate surface
$\mathbf{s} = (\mathbf{x},\psi(\mathbf{x}))$. 
The vertical displacement of $\mathbf{s}$ with respect to the horizontal plane is
\begin{equation}
\label{verticaldisplacement}
\Delta\psi(\mathbf{r}_{\bot}) \equiv  \frac{1}{2}k_{1}x^{2} + \frac{1}{2}k_{2}y^{2}  +\cdots,
\end{equation}
where we have written $\mathbf{r}_{\bot}=\mathbf{x}-\mathbf{x}_{0}\equiv(x,y)$ for the projection of the vector $\mathbf{s}-\mathbf{s}^{\prime}$ onto the horizontal plane $\pi_{0}$, tangent to the graph of $\psi(\mathbf{x})$ in $\mathbf{s}^{\prime}$. With this parametrization, the coefficients  $k_{i}$ are exactly the principal curvatures  
$k_{i}(\mathbf{s}^{\prime})$ evaluated at $\mathbf{s}^\prime$. The ellipsis in Eq.~(\ref{verticaldisplacement}) stands for higher-order terms in $(x,y)$, which are coupled to higher orders of the local curvatures as well. We assume that the Taylor coefficients associated with terms
$x^m y^{n-m}$ (with $0\le m \le n$ and $n>3$) scale as $R^{-n+1}$. With this property, close to the origin 
$\Delta \psi(\mathbf{r}_{\bot}) = R \Delta 
\bar{\psi} (\mathbf{r}_{\bot}R^{-1})$, i.e., $R$ is the only relevant length scale for
the substrate shape. 

Let us consider first the OZ kernel. The two points are separated by the distance $|\mathbf{s}-\mathbf{s}^{\prime}| = \sqrt{\mathbf{r}_{\bot}^{2} + [\Delta \psi(\mathbf{r}_{\bot})]^{2}}$ and for small curvatures we can Taylor expand around the flat configuration to express the OZ kernel as a power series in the curvature:
\begin{equation}
K(\mathbf{s},\mathbf{s}^{\prime}) = \mathcal{K}(r_{\bot}) \Biggl[ 1 - \frac{1}{2}(1+\kappa r_{\bot}) \left(\frac{\Delta\psi(\mathbf{r}_{\bot})}{r_{\bot}}\right)^{2} + {O}(R^{-4}) \Biggr] ,
\label{expk}
\end{equation}
where $\mathcal{K}(x)=\kappa e^{-\kappa x}/2\pi x$. As the kernel $\mathcal{K}$ decays exponentially with
a lengthscale $\kappa^{-1}$, Eq.~(\ref{expk}) is a faithful representation of $K(\mathbf{s},\mathbf{s}^{\prime})$
around $\mathbf{s}^{\prime}$ if $\kappa R \gg 1$. For the normal derivative of the kernel we have
\begin{equation}
\partial_{n} K(\mathbf{s},\mathbf{s}^{\prime}) = \mathbf{n}(\mathbf{s}) \mathbf{\cdot}\boldsymbol \nabla_{\mathbf{s}} K(\mathbf{s},\mathbf{s}^{\prime}) = \mathbf{n}(\mathbf{s}) \mathbf{\cdot} \frac{\mathbf{s}-\mathbf{s}^{\prime}}{|\mathbf{s}-\mathbf{s}^{\prime}|} \, \frac{\partial\mathcal{K}(|\mathbf{s}-\mathbf{s}^{\prime}|)}{\partial|\mathbf{s}-\mathbf{s}^{\prime}|} .
\end{equation}
The Monge parametrization of the normal vector is
\begin{eqnarray}
\mathbf{n}(\mathbf{s}) &=& \frac{(-\boldsymbol\nabla_{\bot}\Delta\psi(\mathbf{r}_{\bot}),1)}{\sqrt{1+\left[\boldsymbol\nabla_{\bot}\Delta\psi(\mathbf{r}_{\bot})\right]^{2}}},
\end{eqnarray}
where $\boldsymbol\nabla_\bot\equiv (\partial_{x},\partial_{y})$. 
Then, following the above ideas, we can show that
\begin{eqnarray}
\partial_{n} K(\mathbf{s},\mathbf{s}^{\prime}) = \frac{1+\kappa r_{\bot}}{r_{\bot}} \mathcal{K}(r_{\bot}) \left(\frac{
\mathbf{r}_{\bot}\mathbf{\cdot}\boldsymbol\nabla_{\bot}\Delta \psi-
\Delta\psi}{r_{\bot}}\right) \nonumber \\ 
+ {O}(R^{-3}) ,
\label{expk2}
\end{eqnarray}
where the term in large parentheses is ${O}\left(R^{-1}\right)$ [see Eq.~(\ref{verticaldisplacement})]. 
The surface element $d \mathbf{s}=\sqrt{1+(\boldsymbol\nabla_{\bot}\psi)^{2}}d \mathbf{r}_{\bot}=
[1+(\boldsymbol\nabla_{\bot}\Delta \psi)^{2}/2]d \mathbf{r}_{\bot} + {O}(R^{-4})$. The normal derivative of the order parameter can be expanded in a similar way; thus, $q(\mathbf{s}) = \sum_{n=1}^{\infty} q_{n}(\mathbf{s})$, where $q_{n} = {O}(R^{-n})$. Plugging the above relations into Eq.~(\ref{inteq1}) and identifying the corresponding terms, order by order, we find a recursive chain of integral equations for $q_{n}(\mathbf{s})$ of the form (see Appendix~\ref{appendix_A})
\begin{equation}
\int_{\mathbb{R}^{2}} d \mathbf{r}_{\bot} \, q_{n}(\mathbf{r}_\bot) \left[\mathcal{K}(r_{\bot}) - \frac{\kappa}{g}\delta
(\mathbf{r}_{\bot})\right] = f_n[q_0,\ldots,q_{n-1}],
\end{equation}
where we have extended the integral to $\mathbb{R}^{2}$, ignoring terms exponentially decaying in $\kappa R$.
In general, $f_n$ is a functional of $q_i$ for $i<n$. For $n=0$, $f_0=\kappa(h_1+gm_b)/g$, which is independent of
$\mathbf{s}^\prime$, so $q_0(\mathbf{s})$ is given by Eq.~(\ref{qflat}) everywhere on the substrate.
Following the procedure outlined in Appendix~\ref{appendix_A}, we find for the next-order terms
\begin{eqnarray}
q_1&=&q_0 \frac{g}{\kappa - g}\frac{H}{\kappa}, \label{q1}\\
q_2&=& q_0 \frac{g}{\kappa-g} \left(\frac{H^2}{2\kappa^2}\left(1+\frac{2\kappa}{\kappa-g}\right)-\frac{K_G}{2\kappa^2}\right).
\label{q2}
\end{eqnarray}

We are now in the position to estimate the interfacial thermodynamic properties and the order-parameter profile.
The interfacial free energy $\mathcal{F}_{wl}$ of the wall-liquid interface is obtained from Eq.~(\ref{mfHLGW}) as
\begin{eqnarray}
{\mathcal{F}}_{wl}&=&\frac{h_1+gm_b}{2g}\int_\psi d\mathbf{s}\ q(\mathbf{s})  
=\sigma_{wl} {\cal A} + \Delta \mathcal{F}_{wl}[\psi],
\label{fsw}
\end{eqnarray} 
where $\sigma_{wl}\equiv(\kappa/2)(h_1+gm_b)^2/g(g-\kappa)$ is the surface tension defined for
a planar wall-liquid interface 
and ${\cal A}$ is the total substrate area. Thus, the increment 
$\Delta \mathcal{F}_{wl}[\psi]$ accounts for all the curvature-related terms.
For large $\kappa R$, $\Delta \mathcal{F}_{wl}$ can be expressed as
\begin{eqnarray}
\frac{\Delta \mathcal{F}_{wl}}{\sigma_{wl} {\cal A}} &=& \frac{g}{\kappa-g} \frac{\overline{H}}{\kappa} + 
\frac{1}{2}\left(\frac{g}{\kappa-g}\right)\left(1+\frac{2\kappa}{\kappa-g}\right)\frac{\overline{H^2}}{\kappa^2}
\nonumber\\&-&
\frac{1}{2}\left(\frac{g}{\kappa-g}\right)\frac{\overline{K_G}}{\kappa^2} + \cdots,
\label{deltafwlexpansion}
\end{eqnarray}
where $\overline{H}$, $\overline{H^2}$ and $\overline{K_G}$ are the averages over the substrate of the mean 
curvature, its square, and the Gaussian curvature, respectively. The leading order 
is consistent with the expression obtained in Refs.~\cite{parry2,parry3}. Finally, the ellipsis corresponds to 
higher-order curvature contributions which, in general, are nonvanishing. This feature, as well as the 
contribution being proportional to $\overline{H^2}$, is nonzero, which implies that the DP model does not 
satisfy the morphological thermodynamics hypothesis for confined fluids of hard bodies \cite{konig} (see
also Ref. \cite{blokhuis} for a critical review of this proposal).   

Diagrammatically, the interfacial free energy can be represented as
\begin{eqnarray}
\label{free_energy_sub}
\mathcal{F}_{wl}=\sigma_{wl}\Bigg[ \fig{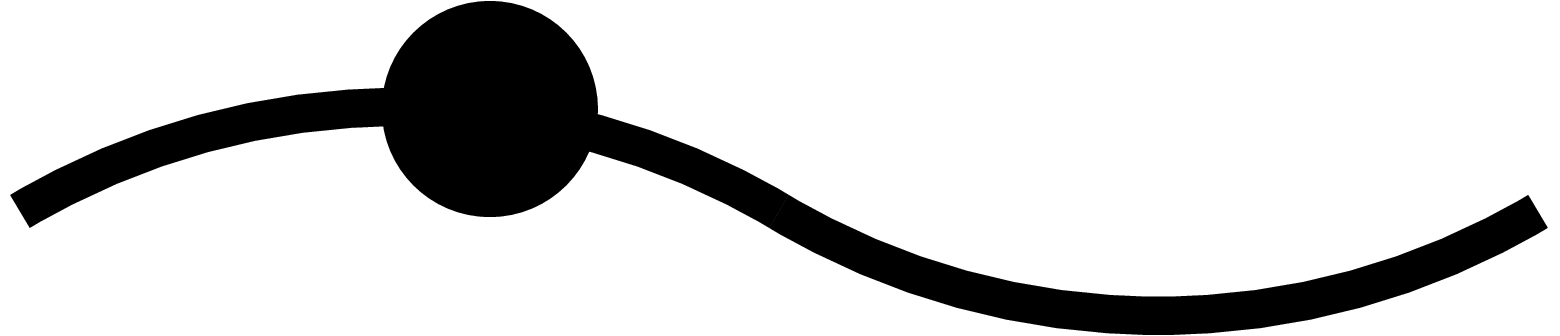}+\left(\frac{g}{\kappa-g}\right) \fig{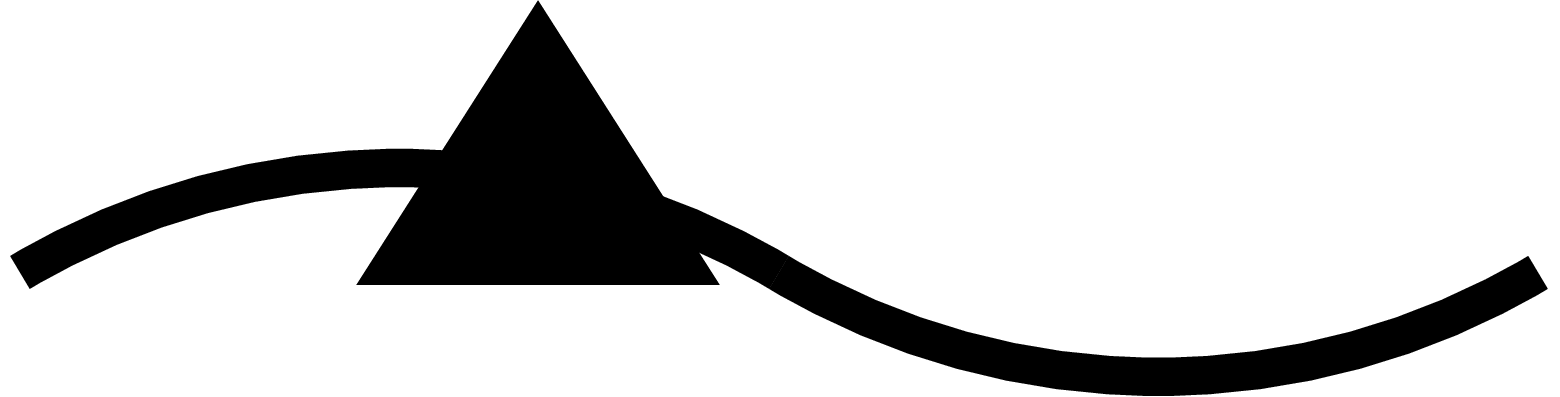}\nonumber\\ +
\frac{1}{2}\left(\frac{g}{\kappa-g}\right)\left(1+\frac{2\kappa}{\kappa-g}\right) \fig{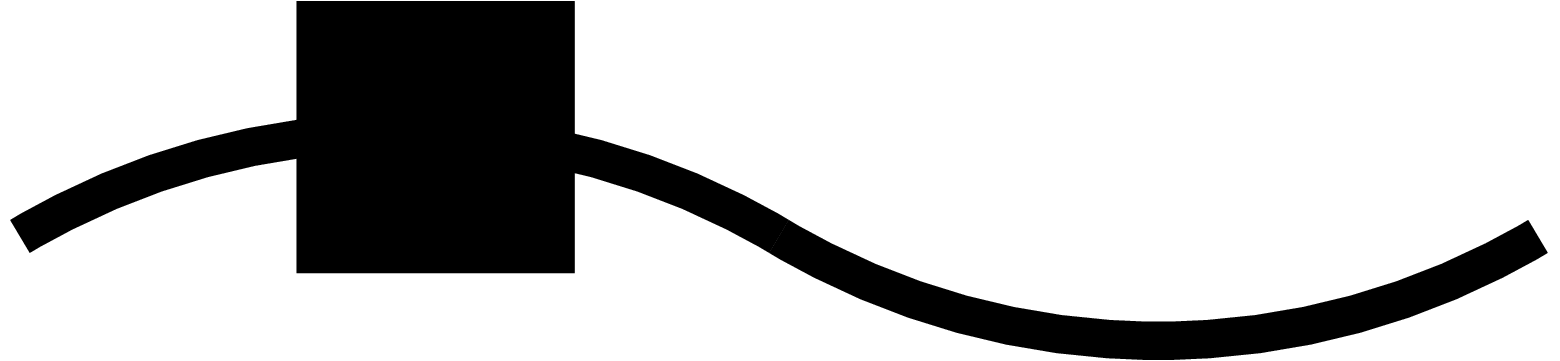} \nonumber\\
- 
\frac{1}{2}\left(\frac{g}{\kappa-g}\right) \fig{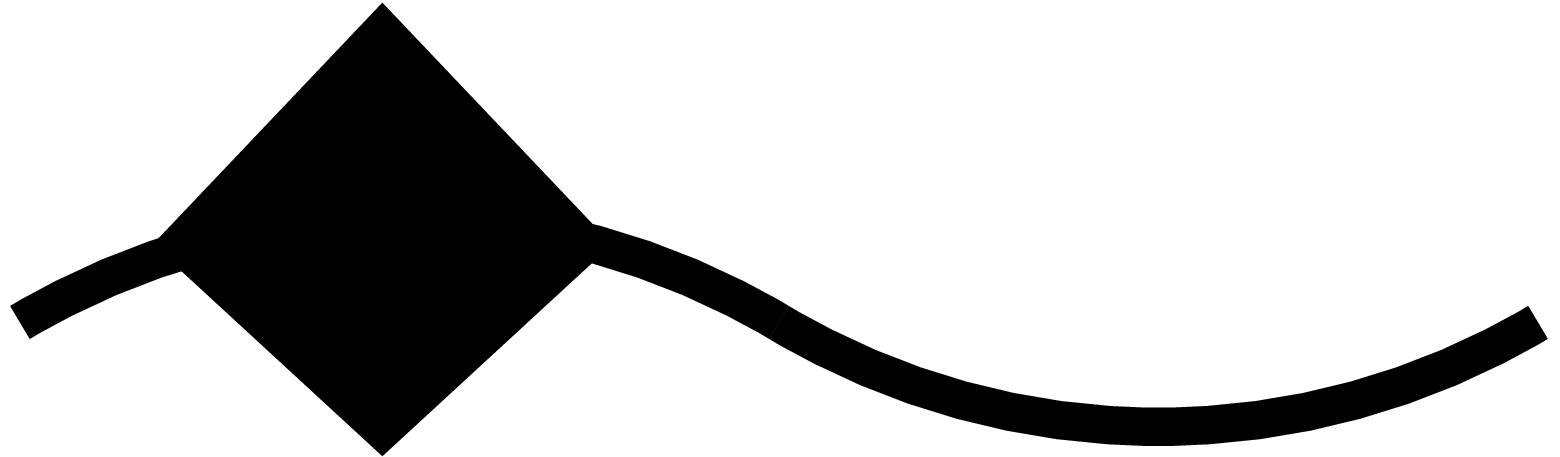}+ \cdots \Bigg] ,
\end{eqnarray}
where the wavy line corresponds to the substrate surface and the closed circle means that one must integrate
over all the positions on that surface with the appropriate infinitesimal area element. The closed triangle 
corresponds to integration over the surface, weighted by the local mean curvature in units of $\kappa$ (our
notation for this symbol differs from that used in Ref. \cite{parry3} by a factor of $1/2$). Finally, for the
closed square and the rhombus the weight function for the surface integrations are the squared mean curvature
and the Gaussian curvature, respectively, in units of $\kappa^2$. The present treatment highlights nonzero 
bending 
rigidity and saddle-splay coefficients, which were missing in the original formulation of the 
nonlocal model \cite{parry2,parry3}. The values of these are in agreement with those obtained from the 
exact 
solutions for the free energy of a fluid outside or inside a spherical or a cylindrical surface of radius $R$ 
within the DP model \cite{laura} (see also Appendix~\ref{appendixe}). 

\subsection{General diagrammatic approach}
We can go beyond the approach presented in the preceding section 
and obtain formally the full expansion of the interfacial free energy in powers of 
substrate curvatures. For this purpose, we return to Eq.~(\ref{inteq1}). The integral-equation kernel can
be formally inverted as
\begin{eqnarray} 
X(\mathbf{s},\mathbf{s}')\equiv\left(K(\mathbf{s},\mathbf{s}')+\frac{1}{g}\partial_{n} K(\mathbf{s},\mathbf{s}')-\frac{\kappa}{g}
\delta(\mathbf{s}-\mathbf{s}')\right)^{-1}\nonumber\\ = \frac{g}{g-\kappa} \delta(\mathbf{s}-\mathbf{s}')
\\-\frac{g}{g-\kappa} \int_{\psi} d\mathbf{s}_1 X(\mathbf{s},\mathbf{s}_1)\left(U(\mathbf{s}_1,\mathbf{s}')+
\frac{1}{g}\partial_{n_1} K(\mathbf{s}_1,\mathbf{s}')\right),
\nonumber
\label{seriesx}
\end{eqnarray}
where $U(\mathbf{s},\mathbf{s}')\equiv K(\mathbf{s},\mathbf{s}')-\delta(\mathbf{s}-\mathbf{s}')$ is the barred kernel 
introduced in Ref. \cite{PR} and $\partial_{n_1} \equiv \mathbf{n}(\mathbf{s}_1)\mathbf{\cdot}
\boldsymbol{\nabla}_{\mathbf{s}_1}$. This expression can be iterated, so we obtain the following formal 
expansion for $X$:
\begin{eqnarray} 
X(\mathbf{s},\mathbf{s}')= \frac{g}{g-\kappa} \delta(\mathbf{s}-\mathbf{s}')
\label{seriesx2}
\\-\left(\frac{g}{g-\kappa}\right)^2 \left(U(\mathbf{s},\mathbf{s}')+
\frac{1}{g}\partial_{n} K(\mathbf{s},\mathbf{s}')\right)
\nonumber\\ 
+\left(\frac{g}{g-\kappa}\right)^3 \int_{\psi} d\mathbf{s}_1  \left(U(\mathbf{s},\mathbf{s}_1)+
\frac{1}{g}\partial_{n} K(\mathbf{s},\mathbf{s}_1)\right)
\nonumber\\\times \left(U(\mathbf{s}_1,\mathbf{s}')+
\frac{1}{g}\partial_{n_1} K(\mathbf{s}_1,\mathbf{s}')\right)+\cdots.
\nonumber
\end{eqnarray}
The solution to Eq.~(\ref{inteq1}) can be expressed as
\begin{eqnarray}
&&q(\mathbf{s})=-\frac{\kappa(h_1+gm_b)}{\kappa-g}\Bigg[1-\int_\psi d\mathbf{s}_1 d\mathbf{s}_2 
X(\mathbf{s},\mathbf{s}_1) U(\mathbf{s}_1,\mathbf{s}_2)\nonumber\\
&&-\int_\psi d\mathbf{s}_1 d\mathbf{s}_2 
X(\mathbf{s},\mathbf{s}_1) 
\frac{1}{\kappa}\partial_{n_1} K(\mathbf{s}_1,\mathbf{s}_2) \Bigg] .
\label{qseries}
\end{eqnarray}
When the expansion ~(\ref{seriesx2}) is introduced in the expression (\ref{qseries}), we arrive at a formal series
for $q$, where each term is proportional to the convolution of $n$ functions, each being either $U$ or
$\partial_{n} K/\kappa$. We introduce the diagrammatic representation
\begin{equation}
U(\mathbf{s},\mathbf{s}')=\fig{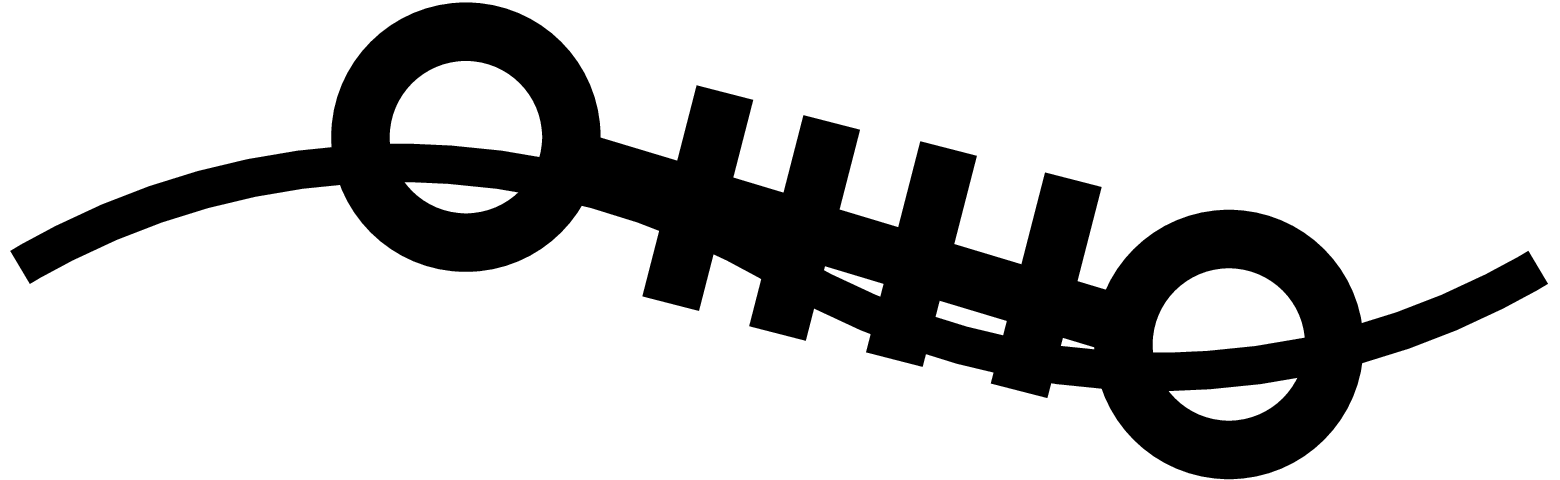} \quad,\quad \frac{1}{\kappa}\partial_{n} K(\mathbf{s},\mathbf{s}')
=\fig{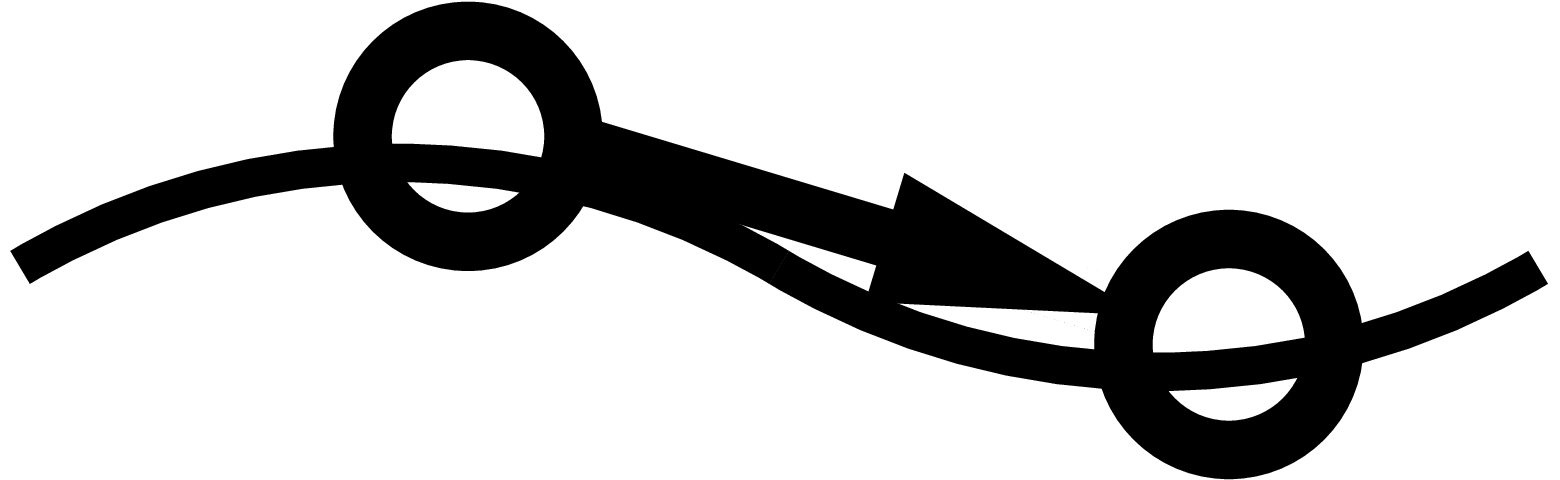} ,
\end{equation}
where in the latter diagram the arrow points to the position where the normal derivative is taken. In
this way, the 
expansion terms appearing in $q$ can be represented as chainlike diagrams. For example,
\begin{equation}
\int_\psi d\mathbf{s}_1 d\mathbf{s}_2 
U(\mathbf{s},\mathbf{s}_1)\frac{1}{\kappa}\partial_{n_1} K(\mathbf{s}_1,\mathbf{s}_2)
=\fig{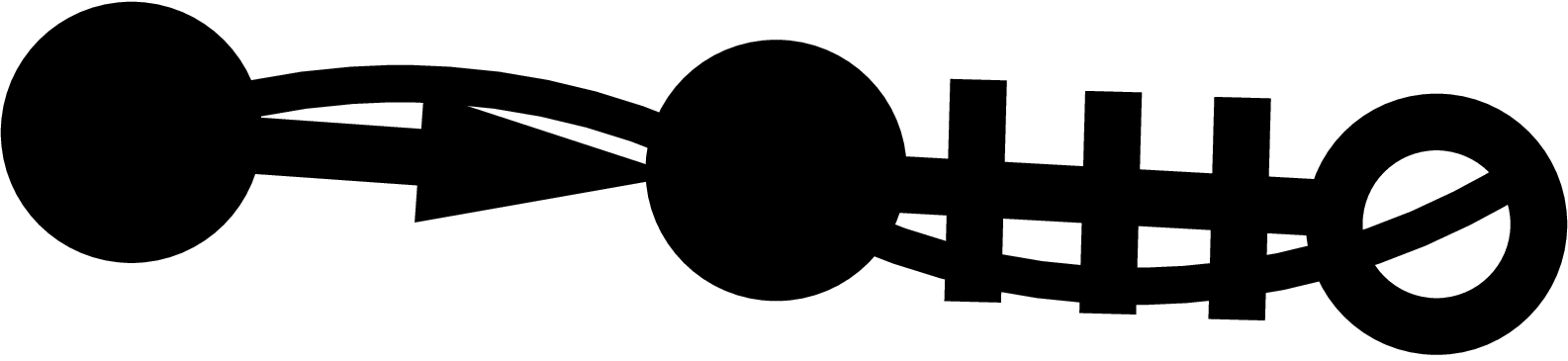},
\end{equation}
where a closed circle corresponds to an integrated position and the open circle represents the 
point $\mathbf{s}$ where $q$ is evaluated. Thus, from Eq.~(\ref{qseries}) we have
\begin{eqnarray}
&& q(\mathbf{s})=-\frac{\kappa(h_1+gm_b)}{\kappa-g}\Bigg[1+\frac{g}{\kappa-g}\fig{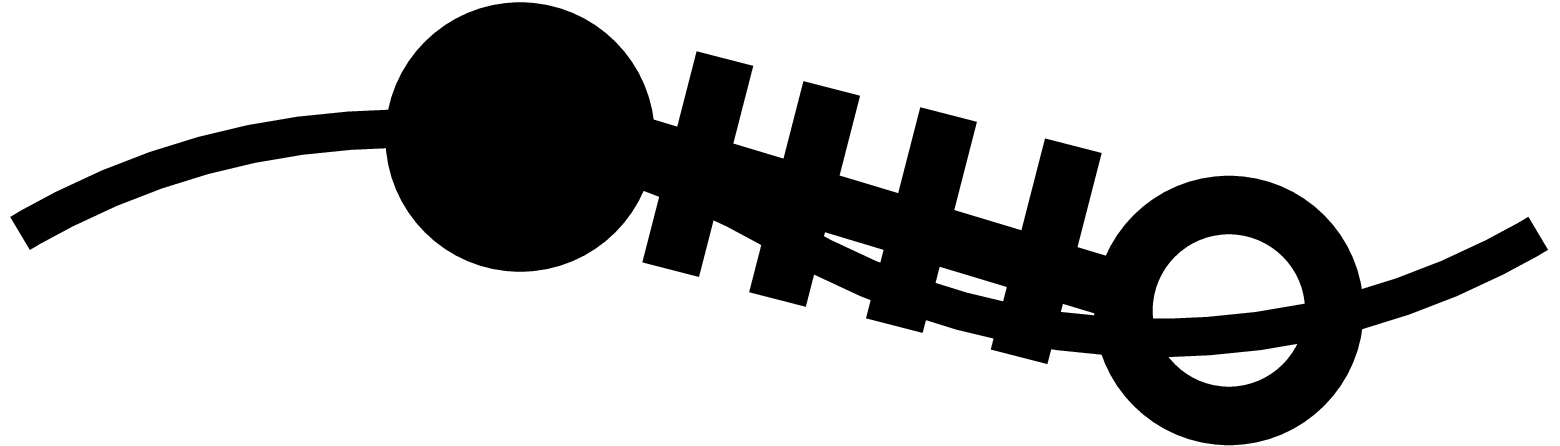}\nonumber\\
&&+
\frac{g}{\kappa-g}\fig{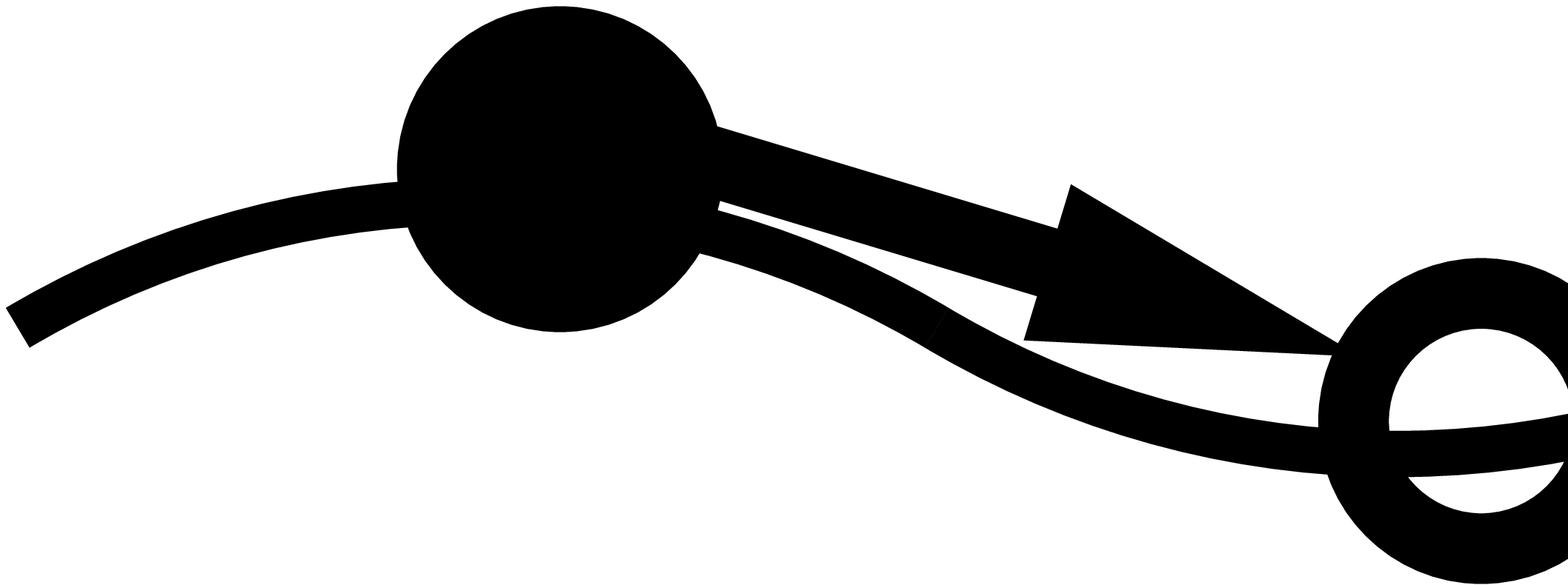}+\frac{g}{\kappa-g}\frac{\kappa}{\kappa-g} \fig{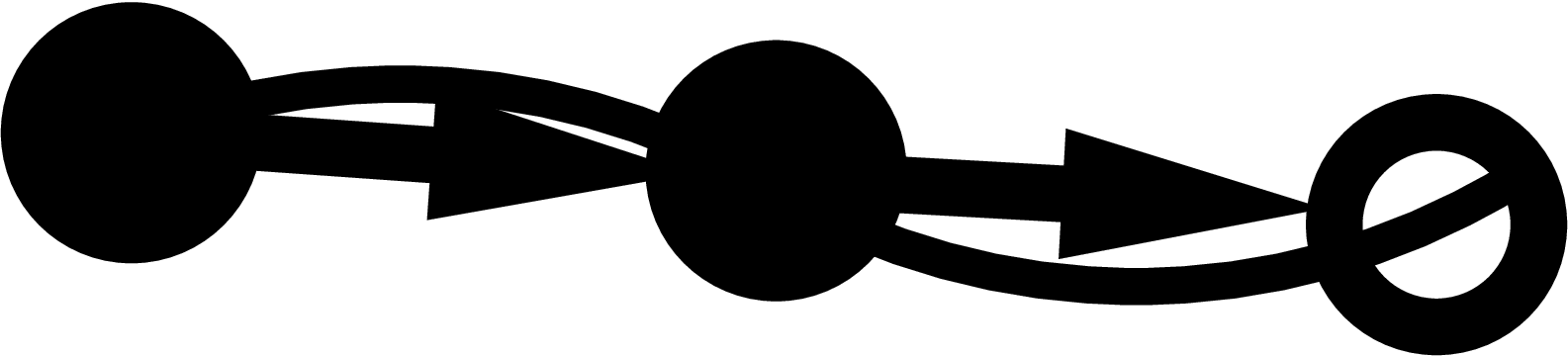}
\nonumber\\&&+\frac{g}{\kappa-g}\frac{\kappa}{\kappa-g} \fig{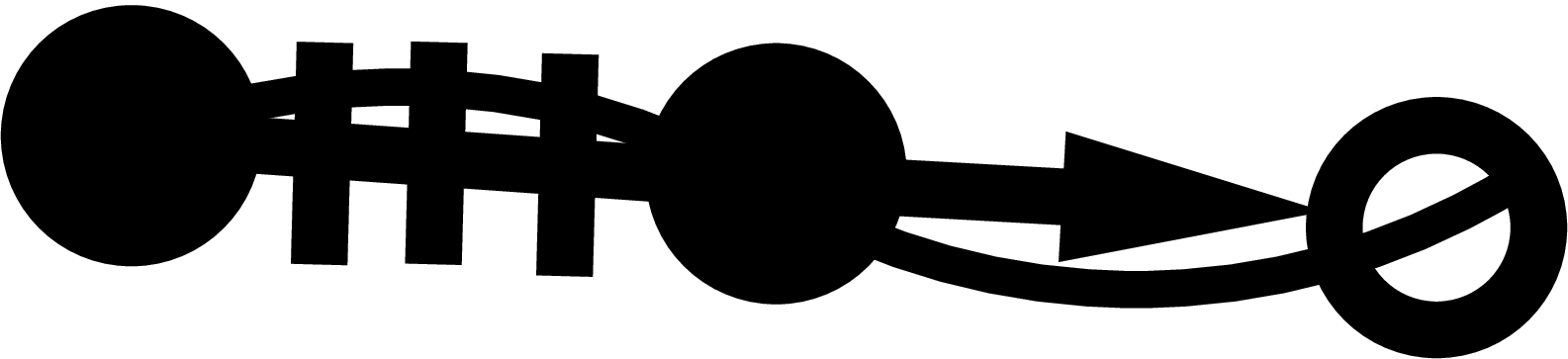}
\nonumber\\&&+\left(\frac{g}{\kappa-g}\right)^2 \left(\fig{diagram7}+\fig{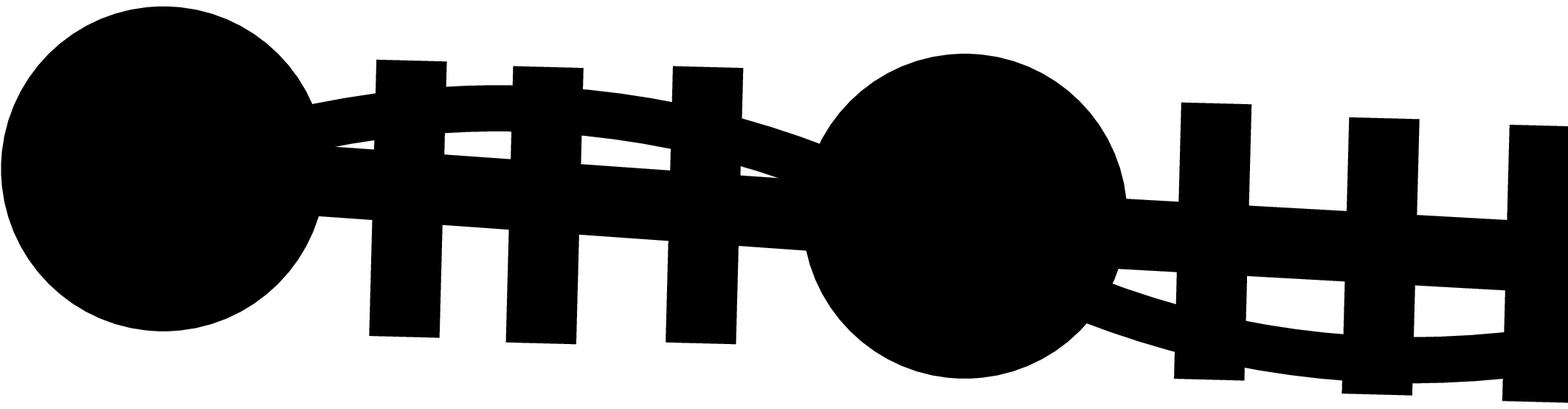}\right) + \cdots\Bigg] .
\label{qpsidiagram}
\end{eqnarray}
Here, each diagram with $n$ bonds (of which $m$ are of $\partial_{n} K/\kappa-$type) must be multiplied by a factor
\begin{equation}
\left(\frac{g}{\kappa-g}\right)^n \left(\frac{\kappa}{g}\right)^{m-m_0},
\label{factor}
\end{equation}
where the index $m_0$ is either $0$ or $1$, depending on whether the first bond on the left (i.e., the 
one that emerges from the closed extreme circle) is of $U$ or $\partial_{n} K/\kappa$ type, respectively. 
The connection with the curvature expansion is evident as, taking into account Eqs. (\ref{expk}) and
(\ref{expk2}), we find that
\begin{eqnarray}
\label{dashedk}
\fig{diagram8}&\equiv& \fig{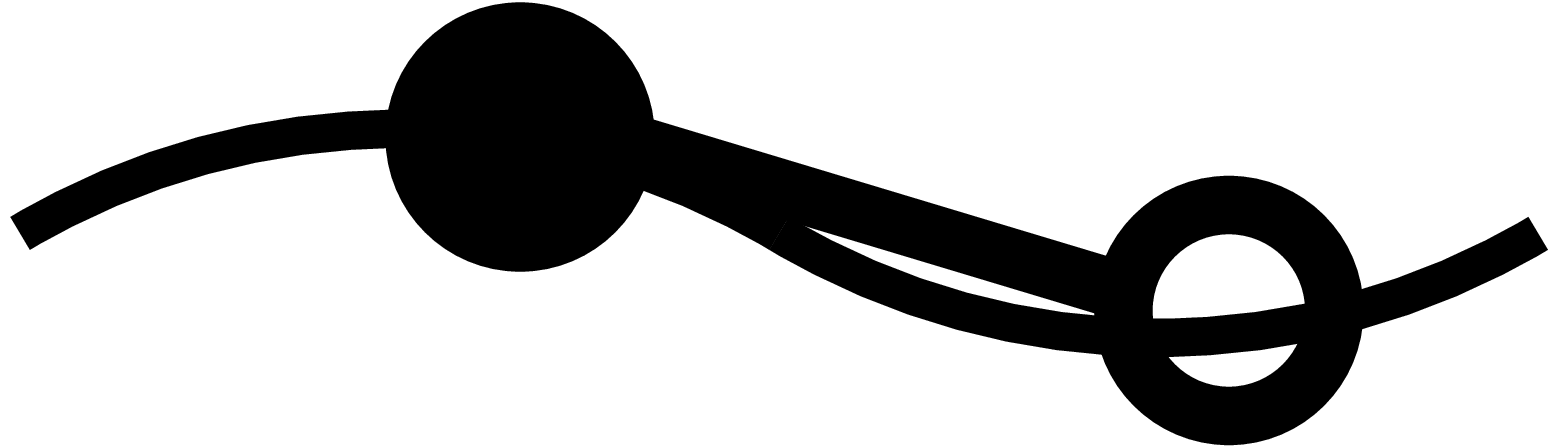}-1\\&=&
\frac{1}{2} \biggl[ \fig{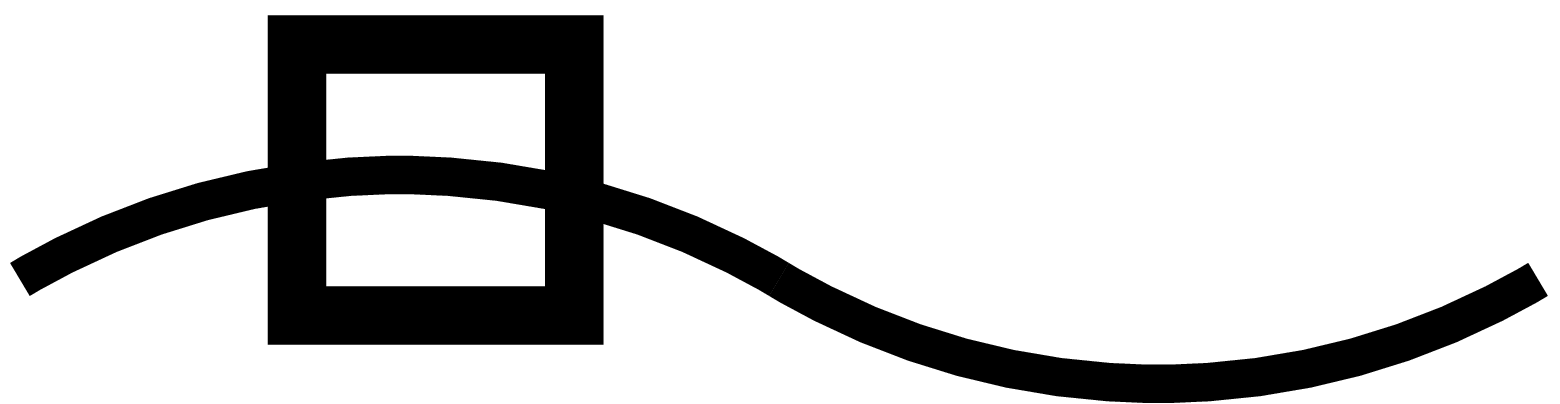}-\fig{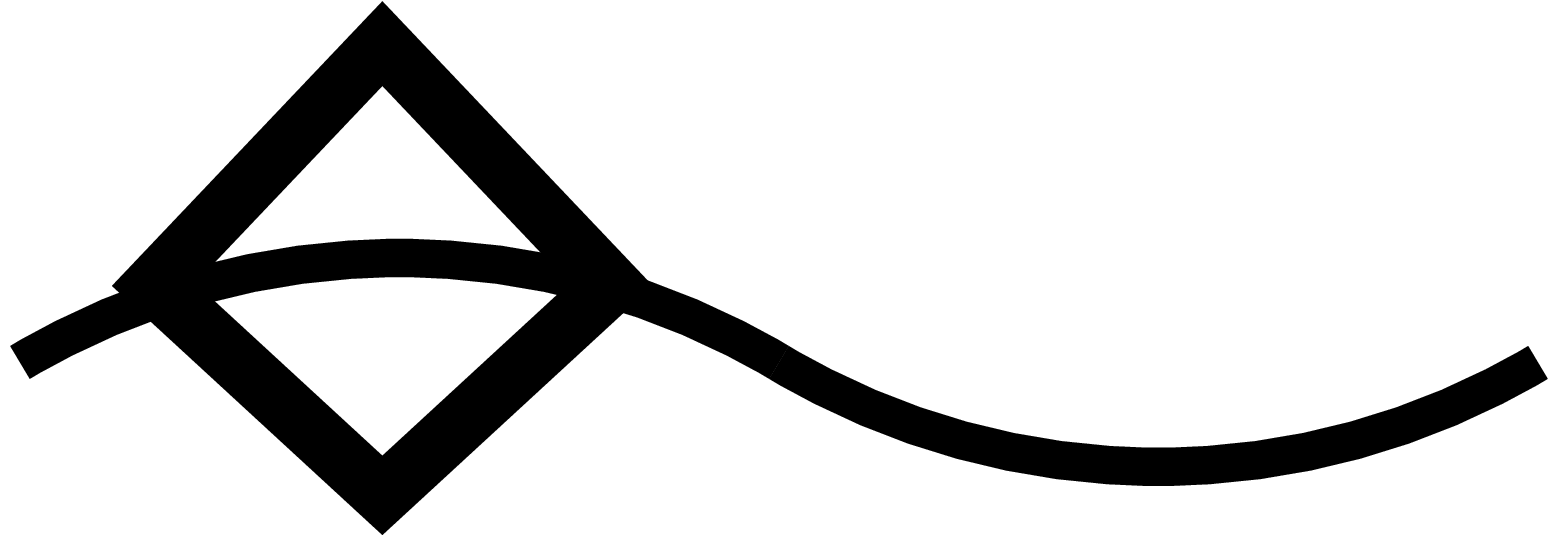} \biggr] 
+{O}(R^{-4}),
\nonumber\\
\label{dashedk2}
\fig{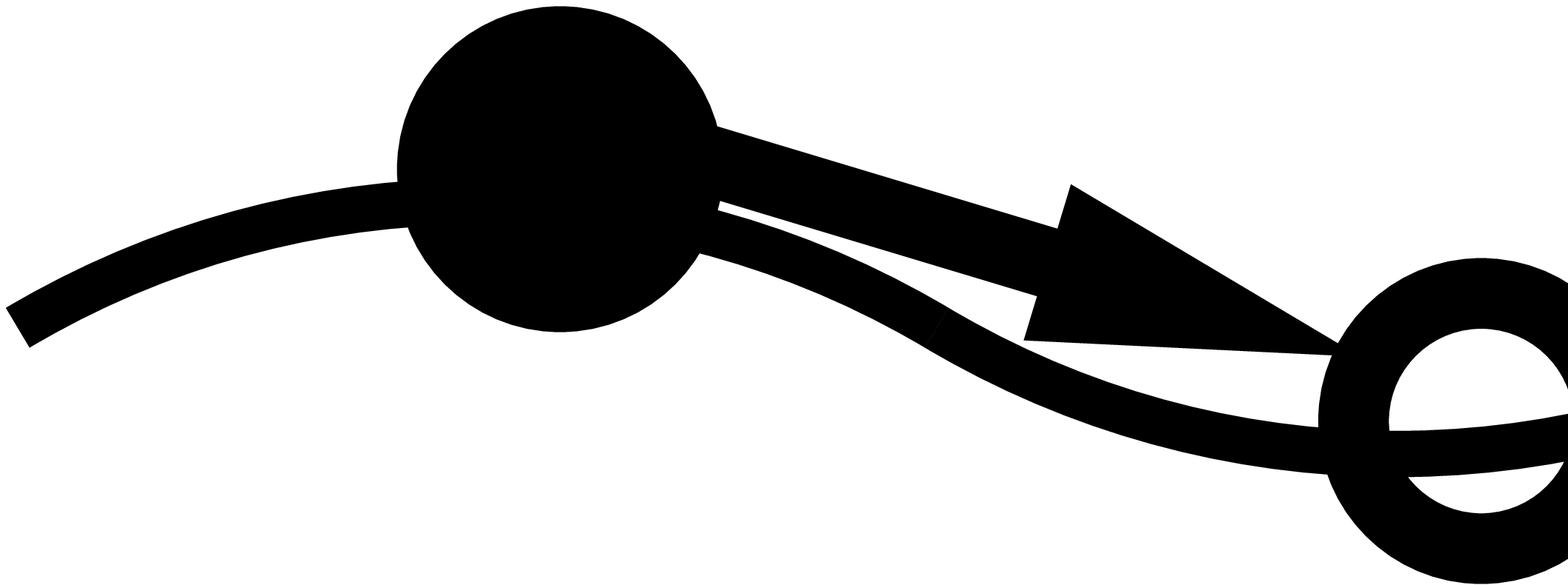}&=& \fig{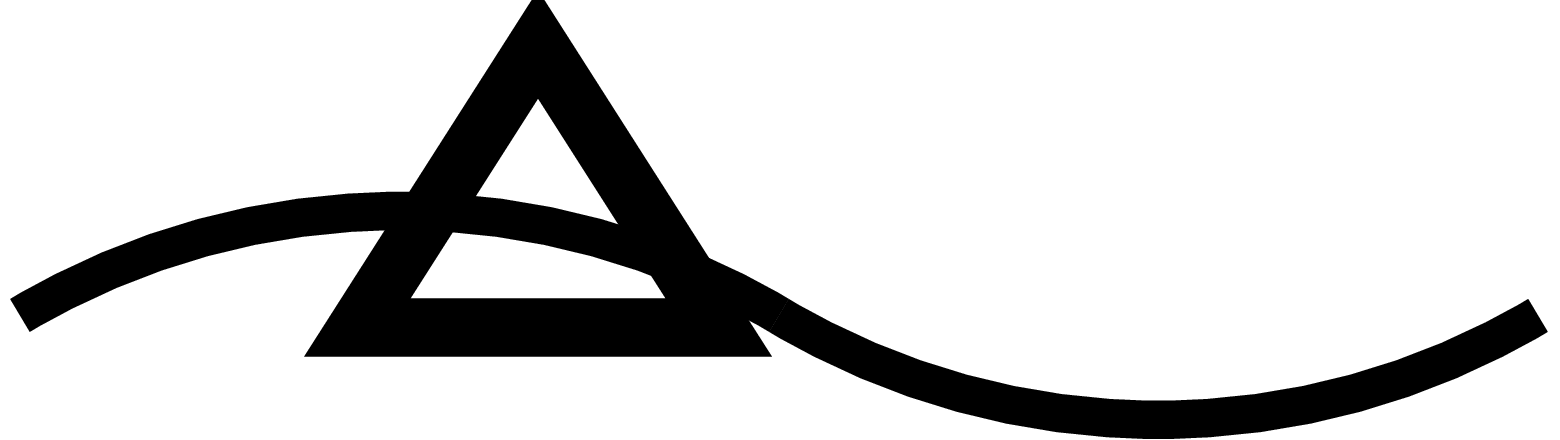} +{O}(R^{-3}),
\end{eqnarray} 
where by the open symbols we denote the evaluation of the corresponding weight functions at $\mathbf{s}$. 
Thus, a diagram with $n$, $U-$type and $m$, $\partial_{n} K/\kappa-$type bonds is of order of $R^{-(2n+m)}$.
This demonstrates that in order to obtain the corrections to $q$ to order $R^{-2}$, only the first three diagrams in Eq.~(\ref{qpsidiagram}) are needed.

By substituting Eq.~(\ref{qpsidiagram}) into Eq.~(\ref{fsw}) we obtain the diagrammatic expansion of the interfacial free energy  
\begin{eqnarray}
\label{F_chain_1}
&&\mathcal{F}_{wl}=\sigma_{wl}\Bigg[ \fig{diagram1}+\left(\frac{g}{\kappa-g}\right) (\fig{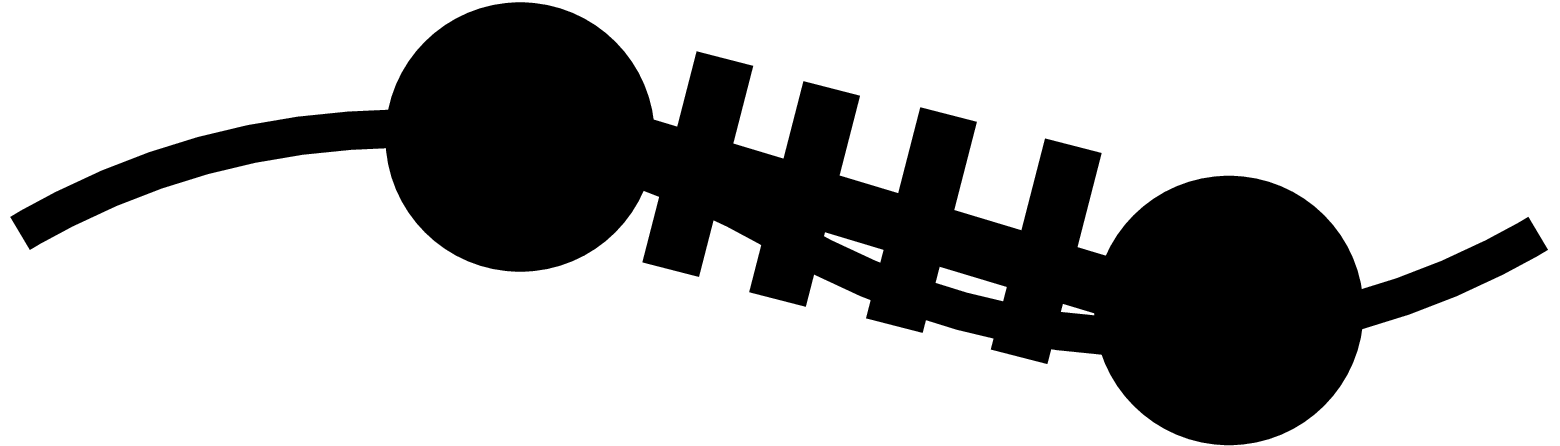}+
\fig{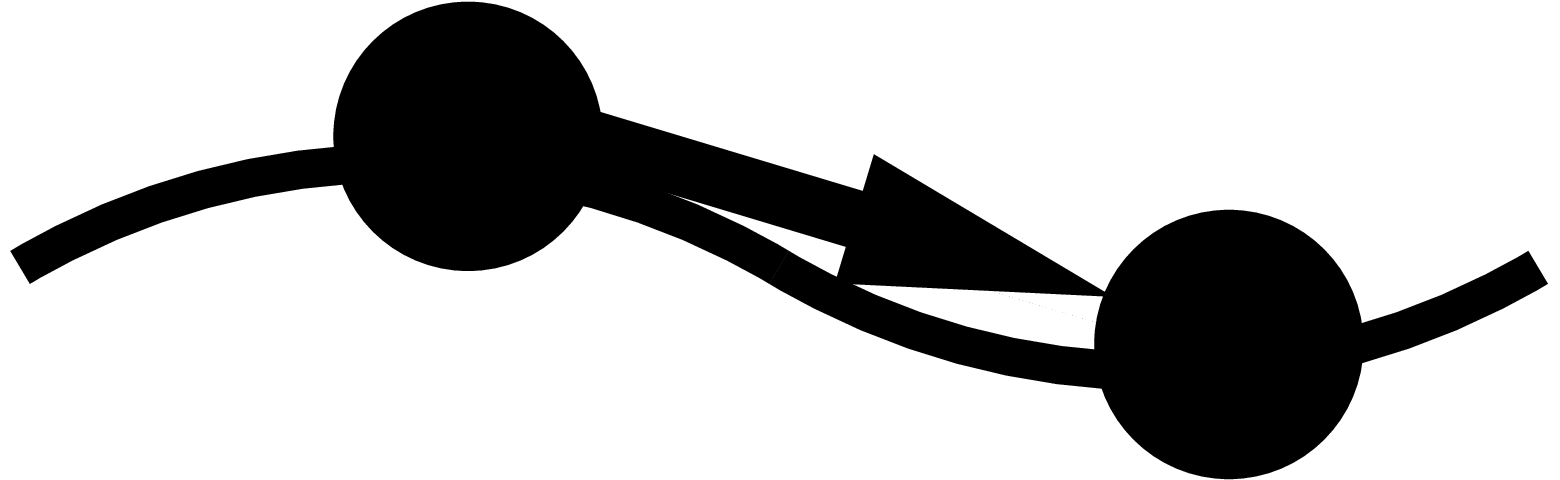}) \nonumber\\
&&+ 
\frac{g}{\kappa-g}\frac{\kappa}{\kappa-g} \fig{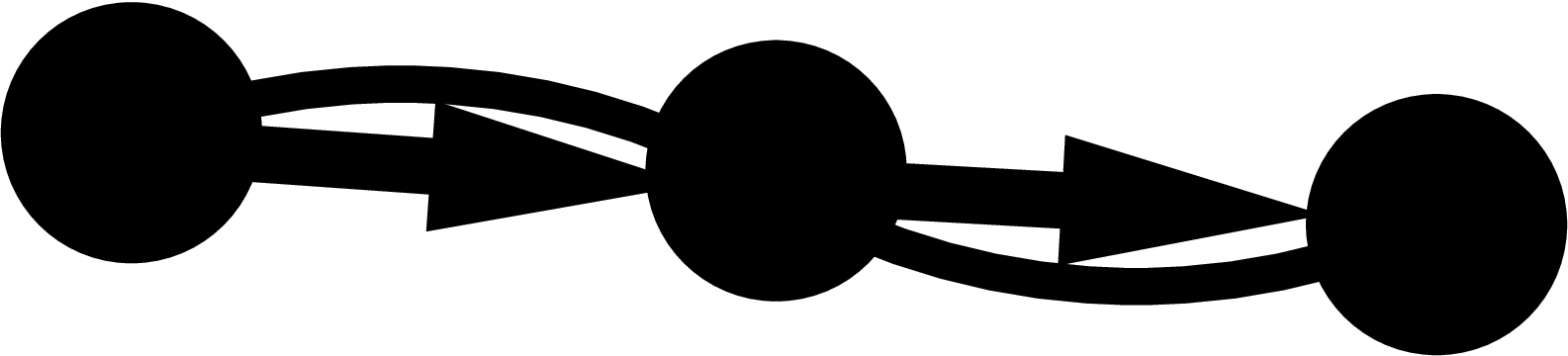} +\cdots \Bigg],
\end{eqnarray}
which coincides with our previous result ~(\ref{free_energy_sub}) up to $R^{-2}$ corrections.
In this expansion, the factors that multiply each diagram are the same as those that multiply the 
corresponding diagrams in Eq.~(\ref{qpsidiagram}).

\subsection{Evaluation of the order-parameter profile}
As for the interfacial free energy, we can obtain a formally exact expression for 
the order-parameter profile
from Eq.~(\ref{greenidentity3}),
\begin{eqnarray}
\delta m_\Xi(\mathbf{r})&=&
-\frac{1}{2\kappa}\int_{\psi} d\mathbf{s}
K(\mathbf{s},\mathbf{r}) q(\mathbf{s})
\label{profileprev}
\\
&-&\frac{1}{2\kappa}\int_{\psi} d\mathbf{s} \left(\frac{h_1}{g}+m_b+\frac{q(\mathbf{s})}{g}\right)
\partial_{n} K(\mathbf{s},\mathbf{r}),
\nonumber
\end{eqnarray}
or, equivalently,
\begin{eqnarray}
\delta m_\Xi(\mathbf{r})
&=&-\frac{1}{2}\int_{\psi} d\mathbf{s}
K(\mathbf{s},\mathbf{r}) \left(\frac{h_1}{g}+m_b+\frac{\kappa+g}{\kappa g}q(\mathbf{s})\right)
\nonumber\\
&-&\frac{1}{2\kappa}\int_{\psi} d\mathbf{s} \left(\frac{h_1}{g}+m_b+\frac{q(\mathbf{s})}{g}\right)
\nonumber\\
&\times&\left[\partial_{n} K(\mathbf{s},\mathbf{r})-\kappa K(\mathbf{s},\mathbf{r})\right]
\label{profile}
\end{eqnarray}
by substitution of the expansion (\ref{qpsidiagram}) for $q(\mathbf{s})$.
However, it is more convenient to use 
the single-layer potential representation ~(\ref{singlelayer}). Note that, once we know $q$, the
order parameter on the substrate can be obtained from Eq.~(\ref{boundary2}) as $-h_1/g-m_b - q/g$. 
It is instructive to derive the order parameter starting from the perturbative approach by expanding $\Psi$ in powers of 
the local curvature. The derivation proceeds 
along the same steps as the previous sections (see Appendix~\ref{appendix_A}).
After substitution of the expansion of $\Psi$ into Eq.~(\ref{singlelayer}), we 
find the following diagrammatic representation of the order parameter:
\begin{eqnarray}
\label{OP_curv_exp}
\delta m_\Xi\approx \frac{h_1+gm_b}{\kappa-g}\Bigg[\fig{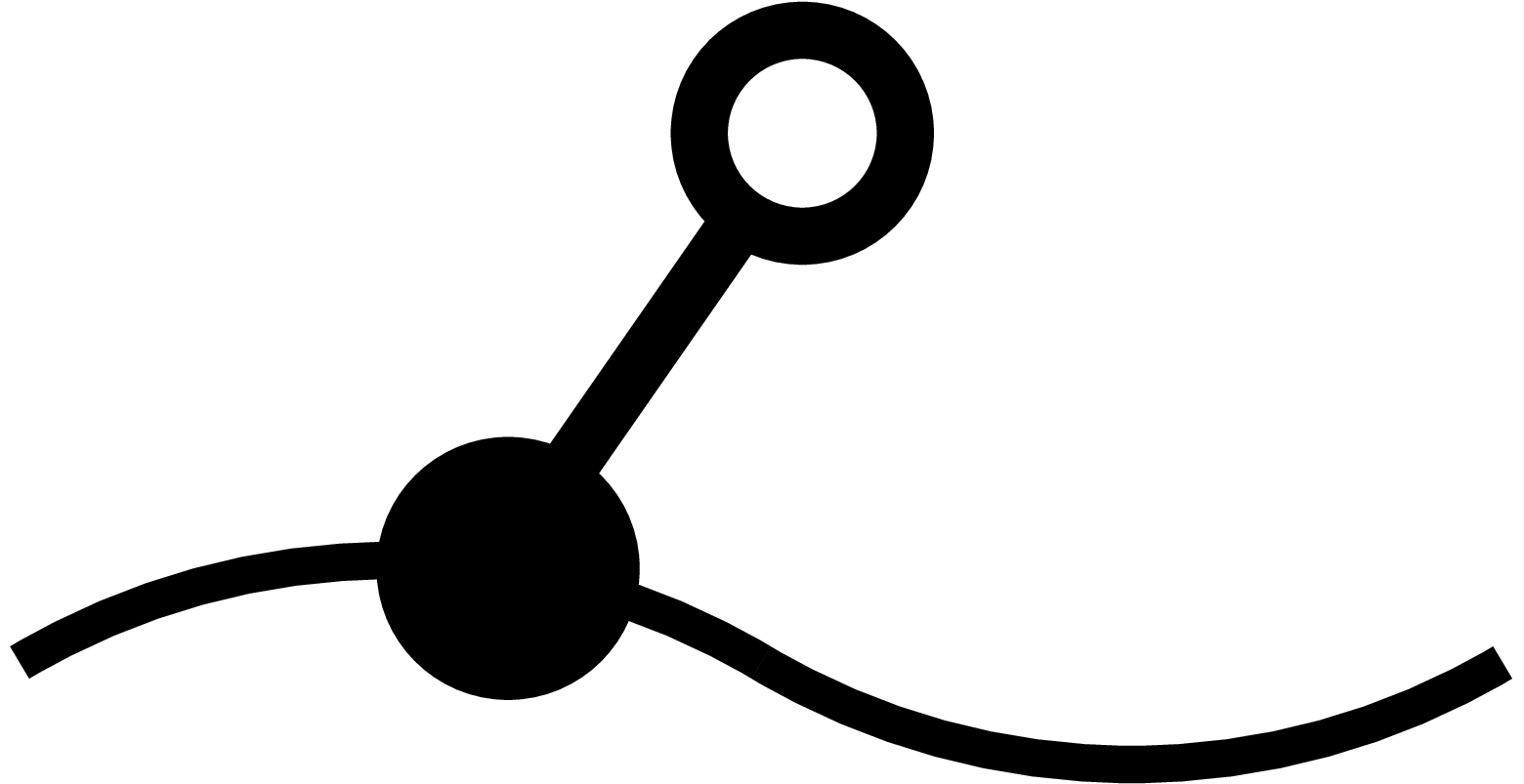}+
\frac{\kappa}{\kappa-g}
\fig{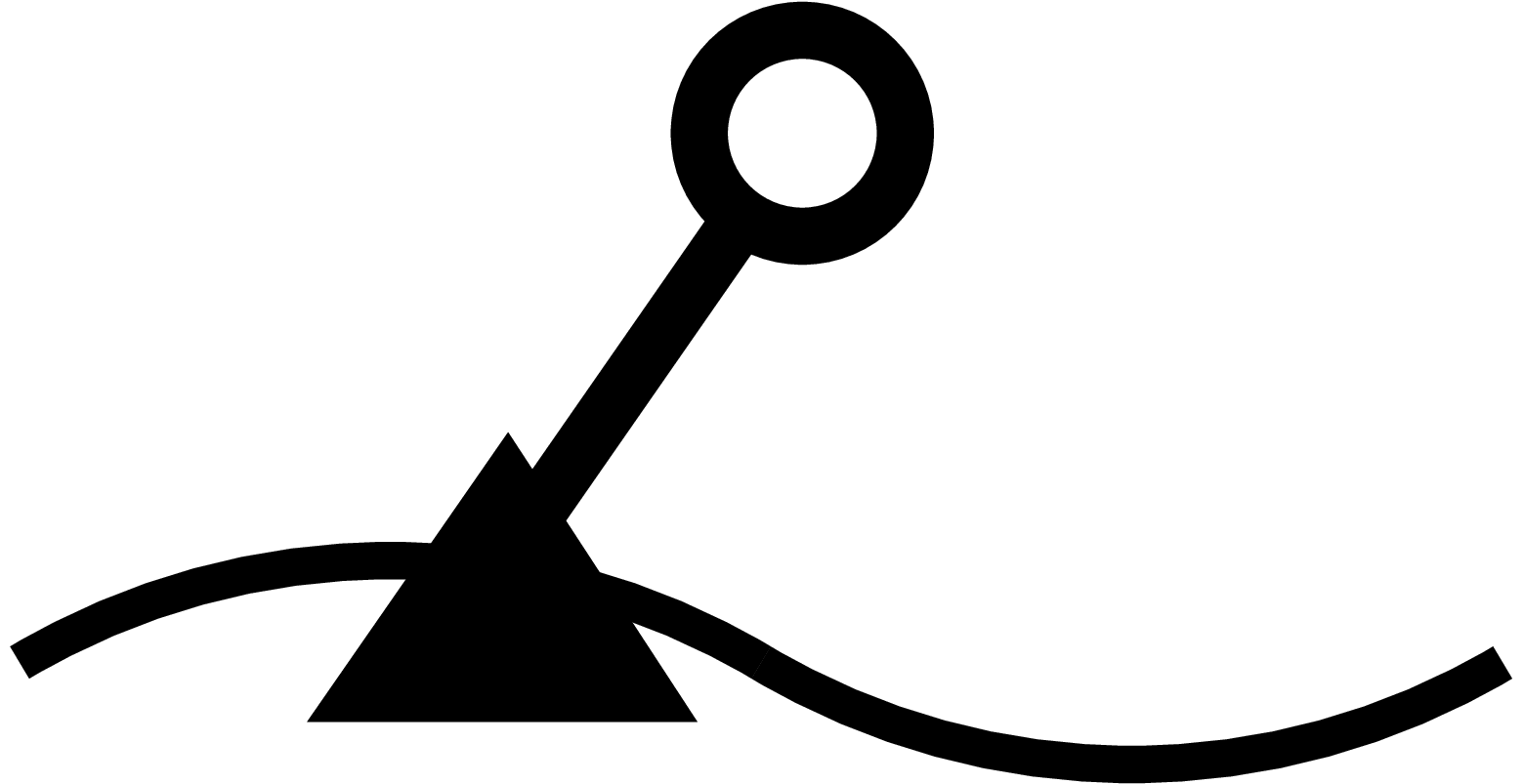}\nonumber\\
+\left(\frac{1}{2}\frac{g}{\kappa-g}+\frac{\kappa^2}{(\kappa-g)^2}\right)\fig{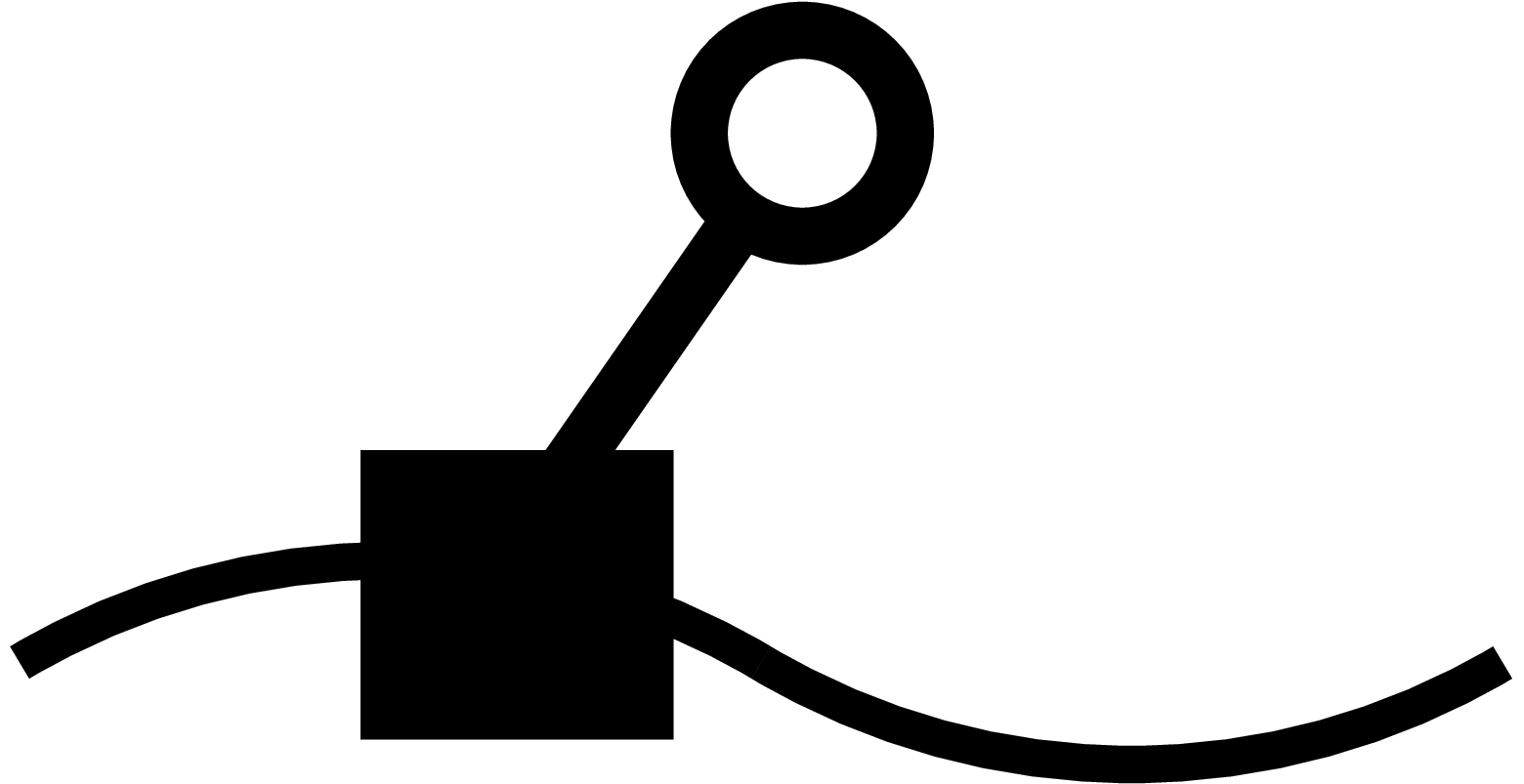}\nonumber\\
-\frac{1}{2}\frac{g}{\kappa-g}\fig{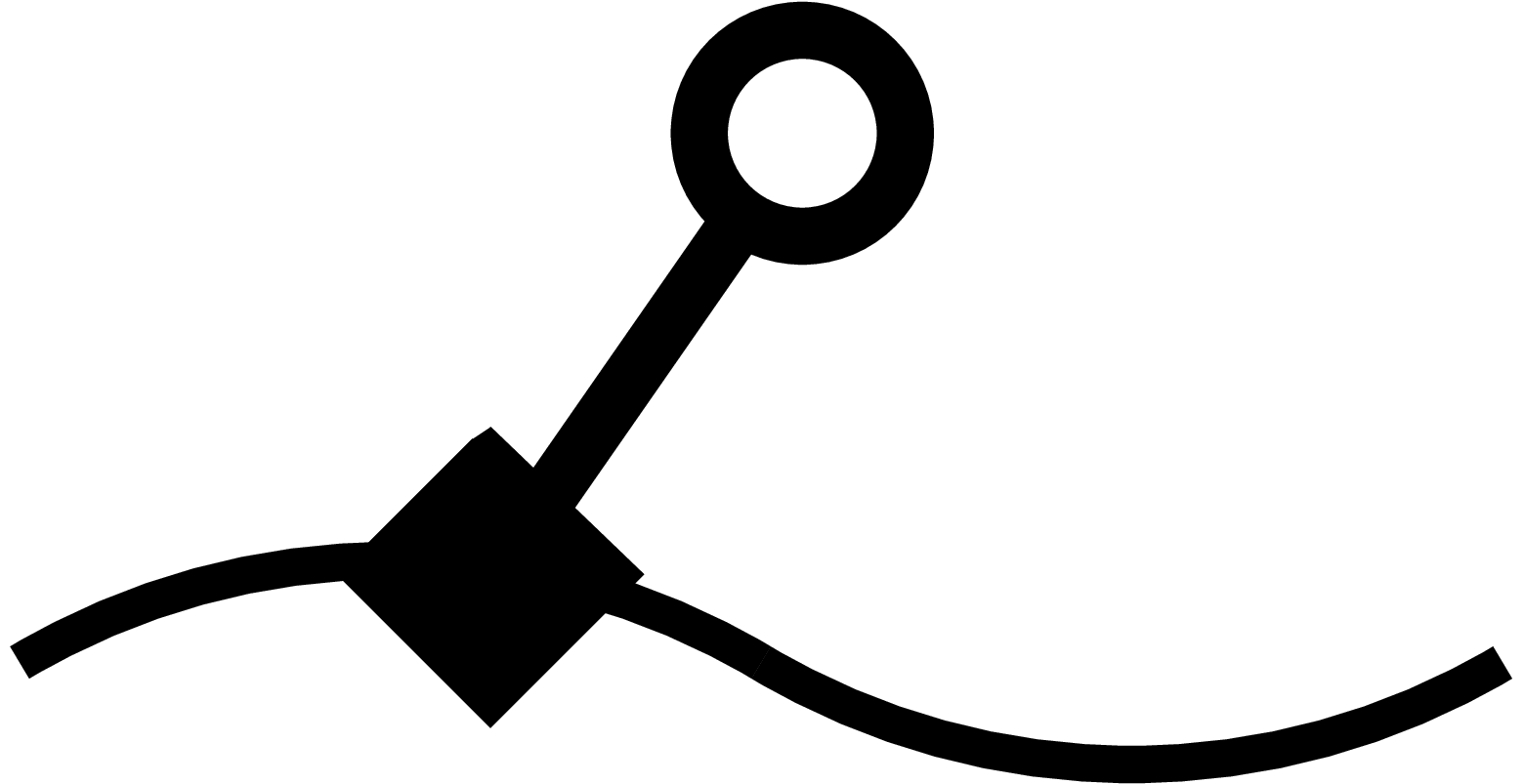}\Bigg] .
\end{eqnarray}
The presence of the curvature correction terms, which are not accounted for in the original
nonlocal ansatz, can be checked again with the exact solutions known for spherical and cylindrical substrates 
(see Appendix~\ref{appendixe}). 

As for the free energy, we can also generate a general diagrammatic approach to the curvature corrections. For this purpose,
Eq.~(\ref{singlelayer2}) can be formally solved as
\begin{equation}
\Psi(\mathbf{s})=2\kappa \int_\psi d\mathbf{s}_0 K^{-1}(\mathbf{s},\mathbf{s}_0)\left(-\frac{h_1+gm_b}{g}-
\frac{q(\mathbf{s}_0)}{g}\right),
\label{defPsi}
\end{equation} 
where $K^{-1}(\mathbf{s},\mathbf{s}_0)=\lim_{g\to\-\infty} X(\mathbf{s},\mathbf{s}_0)$, i.e.,
\begin{equation}
K^{-1}(\mathbf{s},\mathbf{s}_0)=\delta(\mathbf{s}-\mathbf{s}_0)-U(\mathbf{s},\mathbf{s}_0)+\int_\psi d\mathbf{s}_1
U\mathbf{s},\mathbf{s}_1)U(\mathbf{s}_1,\mathbf{s}_0)-\cdots .
\label{definvk}
\end{equation}
Substituting the expansions (\ref{qpsidiagram}) and (\ref{definvk}) into Eq.~(\ref{defPsi}), and back
into Eq.~(\ref{singlelayer}), we obtain the following expansion for the order-parameter profile: 
\begin{eqnarray}
&&\delta m_\Xi=\frac{h_1+gm_b}{\kappa-g}\Bigg[\fig{diagram21}+\frac{\kappa}{\kappa-g} 
\fig{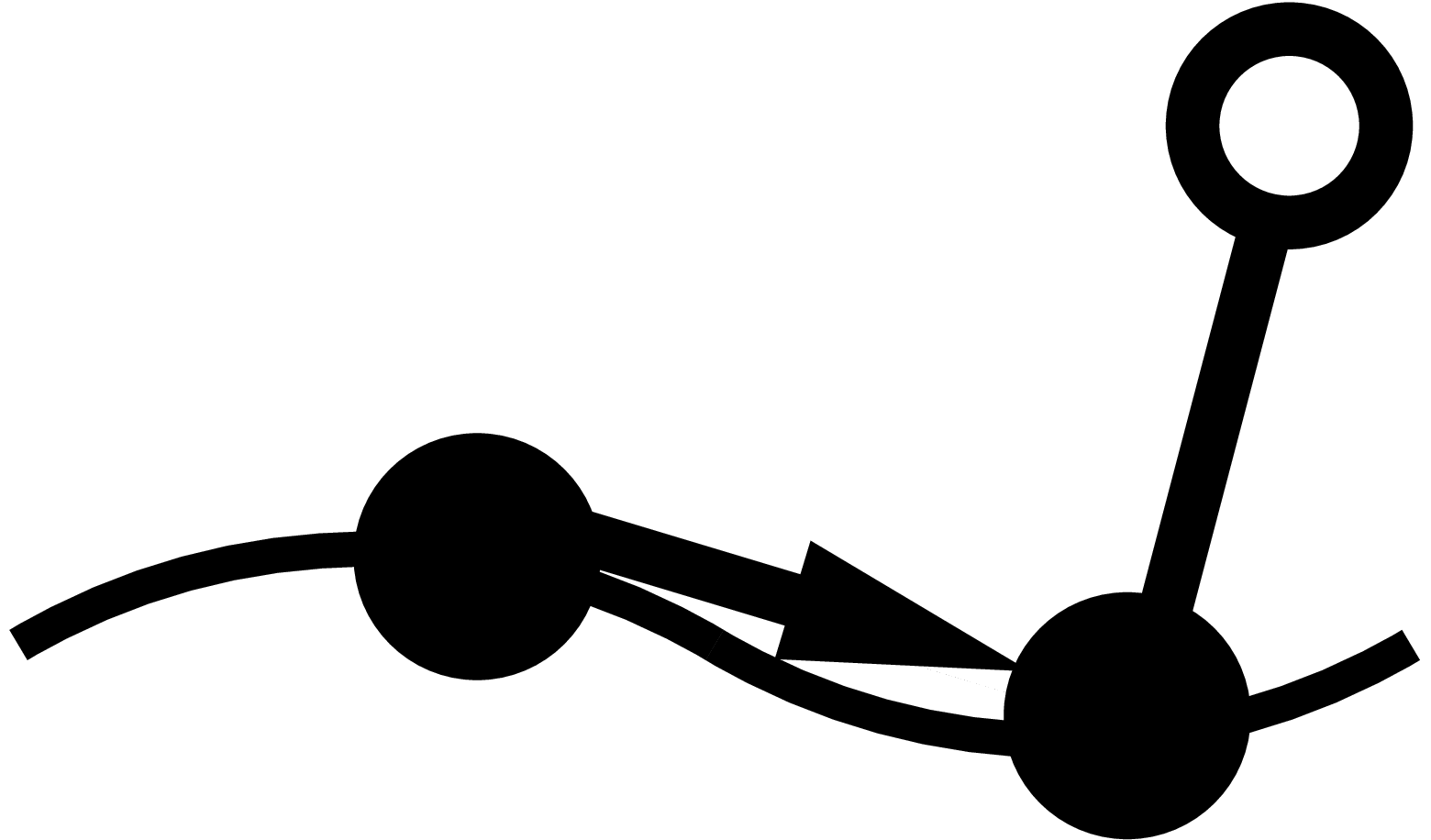}\nonumber\\&+
&\frac{g}{\kappa-g}\fig{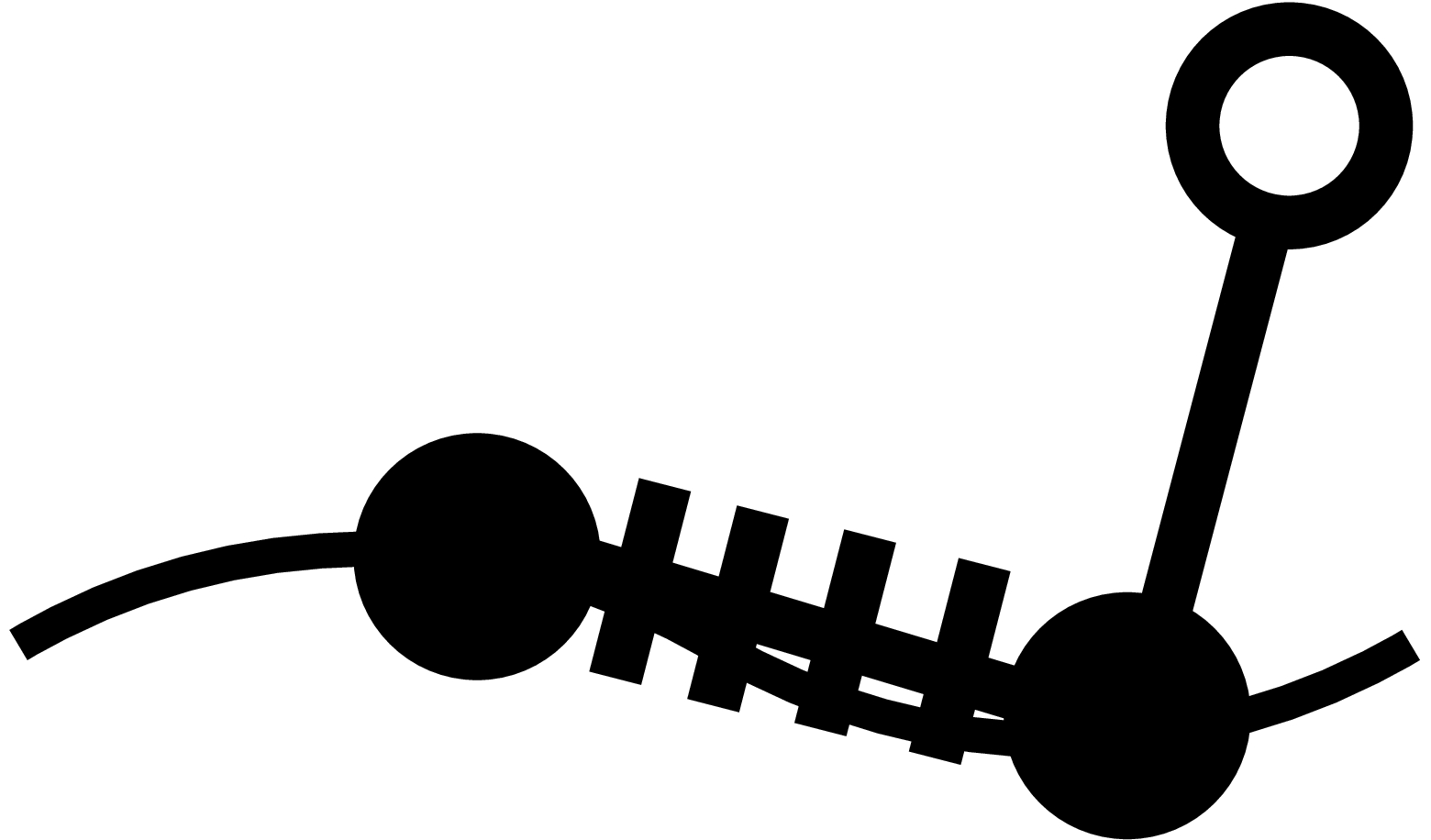}
+\left(\frac{\kappa}{\kappa-g}\right)^2
\fig{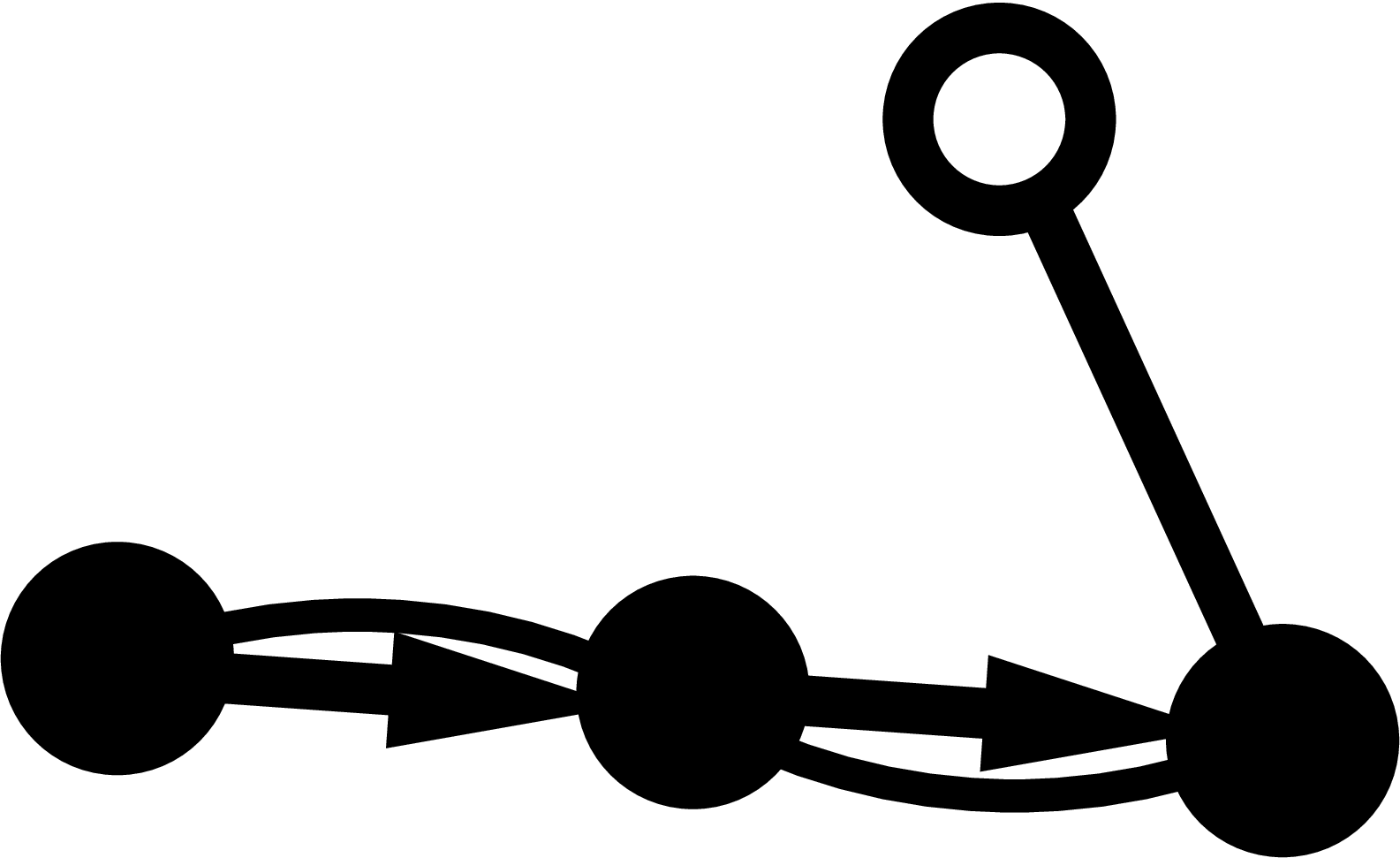}
+\cdots\Bigg] .
\label{deltamdiagram}
\end{eqnarray}
The diagrams of this expansion are obtained by convolution of a chainlike diagram of $q$ from 
Eq.~(\ref{qpsidiagram}) and the Green's function $K(\mathbf{s},\mathbf{r})$. The factor associated with each 
diagram [except for the first one in Eq.~(\ref{deltamdiagram})] is given by
the product of two terms: the factor associated with the $q$ diagram, Eq.~(\ref{factor}), and either 
$1$ (if the chainlike diagram has only $U-$type bonds) or $1-(1-\kappa/g)^{l+1}$ otherwise, with $l$ being
the number of $U-$type bonds from the last $\partial_n K/\kappa-$type bond to the extreme where the Green's 
function $K(\mathbf{s},\mathbf{r})$ emerges. By using (\ref{dashedk}) and (\ref{dashedk2}), the diagrammatic expansion
(\ref{deltamdiagram}) reduces to (\ref{OP_curv_exp}) up to $R^{-3}$ corrections.

\subsection{Summary and remarks}
So far we have done two things. First, we
devised a perturbative approach based on a small-curvature expansion of surfaces and we have applied this to a (nonwetting) bulk phase in contact with a substrate. The free energy (\ref{free_energy_sub}) contains curvature corrections that we have identified exactly at the leading (nontrivial) order. The curvature expansions are, however, quite cumbersome, and to this end we developed a more fundamental approach, based on the formal inversion of integral equations satisfied by the order-parameter field. By using a diagrammatic approach we have found a formal expansion of the interfacial free energy [see Eq.~(\ref{F_chain_1})]. By following this approach we have also derived the order-parameter profile in the bulk phase [see Eq.~(\ref{deltamdiagram})]. For small interfacial or surface curvatures, the wetting diagrams entering into the formal expansions simplify and they reveal the curvature corrections in terms of local Gaussian and average curvatures; this property will be analyzed in detail in the next section for the case of an isolated liquid-gas interface.

\section{The free interface and its self-interaction energy functional}
Next we turn our attention to the liquid-gas interface. This is free in the sense that it is isolated, 
infinitely far from any confining walls but \emph{constrained} so that the surface of zero magnetization adopts a given, smooth,
 nonplanar configuration $\ell(\bf{x})$. Overhangs are excluded and again we will suppose that the curvature 
is small everywhere. We follow the prescription set out by Fisher and Jin \cite{Jin1,Jin3}, whereby the 
effective interfacial model is identified as the minimum of the LGW model subject to this cross-criterion constraint 
together with the appropriate bulk boundary conditions, viz., that $m_\Xi(\mathbf{r})\to \mp m_0$ as $z\to\pm\infty$, i.e., gas 
is above and liquid is below the interface corresponding to the  two domains $\Omega_1$ and $\Omega_2$. 
These regions are uncoupled, and in each the equilibrium constrained profile satisfies the Helmholtz 
equation (\ref{pde}) subject to the above boundary conditions. In this case, the solution to 
Eq.~(\ref{greenidentity6}) can be written as
\begin{equation}
\label{consistent}
\int_{\ell} d \mathbf{s} \, K(\mathbf{s},\mathbf{s}^\prime) q^\pm(\mathbf{s}) = -m_{0} \Biggl( \kappa \mp \int_{\ell} d \mathbf{s} \, \partial_{n} K(\mathbf{s},\mathbf{s}^\prime) \Biggr) ,
\end{equation}
where $\mathbf{n}$ is the interface normal towards the gas phase.
This relation tells us that, given the interfacial profile $\ell(\mathbf{x})$ as a background, the order parameter can be found from knowledge of $q^\pm(\mathbf{s}_{\ell}) = \partial_{n_{\ell}}\delta m_{\Xi}(\mathbf{s_{\ell}^{\pm}})$, where $\delta m_\Xi=m(\mathbf{r})\pm m_0$ for $\mathbf{r}$ lying above or below the interface. Notice that the 
order parameter is a function of the position but also a \emph{functional} of the interfacial shape $\ell(\mathbf{x})$. Our main goal is to determine this functional dependence.  

\subsection{Perturbative approach}
If we assume that the local interfacial curvatures are small, we can proceed in a similar way to the preceding 
section. The normal derivative of the order parameter can be expanded in powers of the minimum local curvature 
radius $R$; thus $q^{\pm}(\mathbf{s}_{\ell}) = \sum_{n=1}^{\infty} q_{n}^{\pm}(\mathbf{s}_{\ell})$, where $q_{n} = {O}(R^{-n})$. It follows that for a flat interface $q_{n}=0$ (for $n\geqslant1$), and the series reduces to $q^{\pm}(\mathbf{s}_{\ell})=q_{0}(\mathbf{s}_{\ell})$. Inserting the above relations into Eq.~(\ref{consistent}) and identifying the corresponding terms, order by order, we find a recursive chain of equations for the $q_{n}$'s. Following the scheme used in the preceding section, we can solve up to ${O}(R^{-3})$ and obtain the desired $q$'s (see Appendix~\ref{appendix_A}). The DP potential allows us to write the LGW Hamiltonian in terms of surface integrals only, viz.,
\begin{equation}
\mathcal{H}_{LGW}[m_{\Xi}] \equiv H[\ell] = -\frac{m_{0}}{2} \int_{\ell} d \mathbf{s}_{\ell} \left[ q^{+}(\mathbf{s}_{\ell}) + q^{-}(\mathbf{s}_{\ell}) \right] ,
\label{finterface}
\end{equation}
where $H[\ell]$ is the interfacial Hamiltonian. This functional can be evaluated with the perturbative expansion for $q^{\pm}$ mentioned above and, consequently, it leads to a similar expansion of the Hamiltonian that we cast in the form $H[\ell]=\sigma \mathcal{A}_{lg}+\Delta H[\ell]$, where
\begin{equation}
\Delta H[\ell] \equiv -\sigma \sum_{n=2}^{\infty} (-1)^{n} \omega_{n}[\ell] ,
\end{equation}
is the self-interaction contribution \cite{PR}, $\sigma=\kappa m_0^2$ is the surface tension, and $\mathcal{A}_{lg}$ is the interfacial area. The functionals $\omega_{n}[\ell]$ are (ensured to be) of order ${O}(R^{-2(n-1)})$. Having in mind the approximate solutions for $q$ just derived, we find that
\begin{equation}
H[\ell] = \kappa m_{0}^{2} \int_{\ell} d \mathbf{s}_{\ell} - \frac{m_{0}}{2} \int_{\ell} d \mathbf{s}_{\ell} \left[ q_{2}^{+}(\mathbf{s}_{\ell}) + q_{2}^{-}(\mathbf{s}_{\ell}) \right] + {O}(R^{-3}) .
\end{equation}
Thus we immediately recover that $H[\ell]=\sigma\int d{\mathbf{s}}_\ell+\cdots$, which is just the standard capillary-wave Hamiltonian.
The ellipsis stands for the energy corrections due to the self-interaction: The first of these corrections is
\begin{equation}
\label{omega2}
-\sigma \omega_{2}[\ell]=-\frac{m_{0}}{2} \int_{\ell} d \mathbf{s}_{\ell} \left[ q_{2}^{+}(\mathbf{s}_{\ell}) + q_{2}^{-}(\mathbf{s}_{\ell}) \right]
\end{equation}
or, more explicitly, using Eq.~(\ref{exph22}),
\begin{equation}
\label{ham1}
H[\ell] \approx \sigma \int_{\ell} d \mathbf{s}_{\ell} - \frac{\sigma}{8} \int_{\ell} d \mathbf{s}_{\ell} \left( \frac{k_{1}(\mathbf{s}_{\ell})-k_{2}(\mathbf{s}_{\ell})}{\kappa} \right)^{2}.
\end{equation}
We use the symbol $\approx$ to mean that relationships hold up to ${O}(R^{-3})$ corrections. The
absence of a $1/R$ contribution, i.e., the vanishing of the Tolman length, is due to the Ising symmetry of the 
DP model. In the theory of lipid membranes the above functional (\ref{ham1}) is commonly expressed in terms of the Gaussian and extrinsic curvatures of the interface, following Helfrich \cite{Helfrich},
\begin{equation}
\label{Helfrich}
H[\ell] \approx \sigma \mathcal{A}_{\ell} + \int_{\ell} d \mathbf{s}_{\ell} \left(\kappa_B H^{2} + \kappa_{G} K_G \right),
\end{equation}
where the coupling constants are the bending rigidity
\begin{equation}
\label{bendrig}
\kappa_{B}=-\frac{\sigma}{2\kappa^{2}}
\end{equation}
and the saddle-splay rigidity
\begin{equation}
\label{saddlesplayrig}
\kappa_{G}=\frac{\sigma}{2\kappa^{2}}.
\end{equation}
Using the diagrammatic notation introduced in the preceding section, the full $H[\ell]$ can be expressed as
\begin{equation}
H[\ell]=\sigma\Bigg[ \fig{diagram1}-\frac{1}{2}\left(\fig{diagram3}-\fig{diagram4}\right) + \cdots \Bigg] .
\end{equation}
The constrained order-parameter profiles in the gas and liquid phases can be obtained from 
Eq.~(\ref{greenidentity3}) as
\begin{eqnarray}
\delta m_\Xi(\mathbf{r})\equiv m_\Xi(\mathbf{r})\mp m_0 =
\mp\frac{1}{2\kappa}\int_{\ell} d\mathbf{s}
K(\mathbf{s},\mathbf{r}) q^{\pm}(\mathbf{s})
\nonumber\\
+
\frac{m_0}{2\kappa}\int_{\ell} d\mathbf{s} \ 
\partial_{n} K(\mathbf{s},\mathbf{r}) ,
\label{profile2}
\end{eqnarray}
where the upper (lower) sign must be selected when $\mathbf{r}$ is in the gas (liquid) region.
Proceeding in a similar way as in the preceding section, the constrained order-parameter profiles can be expressed in the gas phase as
\begin{eqnarray}
\label{OP_vapor}
\delta m_\Xi&\approx& m_0 \Bigg[\fig{diagram21}
-\frac{1}{2}\left(\fig{diagram23}-\fig{diagram24}\right)\Bigg]
\nonumber\\
\end{eqnarray}
and in the liquid phase as
\begin{eqnarray}
\label{OP_liquid}
\delta m_\Xi&\approx& -m_0 \Bigg[\fig{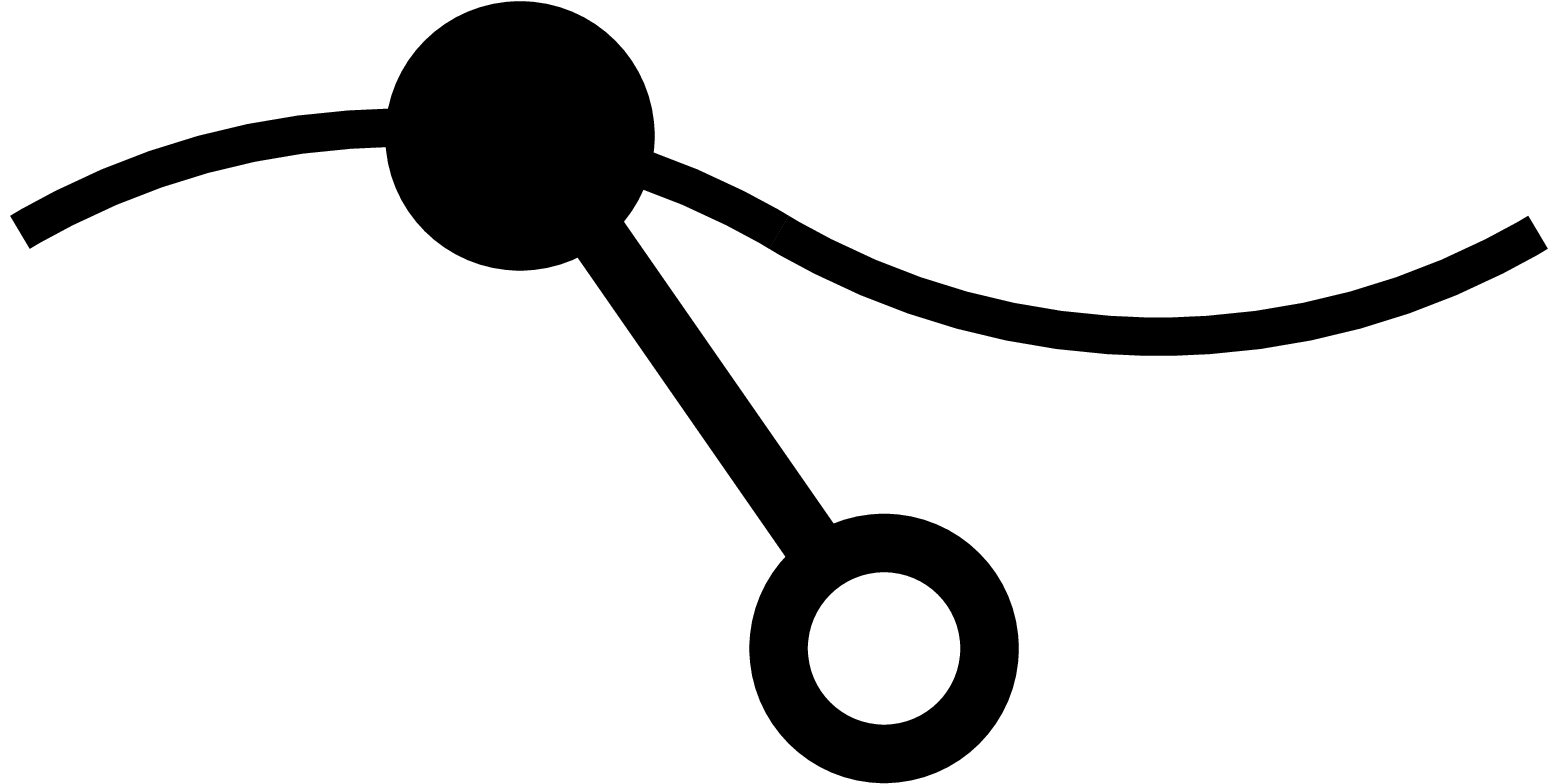}
-\frac{1}{2}\left(\fig{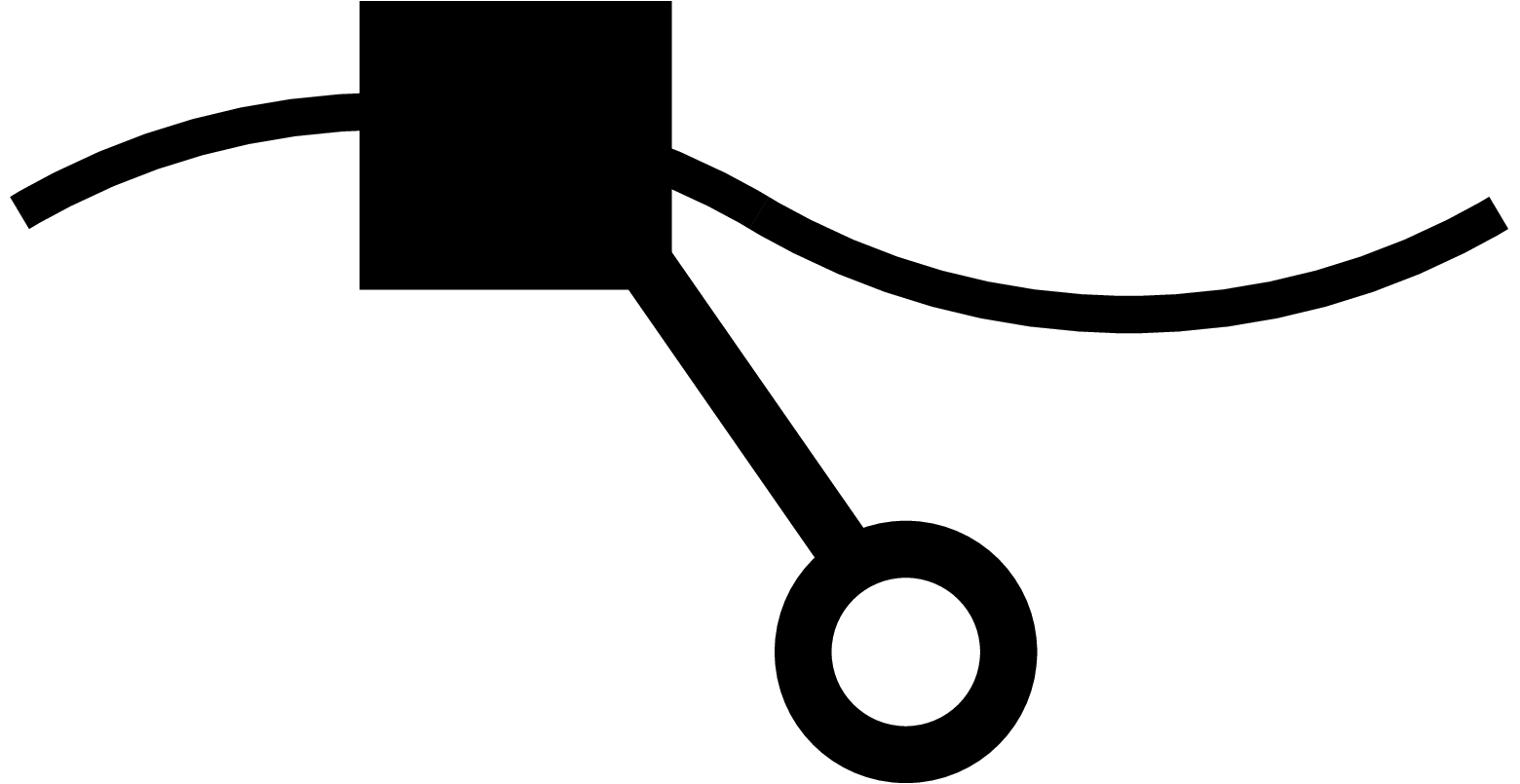}-\fig{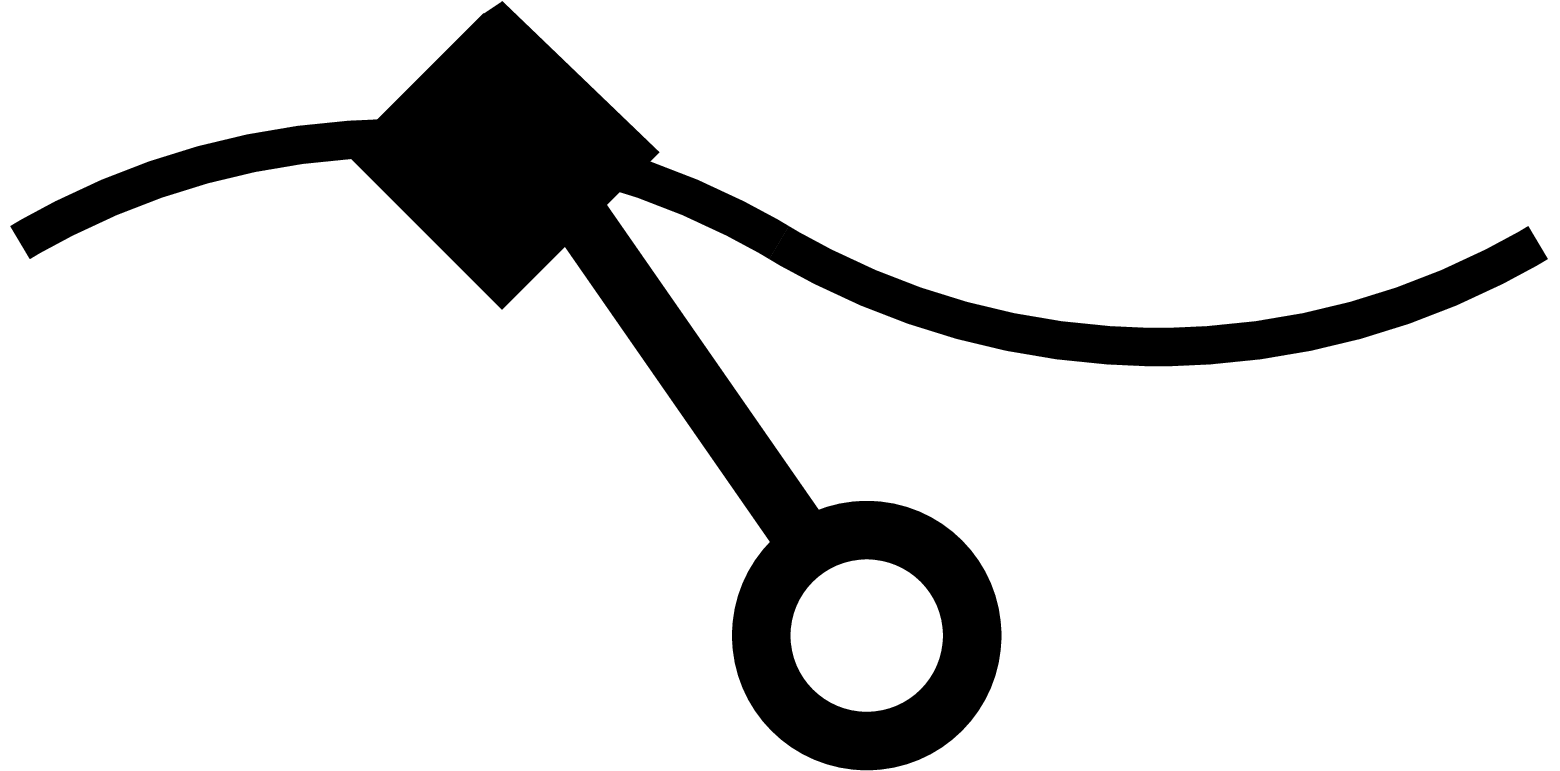}\right)
\Bigg].\nonumber\\
\end{eqnarray}

\subsection{General diagrammatic approach}
We can go beyond this perturbative approach and reobtain the full set of functionals $\omega_n$. For this purpose, we define
$\overline{q}(\mathbf{s})$ to be $[q^+(\mathbf{s})+q^-(\mathbf{s})]/2$. Note that the interfacial Hamiltonian 
~(\ref{finterface}) 
is proportional to the surface integral of $\overline{q}$. From Eq.~(\ref{consistent}) it follows
that $\overline{q}$ satisfies the
integral equation
\begin{equation}
\overline{q}(\mathbf{s}_\ell)=-\kappa m_0 - 
\int_{\ell} d \mathbf{s}\, U(\mathbf{s}_{\ell},\mathbf{s}) \overline{q}(\mathbf{s}). 
\end{equation}
Formally, this equation can be solved iteratively as
\begin{eqnarray}
\overline{q}(\mathbf{s}_\ell)=-\kappa m_0 \Bigg(1-\int_{\ell} d \mathbf{s}\, U(\mathbf{s}_{\ell},\mathbf{s})
\nonumber\\
+\int_{\ell} d \mathbf{s}\, d \mathbf{s}^\prime\, U(\mathbf{s}_{\ell},\mathbf{s}) U(\mathbf{s},\mathbf{s}^\prime)
-\cdots
\Bigg),
\end{eqnarray}
where the $n^{\textrm{th}}$ term involves the convolution of $n$ $U-$type functions on the interface. Upon substituting this expression 
into Eq.~(\ref{finterface}) we are lead to the expansion
\begin{eqnarray}
&&H[\ell]=\sigma\Bigg(1-\int_{\ell} d \mathbf{s}_1 \, d \mathbf{s}_2  U(\mathbf{s}_1,\mathbf{s}_2) \nonumber\\
&&+ \int_{\ell} d \mathbf{s}_1 \, d \mathbf{s}_2 \, d \mathbf{s}_3\,  U(\mathbf{s}_1,\mathbf{s}_2)
U(\mathbf{s}_2,\mathbf{s}_3)-\cdots
\Bigg)
\end{eqnarray}
or, diagrammatically,
\begin{equation}
\label{F_chain}
H[\ell]=\sigma\Bigg[ \fig{diagram1}-\fig{diagram18} + \fig{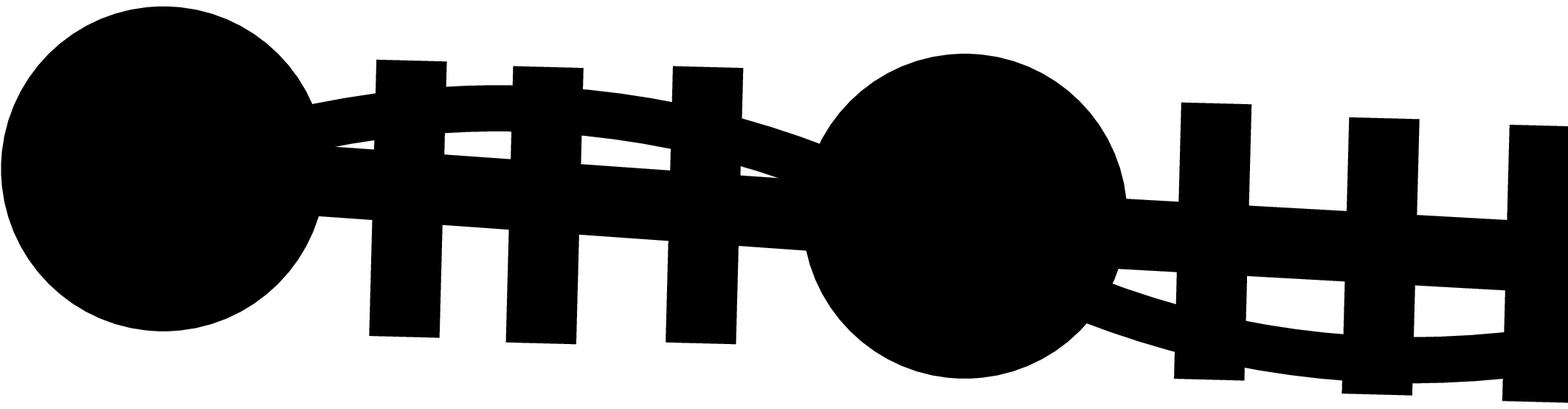} + \cdots \Bigg] .
\end{equation}
From this expression, we obtain that
\begin{equation}
\omega_n = \int_\ell d \mathbf{s}_1 \, \cdots \, d \mathbf{s}_{n} \, U(\mathbf{s}_1,\mathbf{s}_2)
U(\mathbf{s}_2,\mathbf{s}_3)\cdots U(\mathbf{s}_{n-1},\mathbf{s}_n)
\end{equation}
which is the expression obtained in Ref. \cite{PR}. In general, from Eq.~(\ref{dashedk}) we get that
$\omega_n \sim R^{-2(n-1)}$, which connects the self-interaction contributions to the curvature corrections to the
free energy. We note that this expression is general, so it is also valid for spherical
bubbles, for which $\omega_n = \exp[2(n-1)\kappa R]$ \cite{PR} because $H^2=K_G$.  
We can also obtain the order-parameter profiles by considering in 
Eq.~(\ref{deltamdiagram}) the limit $h_1=0$, $g\to -\infty$, and $m_b=-m_0$ (or $+m_0$) for the gas (or liquid) 
phase, respectively. Thus, the order-parameter profile in the gas phase has the expansion
\begin{eqnarray}
\delta m_\Xi = m_0 \Bigg[\fig{diagram21}-\fig{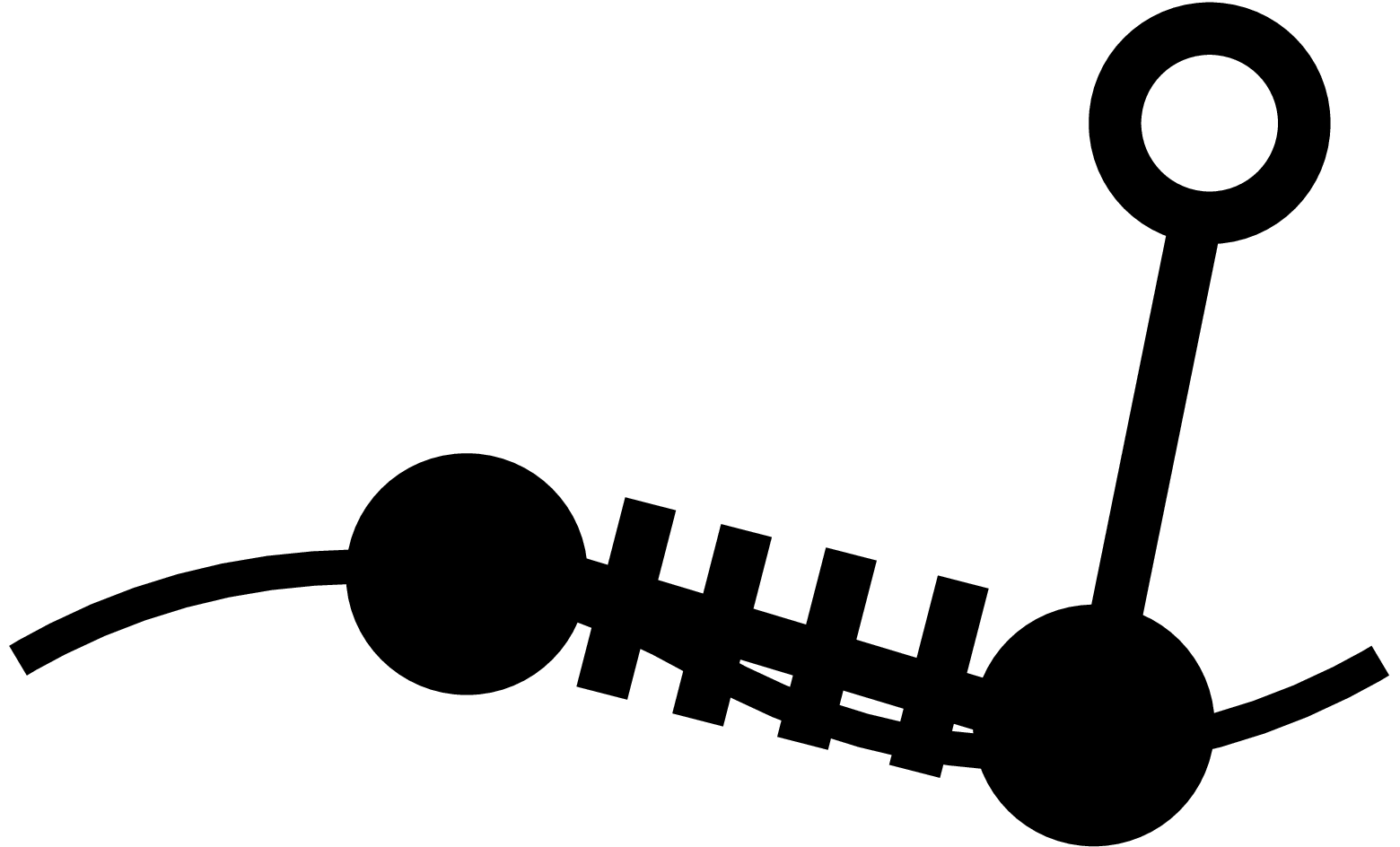}+\fig{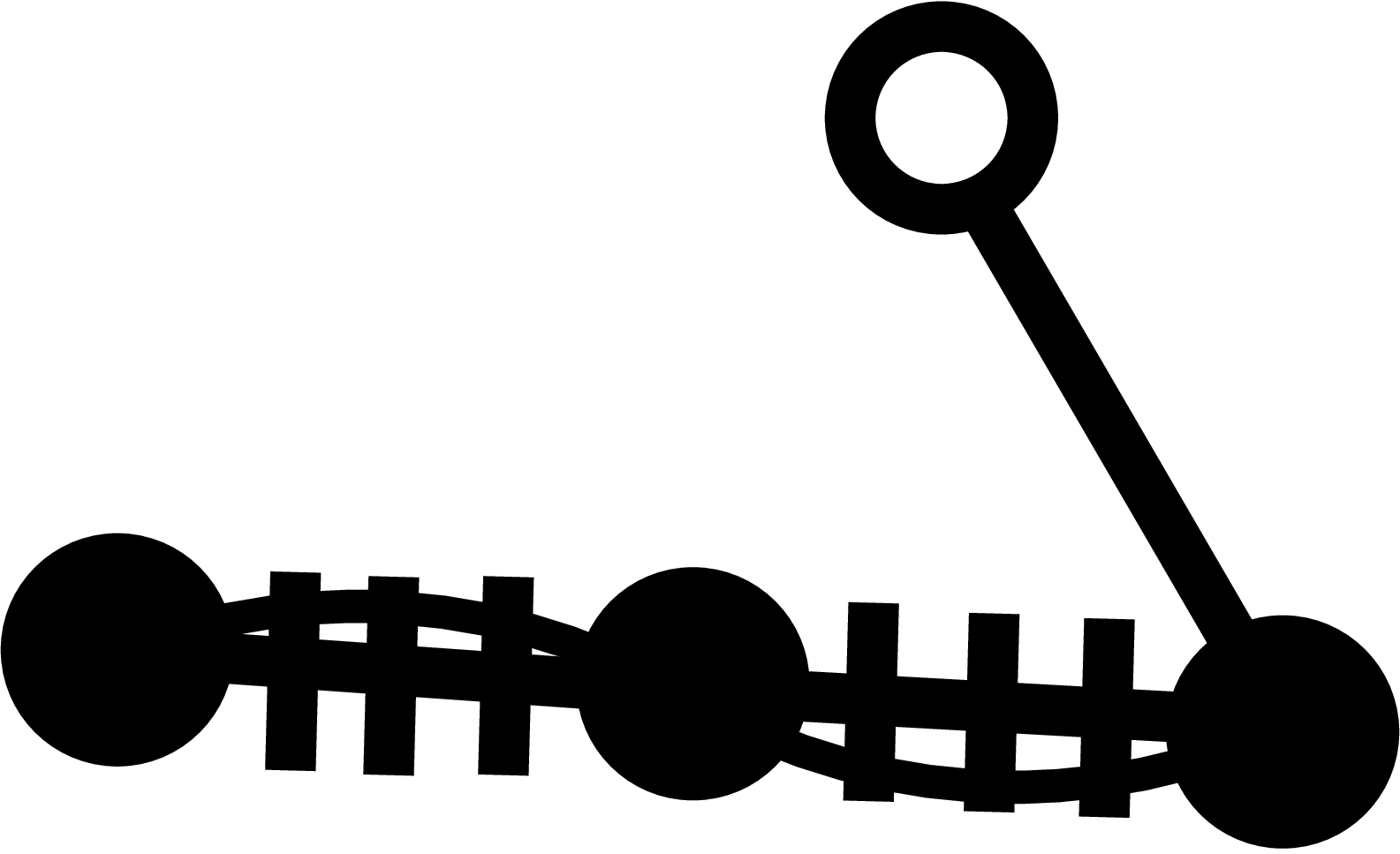}-\cdots\Bigg],
\label{deltamdiagram2}
\end{eqnarray}
and in the liquid phase has the expansion
\begin{eqnarray}
\delta m_\Xi
= -m_0 \Bigg[\fig{diagram28}-\fig{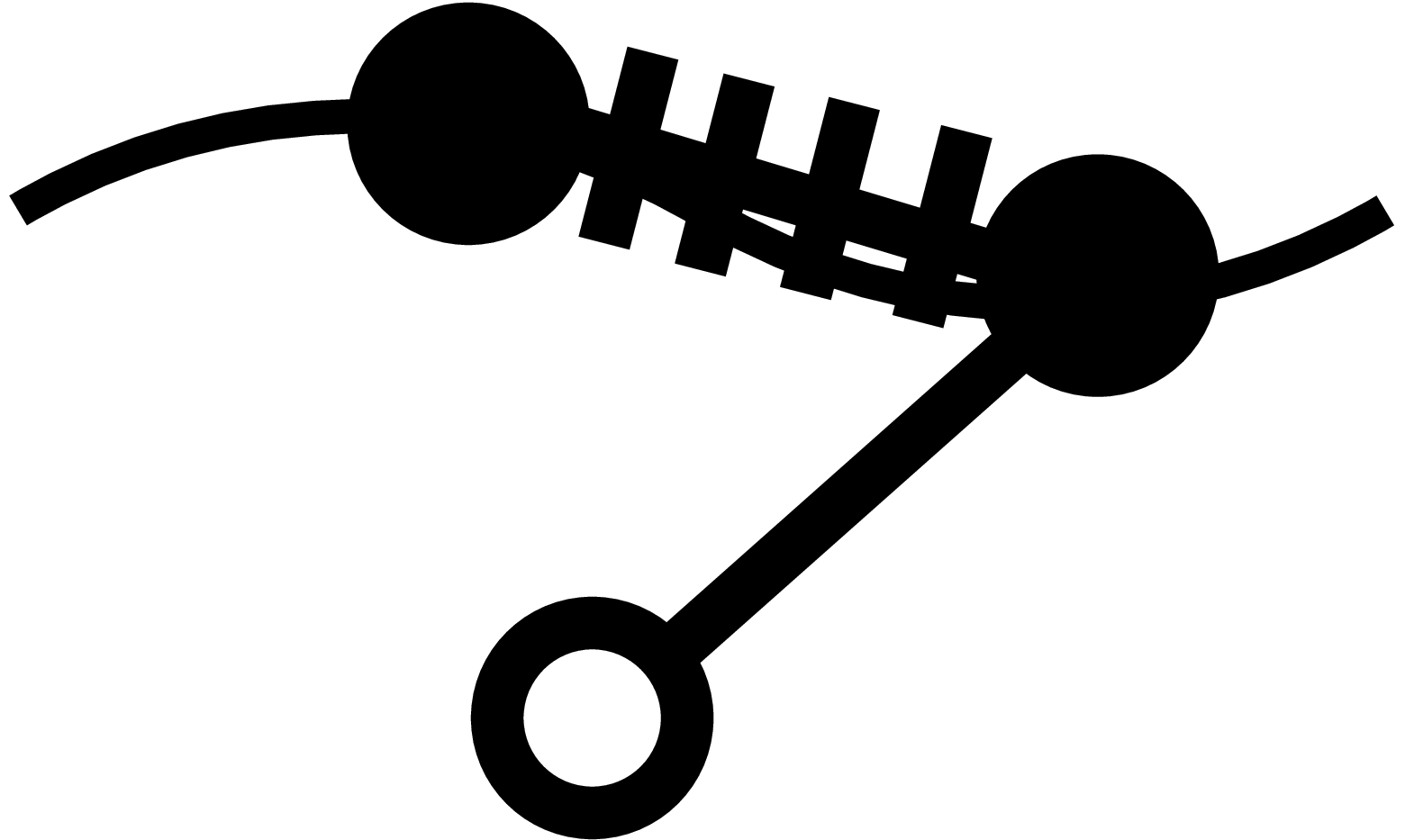}+\fig{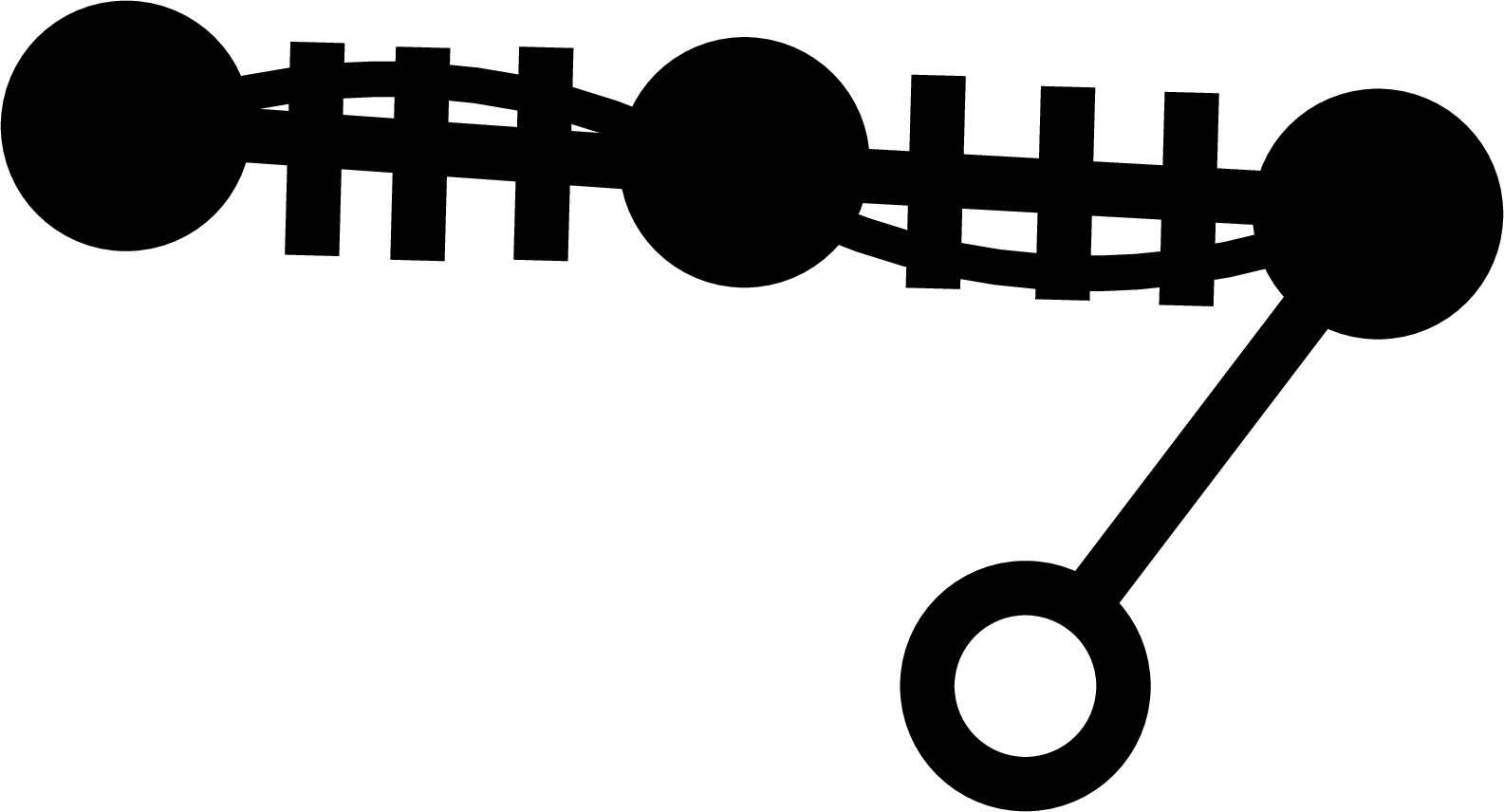}-\cdots \Bigg].
\label{deltamdiagram3}
\end{eqnarray}
Again, these equations reduce to Eqs.~(\ref{OP_vapor}) and (\ref{OP_liquid}) upon using Eq.~(\ref{dashedk}) up to $R^{-3}$ 
corrections. 

Once we have established the connection between the self-interaction of the fluid interface and the curvature
corrections to the interfacial free energy, we can find the full functional  
of the interfacial shape. 
This task can be pursued to any desired accuracy in the curvature, which we leave at ${O}(R^{-3})$. 
Leaving the technical aspects aside here (see Appendix~\ref{appendix_B} for details) we find that Eq.~(\ref{Helfrich}) reduces to
the interfacial Hamiltonian
\begin{equation}
\label{ham4}
H[\ell] \approx \sigma \mathcal{A}_{\pi} + \frac{\sigma}{2} \int d \mathbf{x}_{1} d \mathbf{x}_{2} \, \mathcal{W}(x_{12}) \bigl[\ell(\mathbf{x}_{2}) - \ell(\mathbf{x}_{1})\bigr]^{2} ,
\end{equation}
where the self-interaction is described by the function
\begin{equation}
\mathcal{W}(x) \equiv \frac{\kappa}{2\pi} \frac{1+\kappa x}{x^{3}} \textrm{e}^{-\kappa x}
=\frac{1+\kappa x}{x^2}\mathcal{K}(x) ,
\end{equation}
thus rigorously rederiving the result first obtained in Ref.~\cite{PR}. Clearly, for a flat interface 
$H[\ell]=\sigma \mathcal{A}_\pi$, with $\mathcal{A}_\pi$ being the planar (projected) area.  As shown in 
Ref.~\cite{PR}, when the gradient is small we can expand as $\ell(\mathbf{x}_{2}) \simeq \ell(\mathbf{x}_{1}) + \mathbf{x}_{21} \mathbf{\cdot}\boldsymbol \nabla\ell(\mathbf{x}_{1})+\cdots$,  in which case 
Eq.~(\ref{ham4}) reduces to
\begin{equation}
H[\ell] \approx \sigma \mathcal{A}_{\pi}+\frac{\sigma}{2} \int d \mathbf{x} \, \bigl[\boldsymbol\nabla\ell(\mathbf{x})\bigr]^{2},
\end{equation}
thus recovering the standard mesoscopic capillary-wave Hamiltonian, which can now be seen as a particular local small-gradient limit of the nonlocal functional (\ref{ham4}). The present derivation of the nonlocal self-interaction improves on that presented in Ref.~\cite{PR} inasmuch as it systematically and rigorously accounts for all curvature corrections.

\subsection{Summary and remarks}
In this section we considered an isolated liquid-gas interface and solved the 
Helmholtz equations required to identify the free energy of a constrained interfacial configuration defined by a 
crossing criterion. We first implemented a direct perturbation expansion in the local curvature, obtaining the 
Helfrich-like corrections to the surface tension term and identifying the values of bending and saddle-splay rigidities 
for the DP potential. We then refined this expansion by considering the order-parameter profile around the 
interface in which the curvature corrections are explicit [see Eqs.~(\ref{OP_vapor}) and (\ref{OP_liquid})]. This leads us naturally to express the free energy as an interfacial self-interaction that can be neatly expressed that involves a formally exact way using a diagrammatic expansion. Finally, in the limit of small curvatures this nonlocal interaction recovers the standard local capillary-wave model.

\section{Binding energy for a wetting film configuration}
Having examined the wall-liquid and free (but constrained) liquid-gas interfaces, we turn to the case of a wall-gas interface where an intruding wetting layer of liquid, with positive order parameter (i.e., $m>0$), intrudes between the substrate and the bulk gas (where the order parameter is set to $-m_0$). The wall is again described by a height function $\psi(\mathbf{x})$ and the liquid-gas interface (i.e., the surface on which $m=0$) is constrained to lie along $\ell(\mathbf{x})$. No overhangs of either the interface or substrate occur; nor do these two surfaces touch. The minimum of the LGW Hamiltonian (\ref{HLGW}) subject to the substrate, bulk, and crossing-criterion boundary conditions 
\begin{eqnarray}
\mathbf{n}_\psi\mathbf{\cdot}\boldsymbol \nabla m_\Xi (\mathbf{r})=-h_1-gm_\Xi(\mathbf{r}), \  
\textrm{for}\ \mathbf{r}=(\mathbf{x},\psi(\mathbf{x})), \label{boundary22}\\
m_\Xi(z)\to -m_0, \qquad 
\textrm{for}\quad z\to +\infty,
\label{boundary2b2}\\
m_\Xi(\mathbf{r})=0,\ \textrm{for}\ \mathbf{r}=(\mathbf{x},\ell(\mathbf{x})) \label{boundary2c2}
\end{eqnarray}
defines a constrained excess free energy for the wall-gas interface, which by Eq.~(\ref{mfHLGW2}) can be recast as
\begin{eqnarray}
\mathcal{F}_{wg}[\ell,\psi]=-\frac{m_0}{2}\int_{\ell} d\mathbf{s}
\left[q^+_\ell(\mathbf{s}) + q^-_\ell (\mathbf{s})\right]\nonumber\\
+\frac{h_1+gm_0}{2g}\int_\psi d\mathbf{s} q_\psi(\mathbf{s}),
\label{Hmin}
\end{eqnarray}
where $q_\ell^{\pm}(\mathbf{s})\equiv \mathbf{n}_\ell(\mathbf{s})\mathbf{\cdot}\boldsymbol \nabla_\mathbf{s} 
\delta m^{\pm}(\mathbf{s})$ for $\mathbf{s}$ on the liquid-gas interface, and 
$q_\psi\equiv \mathbf{n}_\psi(\mathbf{s})\mathbf{\cdot}\boldsymbol \nabla_\mathbf{s} 
\delta m(\mathbf{s})$ on the substrate. 

The next step is to define and identify the binding potential $W[\ell,\psi]$. By analogy with isolated interfaces, the free energy of a wetting layer can be expressed as a functional of the normal derivatives of the order parameter computed at the layer boundaries. The binding potential takes into account the interaction of the interface with the wall and is determined by subtracting the contributions arising from the isolated wall-liquid and constrained but free liquid-gas interfaces, which we have already determined. Therefore, before presenting the final result for $W[\ell,\psi]$ and its diagrammatic formulation, we need to consider the fundamental relations obeyed by the order parameter in a wetting layer. In order to do so we need some technical preliminaries.

\subsection{Technical preliminaries}
From Eqs.~(\ref{greenidentity5}) and (\ref{greenidentity6}) it follows that 
the functions $q^+_\ell(\mathbf{s})$, $q^-_\ell(\mathbf{s})$,
and $q_\psi(\mathbf{s})$ satisfy the coupled integral equations
\begin{eqnarray}
\int_\psi d\mathbf{s} \Bigg[K(\mathbf{s}_\psi,\mathbf{s})
+\frac{1}{g}\partial_{n} K(\mathbf{s}_\psi,\mathbf{s}) 
-\frac{\kappa}{g}\delta(\mathbf{s}-\mathbf{s}_\psi)\Bigg]q_\psi(\mathbf{s})
\nonumber\\-\int_\ell d\mathbf{s} K(\mathbf{s}_\psi,\mathbf{s})q_\ell^-(\mathbf{s})
\nonumber\\=\left(\frac{-h_1}{g}-m_0\right)\left(-\kappa+\int_\psi d\mathbf{s}\  
\partial_n K(\mathbf{s}_\psi,\mathbf{s})\right)
\nonumber\\
+m_0\int_\ell d\mathbf{s}\ \partial_n K(\mathbf{s}_\psi,\mathbf{s});
\label{inteq1a}\\
\int_\psi d\mathbf{s} \Bigg[K(\mathbf{s}_\ell,\mathbf{s})
+\frac{1}{g}\partial_n K(\mathbf{s}_\ell,\mathbf{s})
\Bigg]q_\psi
(\mathbf{s})\nonumber\\
-\int_\ell d\mathbf{s} K(\mathbf{s}_\ell,\mathbf{s}) q^-_\ell
(\mathbf{s})
=m_0\left(\kappa+\int_\ell d\mathbf{s}\ 
\partial_n K(\mathbf{s}_\ell,\mathbf{s})\right)
\nonumber\\
+\left(\frac{-h_1}{g}-m_0\right)\int_\psi d\mathbf{s}\ 
\partial_n K(\mathbf{s}_\ell,\mathbf{s})
\label{inteq1b}
\end{eqnarray}
and
\begin{eqnarray}
\int_\ell d\mathbf{s} K(\mathbf{s}_\ell,\mathbf{s})q_\ell^+
(\mathbf{s})=-m_0\left(\kappa-
\int_\ell d\mathbf{s}\ 
\partial_n K(\mathbf{s}_\ell,\mathbf{s})\right),
\label{inteq2}
\end{eqnarray}
where $\mathbf{s}_\ell$ and $\mathbf{s}_\psi$ are on the liquid-gas interface and on the substrate, 
respectively. Note that these equations are linear in $q$. In order to extract the interaction between the surfaces, we obtain the equations in terms of the new fields
$\Delta q^\pm_\ell(\mathbf{s})\equiv q^\pm_\ell(\mathbf{s})-q^{0,\pm}_\ell(\mathbf{s})$, and $
\Delta q_\psi(\mathbf{s})\equiv q_\psi(\mathbf{s})-q^{0}_\psi(\mathbf{s})$,
where the $0$ superscript means that the corresponding normal derivative is evaluated on its isolated interface. In addition,
$q^0_\psi$ and $q^{0,\pm}_\ell$ satisfy Eqs.~(\ref{inteq1}) and (\ref{consistent}), respectively. 
Note that $\Delta q^{0,+}_\ell\equiv 0$ (because the gas domain is shielded from influence of the wall) 
and Eqs. (\ref{inteq1a}) and (\ref{inteq1b}) can be recast as
\begin{eqnarray}
& &\int_\psi d\mathbf{s} \Bigg[K(\mathbf{s}_\psi,\mathbf{s})
+\frac{1}{g}\partial_n K(\mathbf{s}_\psi,\mathbf{s})
\nonumber\\&-&\frac{\kappa}{g}\delta(\mathbf{s}_\psi-\mathbf{s})
\Bigg]\Delta q_\psi(\mathbf{s})-\int_\ell d\mathbf{s} 
K(\mathbf{s}_\psi,\mathbf{s})\Delta q_{\ell}^-(\mathbf{s})\nonumber\\
&=&m_0\int_\ell d\mathbf{s} \ \partial_n K(\mathbf{s}_\psi,\mathbf{s})
+\int_\ell d\mathbf{s} K(\mathbf{s}_\psi,\mathbf{s}) q_\ell^{0,-} (\mathbf{s})
\nonumber\\
&\equiv& 2\kappa \delta m_\Xi^{0,\ell}(\mathbf{s}_\psi) 
\label{inteq6}
\end{eqnarray}
and
\begin{eqnarray}
& &\int_\psi d\mathbf{s} \Bigg[K(\mathbf{s}_\ell,\mathbf{s})
+\frac{1}{g}\partial_n K(\mathbf{s}_\ell,\mathbf{s})\Bigg]
\Delta q_\psi (\mathbf{s})\nonumber\\
&-&\int_\ell d\mathbf{s} K(\mathbf{s}_\ell,\mathbf{s})
\Delta q_\ell(\mathbf{s})
\nonumber\\
&=&\left(\frac{-h_1}{g}-m_0\right)\int_\psi d\mathbf{s}
\ \partial_n K(\mathbf{s}_\ell,\mathbf{s})\nonumber\\
&-&\int_\psi d\mathbf{s} \Bigg[K(\mathbf{s}_\ell,\mathbf{s})
+\frac{1}{g}\partial_n K(\mathbf{s}_\ell,\mathbf{s})\Bigg]
q^0_\psi(\mathbf{s})\nonumber\\
&\equiv& 2\kappa \delta m_\Xi^{0,\psi}(\mathbf{s}_\ell),
\label{inteq7}
\end{eqnarray}
where we have identified the right-hand side of both equations as the 
order-parameter profile $\delta m_\Xi^{0,\ell(\psi)}(\mathbf{s})$ at
the boundary point $\mathbf{s}$ due to the
presence of an isolated liquid-gas (wall) interface, respectively 
[see Eqs.~(\ref{profile}) and (\ref{profile2})].
Equations~(\ref{inteq6}) and (\ref{inteq7}) are the basis of our perturbative
approach, as we can expand $\Delta q_\psi$ and $\Delta q_\ell^-$ in 
powers of $K(\mathbf{s}_\ell,\mathbf{s}_\psi)$ as $\Delta q_\psi =\sum_{i=1}^\infty \Delta q_{i,\psi}$
and $\Delta q_\ell =\sum_{i=1}^\infty \Delta q_{i,\ell}$. Each term of this expansion can be
formally solved as follows. At the wall,
\begin{eqnarray}
&&\Delta q_{1,\psi}(\mathbf{s}_\psi)=\int_\psi d\mathbf{s} X_\psi(\mathbf{s}_\psi,\mathbf{s}) 
2\kappa \delta m_\Xi^{0,\ell}
(\mathbf{s}),
\label{formalsol}
\end{eqnarray}
and otherwise
\begin{eqnarray}
&&\Delta q_{i>1,\psi}(\mathbf{s}_\psi)=\int_\psi d\mathbf{s} \int_\ell d\mathbf{s}' 
X_\psi(\mathbf{s}_\psi,\mathbf{s}) K(\mathbf{s},\mathbf{s}')\Delta q_{i-1,\ell}(\mathbf{s}')
.
\nonumber\\
\label{formalsol2}
\end{eqnarray}
Similarly, at the interface
\begin{eqnarray}
&&\Delta q_{1,\ell}(\mathbf{s}_\ell)=-\int_\ell d\mathbf{s} K^{-1}_\ell(\mathbf{s}_\ell,\mathbf{s}) 
2\kappa \delta m_\Xi^{0,\psi}
(\mathbf{s})
\label{formalsol3}
\end{eqnarray}
and otherwise
\begin{eqnarray}
\Delta q_{i>1,\ell}(\mathbf{s}_\ell)=\int_\ell d\mathbf{s} \int_\psi d\mathbf{s}' 
K^{-1}_\ell(\mathbf{s}_\ell,\mathbf{s}) \Bigg(K(\mathbf{s},\mathbf{s}')\nonumber\\
+\frac{1}{g}\partial_{n'}K(\mathbf{s},
\mathbf{s}')\Bigg)\Delta q_{i-1,\psi}(\mathbf{s}'),
\label{formalsol4}
\end{eqnarray}
where $X_\psi$ is the operator on the substrate defined by Eq.~(\ref{seriesx}) and $K^{-1}_\ell$ 
is the inverse operator of $K$ on the liquid-gas interface. 

Now using the Green's identities (\ref{greenidentity7})-(\ref{greenidentity9}),
it follows that
\begin{eqnarray}
&&\Delta q_{i>1,\ell}(\mathbf{s}_\ell)= 
\int_\ell d\mathbf{s} \int_\psi d\mathbf{s}' \int_\psi d\mathbf{s}''
K^{-1}_\ell(\mathbf{s}_\ell,\mathbf{s})\nonumber\\
&&\times K(\mathbf{s},\mathbf{s}')\left(\delta(\mathbf{s}'-\mathbf{s}'')+\frac{\kappa}{g}K^{-1}_\psi(\mathbf{s}',\mathbf{s}'')\right)
\Delta q_{i-1,\psi}(\mathbf{s}'')
\nonumber\\
&&+\frac{\kappa}{g}\int_\ell d\mathbf{s} \int_\psi d\mathbf{s}' \int_\psi d\mathbf{s}'' \int_\psi d\mathbf{s}'''
K^{-1}_\ell(\mathbf{s}_\ell,\mathbf{s})K(\mathbf{s},\mathbf{s}')\nonumber\\
&&\times\frac{1}{\kappa}\partial_{n'} K(\mathbf{s}',\mathbf{s}'')
K^{-1}_\psi(\mathbf{s}'',\mathbf{s}''')
\Delta q_{i-1,\psi}(\mathbf{s}'''),
\label{formalsol5}
\end{eqnarray}
where $K^{-1}_\psi$ is the 
inverse operator of $K$ on the substrate, i.e., $X_\psi$ in the limit $g\to -\infty$.  
Taking into account the expansions ~(\ref{seriesx2}), (\ref{definvk}), (\ref{deltamdiagram}),
and (\ref{deltamdiagram3}), we obtain a diagrammatic expansion for $\Delta q_\ell$ and $\Delta q_\psi$,
\begin{eqnarray}
&&\Delta q_\psi(\mathbf{s})=-2\kappa m_0\Bigg[\frac{g}{g-\kappa}\Bigg(\fig{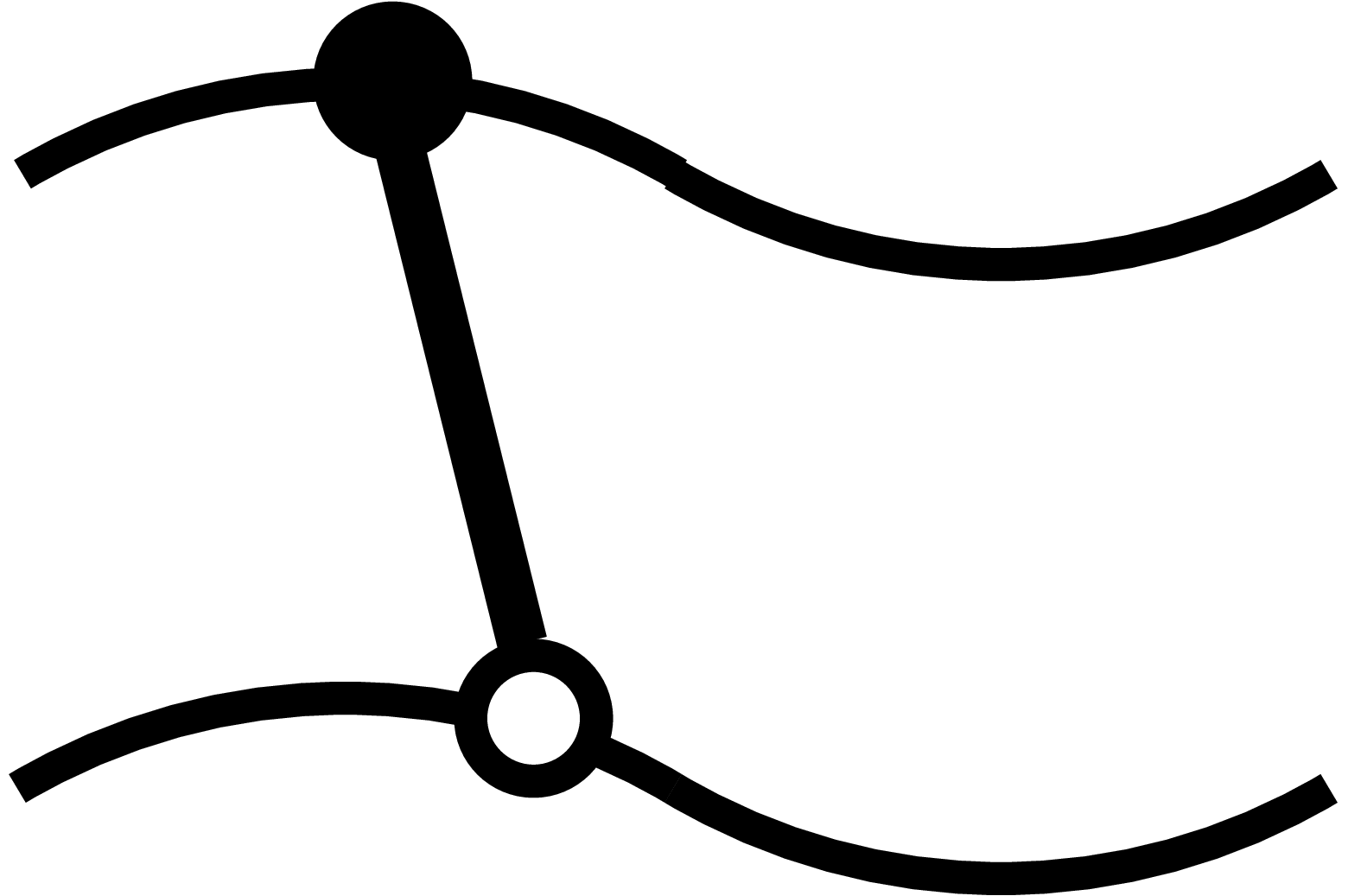}-\fig{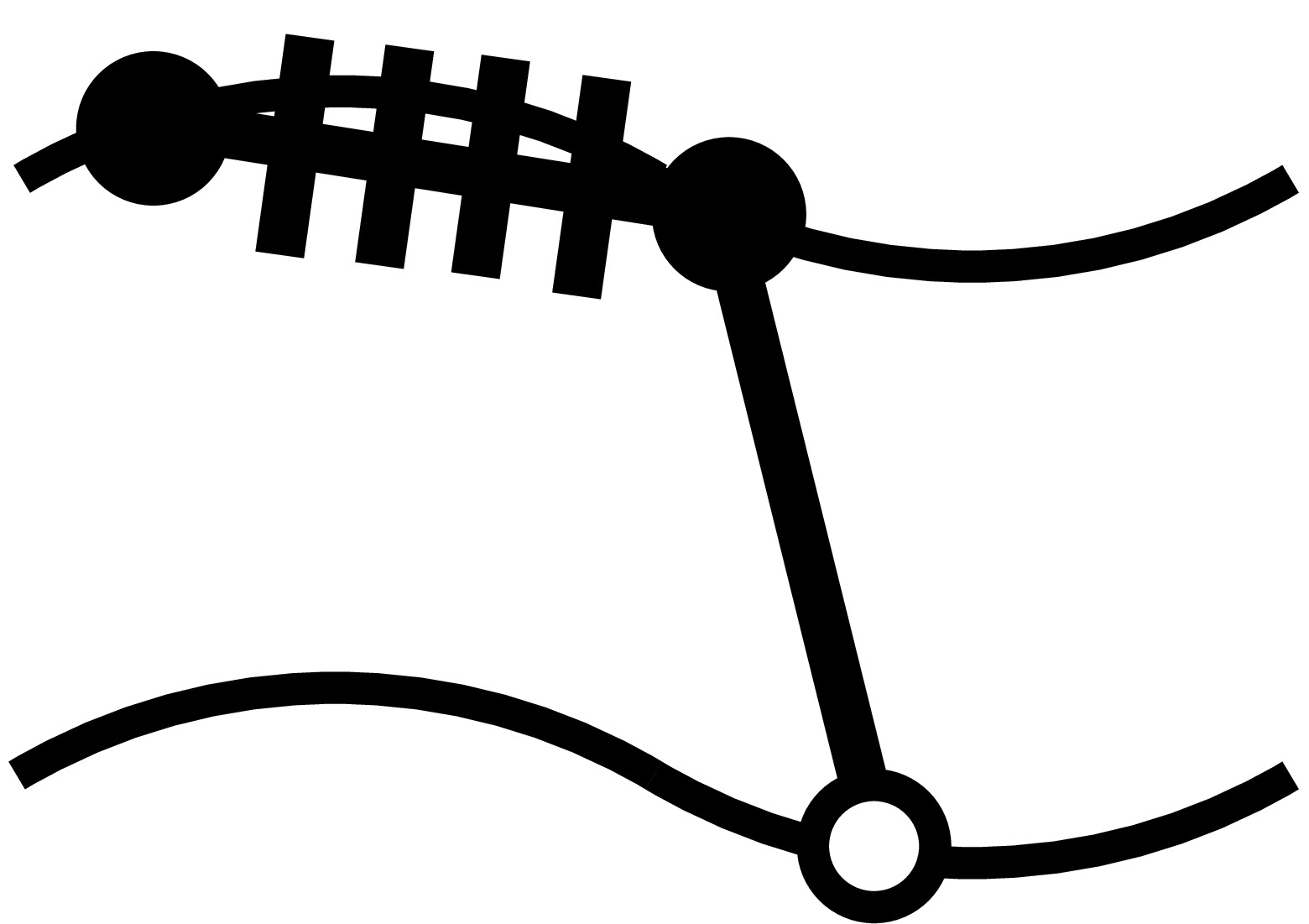}+
\cdots\nonumber\\
&&+\frac{\kappa}{\kappa-g}\fig{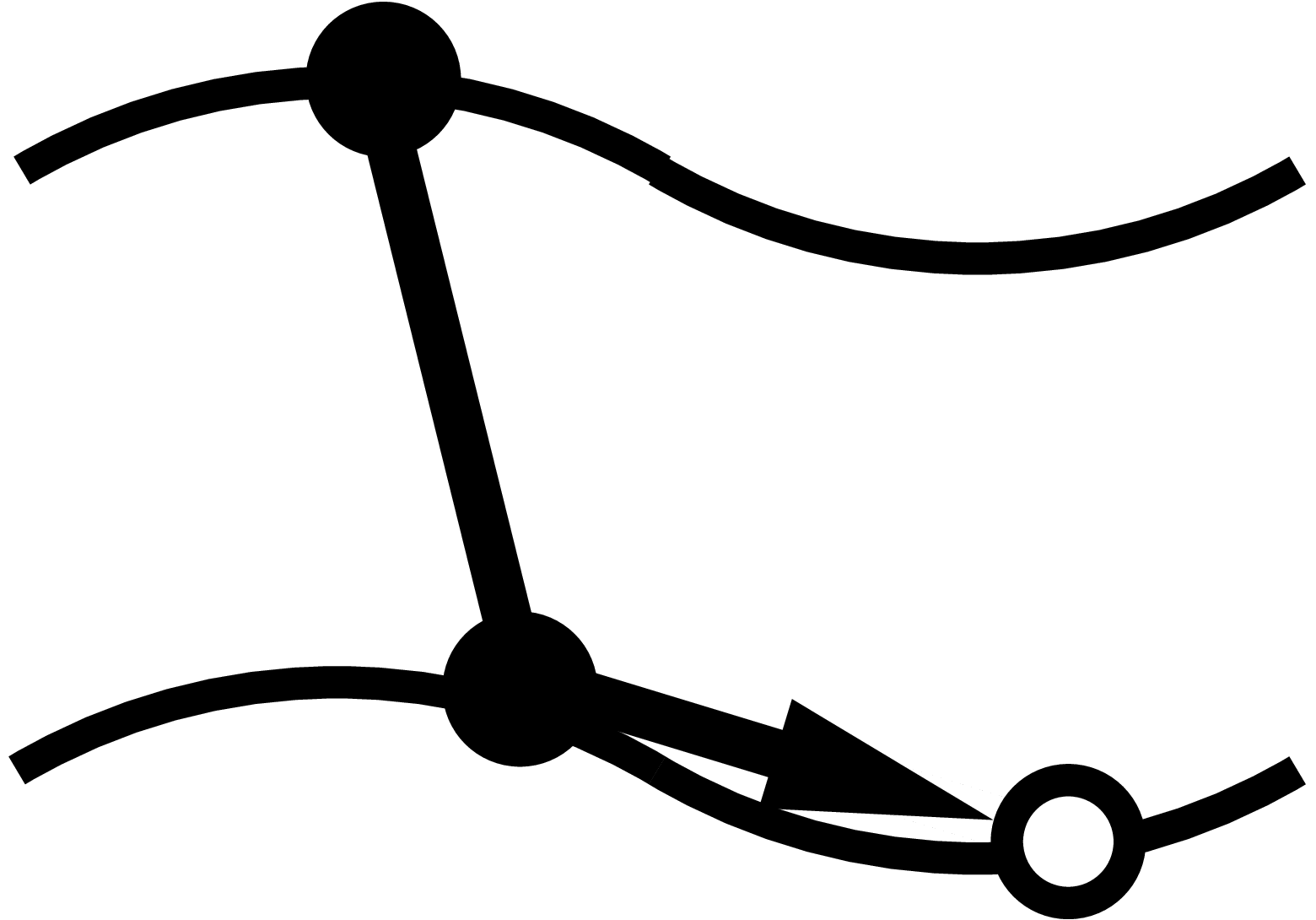}+\cdots-\frac{\kappa}{\kappa-g}\fig{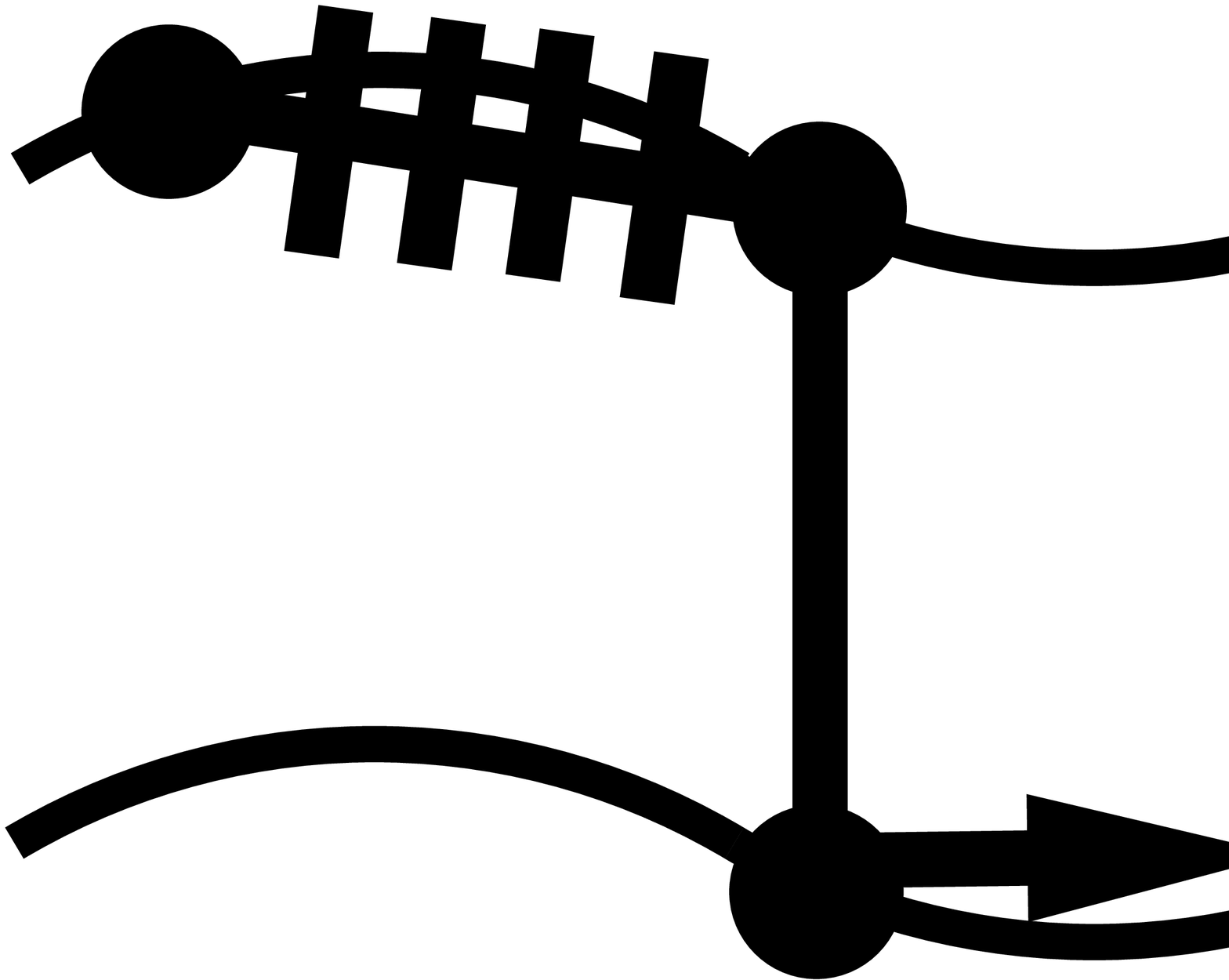}\Bigg)\nonumber\\
&&+\left(\frac{g}{g-\kappa}\right)^2\left(1+\frac{\kappa}{g}\right)\fig{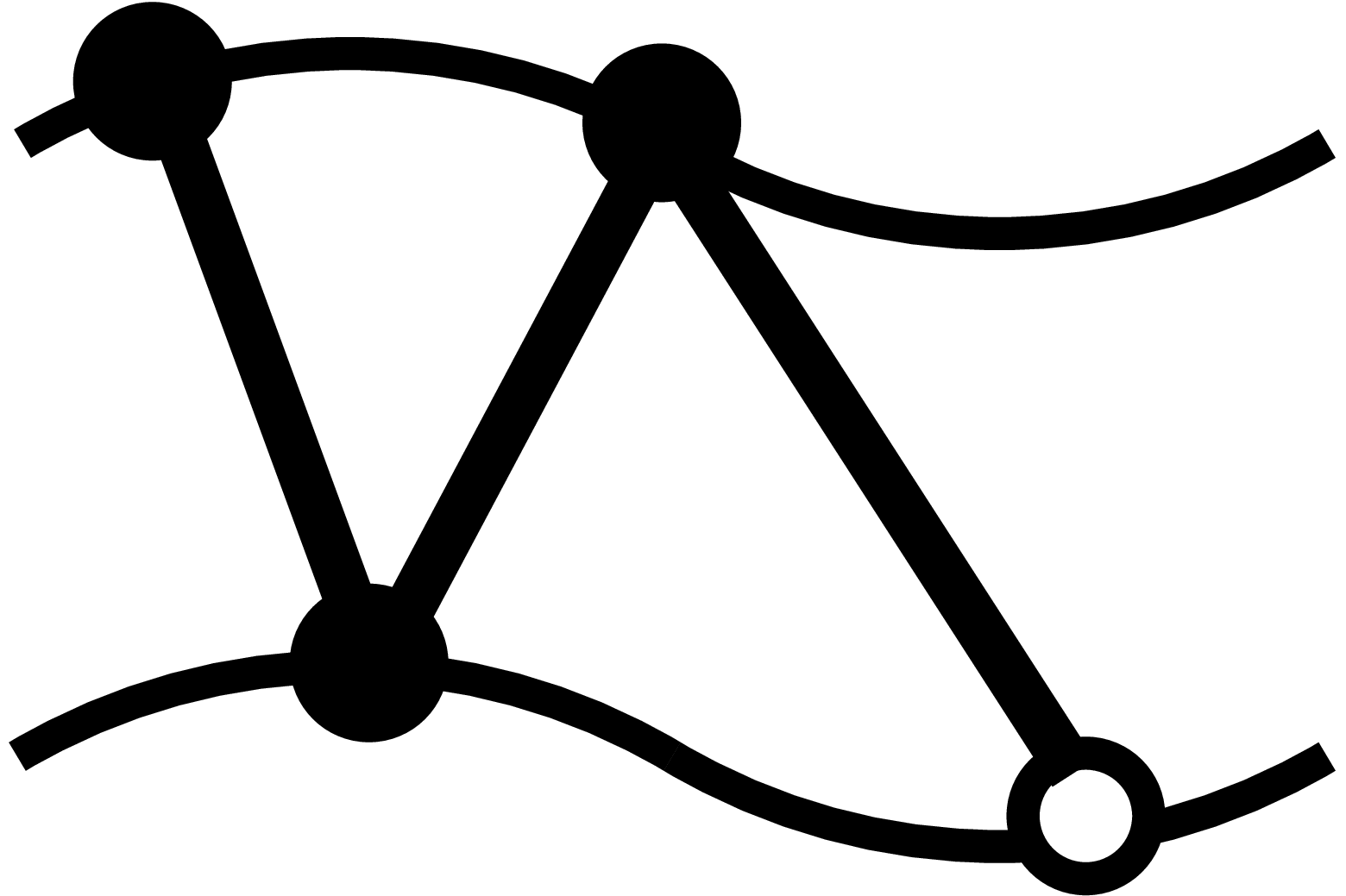}+\cdots\Bigg]
\nonumber\\
&&+2\kappa \frac{h_1+gm_0}{g} \Bigg[\left(\frac{g}{g-\kappa}\right)^2\Bigg(\fig{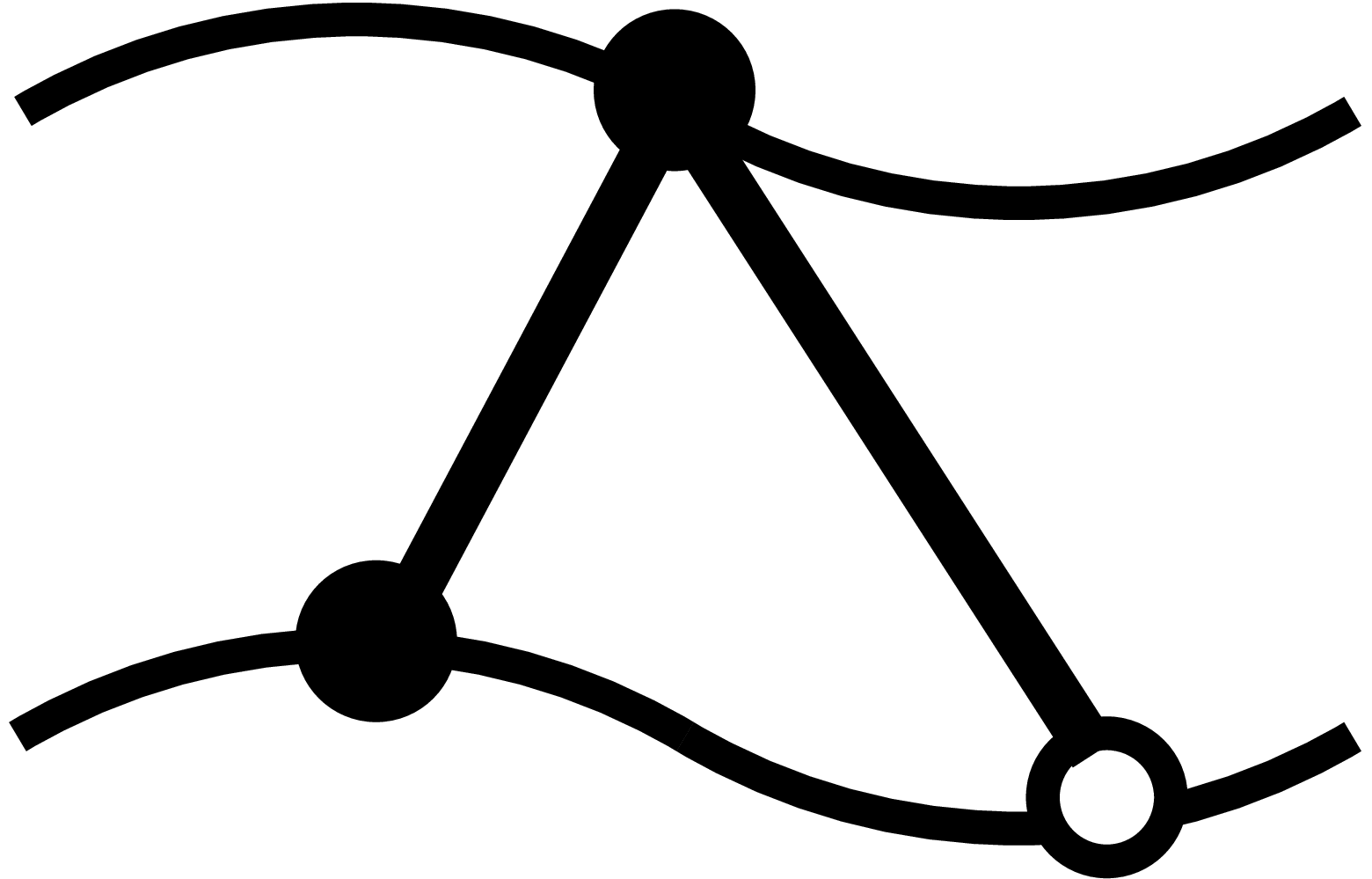}\nonumber\\
&&+\frac{g}{\kappa-g}
\fig{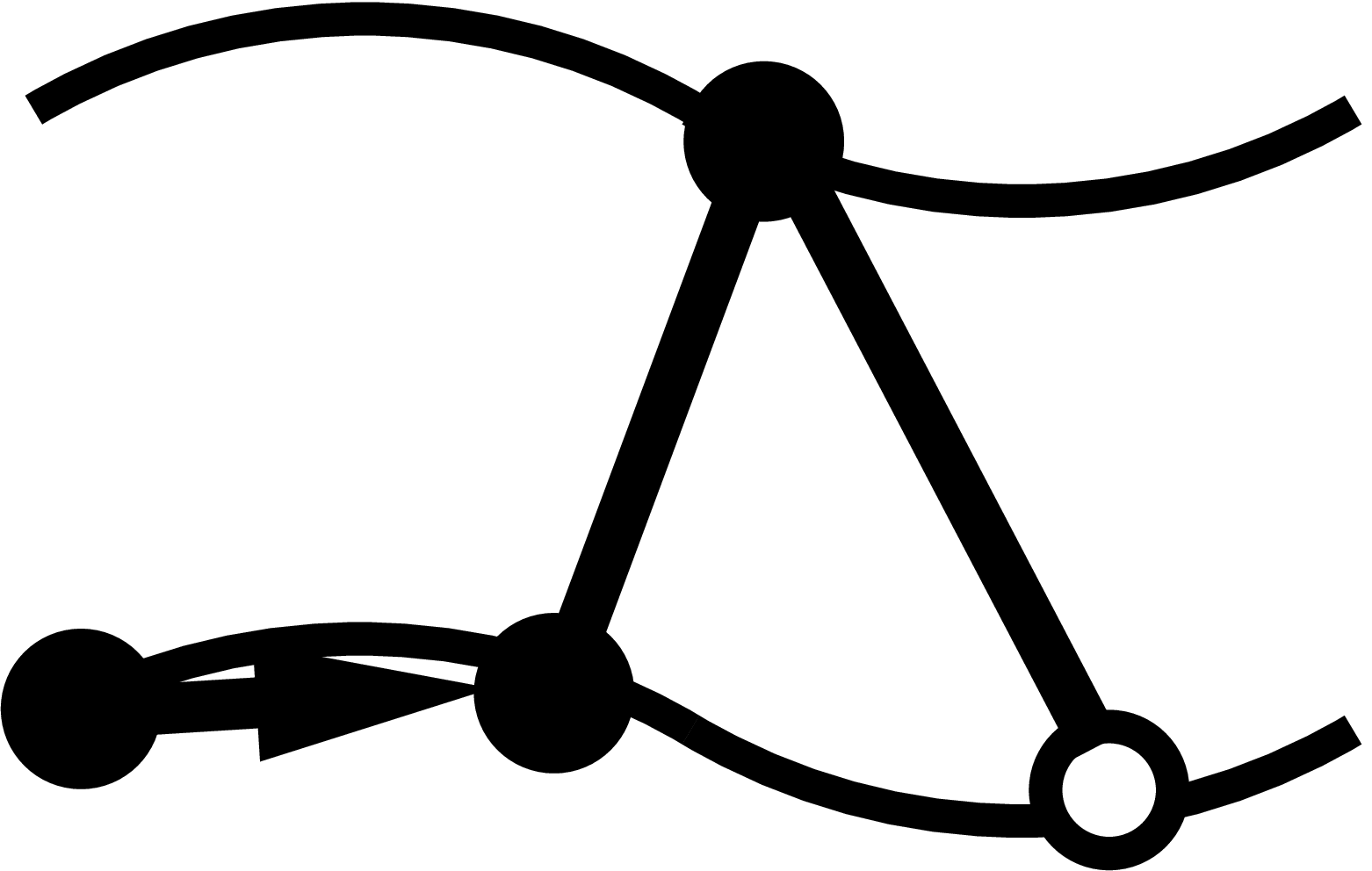}+\cdots\Bigg)+\cdots\Bigg]
\label{deltaqpsi}
\end{eqnarray}
and
\begin{eqnarray}
&&\Delta q_\ell^-(\mathbf{s})=
2\kappa \frac{h_1+gm_0}{g}\Bigg[\frac{g}{g-\kappa}\Bigg(\fig{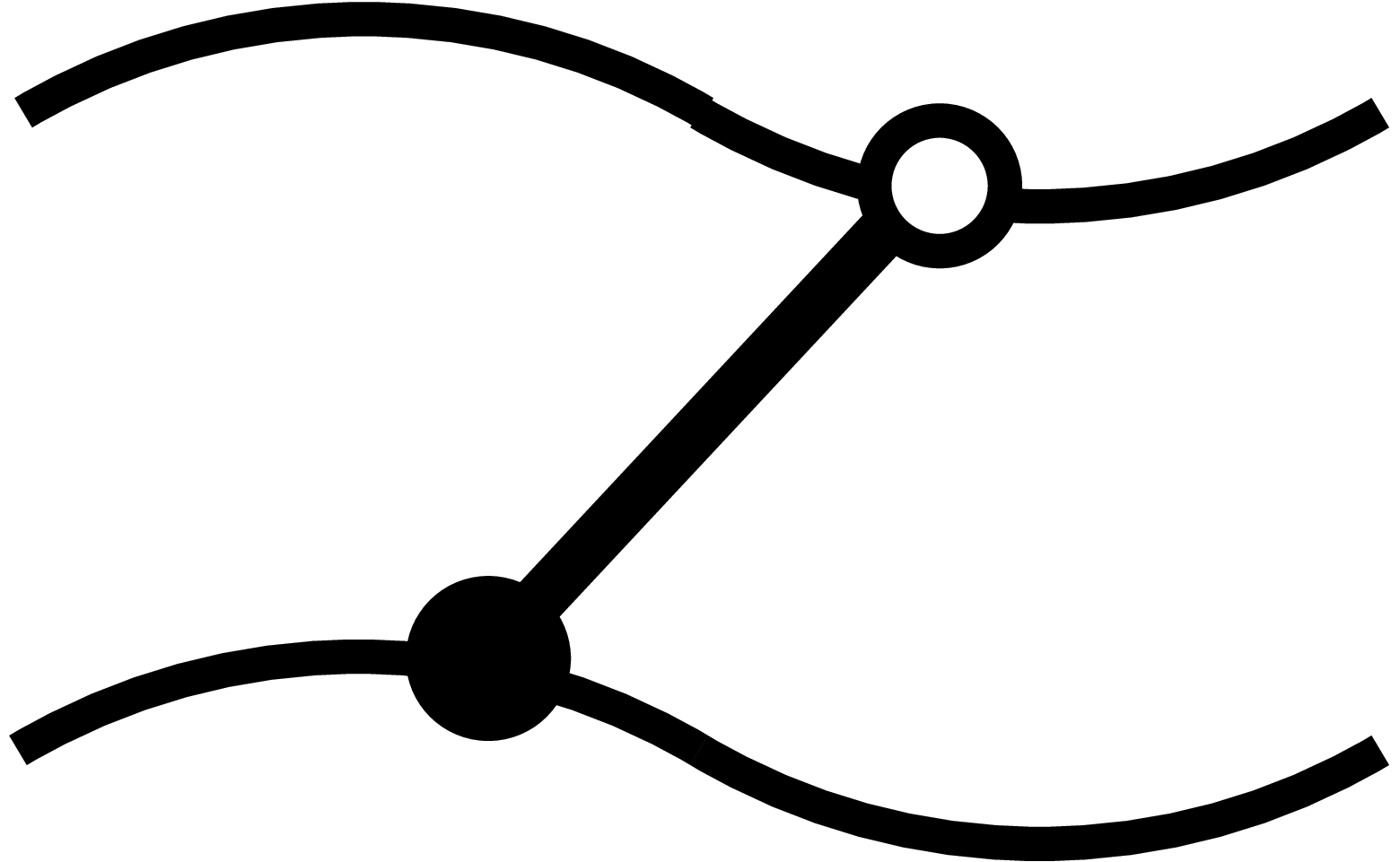}\nonumber\\
&&-\fig{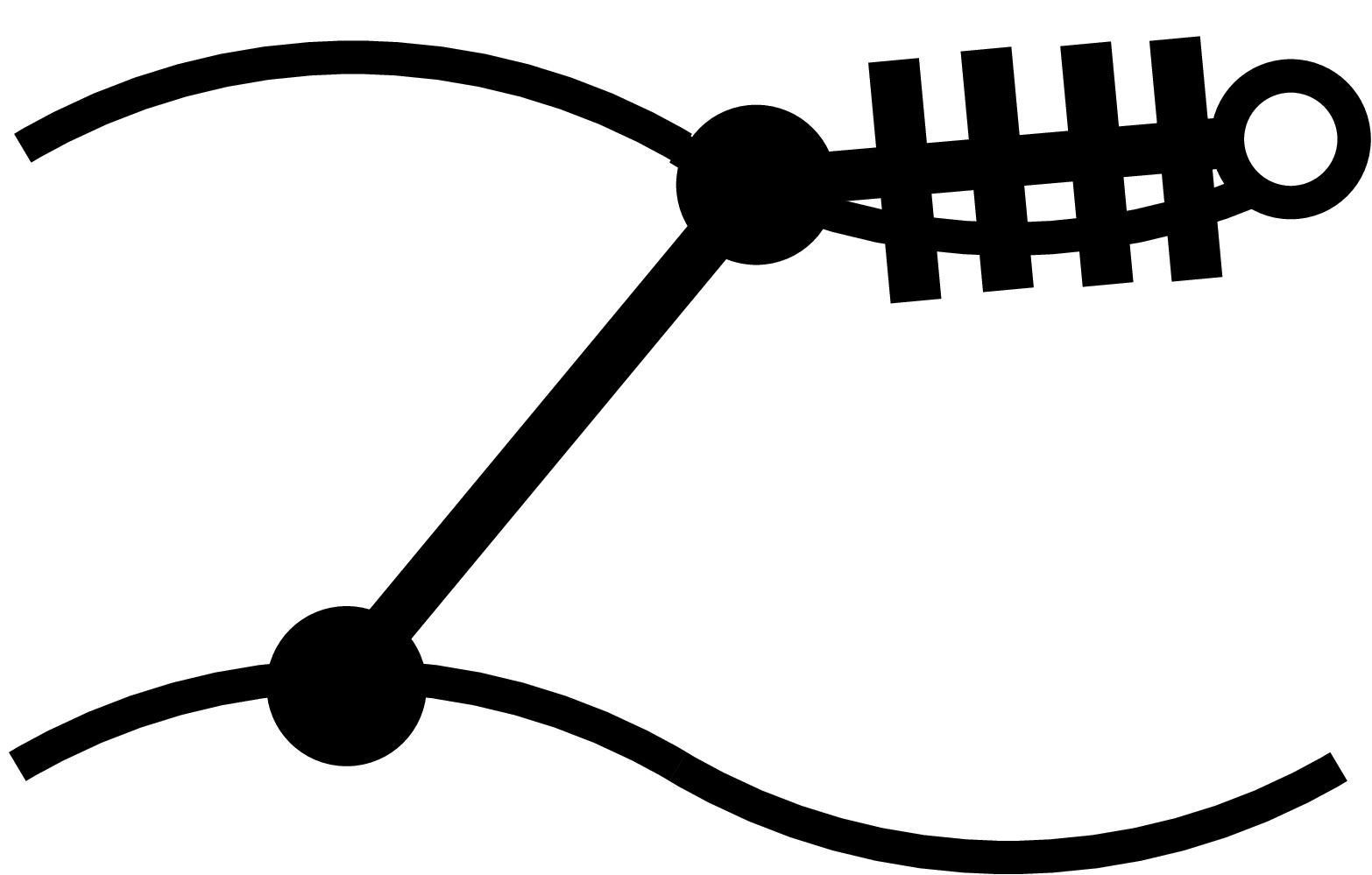}
+\frac{g}{\kappa-g}\fig{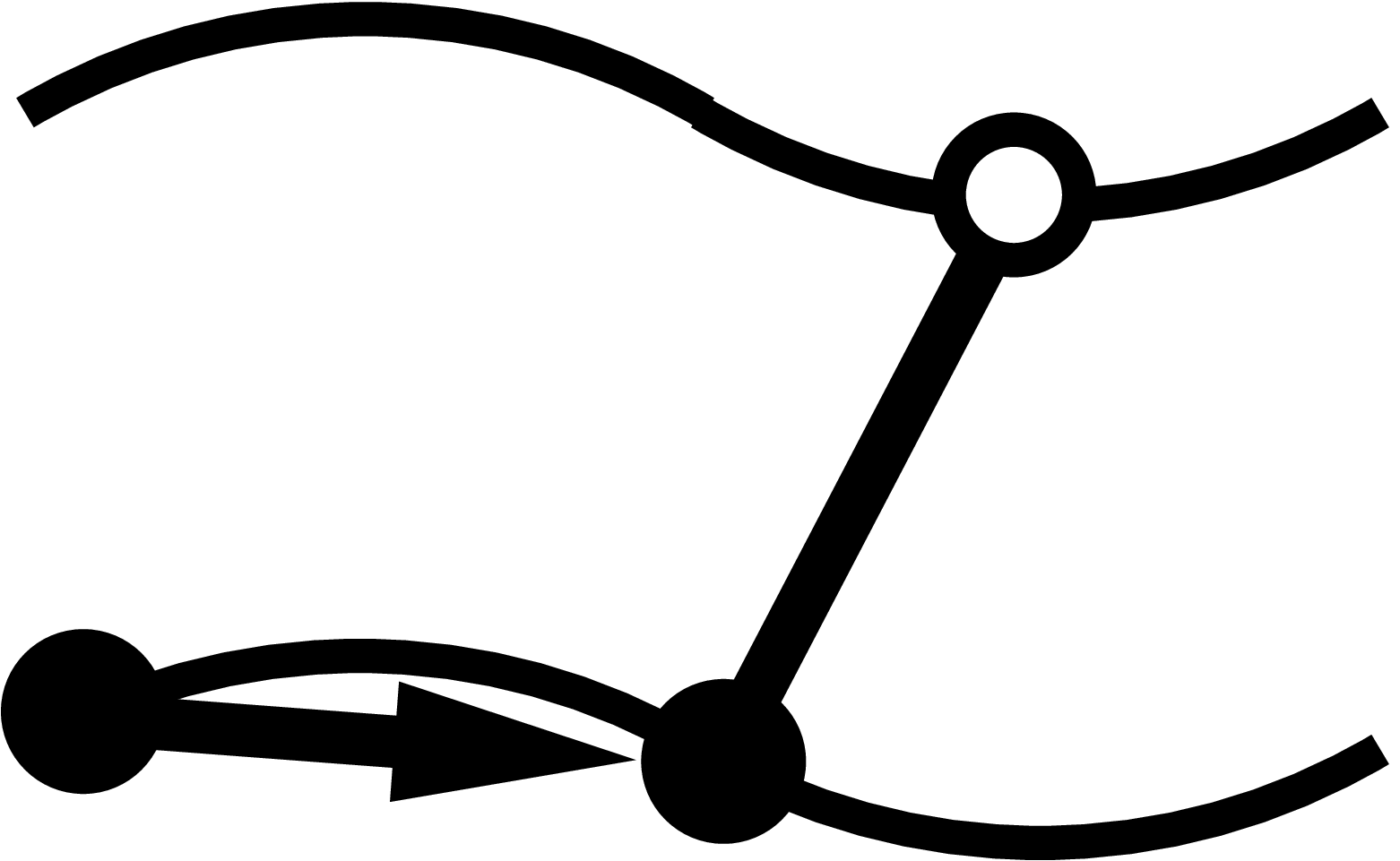}+\cdots\Bigg)\Bigg]\nonumber\\
&&-2\kappa m_0 \Bigg[\frac{g+\kappa}{g-\kappa}\ \Bigg(\fig{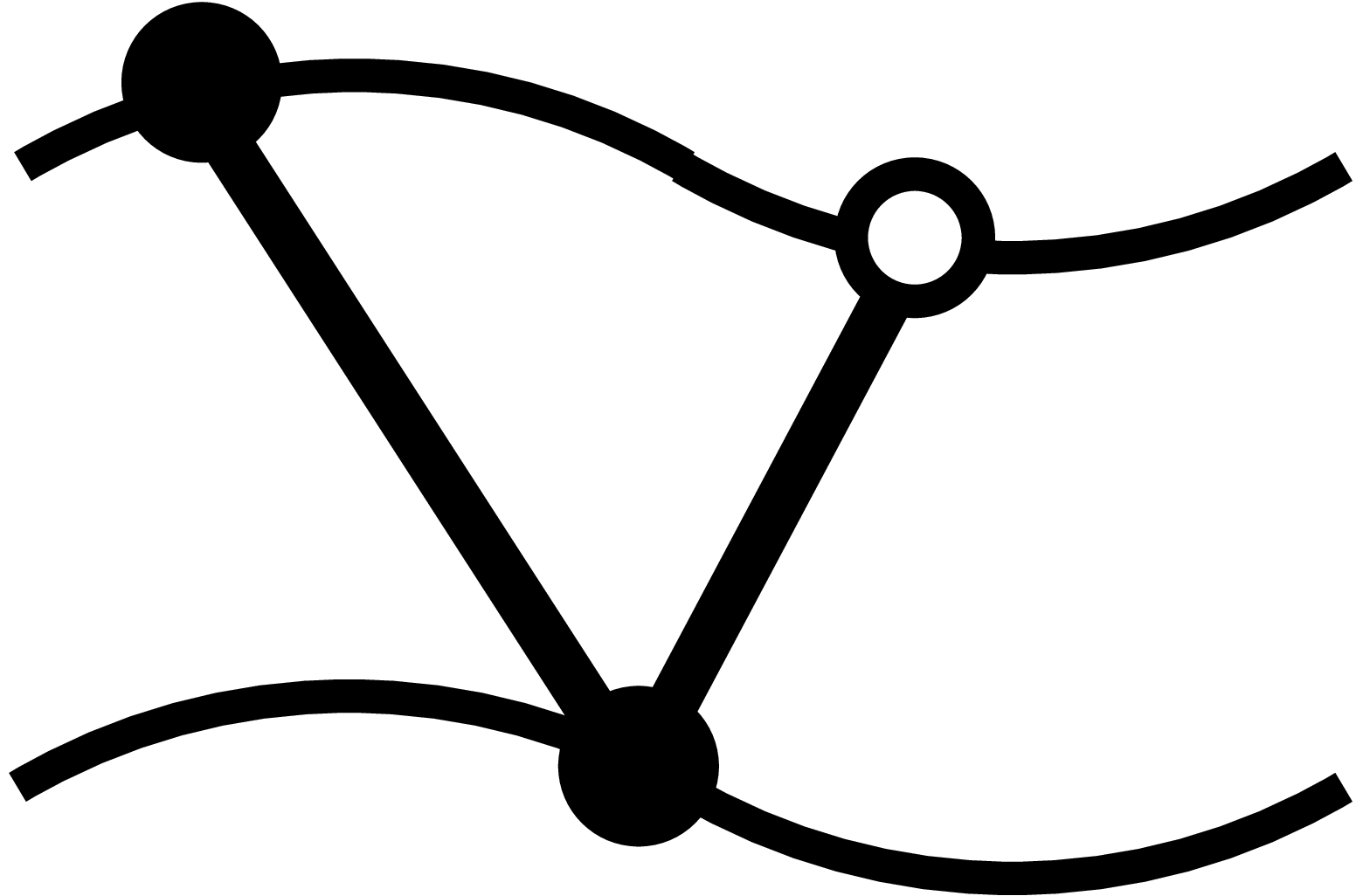}
+\cdots\Bigg)+\cdots\Bigg],
\label{deltaql}
\end{eqnarray}
where the symbols have the meanings as described above.
The diagrams in this expansion have segments on alternating interfaces connected via $K$ kernels, so they 
can be regarded as decorated versions of the zigzag diagrams of the original nonlocal model. 
The segments on the
substrate correspond to convolution products of $U-$ and $\partial_n K/\kappa-$type bonds on this surface, while
on the liquid-gas interface only $U-$type bonds are involved. The closed extreme, which by convention we place on
the left, provides the factor $-2\kappa m_0$ or $-2\kappa (h_1+gm_0)/g$, depending on whether it is located on 
the liquid-gas interface or on the substrate, respectively. On the other hand, the interface on which the
open extreme resides 
indicates whether the diagram contributes to $\Delta q_\psi$ (if it is on the substrate) or 
$\Delta q_\ell^-$ (otherwise). The factor the multiplies each diagram can be obtained as the product of terms
associated with each segment. The segments on the liquid-gas interface have a factor $(-1)^n$, where $n$ is
the number of bonds (of $U$ type) in the segment. The contribution of the segments on the substrate depend on
their positions. Let $n$ be the total number of bonds in the segment, and $m$ be the number
of $\partial_n K/\kappa$ bonds. If the segment contains the closed extreme, its contribution is given by
\begin{eqnarray}
-\left(\frac{g}{\kappa-g}\right)^{n+1}\left(\frac{\kappa}{g}\right)^{m-m_0}\left[1-(1-\delta_{m,0})
\left(1-\frac{\kappa}{g}\right)^{l+1}\right],
\label{factor2}
\end{eqnarray}
where $\delta_{i,j}$ is the Kronecker symbol, the index $m_0$ is either $0$ or $1$ (depending on the first bond 
being of $U$ type or $\partial_n K/\kappa$ type, respectively),
and $l$ is the number of $U-$type bonds after the last $\partial K/\kappa-$type bond. Note that this expression is
the factor that multiplies the diagrams in the expansion (\ref{deltamdiagram}) for the order-parameter profile 
above the substrate, multiplied by $g/(g-\kappa)$. The contribution of a segment on the substrate that contains
the open extreme is
\begin{equation}
-\left(\frac{g}{\kappa-g}\right)^{n+1}\left(\frac{\kappa}{g}\right)^{m}.
\label{factor3}
\end{equation}
Finally, any other segment on the substrate will provide the following factor: either
\begin{equation}
-\left(\frac{g}{\kappa-g}\right)^{n+1}\left(\frac{\kappa}{g}\right)^{m}\left[2-
\left(1-\frac{\kappa}{g}\right)^{l+1}\right],
\label{factor4}
\end{equation}
if the last bond is of $U-$type or 
\begin{eqnarray}
-\left(\frac{g}{\kappa-g}\right)^{n+1}\left(\frac{\kappa}{g}\right)^{m}\Bigg[2-\frac{g}{\kappa}+\frac{\kappa}{g}
\nonumber\\
-\left(1-\frac{g}{\kappa}\right)
\left(1-\frac{\kappa}{g}\right)^{l+1}\Bigg],
\nonumber\\
\label{factor5}
\end{eqnarray}
with $l$ being the number of consecutive $U-$type bonds in the rightmost sequence in the
segment.

\subsection{Binding potential functional and order parameter}
With these preliminaries behind us, we are now in a position to obtain the diagrammatic representation of the binding potential and the order-parameter profile.
The binding potential functional $W[\ell,\psi]$ is defined as the substrate-interface interaction in the excess free energy:
\begin{equation}
\mathcal{F}_{wg}=\mathcal{F}_{wl}[\psi]+H[\ell]+W[\ell,\psi],
\label{bindingpotential}
\end{equation}
where $\mathcal{F}_{wl}[\psi]$ is the free energy of the wall-liquid 
interface and $H[\ell]$ is the free liquid-gas interfacial Hamiltonian, which we have already
determined. So, in terms of $\Delta q_\ell^-$ and $\Delta q_\psi$, we have 
\begin{eqnarray}
W[\ell,\psi]=-\frac{m_0}{2}\int_{\ell} d\mathbf{s} \Delta q_\ell^{-}(\mathbf{s})
+\frac{h_1+gm_0}{2g}\int_\psi d\mathbf{s}\ \Delta q_\psi(\mathbf{s}).
\nonumber\\ 
\end{eqnarray}
Substituting the expansions (\ref{deltaqpsi}) and (\ref{deltaql}) into this expression, we arrive at the 
diagrammatic expansion for the binding potential functional
\begin{eqnarray}
W[\ell,\psi]=\sum_{n=1}^\infty \Bigg(-\kappa m_0\frac{h_1+gm_0}{g} \Omega_n^n+ \kappa m_0^2\Omega_n^{n+1}
\nonumber\\
+\kappa \left(\frac{h_1+gm_0}{g}\right)^2
\Omega_{n+1}^n\Bigg),
\label{Hmin6}
\end{eqnarray}
where $\Omega_i^j$ is the sum of all the independent diagrams that have $i$ segments on the substrate and $j$ 
segments on the liquid-gas interface. Note that these diagrams correspond to those obtained previously
for $\Delta q_\psi$ and $\Delta q_\ell^-$, but integrating over the positions of $\mathbf{s}$, i.e., with 
a closed right extreme. For the first terms we have
\begin{eqnarray}
&&\Omega_1^1=
\frac{g}{g-\kappa}\Bigg(2\fig{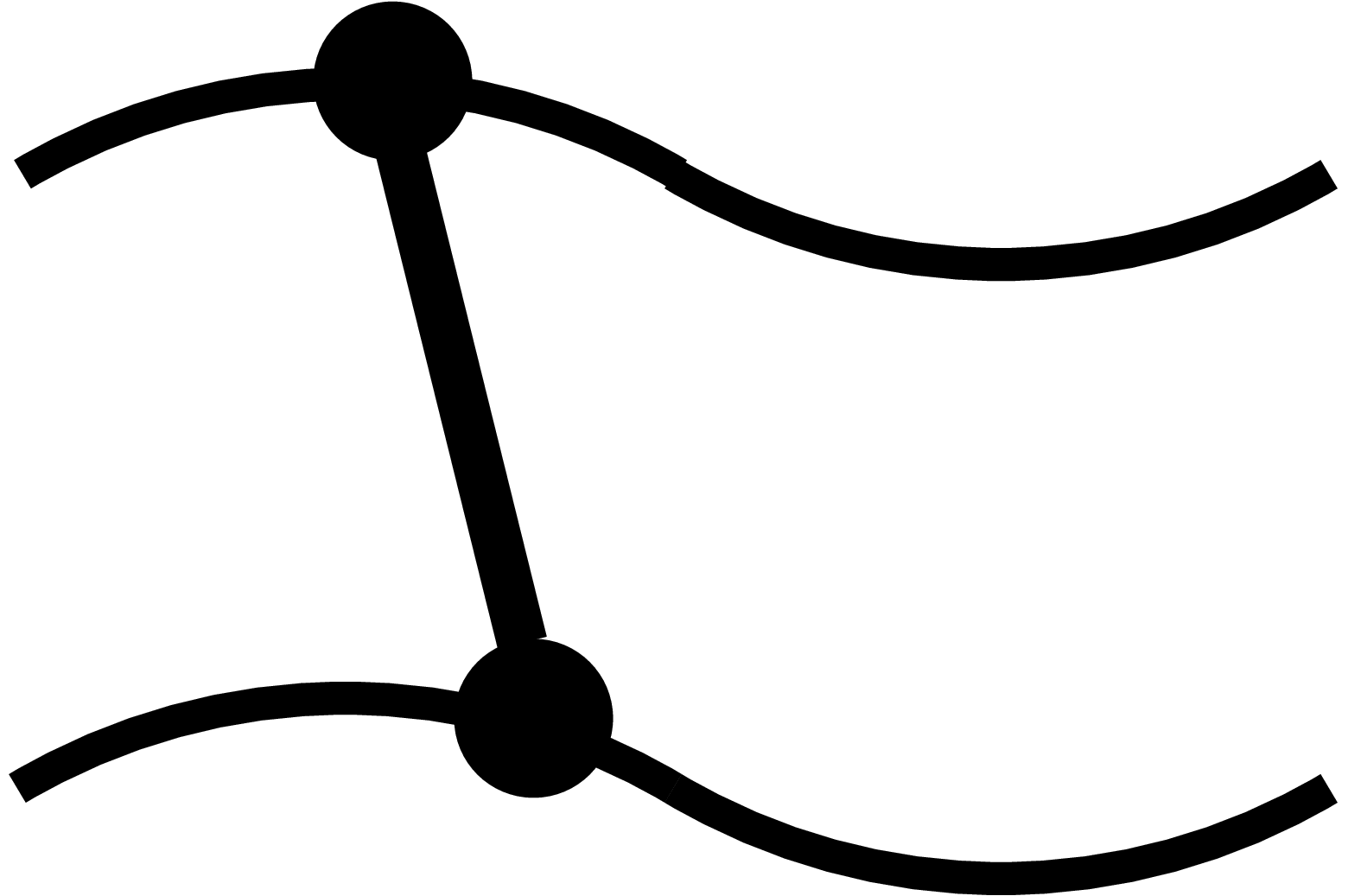}-2\fig{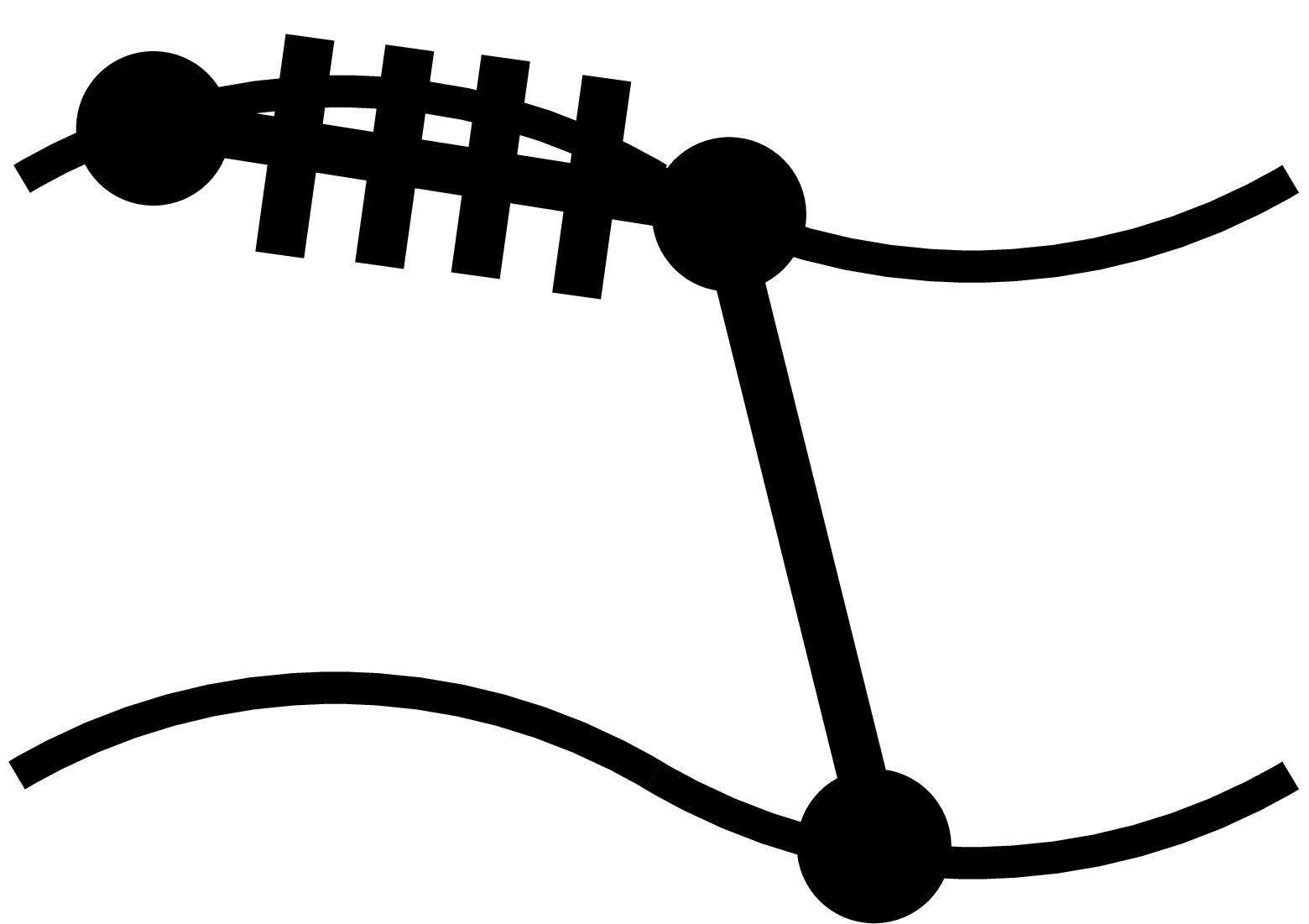}
\nonumber\\
&&+\frac{\kappa}{\kappa-g}\fig{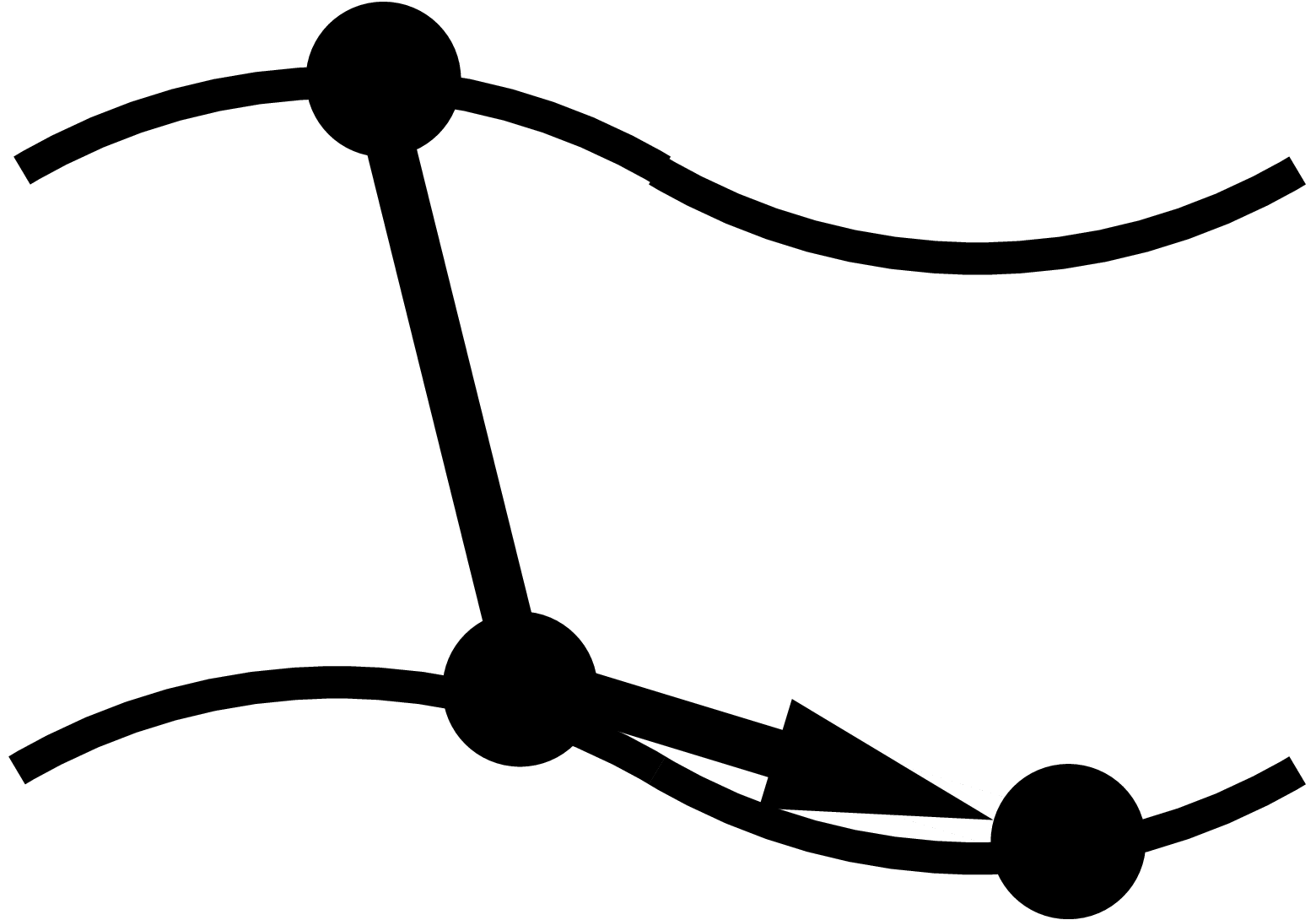}+\frac{g}{\kappa-g}\fig{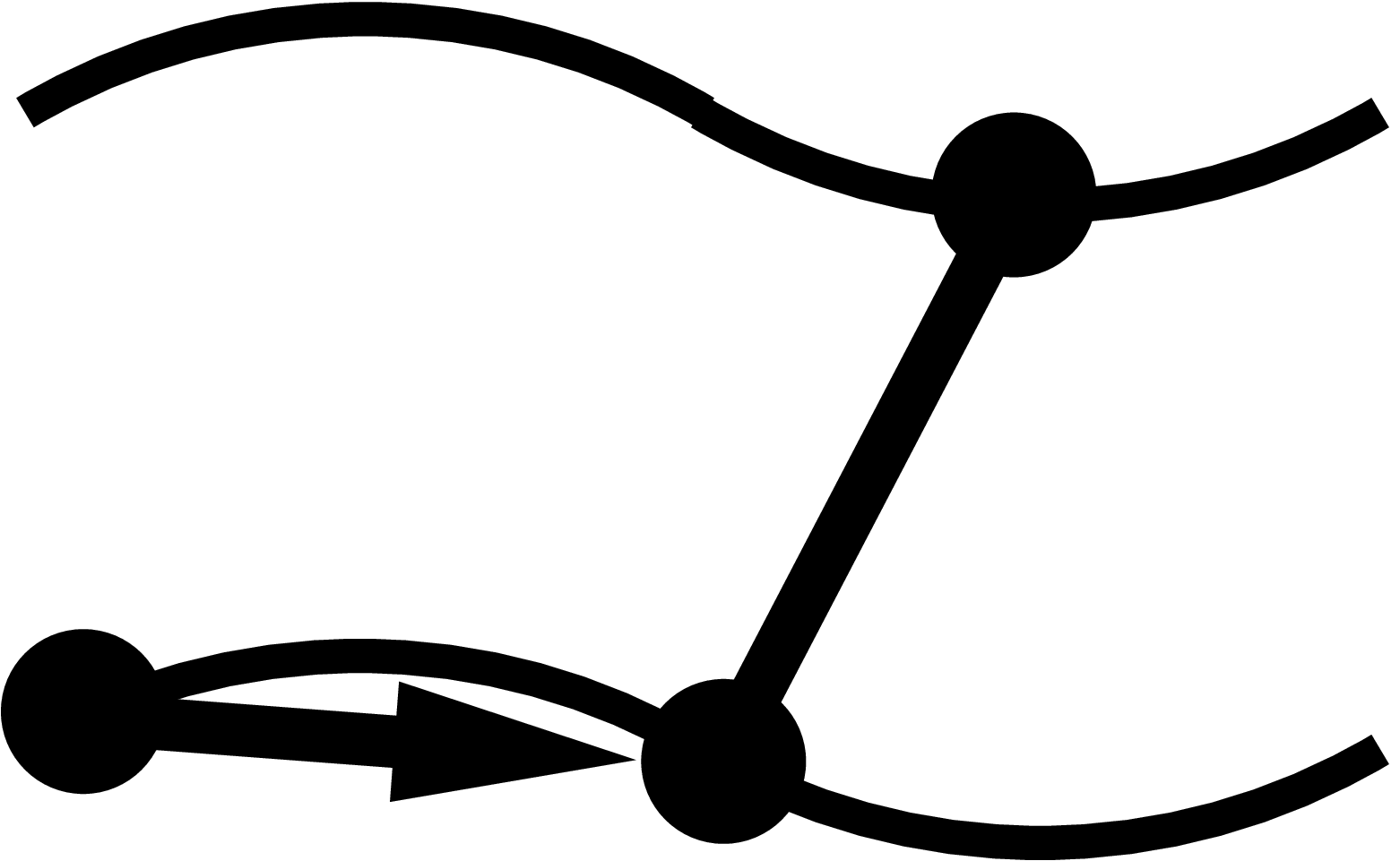}\nonumber\\
&&-\frac{\kappa}{\kappa-g}\fig{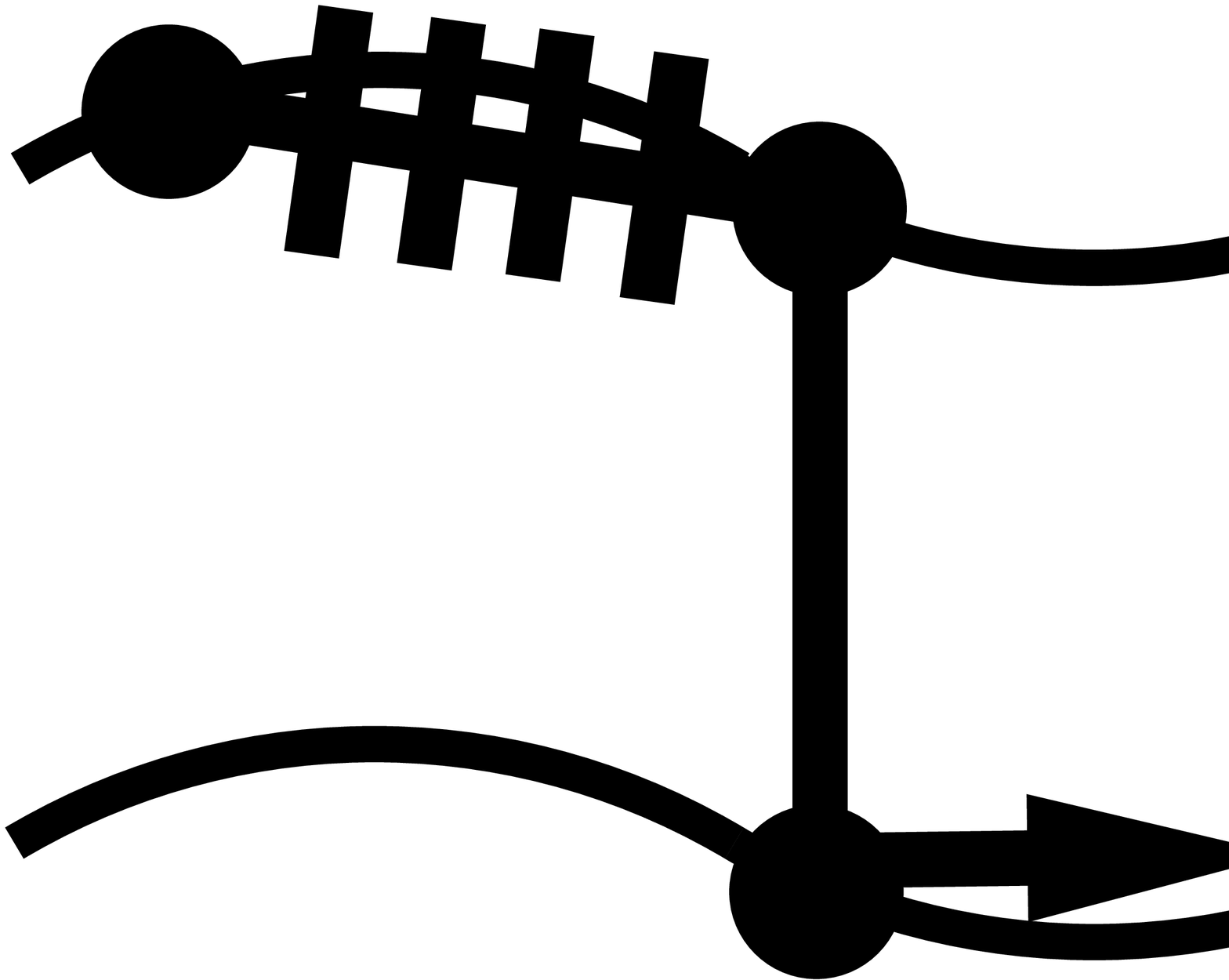}+\cdots\Bigg)\label{omega11}
\end{eqnarray}
while
\begin{eqnarray}
&&\Omega_1^2=
\frac{g}{g-\kappa}\Bigg(\frac{\kappa+g}{g}\fig{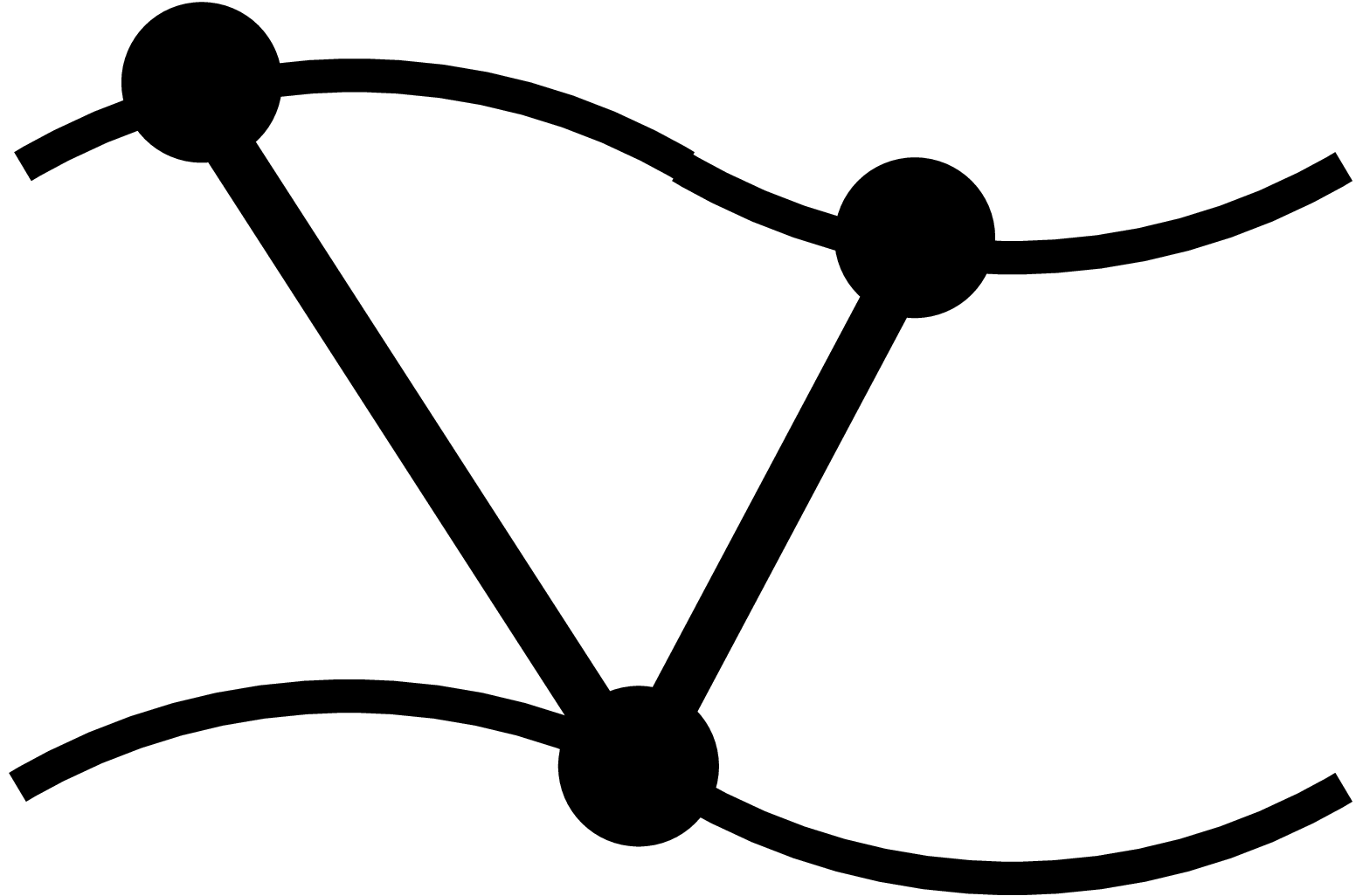}\nonumber\\
&&-2\frac{\kappa+g}{g}\fig{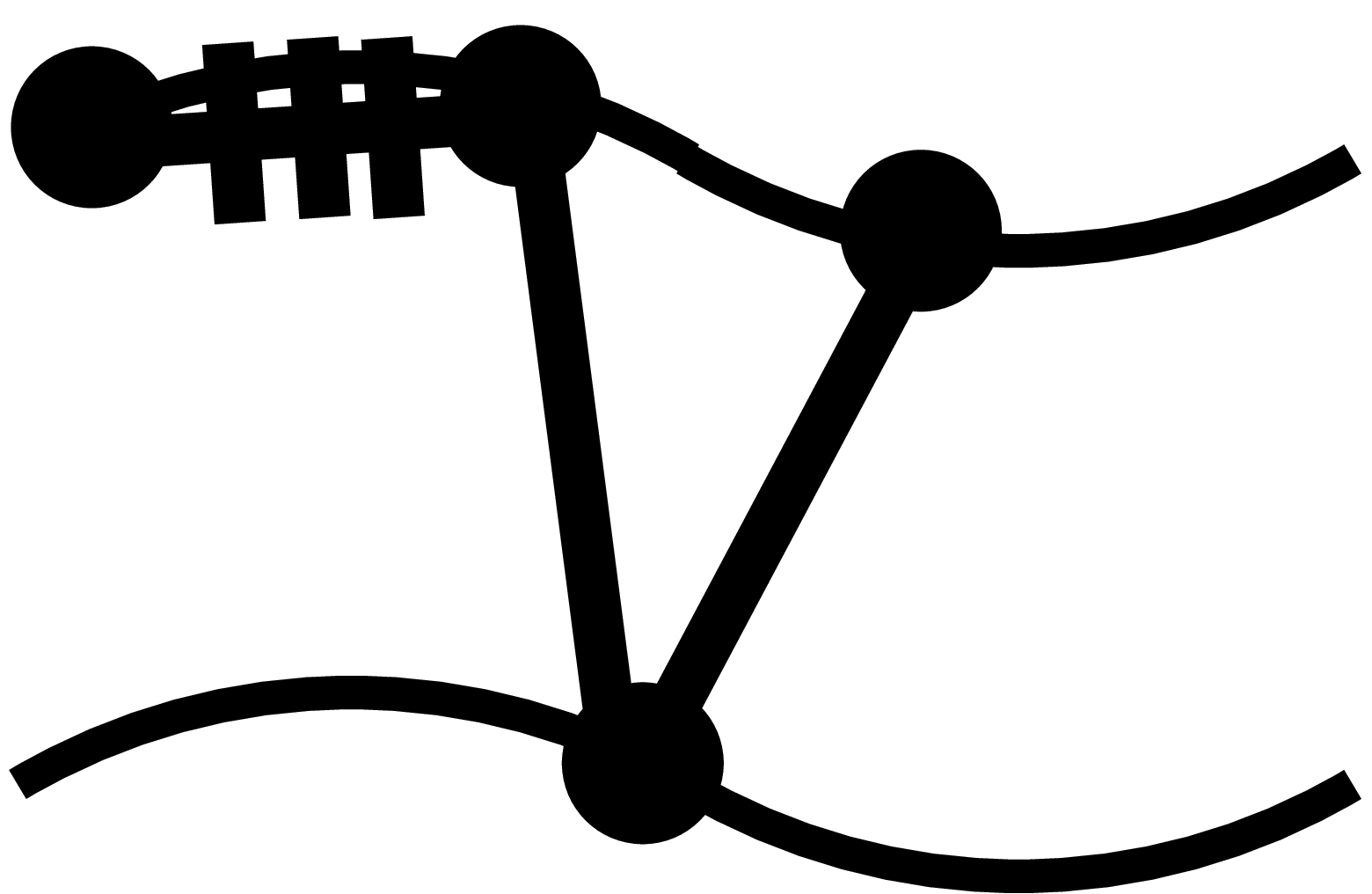}+\frac{\kappa+g}{g}\fig{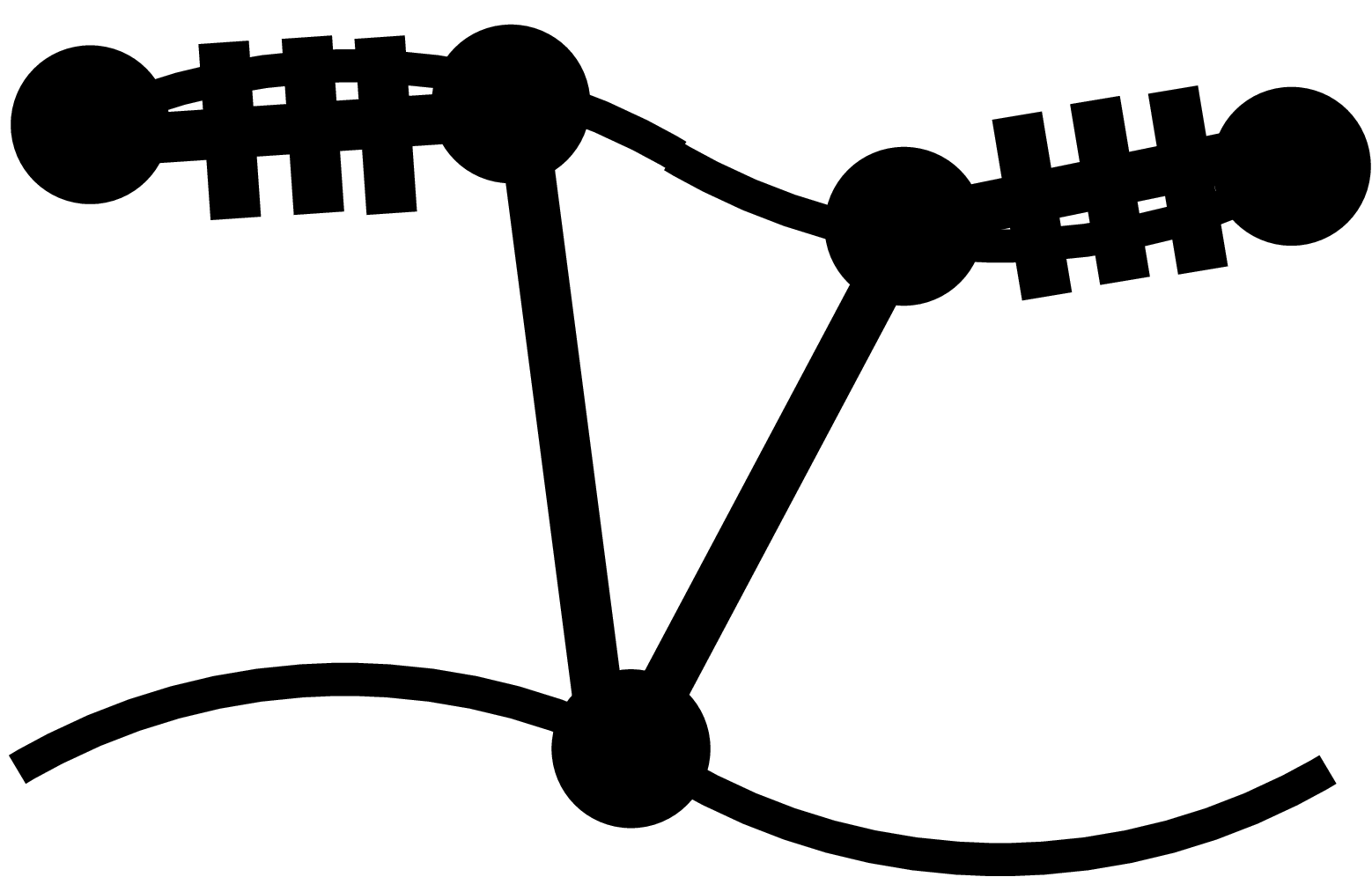}\nonumber\\
&&+\frac{2\kappa}{g}\fig{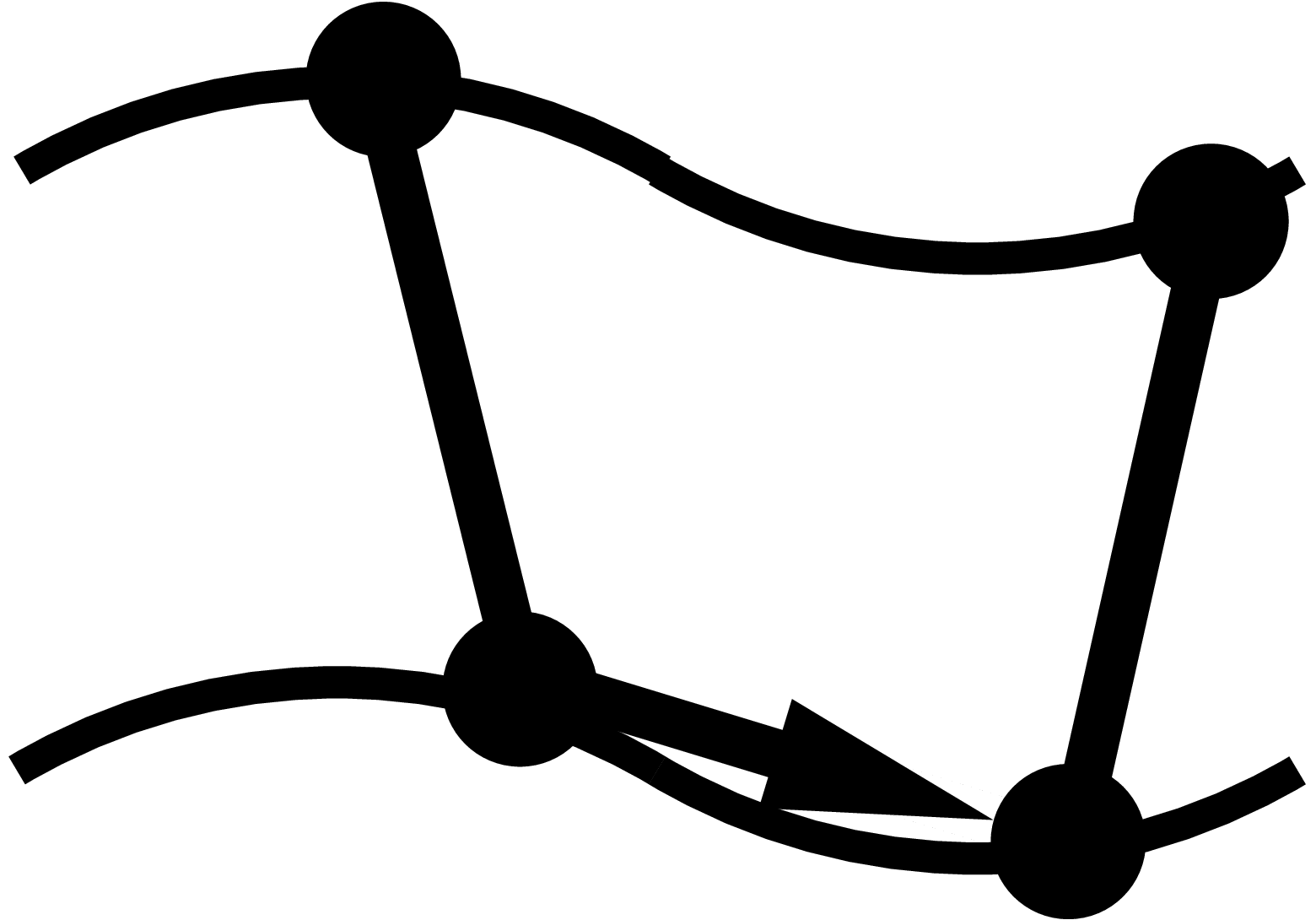}+\cdots\Bigg)
\label{omega12}
\end{eqnarray}
and
\begin{eqnarray}
&&\Omega_2^1=
\left(\frac{g}{g-\kappa}\right)^2\Bigg(\fig{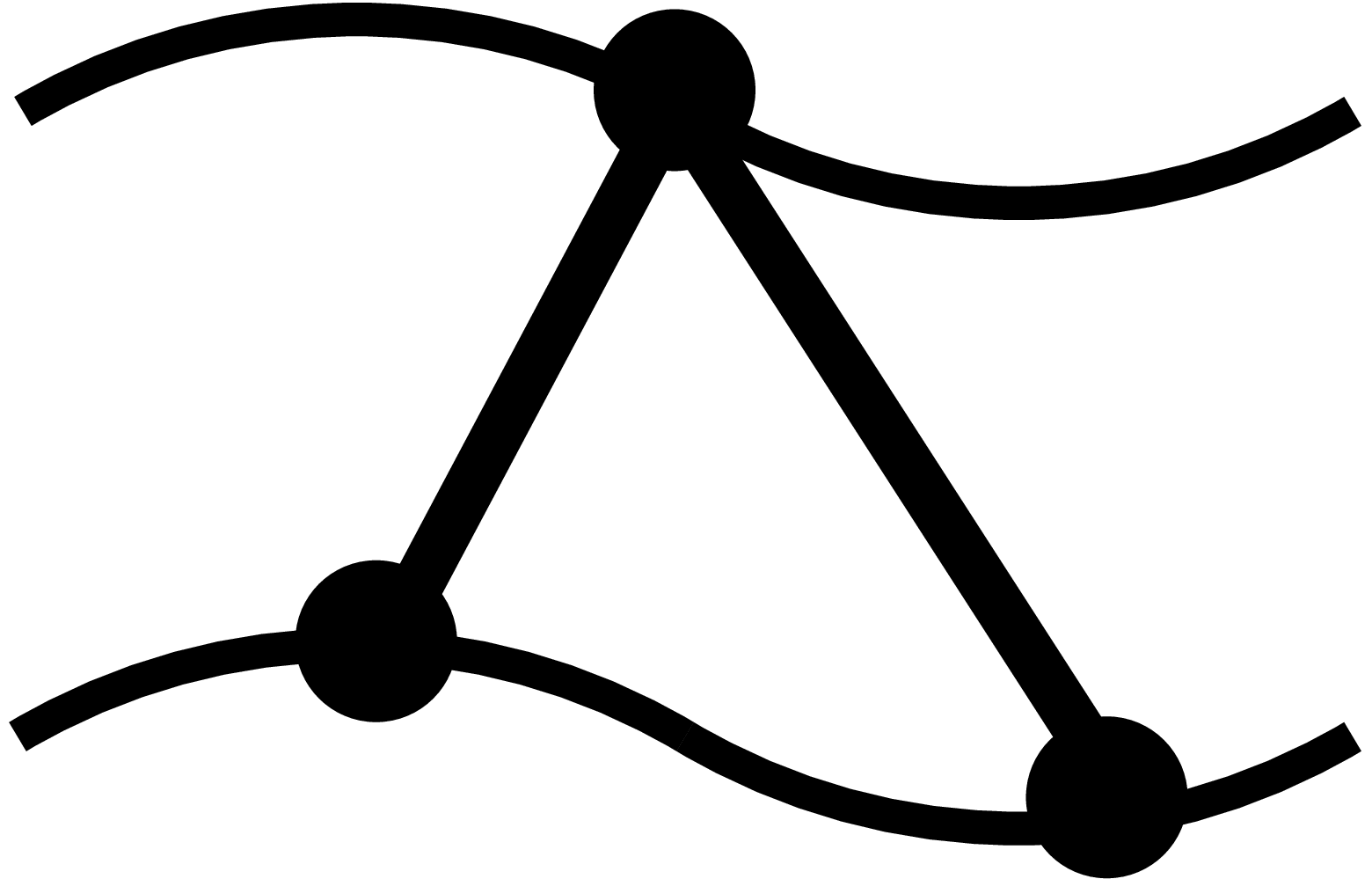}\nonumber\\
&&+\frac{g}{\kappa-g}\fig{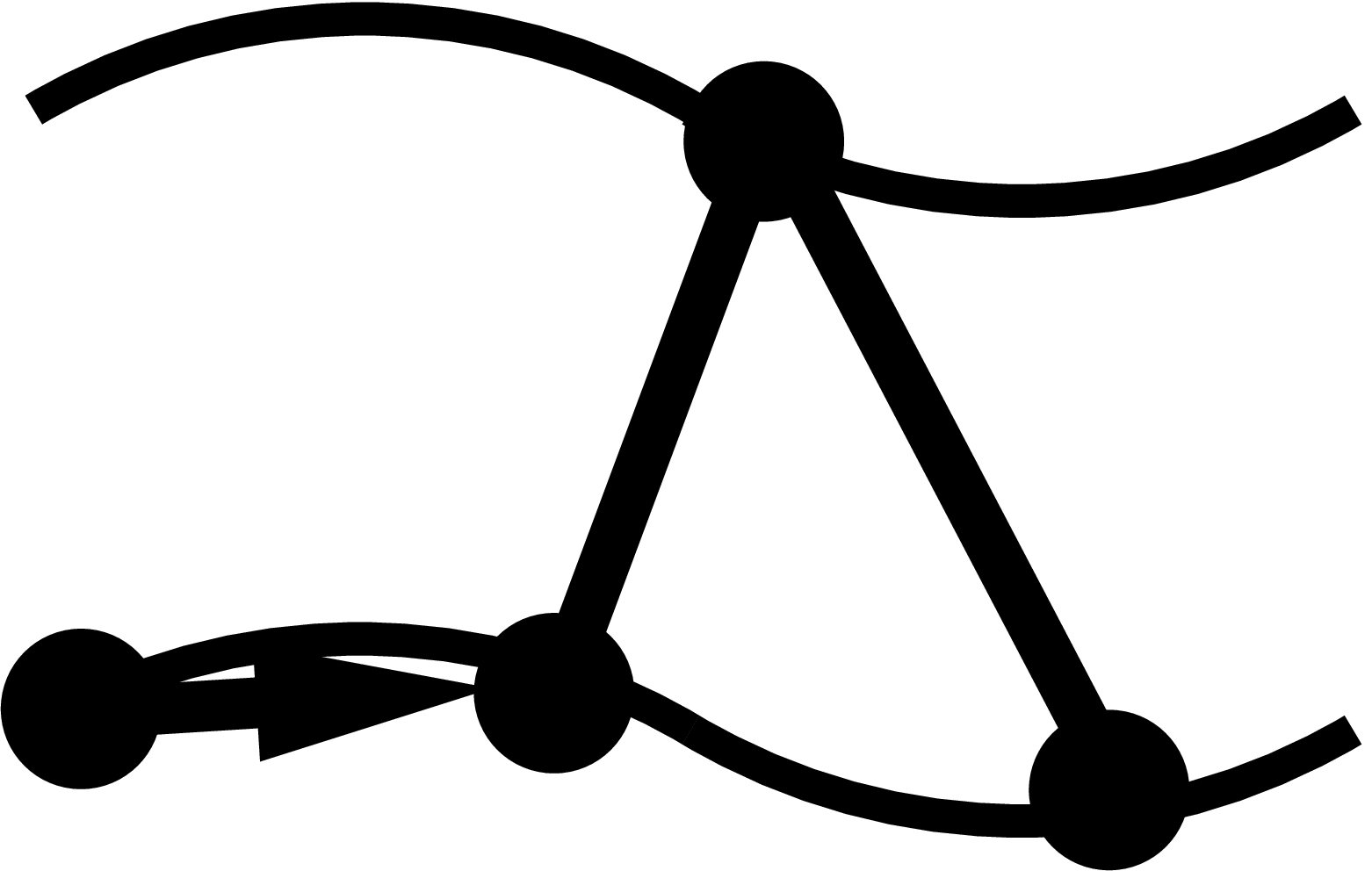}-\fig{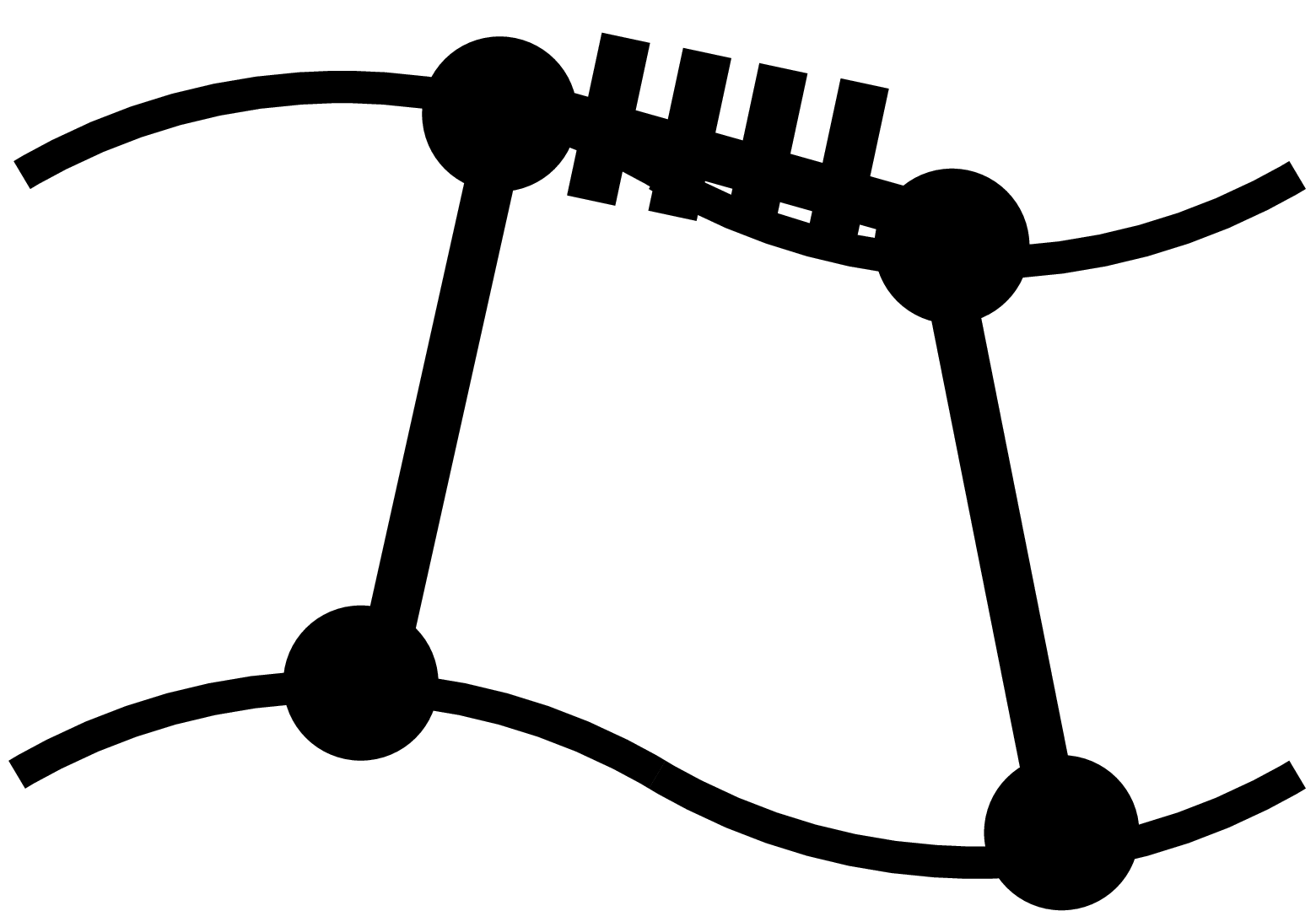}+\cdots\Bigg).
\label{omega21}
\end{eqnarray}
These diagrams are all decorated versions of the diagrams in the original nonlocal model.
They are multiplied by a factor that is the same as the corresponding coefficient for the associated 
$\Delta q$ diagram with open right extreme, provided the diagram either contains $\partial_n K/\kappa-$type 
bonds or
is symmetric under a mirror reflection, i.e., it is the same when read from left to right or the reverse. 
Otherwise, the factor is twice the coefficient for the associated $\Delta q$ diagram. The reason for this is
that two different $\Delta q$ diagrams lead to the same contribution to $W[\ell,\psi]$.
In this sense we mean that only independent diagrams are taken into account in the diagrammatic expansion
of $W[\ell,\psi]$, because we discard one of the two equivalent diagrams, which are related via a mirror reflection.

Now we turn to the order-parameter profile.
Above the liquid-gas interface the profile is uninfluenced by the presence of the substrate, so
it has a diagrammatic expansion given by Eq.~(\ref{deltamdiagram2}).
On the other hand, the order-parameter profile within the adsorbed liquid layer
is influenced by the proximity of both the wall and the liquid-gas interface and has the representation
\begin{eqnarray}
\delta m_\Xi(\mathbf{r})&=&
-\frac{1}{2\kappa}\int_{\psi} d\mathbf{s}
K(\mathbf{s},\mathbf{r}) q_\psi(\mathbf{s})
\nonumber\\
&-&\frac{1}{2\kappa}\int_{\psi} d\mathbf{s} \left(\frac{h_1}{g}+m_b+\frac{q_\psi(\mathbf{s})}{g}\right)
\partial_{n} K(\mathbf{s},\mathbf{r})
\nonumber\\
&+&\frac{1}{2\kappa}\int_\ell d\mathbf{s}
K(\mathbf{s},\mathbf{r}) q_\ell^-(\mathbf{s})
\nonumber\\
&+&\frac{m_0}{2\kappa}\int_\ell d\mathbf{s}\partial_{n} K(\mathbf{s},\mathbf{r}).
\label{profile3}
\end{eqnarray}
Now, writing $q_\psi=q_\psi^0+\Delta q_\psi$ and $q_\ell^-=q_\ell^{0,-}+\Delta q_\ell^-$, and
making use of Eqs.~(\ref{profile}) and (\ref{profile2}), 
we obtain the expression for the order parameter in the
liquid layer
\begin{eqnarray}
\delta m_\Xi(\mathbf{r})&=& \delta m_\Xi^{0,\psi}(\mathbf{r})+\delta m_\Xi^{0,\ell}(\mathbf{r})
\nonumber\\
&+&\frac{1}{2\kappa}\int_\ell d\mathbf{s}
K(\mathbf{s},\mathbf{r}) \Delta q_\ell^-(\mathbf{s})
\label{profile4}
\\
&-&\frac{1}{2\kappa}\int_{\psi} d\mathbf{s}
\left(K(\mathbf{s},\mathbf{r})+\frac{1}{g}\partial_{n} K(\mathbf{s},\mathbf{r})\right) \Delta q_\psi(\mathbf{s})
,
\nonumber
\end{eqnarray}
where the kernel connecting the substrate to the position $\mathbf{r}$ in the last term can be related to 
$K$ using Eq.~(\ref{greenidentity8}). Taking into account the expansions (\ref{deltaqpsi}) and 
(\ref{deltaql}), we find that the order-parameter profile in the liquid layer has the expansion
\begin{eqnarray}
&&\delta m_\Xi=-m_0\Bigg[\fig{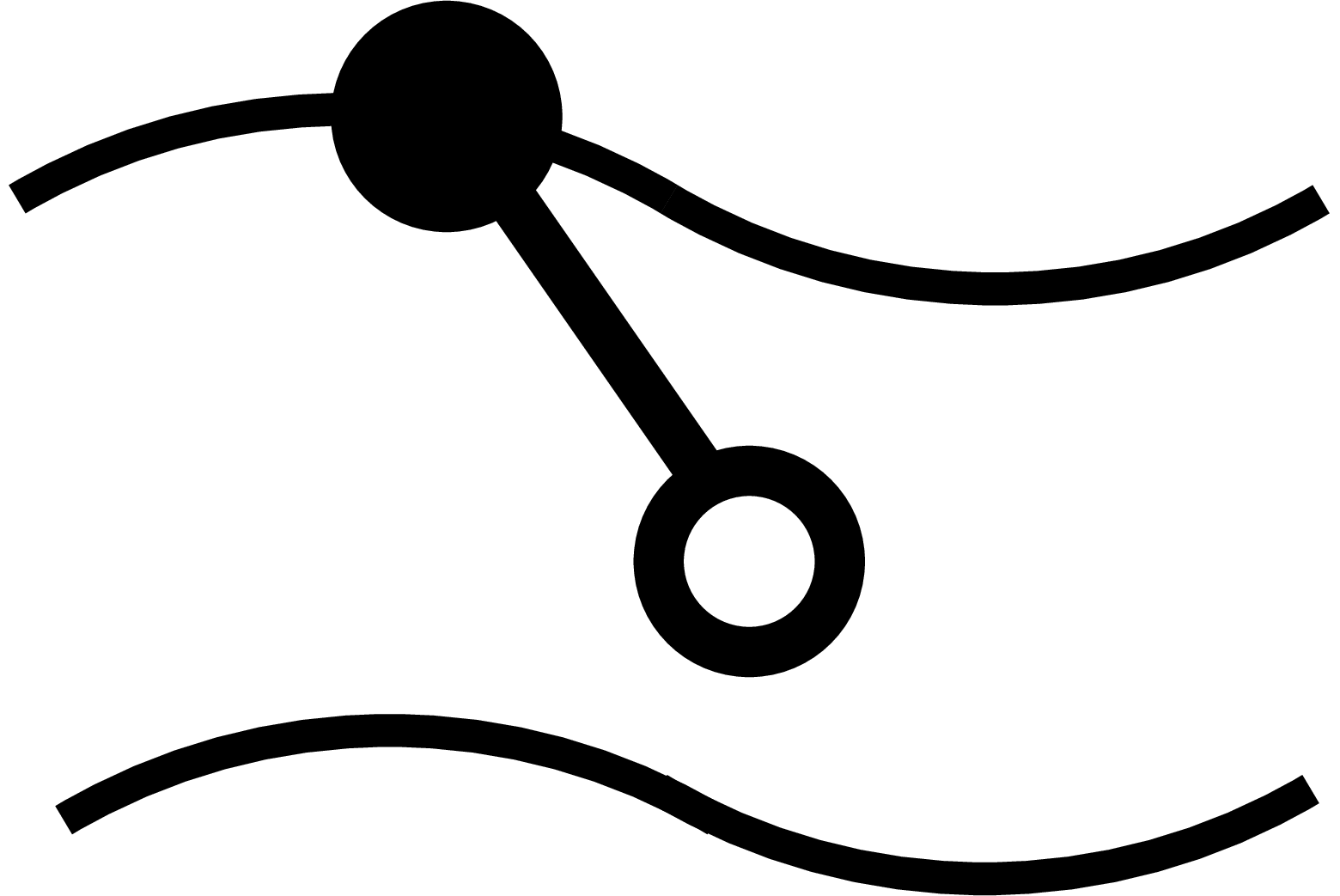}-\fig{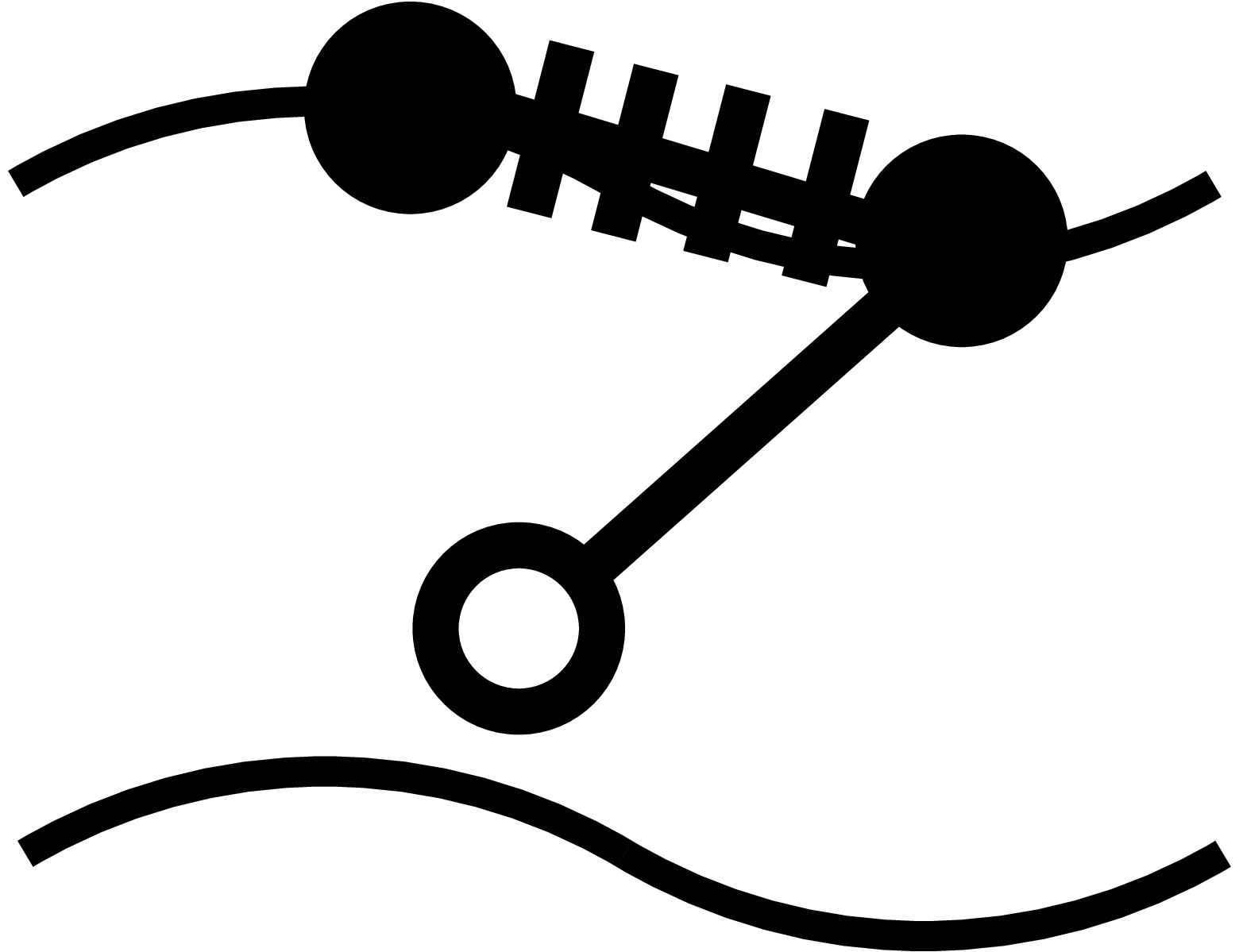}+\cdots\nonumber\\
&&-\frac{\kappa+g}{g-\kappa}\fig{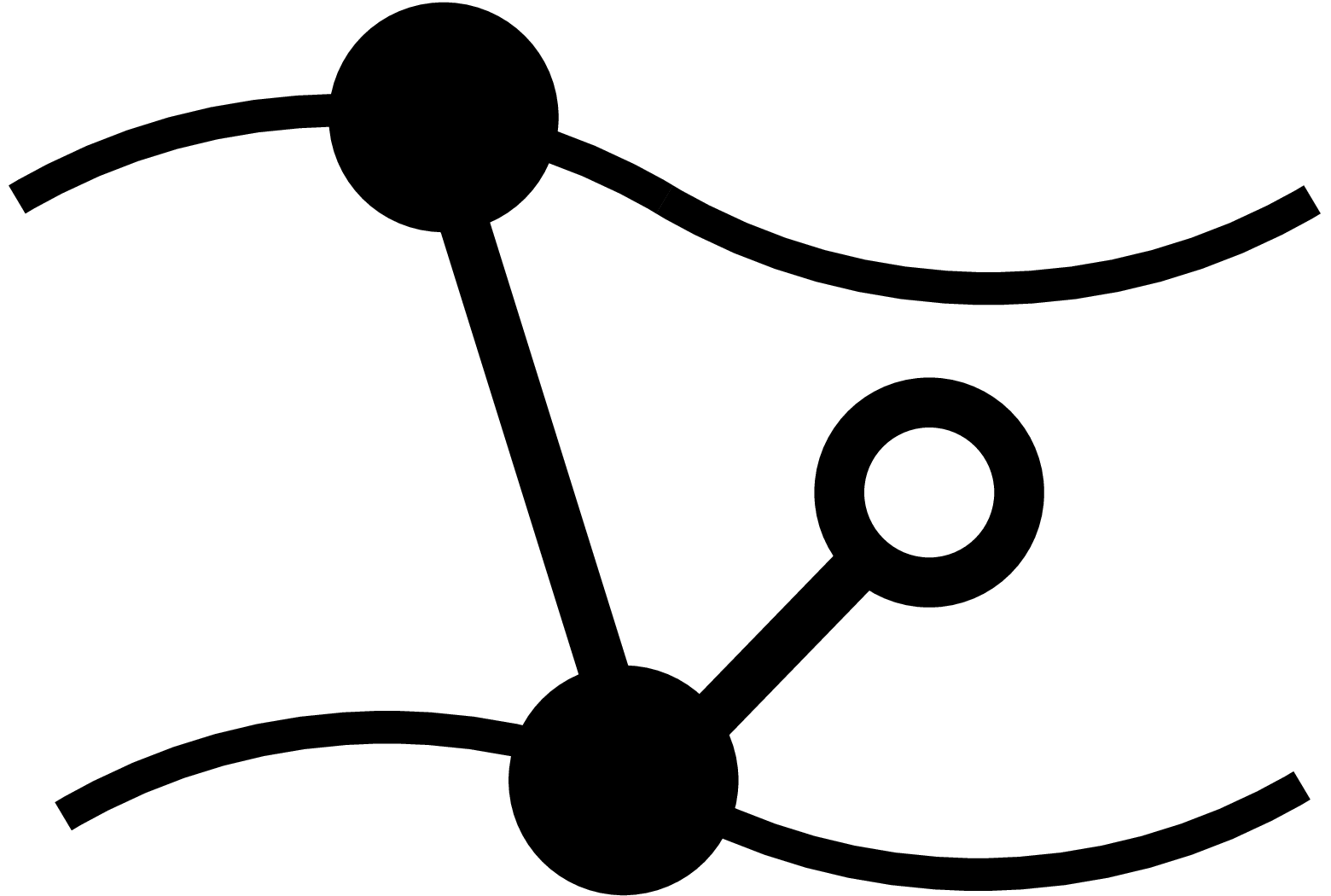}+\frac{\kappa+g}{g-\kappa}\fig{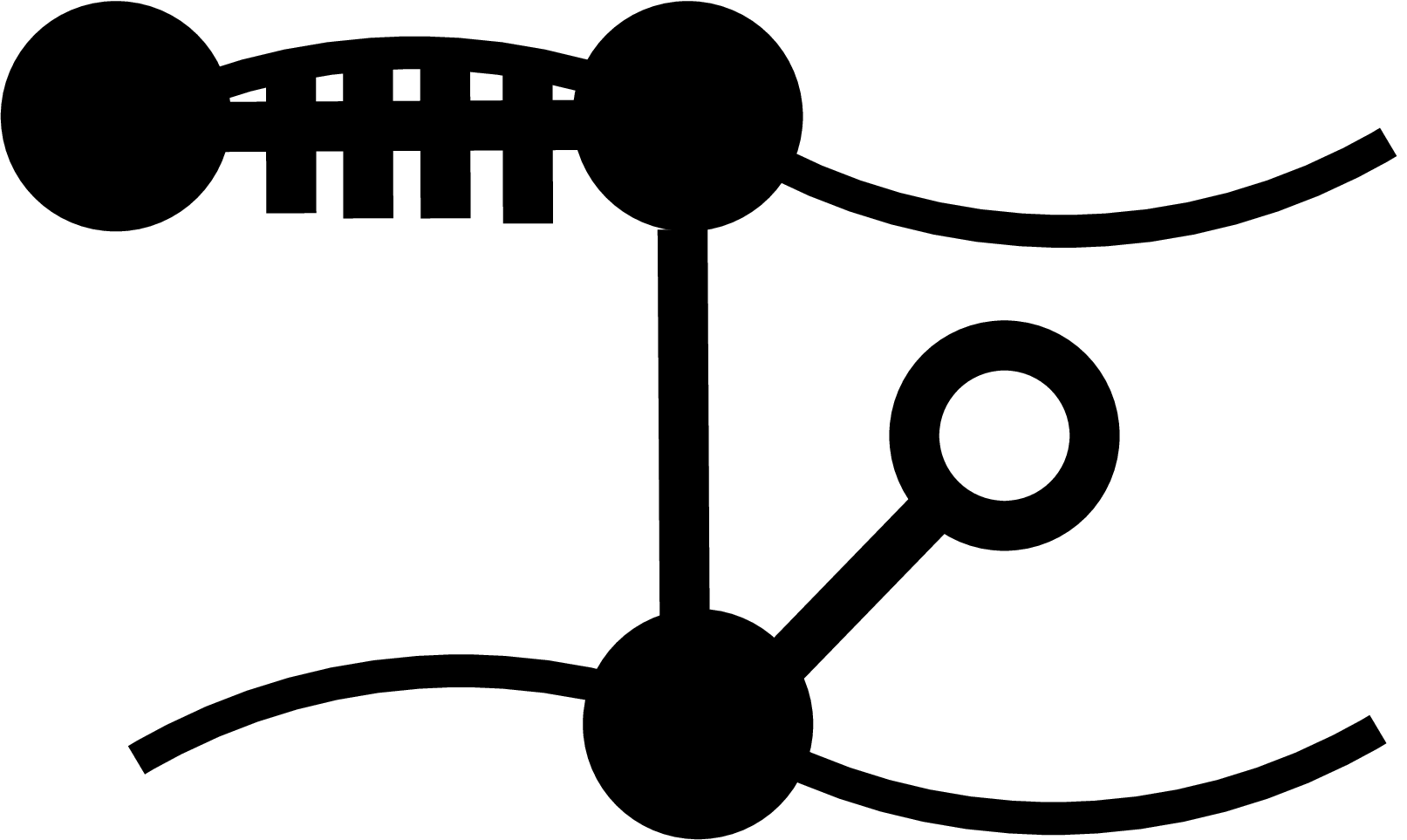}+\cdots\nonumber\\
&&+\frac{\kappa+g}{g-\kappa}\fig{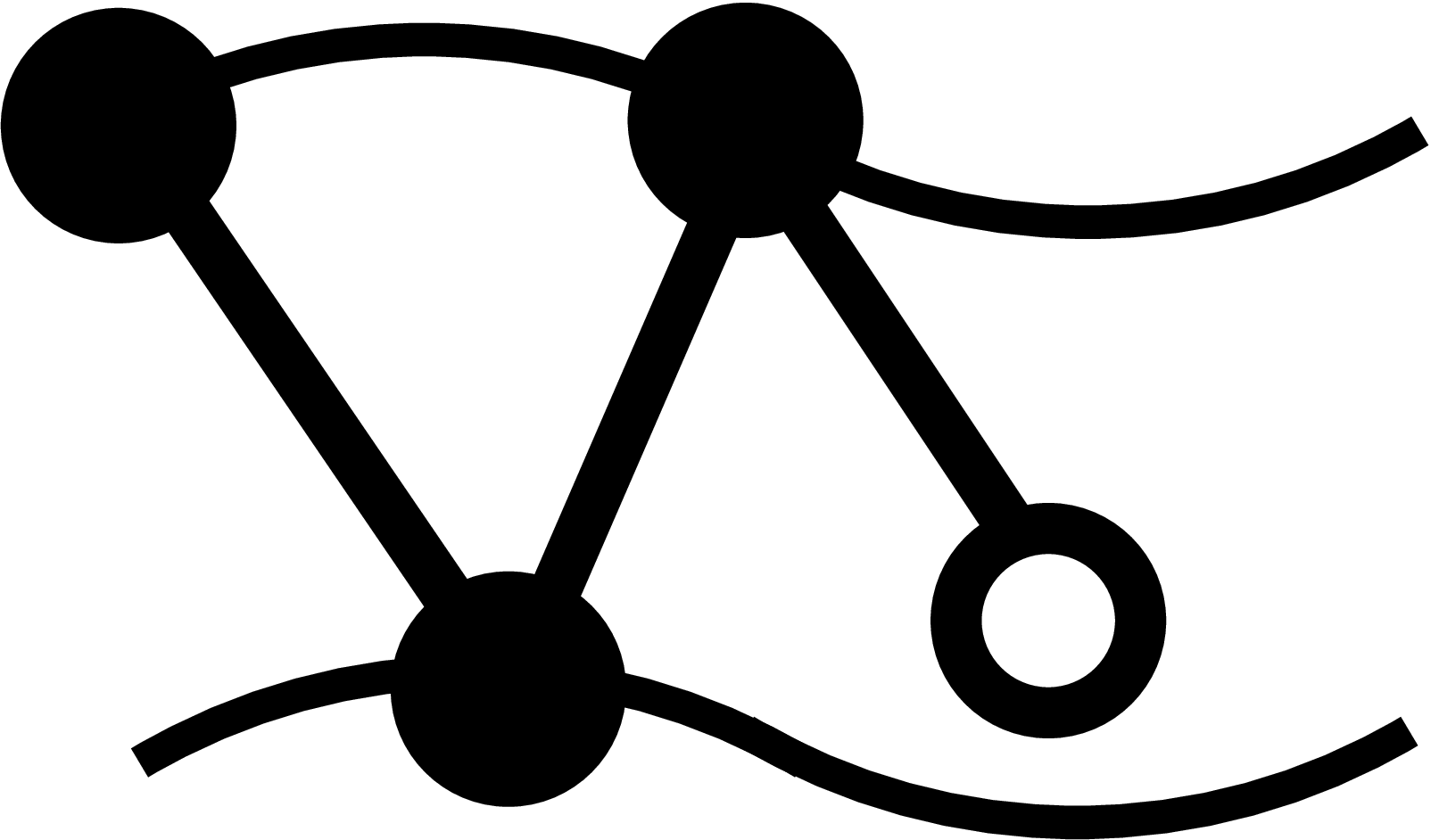}+\cdots\Bigg]\nonumber\\
&&-\frac{h_1+gm_0}{g}\Bigg[\frac{g}{g-\kappa}\fig{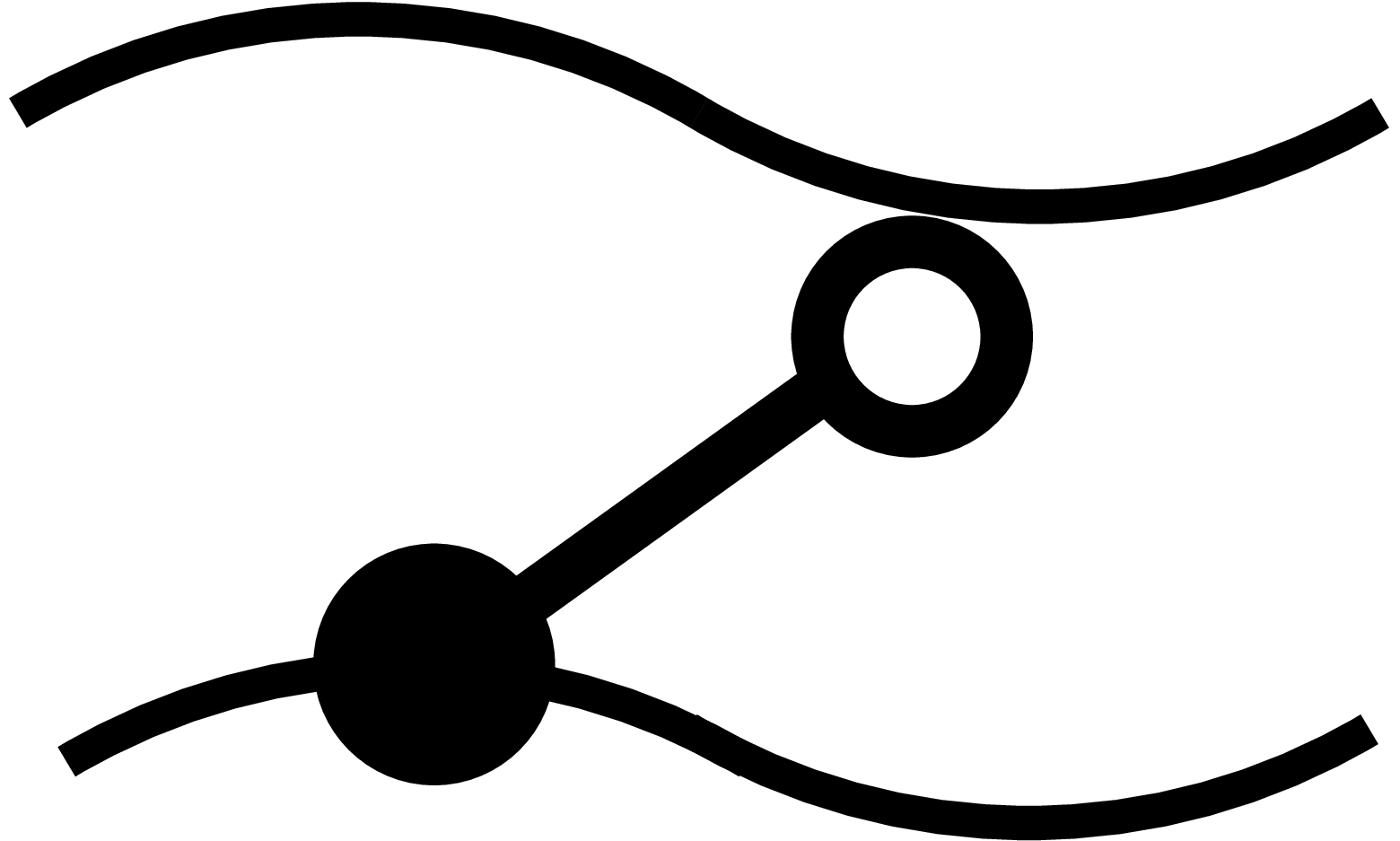}\nonumber\\
&&-\left(\frac{g}{\kappa-g}\right)^2\fig{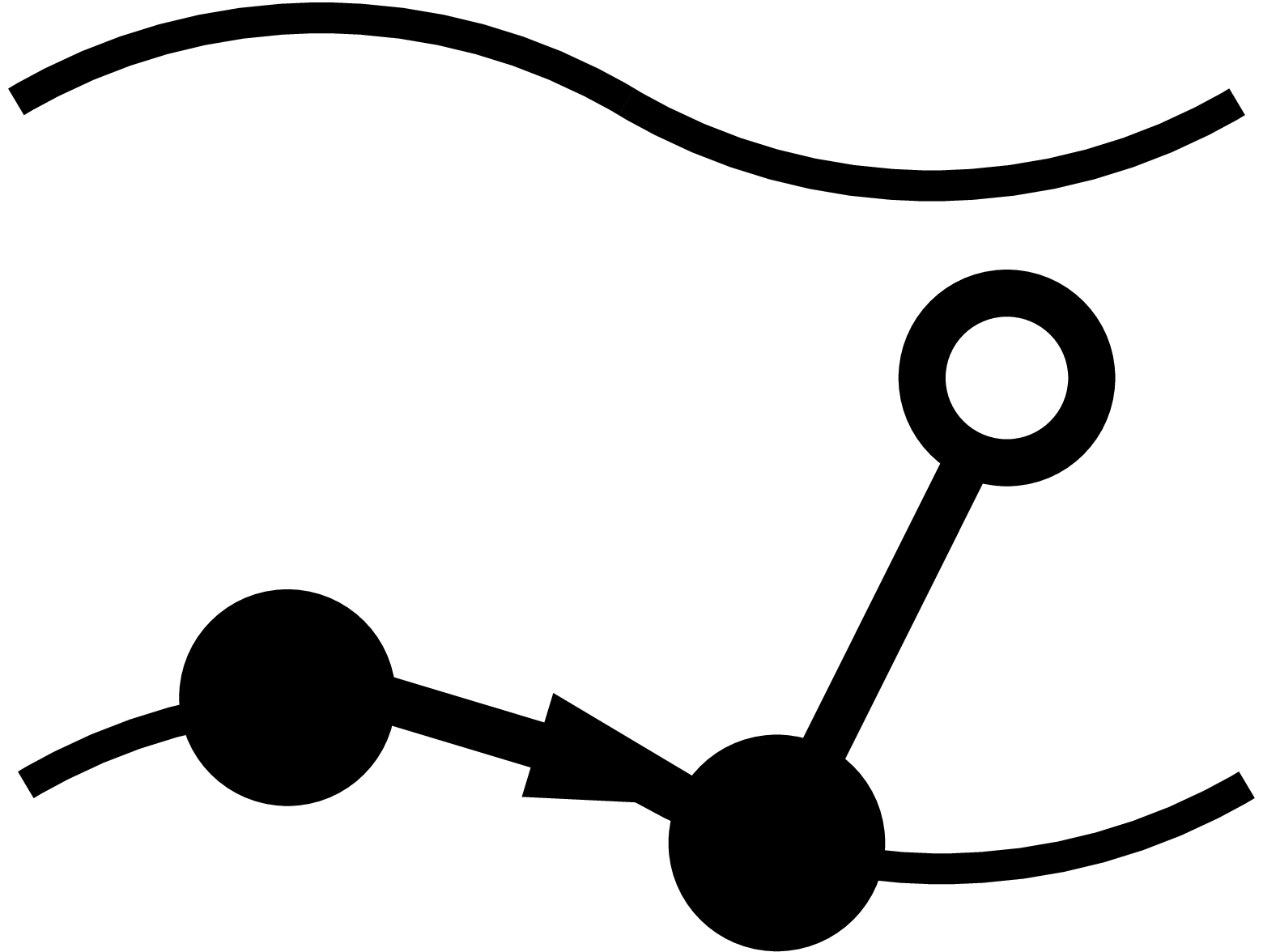}
+\cdots\nonumber\\
&&-\frac{g}{g-\kappa}\fig{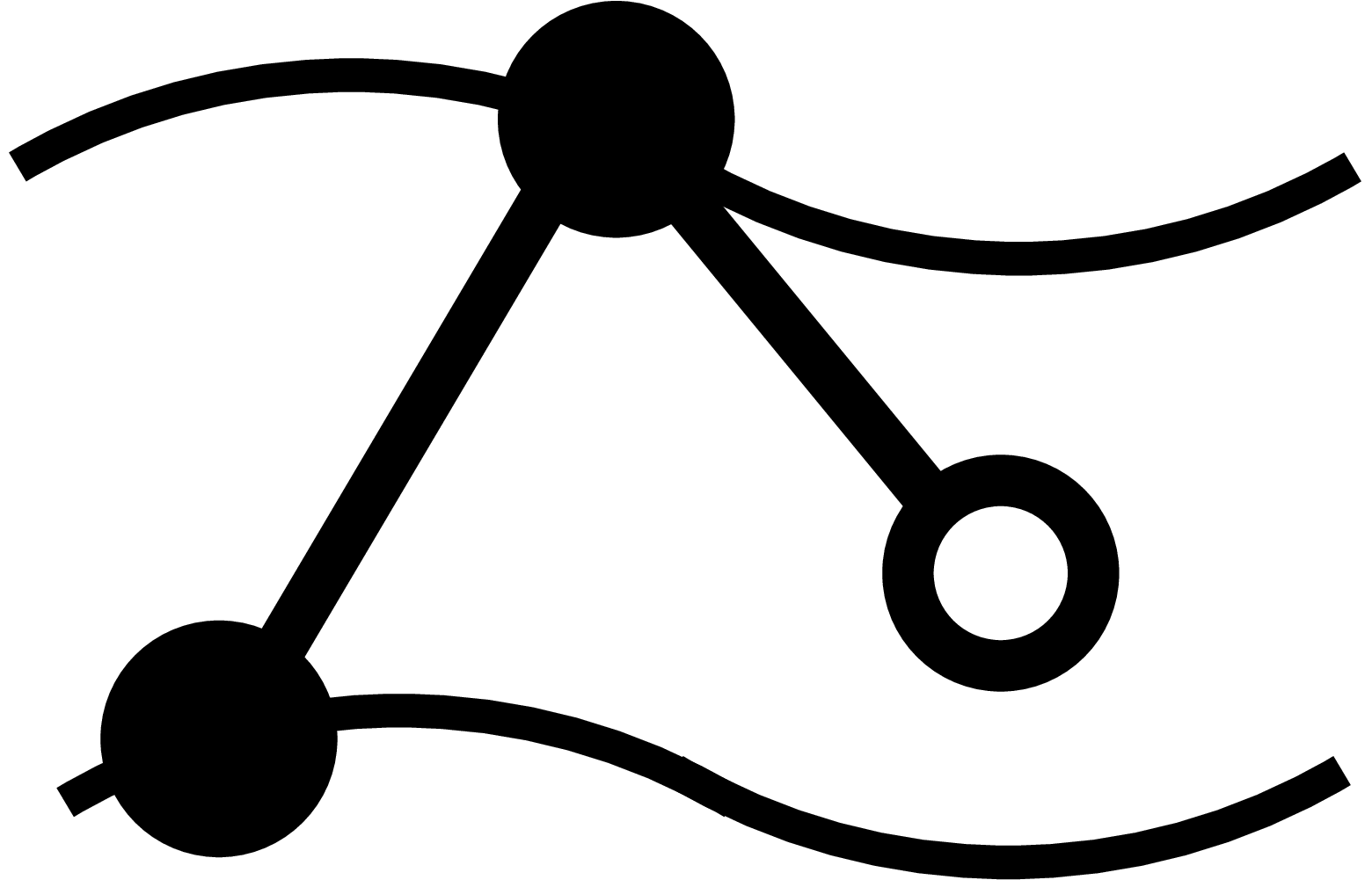}+\frac{g}{g-\kappa}\fig{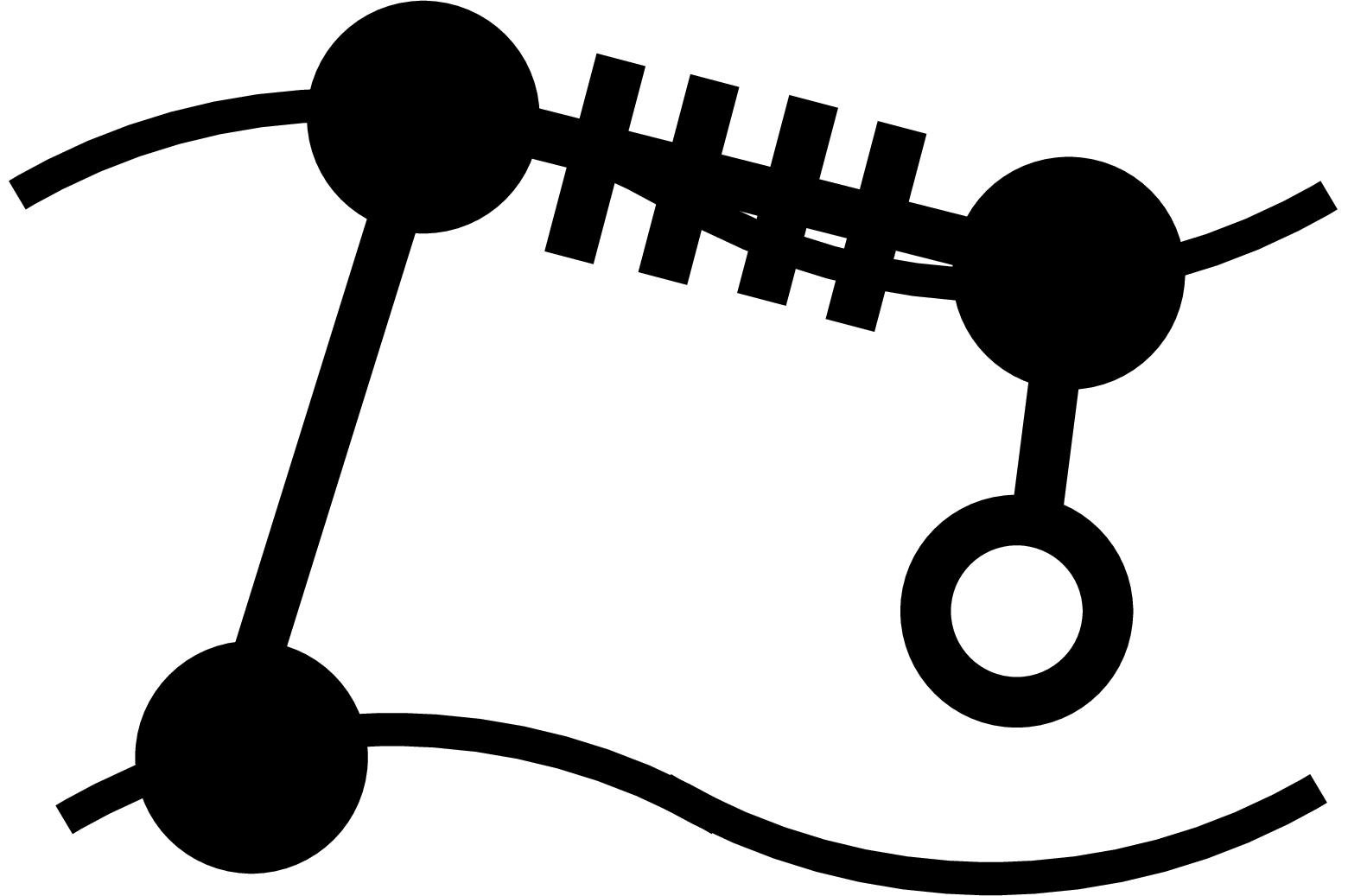}+\cdots \nonumber\\
&&+\left(\frac{g}{g-\kappa}\right)\left(\frac{\kappa+g}{g-\kappa}\right)^2\fig{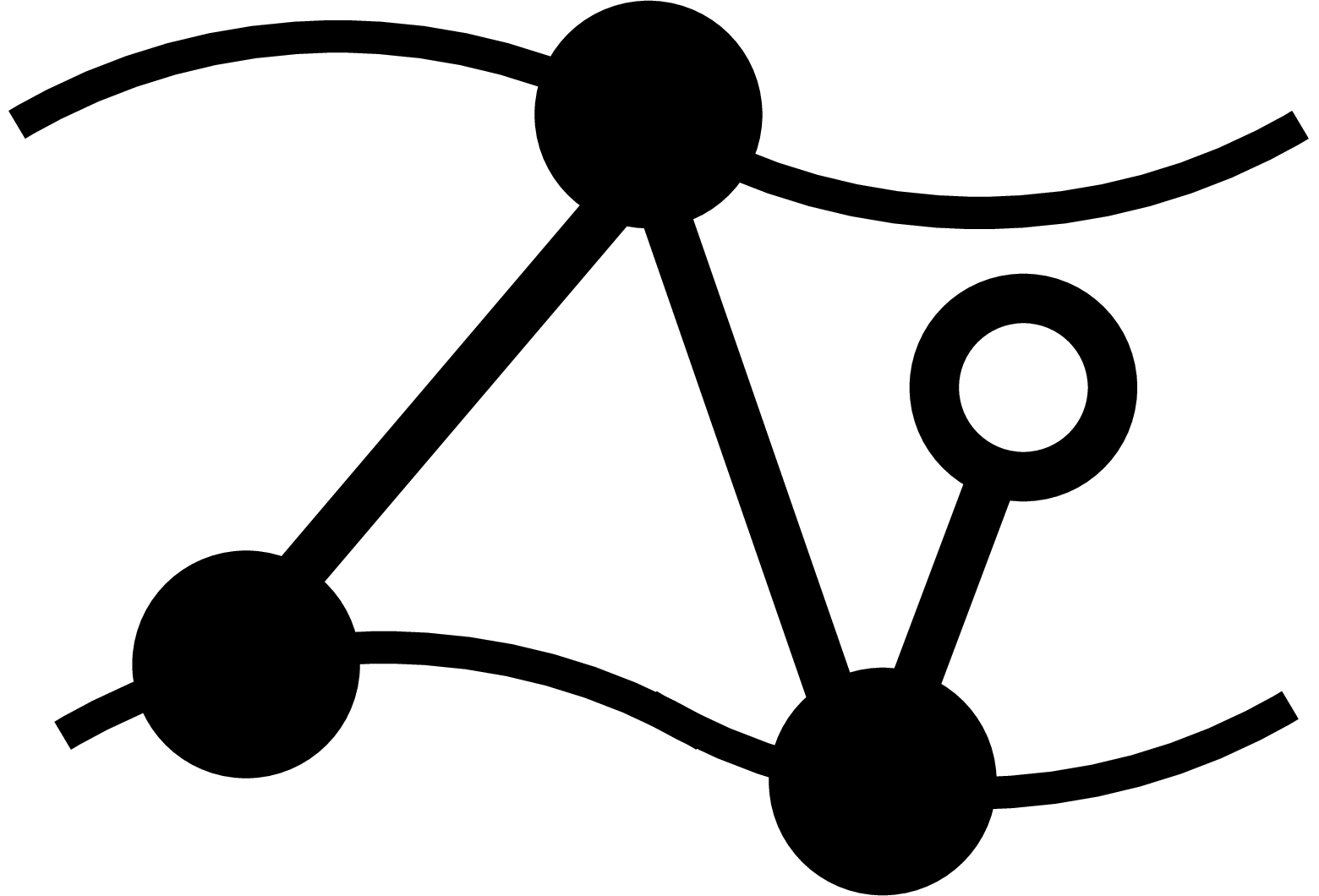}+\cdots\Bigg].
\label{deltamdiagram4}
\end{eqnarray} 
Note that, once again, these diagrams are decorated versions of those obtained in the original nonlocal model.
Their prefactors are either $-m_0$ (if the left extreme is on the liquid-gas interface) or
$-h_1/g-m_0$ (if it is on the substrate). The coefficient that multiplies each diagram is the product of 
$(-1)^{k}$, with $k$ being the number of $K-$type bonds that connect both interfaces, and the factors 
associated with the segments on each substrate. Sections with $n$ of the $U-$type bonds on the liquid-gas interface 
contribute with a factor $(-1)^n$. A segment on the substrate has a factor given by 
Eq.~(\ref{factor2}) if it contains the left diagram extreme and otherwise by Eq.~(\ref{factor4}) or 
(\ref{factor5}), depending on the nature of the rightmost bond in the segment. 

\subsection{Resummation of wetting diagrams}
As we pointed out in the preceding sections, the curvature expansion for isolated interfaces is actually connected to the formal diagrammatic method we have developed. This connection also persists for a wetting film configuration, but it is not at all explicit. The aim of this section is to illustrate how the perturbative scheme emerges from the diagrammatic one. However, 
the connection is actually far from trivial.
The reason is that, although the $\partial K/\kappa-$type
bonds are of order $R^{-1}$ [see Eq.~(\ref{expk2})], this is not the case for the $U-$type bonds: Its 
integral with respect to one argument is of order of $R^{-2}$, but $U$ is of order of unity for 
$\kappa|\mathbf{s}-\mathbf{s}^\prime|\sim 1$. So, if we convolute $U$ with a function that varies on
a length scale much larger than $\kappa^{-1}$, this is not a problem. However, in the case considered in the 
present 
section, we usually convolute $U$ with a kernel $K$ connecting both interfaces, which varies on the same 
length scale (i.e., $\kappa^{-1}$) as $U$. However, we will see that it is possible to resum the diagrams to 
obtain a diagrammatic representation of the zeroth order (in curvature) corrections. By Eq.~(\ref{expk}),
$K(\mathbf{s},\mathbf{s}^\prime)-\mathcal{K}(r_\bot)$ is of order $R^{-2}$, where $\mathbf{r}_\bot$ is the 
projection of $\mathbf{s}-\mathbf{s}^\prime$ on the plane tangent to the interface at $\mathbf{s}^\prime$. Thus,
we can neglect diagrams that present $\partial_n K/\kappa-$type bonds, 
and we replace $K(\mathbf{s},\mathbf{s}^\prime)$ by $\mathcal{K}(r_\bot)$ in the $U-$type bonds.

First, we consider the convolution of $K(\mathbf{r},\mathbf{s})$ with $K(\mathbf{s},\mathbf{s}^\prime)$, 
where $\mathbf{s}$ and $\mathbf{s}^{\prime}$ are on the same interface and $\mathbf{r}$ is either above or 
below this interface. We place the origin at $\mathbf{s}^{\prime}$, neglect curvature corrections, and 
finally assume that $\kappa r \gg 1$. Then we have
\begin{eqnarray}
\int d\mathbf{s} K(\mathbf{s},\mathbf{0})K(\mathbf{s},\mathbf{r}) \approx \left(\frac{\kappa}{2\pi}\right)^2
\int_0^\infty ds \textrm{e}^{-\kappa s}
\\\times
\int_0^{2\pi} d\theta 
\frac{\exp\left[{-\kappa\sqrt{r^2+s^2-2sr\sin\alpha \cos\theta}}\right]}{\sqrt{r^2+s^2-2sr\sin\alpha \cos\theta}}
,
\nonumber
\end{eqnarray}
where $\alpha$ is the angle between $\mathbf{r}$ and the surface normal at the origin $\mathbf{n}(\mathbf{0})$.
As $\kappa r\gg 1$ but $\kappa s\lesssim 1$, we expand the distance between $\mathbf{r}$ and $\mathbf{s}$ in 
powers of $s/r$, so in the first approximation we have
\begin{eqnarray}
&&\int d\mathbf{s} K(\mathbf{s},\mathbf{0})K(\mathbf{s},\mathbf{r}) \approx \\
&&\frac{\kappa \textrm{e}^{-\kappa r}}
{2\pi r}  
\int_0^\infty \kappa ds \textrm{e}^{-\kappa s}\frac{1}{2\pi}
\int_0^{2\pi}d\theta \exp\left[\kappa s \sin\alpha \cos \theta\right].
\nonumber
\end{eqnarray}
The modified Bessel function of zeroth order and the first kind $I_0$ has the integral representation
\begin{equation}
 \label{i0int}
I_0(x)=\frac{1}{2\pi}
\int_0^{2\pi}d\theta \exp\left[x\cos \theta\right].
\end{equation}
Therefore
\begin{eqnarray}
\int d\mathbf{s} K(\mathbf{s},\mathbf{0})K(\mathbf{s},\mathbf{r}) &\approx&
\frac{\kappa \textrm{e}^{-\kappa r}}
{2\pi r}  
\int_0^\infty \kappa ds \textrm{e}^{-\kappa s}I_0(\kappa s \sin\alpha)
\nonumber\\
&=&
\frac{\kappa \textrm{e}^{-\kappa r}}{2\pi r}\frac{1}{|\cos\alpha|},
\end{eqnarray}
and thus
\begin{equation}
\int d\mathbf{s}_0 U(\mathbf{s},\mathbf{s}_0)K(\mathbf{s}_0,\mathbf{r}) \approx K(\mathbf{s},\mathbf{r}) 
\left(\frac{|\mathbf{r}-\mathbf{s}|}{|\mathbf{n}(\mathbf{s})\mathbf{\cdot}(\mathbf{r}-\mathbf{s})|}-1\right),
\label{intuk}
\end{equation}
up to corrections in powers of $(\kappa R)^{-1}$ and $(\kappa r)^{-1}$. 
Equation~(\ref{intuk}) vanishes if $\mathbf{r}$ is on the normal direction to the substrate at $\mathbf{s}$. As a consequence, when 
the liquid-gas interface is parallel to the substrate (e.g., for parallel planes or concentric 
spheres or cylinders), the nonlocal model ansatz is a good approximation to the full solution when curvature
corrections are neglected \cite{laura,parry2} (see also 
Appendix~\ref{appendixe}). On the other hand, a saddle-point analysis of the binding potential shows that
the maximum contribution to the multiple integrals associated with each diagram in Eq.~(\ref{Hmin6}) arises
from the neighborhood of the closest pair of points located on different interfaces, which would lie on a normal
direction common to both substrates. In this sense, the binding-potential representation shown above is extremely 
nonlocal: in leading order only the shape of the substrate and the liquid-gas interfaces around their 
closest positions features. However, this is not true for the order-parameter profile at an arbitrary position 
$\mathbf{r}$ and, in general, corrections beyond $\delta m_\Xi^{0,\ell}+\delta m_\Xi^{0,\psi}$ will be incorrect 
with the original nonlocal model ansatz, even neglecting curvature corrections.

In order to obtain a more local representation of the binding potential and the order-parameter profile, we note that
the structure of the diagrams shows $K$ bonds that connect both interfaces, followed by
a segment of the diagram on one interface. The idea is to resum the convolutions of a $K$ bond connecting
the wall and the gas-liquid interface with all the possible segments either on the liquid-gas interface or on the substrate.
If this is done to zeroth order in the curvature, we obtain renormalized bonds between the wall and the
gas-liquid interface. 
For example, a renormalized bond between a (left) position on the substrate and a (right) position on the 
gas-liquid interface would be
\begin{eqnarray}
K_{\psi\to \ell}
(\mathbf{s},\mathbf{r})=K(\mathbf{s},\mathbf{r})+\sum_{i=1}^\infty (-1)^i\nonumber\\\times \int_\ell d\mathbf{s}_1
\cdots d\mathbf{s}_{i} U(\mathbf{s},\mathbf{s}_1)\cdots U(\mathbf{s}_{i-1},\mathbf{s}_i)
K(\mathbf{s}_i,\mathbf{r}),\label{renormbond1-prevprev}
\end{eqnarray}
which is approximately given by
\begin{eqnarray}
&&K_{\psi\to \ell}
(\mathbf{s},\mathbf{r})\approx K(\mathbf{s},\mathbf{r})\sum_{i=0}^\infty \left(1-\frac{1}{|\cos\alpha|}\right)^i\nonumber\\
&&=K(\mathbf{s},\mathbf{r})|\cos\alpha|
\label{renormbond1-prev}
\end{eqnarray}
and hence
\begin{equation}
K_{\psi\to \ell}
(\mathbf{s},\mathbf{r})
\approx -\frac{1}{\kappa} \partial_{n}K(\mathbf{s},\mathbf{r}), 
\label{renormbond1}
\end{equation}
where $\partial_n K(\mathbf{s},\mathbf{r})=\mathbf{n}(\mathbf{s})\mathbf{\cdot}
\boldsymbol\nabla_{\mathbf{s}} K(\mathbf{s},\mathbf{r})$.
On the other hand, a renormalized bond between a (left) position on the gas-liquid interface and a (right) position on the
substrate would be given by
\begin{eqnarray}
&&K_{\ell\to \psi}
(\mathbf{s},\mathbf{r})=\frac{g+\kappa}{g-\kappa}K(\mathbf{s},\mathbf{r})\nonumber\\
&&-\sum_{i=1}^\infty  
\Bigg[2\left(\frac{g}{\kappa-g}\right)^{i+1}+(-1)^i\Bigg]\nonumber\\
&&\times \int_\psi d\mathbf{s}_1
\cdots d\mathbf{s}_{i} U(\mathbf{s},\mathbf{s}_1)\cdots U(\mathbf{s}_{i-1},\mathbf{s}_i)
K(\mathbf{s}_i,\mathbf{r}),\label{renormbond2-prevprev}
\end{eqnarray}
which is approximately
\begin{eqnarray}
&&K_{\ell\to \psi}
(\mathbf{s},\mathbf{r})\approx-K(\mathbf{s},\mathbf{r})\sum_{i=0}^\infty \left(1-\frac{1}{|\cos\alpha|}\right)^i\nonumber\\ 
&&+2\frac{g}{g-\kappa}K(\mathbf{s},\mathbf{r})\sum_{i=0}^\infty \left[\left(\frac{g}{\kappa-g}\right)
\left(\frac{1}{|\cos\alpha|}-1\right)\right]^i.\label{renormbond2-prev}
\end{eqnarray}
Hence, at leading order,
\begin{equation}
K_{\ell\to \psi}
(\mathbf{s},\mathbf{r})=K(\mathbf{s},\mathbf{r})|\cos\alpha|\frac{g+\kappa|\cos\alpha|}{g-\kappa|\cos\alpha|}
\label{renormbond2-prevpost}
\end{equation}
or, equivalently,
\begin{equation}
K_{\ell\to \psi}
(\mathbf{s},\mathbf{r})\approx \frac{1}{\kappa}\partial_n K(\mathbf{s},\mathbf{r})
\frac{g+\kappa|\cos\alpha|}{g-\kappa|\cos\alpha|},
\label{renormbond2}
\end{equation}
if a new $K-$type bond between the wall and the gas-liquid interface emerges from the right extreme of the diagram on the substrate segment. 
Otherwise
\begin{eqnarray}
&&K_{\ell\to \psi}^\prime
(\mathbf{s},\mathbf{r})=\frac{g}{g-\kappa}\Bigg[K(\mathbf{s},\mathbf{r})\nonumber+\sum_{i=1}^\infty  
\left(\frac{g}{\kappa-g}\right)^i\nonumber\\
&&\times\int_\psi d\mathbf{s}_1
\cdots d\mathbf{s}_{i} U(\mathbf{s},\mathbf{s}_1)\cdots U(\mathbf{s}_{i-1},\mathbf{s}_i)
K(\mathbf{s}_i,\mathbf{r})\Bigg],\label{renormbond3-prevprevprev}
\end{eqnarray}
which is approximately given by
\begin{eqnarray}
&&K_{\ell\to \psi}^\prime
(\mathbf{s},\mathbf{r})
\nonumber\\
&&\approx \frac{g}{g-\kappa}K(\mathbf{s},\mathbf{r})\sum_{i=0}^\infty \left[\left(\frac{g}{\kappa-g}\right)
\left(\frac{1}{|\cos\alpha|}-1\right)\right]^i.\label{renormbond3-prevprev}
\end{eqnarray}
Hence, we have
\begin{equation}
K_{\ell\to \psi}^\prime
(\mathbf{s},\mathbf{r})
\approx K(\mathbf{s},\mathbf{r})\frac{g|\cos\alpha|}{g-\kappa|\cos\alpha|}
\label{renormbond3-prev}
\end{equation}
and finally
\begin{equation}
K_{\ell\to \psi}^\prime
(\mathbf{s},\mathbf{r})\approx \frac{1}{\kappa}\partial_n K(\mathbf{s},\mathbf{r})
\frac{g}{g-\kappa|\cos\alpha|} .
\label{renormbond3}
\end{equation}
It follows that in the limit of small curvatures we can perform a resummation of a rather generic convolution of wetting diagram, the ones above providing the most important examples. The resulting diagrams, which are proportional to $\kappa^{-1}(\partial_{n} K)$, are the basic ingredients entering in the binding potential for fixed boundary conditions (i.e., where $g\rightarrow\infty$); this is what we are going to prove in the next section.

\subsection{Alternative representation of the binding potential functional for fixed boundary conditions}
It is possible to systematically explore the curvature corrections to the binding potential through the consideration of connecting the 
interface and those interfacial segments involving three types of bonds:
$U_\pi\equiv \mathcal{K}(\mathbf{r}_\bot)-\delta(\mathbf{r}_\bot)$, $\partial_n K/\kappa$ (only on the substrate),
and $\tilde U=K(\mathbf{s},\mathbf{s}^\prime)-\mathcal{K}(\mathbf{r}_\bot)$, which are of order of $1$, $(\kappa R)^{-1}$,
and $(\kappa R)^{-2}$, respectively. However, to simplify the discussion we restrict ourselves to the 
case in which the order parameter is  fixed, to a value $m_1$, on the substrate. This will allow us to make the 
connection with 
the original nonlocal model formulation more easily. This case corresponds to the limit 
$g\to -\infty$ and $-h_1/g-m_0\to \delta m_1\equiv m_1-m_0$. Thus, the 
expansions (\ref{omega11}), (\ref{omega12}), (\ref{omega21})
and (\ref{deltamdiagram4}) only include diagrams that do not present
$\partial_n K/\kappa$ bonds. On the other hand, the contributions to the 
coefficients of the segments on the interfaces now become $(-1)^n$, 
with $n$ being the number of $U-$type bonds of the diagram segment, 
regardless of whether or not it lies on the liquid-gas interface or substrate.
This diagrammatic representation presents the same problems as mentioned
above for the finite-$g$ case. However, we can rationalize them using the
identities
\begin{eqnarray}
\label{id_1a}
&&\int_\psi d\mathbf{s}_0 K_\psi^{-1}(\mathbf{s},\mathbf{s}_0)K(\mathbf{s}_0,\mathbf{r})\label{renormbond4}\\
&&=\int_\psi d\mathbf{s}_0 \left(\delta(\mathbf{s}-\mathbf{s}_0)+\frac{1}{\kappa}\partial_n K(\mathbf{s},
\mathbf{s}_0)\right)^{-1}\frac{1}{\kappa}\partial_{n_0}K(\mathbf{s}_0,\mathbf{r}),
\nonumber
\end{eqnarray}
and
\begin{eqnarray}
\label{id_2a}
&&\int_\ell d\mathbf{s}_0 K_\ell^{-1}(\mathbf{s},\mathbf{s}_0)K(\mathbf{s}_0,\mathbf{r})\label{renormbond5}\\
&&=-\int_\ell d\mathbf{s}_0 \left(\delta(\mathbf{s}-\mathbf{s}_0)-\frac{1}{\kappa}\partial_n K(\mathbf{s},
\mathbf{s}_0)\right)^{-1}\frac{1}{\kappa}\partial_{n_0}K(\mathbf{s}_0,\mathbf{r}),
\nonumber
\end{eqnarray}
where $\partial_n K(\mathbf{s},\mathbf{s}_0)\equiv \mathbf{n}(\mathbf{s})\mathbf{\cdot}\boldsymbol
{\nabla}_{\mathbf{s}}K(\mathbf{s},\mathbf{s}_0)$ and $\partial_{n_0} K(\mathbf{s}_0,\mathbf{r})\equiv 
\mathbf{n}(\mathbf{s}_0)\mathbf{\cdot}\boldsymbol {\nabla}_{\mathbf{s}_0}K(\mathbf{s}_0,\mathbf{r})$.
These identities arise from Eq.~(\ref{greenidentity8}) or, alternatively, from 
the equivalence of the single- and double-layer potentials ~(\ref{singlelayer})
and (\ref{doublelayer}) for given Dirichlet boundary conditions [see Eqs.~(\ref{singlelayer2}) and
(\ref{doublelayer2})]. By using the fundamental relations $\int \mathcal{O}\mathcal{O}^{-1}=\delta$ 
for the inverse operators $(\delta \pm \partial K/\kappa)^{-1}$ we obtain 
\begin{eqnarray}
\label{inverseops1}
&&\left(\delta(\mathbf{s}-\mathbf{s}_0)\pm \frac{1}{\kappa}\partial_n K(\mathbf{s},
\mathbf{s}_0)\right)^{-1}=\delta (\mathbf{s}-\mathbf{s}_0) \\&&\mp \int d\mathbf{s}_1
\left(\delta(\mathbf{s}-\mathbf{s}_1)\pm \frac{1}{\kappa}\partial_n K(\mathbf{s},
\mathbf{s}_1)\right)^{-1}\frac{1}{\kappa}\partial_{n_1}K(\mathbf{s}_1,\mathbf{s}_0),
\nonumber
\end{eqnarray}
which formally can be represented as
\begin{eqnarray}
\label{inverseops2}
&&\left(\delta(\mathbf{s}-\mathbf{s}_0)\pm \frac{1}{\kappa}\partial_n K(\mathbf{s},
\mathbf{s}_0)\right)^{-1}=\delta (\mathbf{s}-\mathbf{s}_0) \\&&\mp \frac{1}{\kappa}\partial_n K(\mathbf{s},
\mathbf{s}_0)+ \int d\mathbf{s}_1 \frac{1}{\kappa}\partial_n K(\mathbf{s},
\mathbf{s}_1) \frac{1}{\kappa}\partial_{n_1} K(\mathbf{s}_1,
\mathbf{s}_0) + \cdots .
\nonumber
\end{eqnarray}
It is straightforward to recognize that Eqs.~(\ref{renormbond4}) and (\ref{renormbond5}) can be represented diagrammatically as
\begin{eqnarray}
\label{id_1}
&&\fig{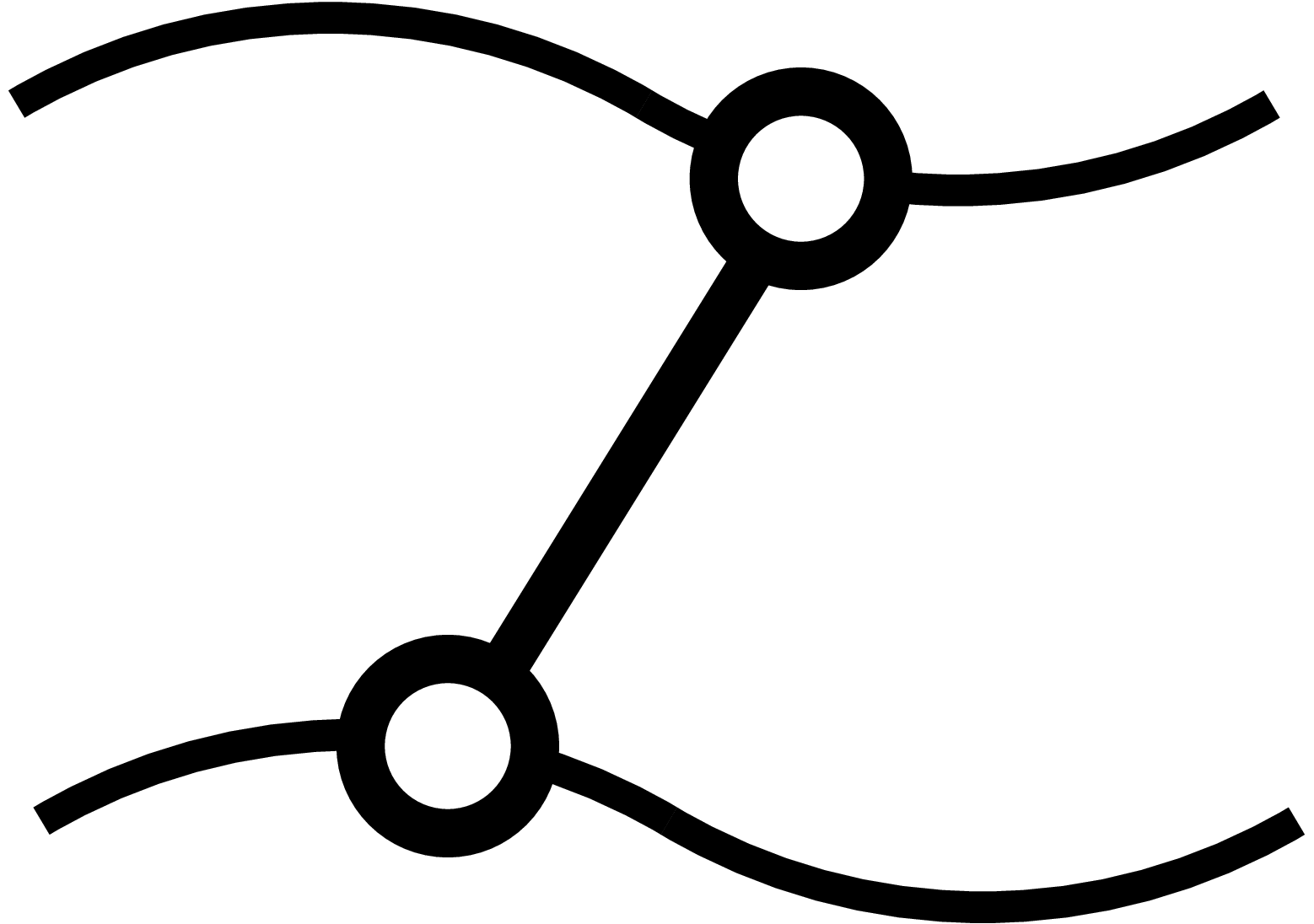}-\fig{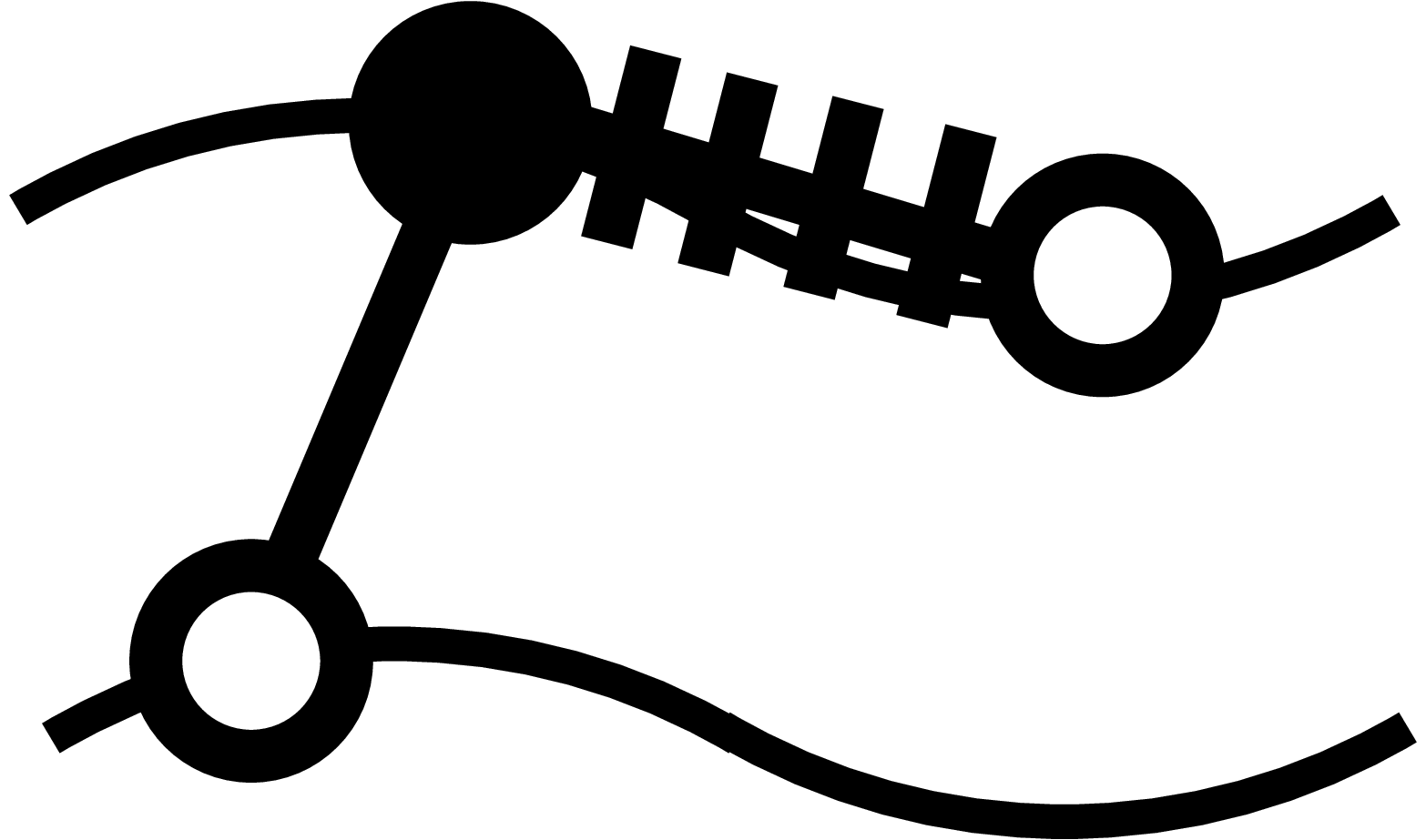}+\fig{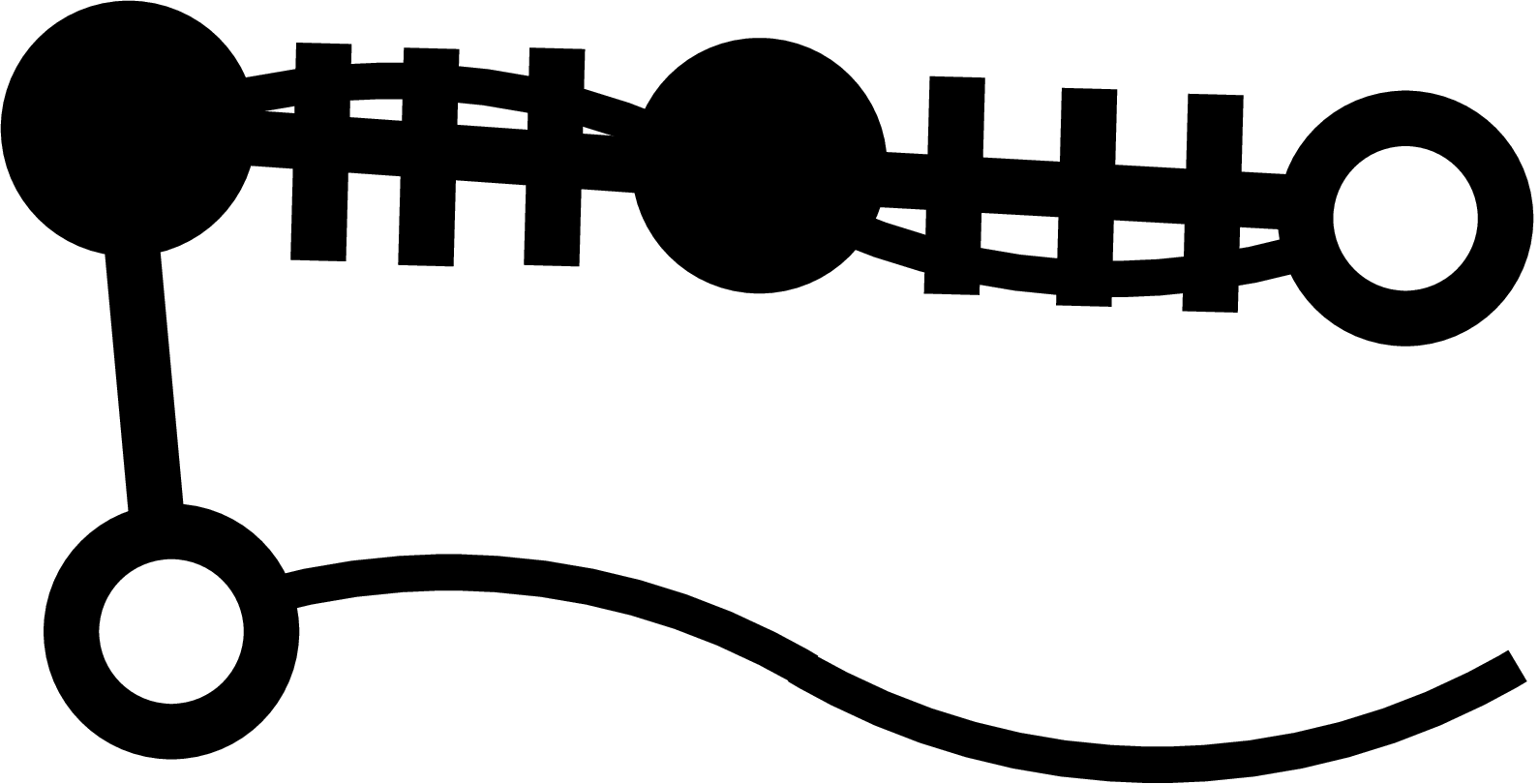}-\cdots\nonumber\\
&&=-\fig{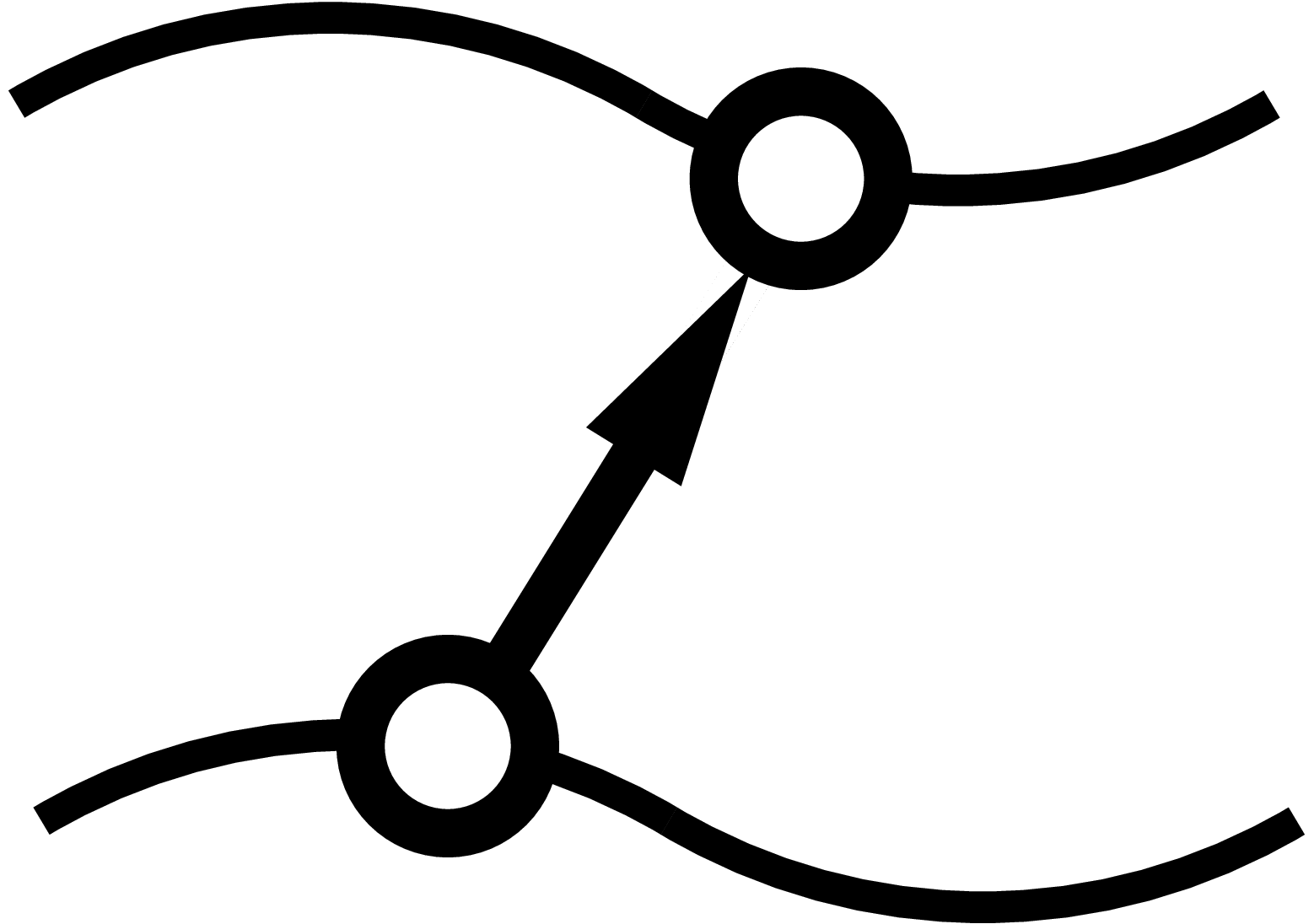}-\fig{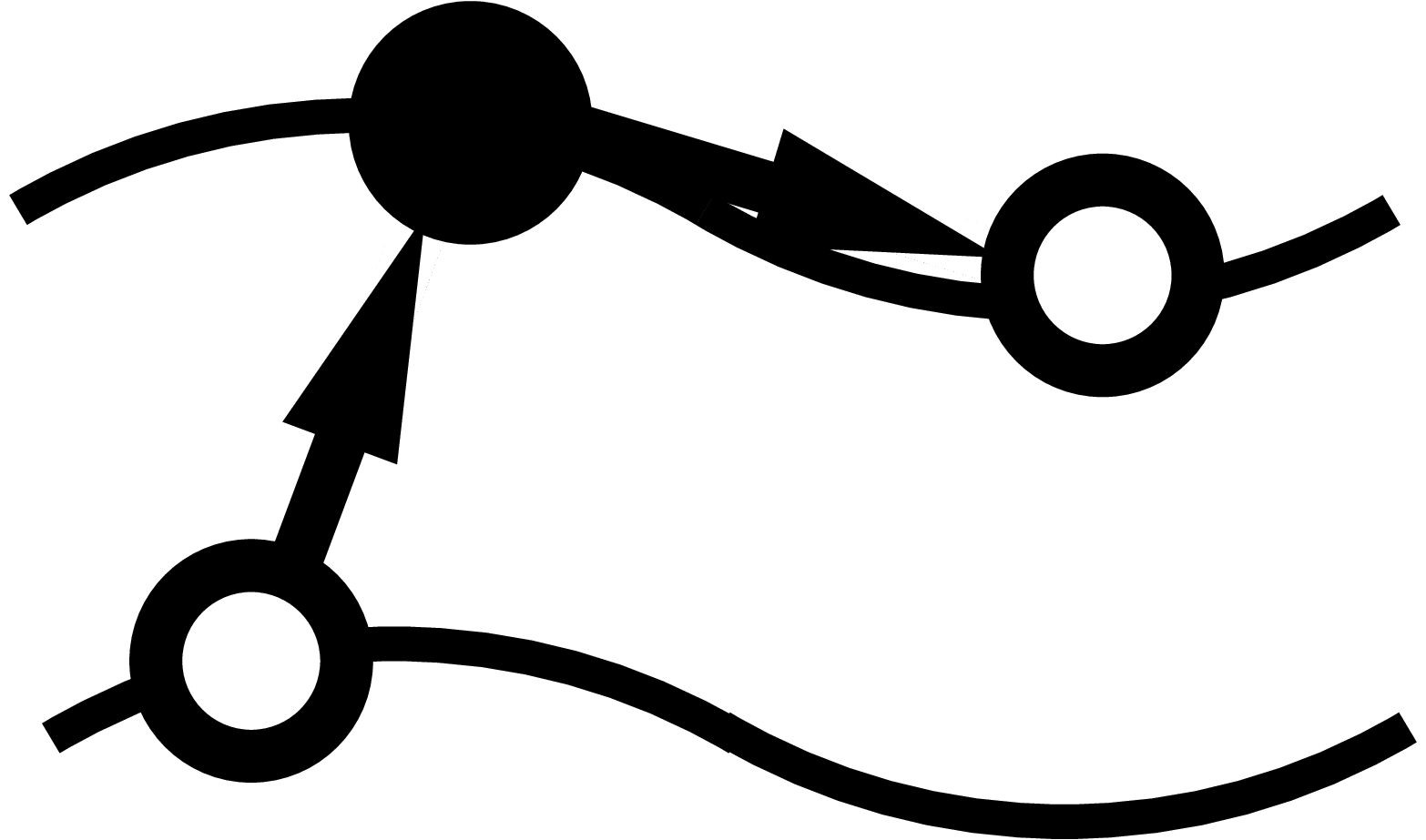}-\fig{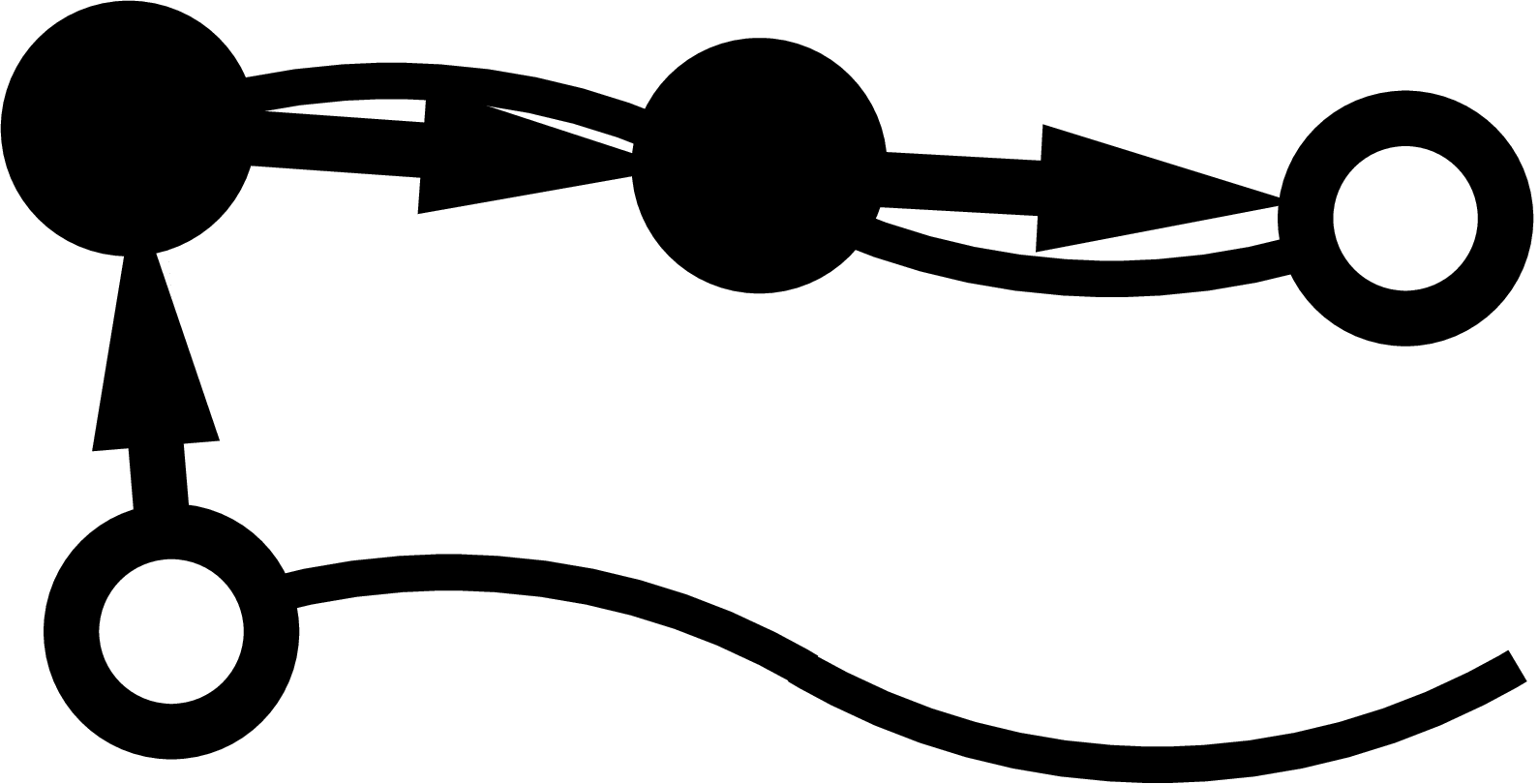}-\cdots\label{identity1}
\\
\label{id_2}
&&\fig{diagram69}-\fig{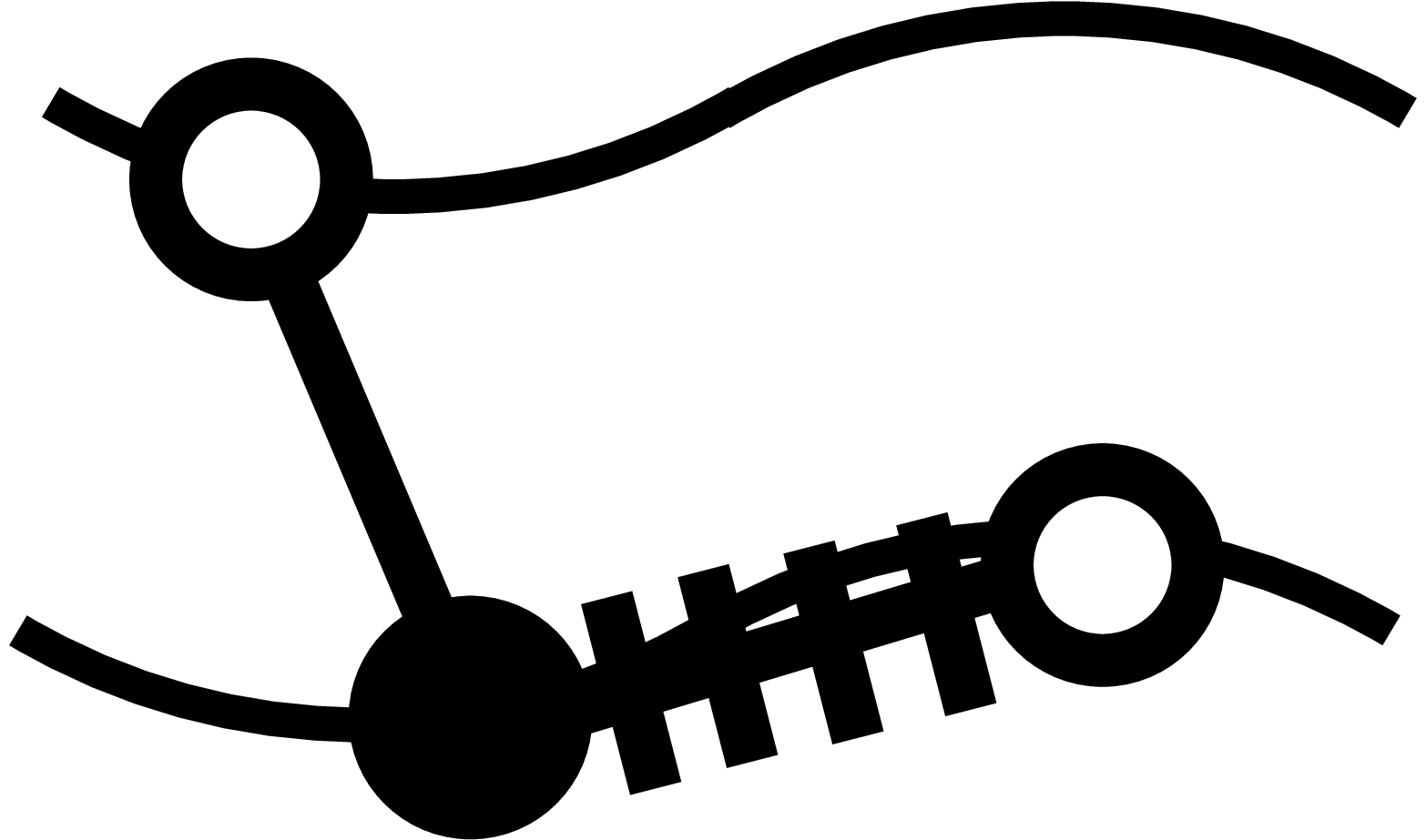}+\fig{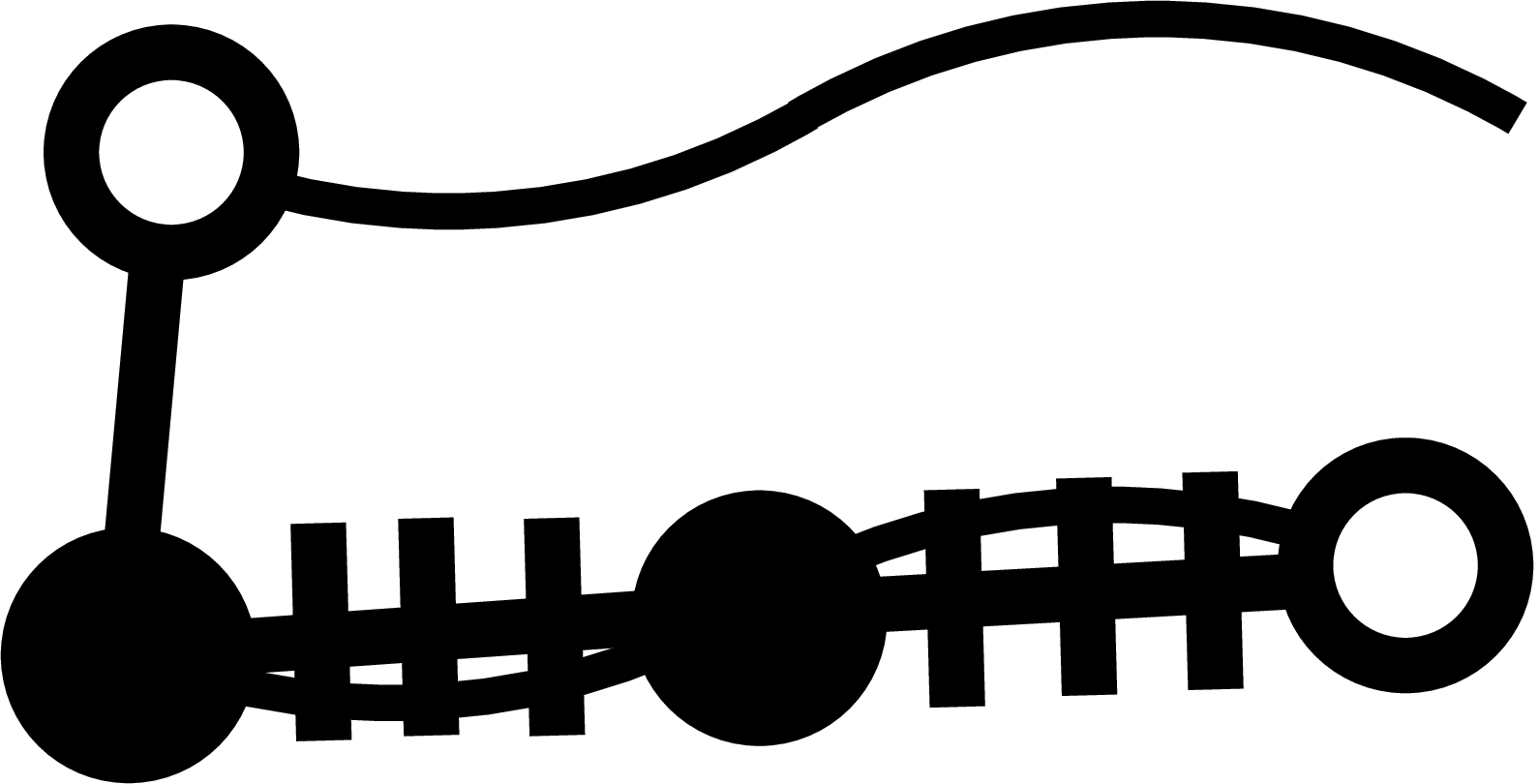}-\cdots\nonumber\\
&&=\fig{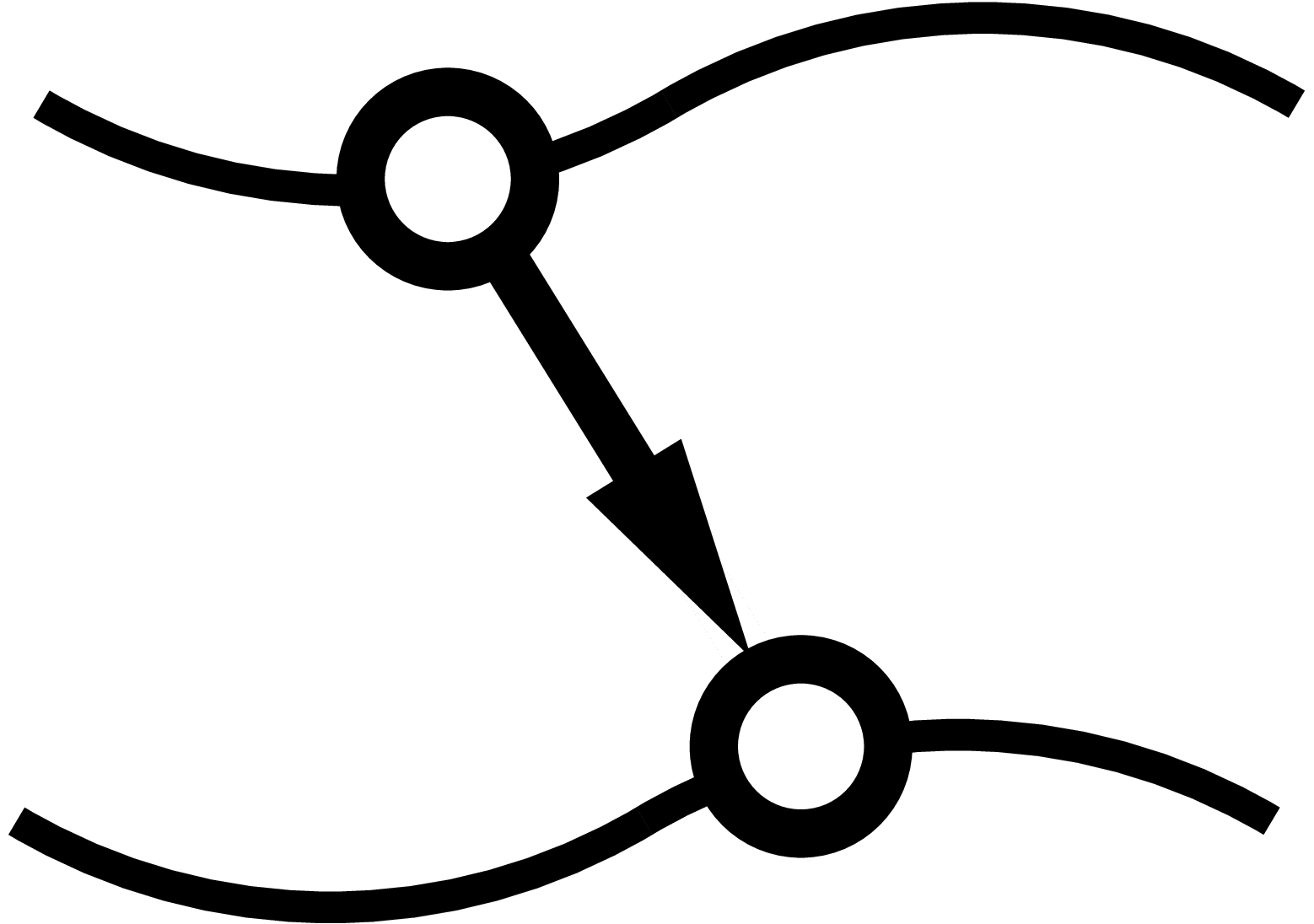}-\fig{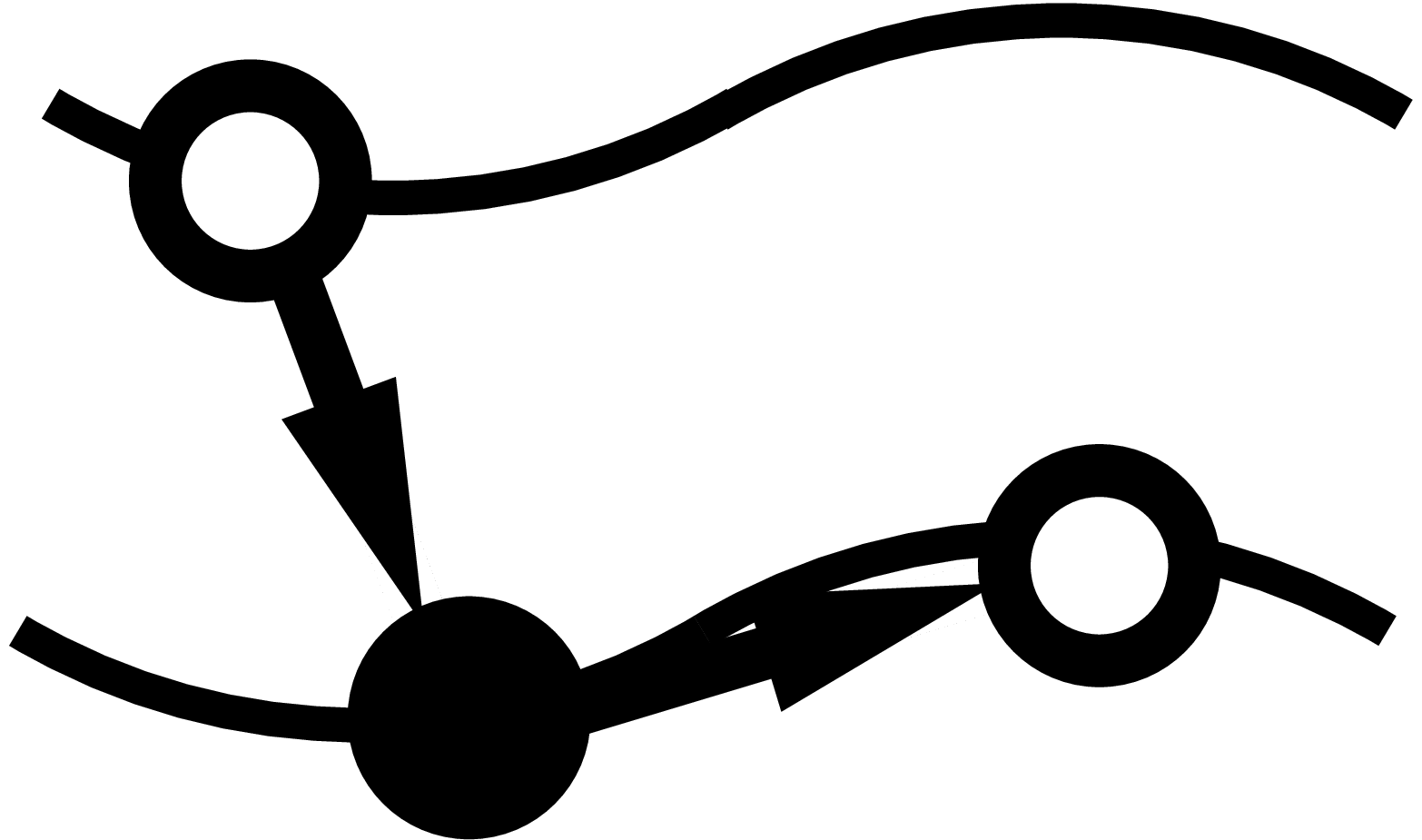}+\fig{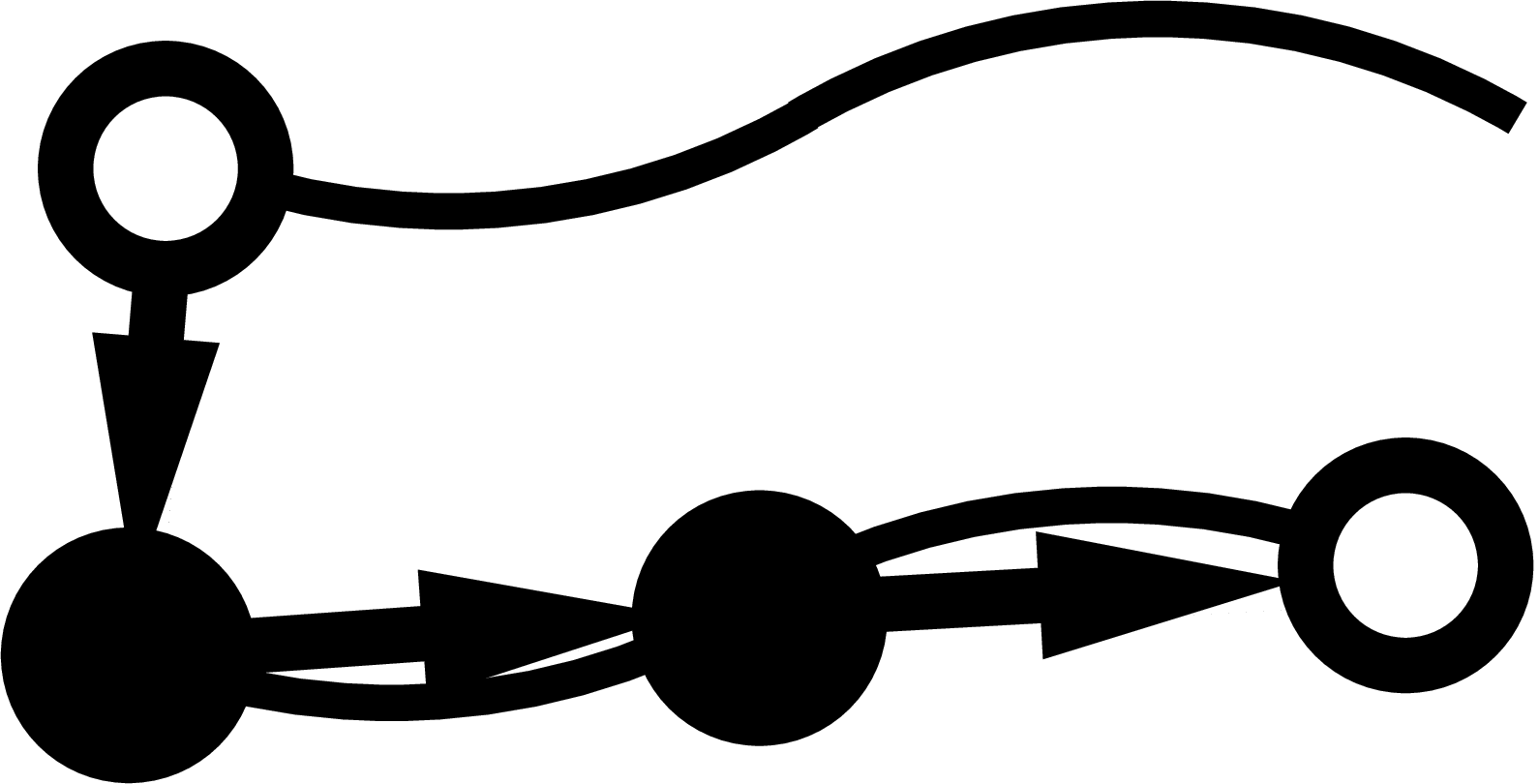}-\cdots,\label{identity2}
\end{eqnarray}
where the bonds carrying arrows linking both interfaces are $\partial_n K/\kappa$ functions, and the arrow
indicates the position where the normal derivative is applied. Note that the right-hand sides of these equations
correspond to an expansion in powers of $(\kappa R)^{-1}$, as $\partial_n K/\kappa\sim (\kappa R)^{-1}$. On the
other hand, the leading-order contributions are consistent with the renormalized bonds obtained previously.
[Consider the limit $g\to -\infty$ in Eqs. (\ref{renormbond1}), (\ref{renormbond2}), and (\ref{renormbond3}).] 
On using Eqs.~(\ref{identity1}) and (\ref{identity2}), we obtain the alternative representation of the
binding potential
\begin{eqnarray}
&&\frac{W[\ell,\psi]}{\kappa m_0^2}=\sum_{n=1}^\infty \Bigg(\frac{\delta m_1}{m_0} \Omega_n^n+ \Omega_n^{n+1}
+\left(\frac{\delta m_1}{m_0}\right)^2
\Omega_{n+1}^n\Bigg),
\nonumber\\
\label{Hmin7}
\end{eqnarray}
which is now similar to the structure of the original nonlocal treatment, for example, the leading-order 
contribution, viz.,
\begin{eqnarray}
\Omega_1^1=&&
-\fig{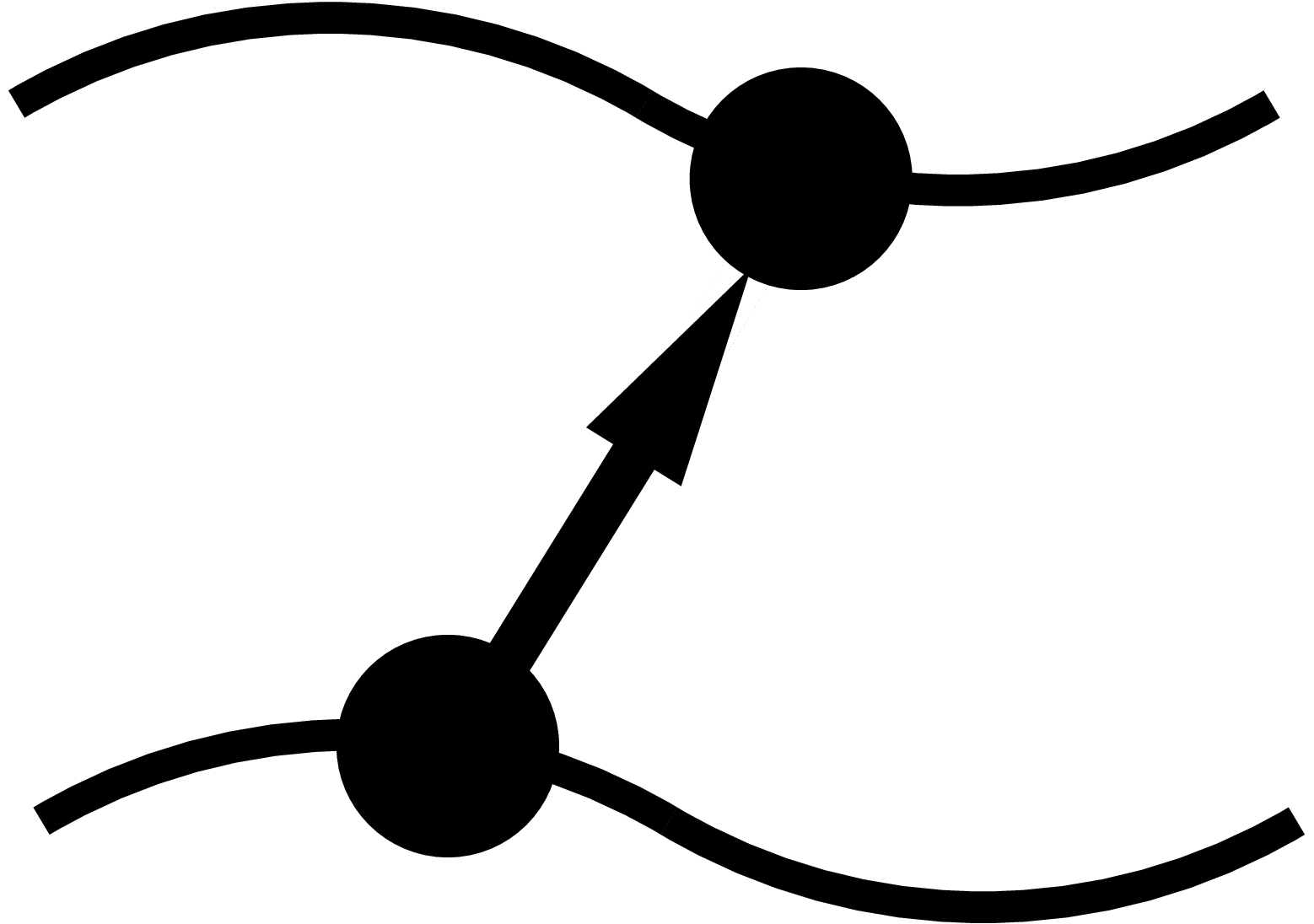}+
\fig{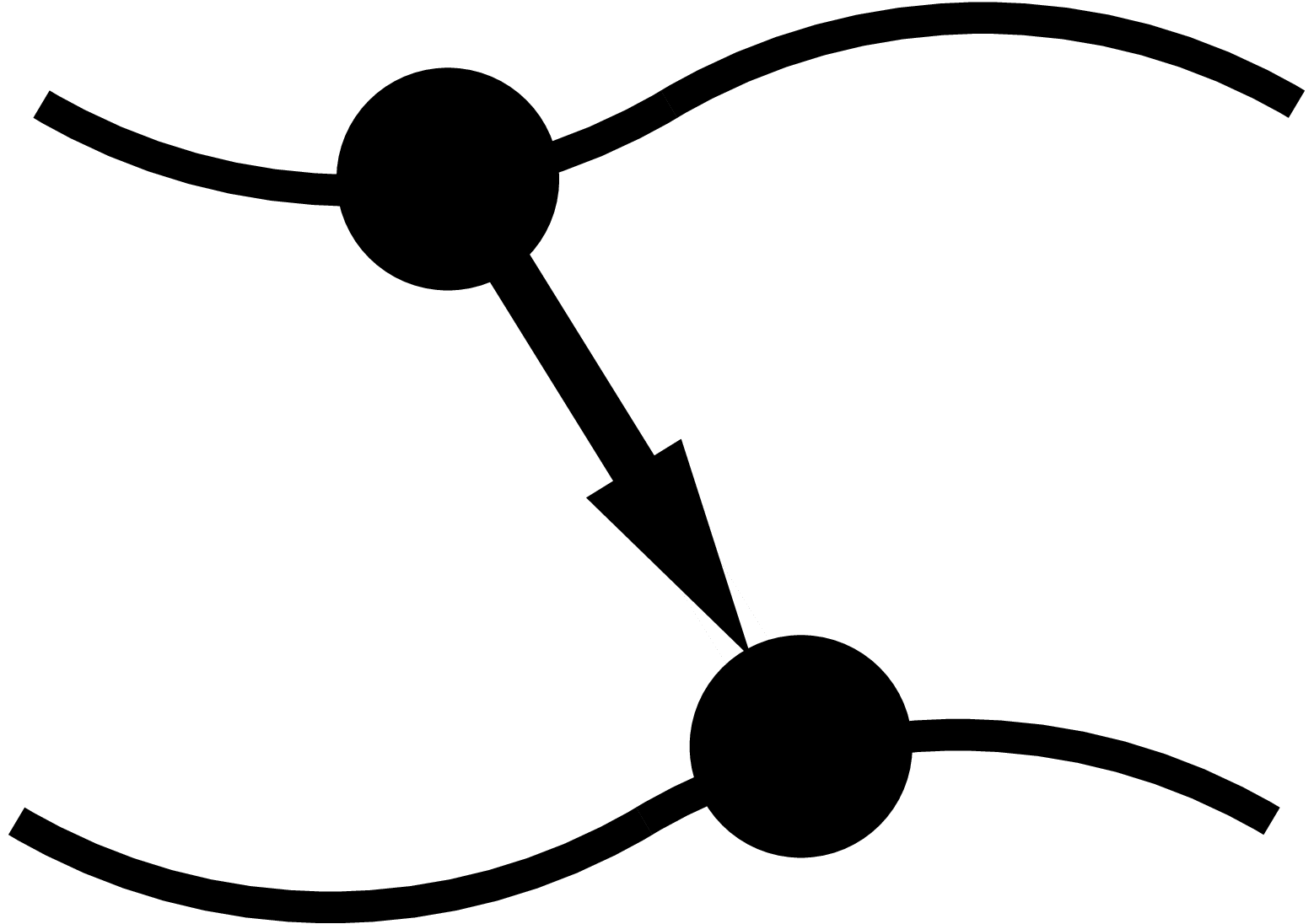}\nonumber\\
&&+\fig{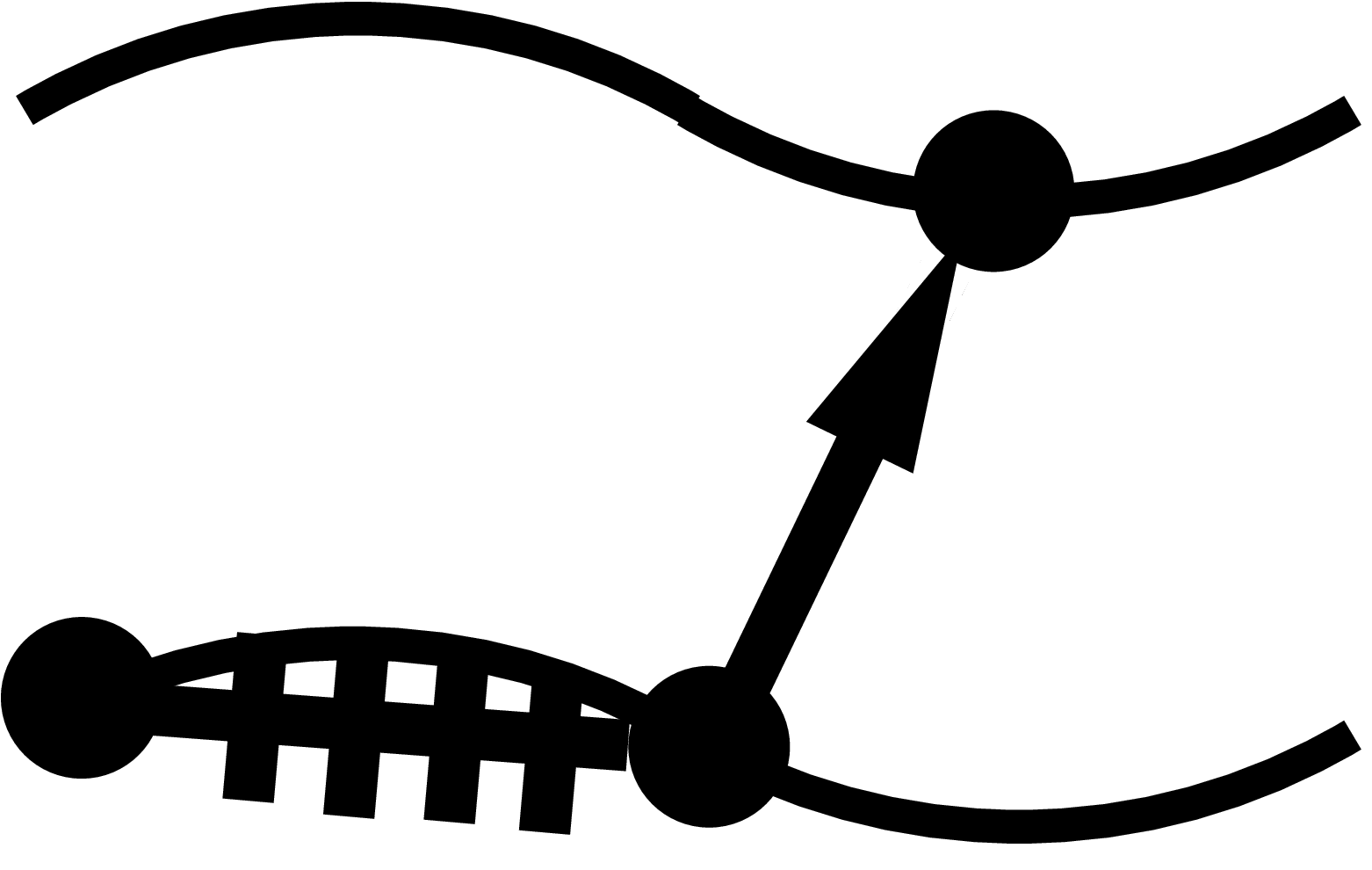}
-\fig{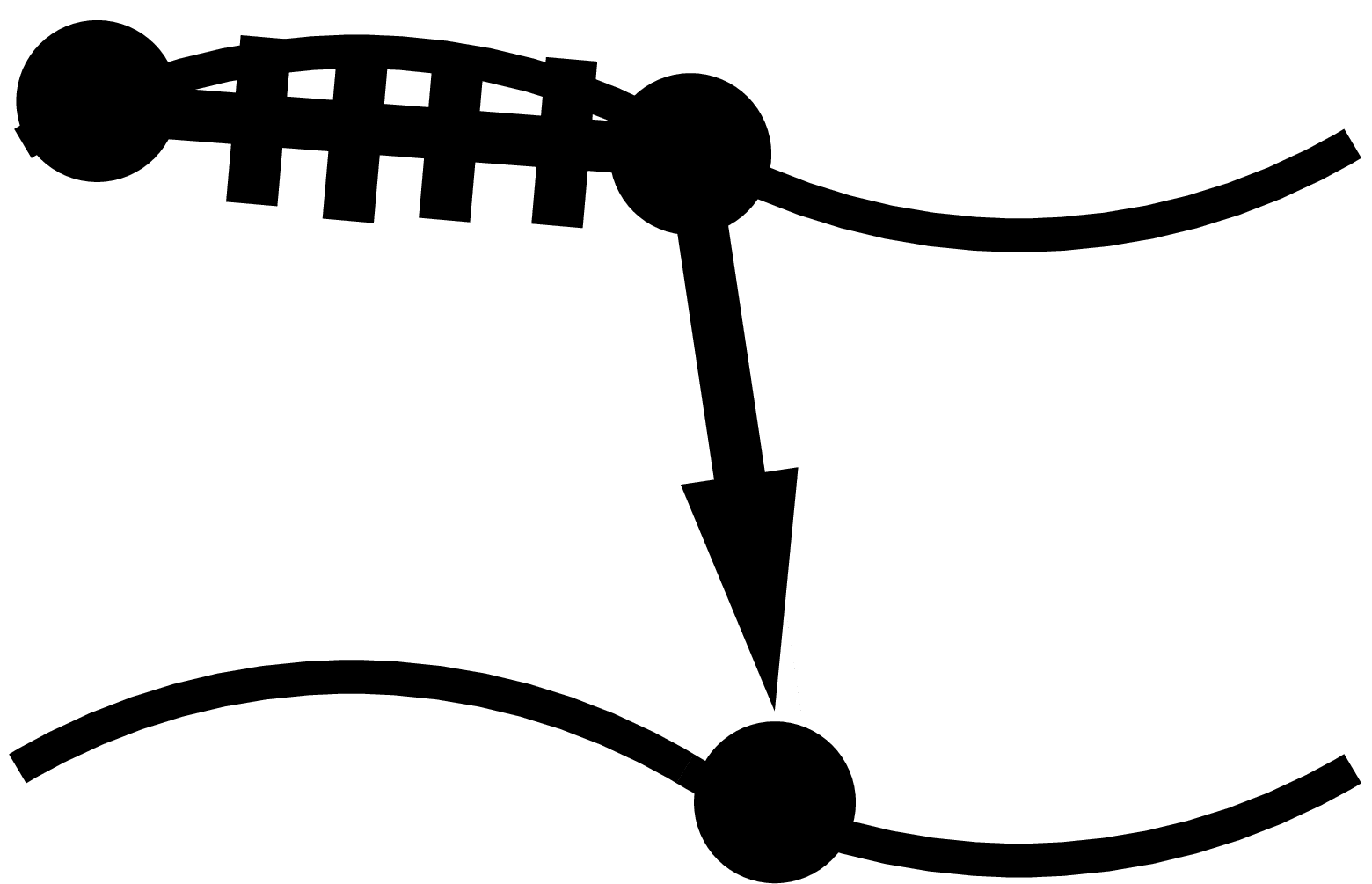}\nonumber\\
&&+\fig{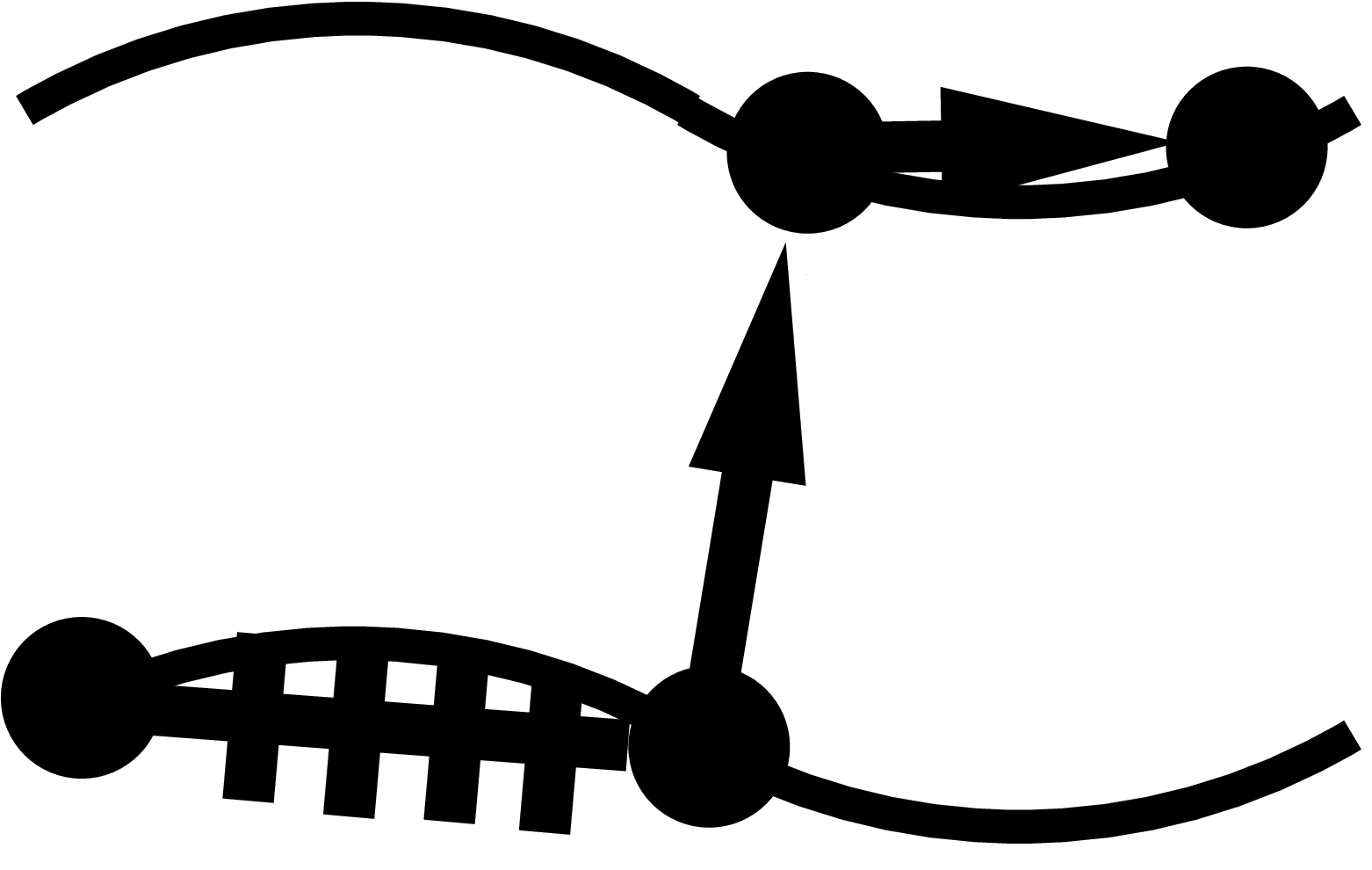}+\fig{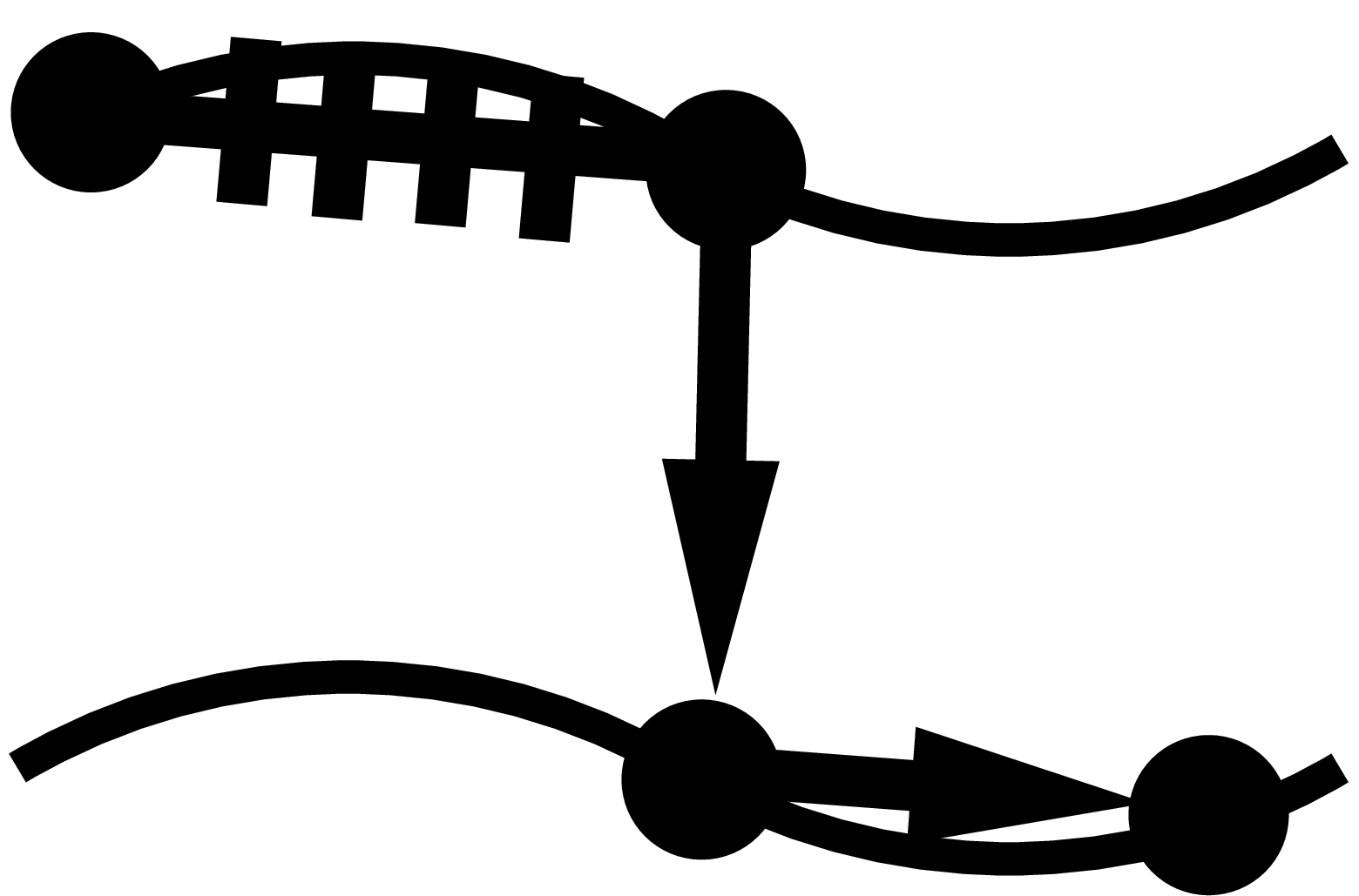}+\cdots,\label{omega11-2}
\end{eqnarray}
while
\begin{eqnarray}
\Omega_1^2=&&
-\fig{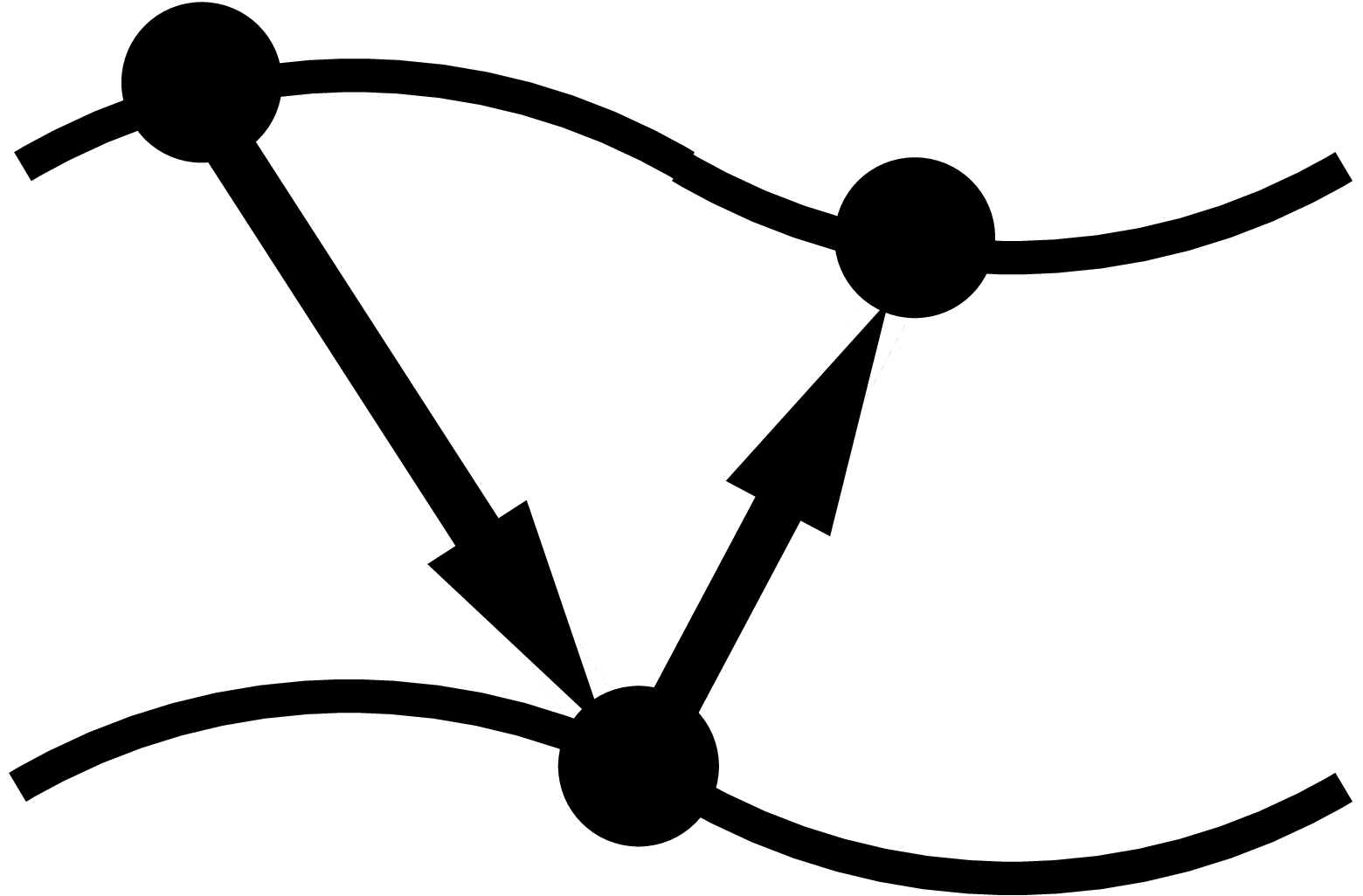}+\fig{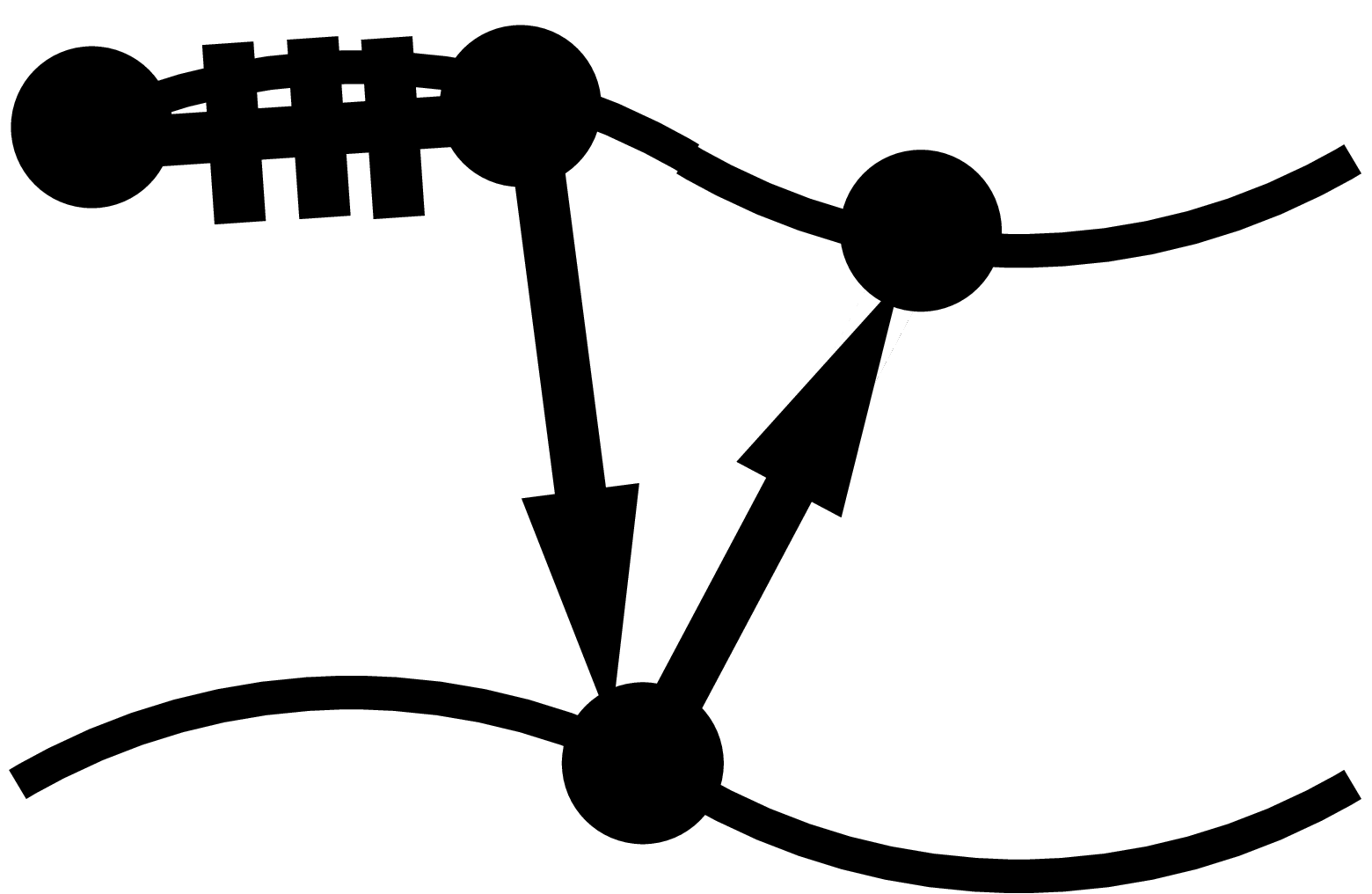}\nonumber\\
&&+\fig{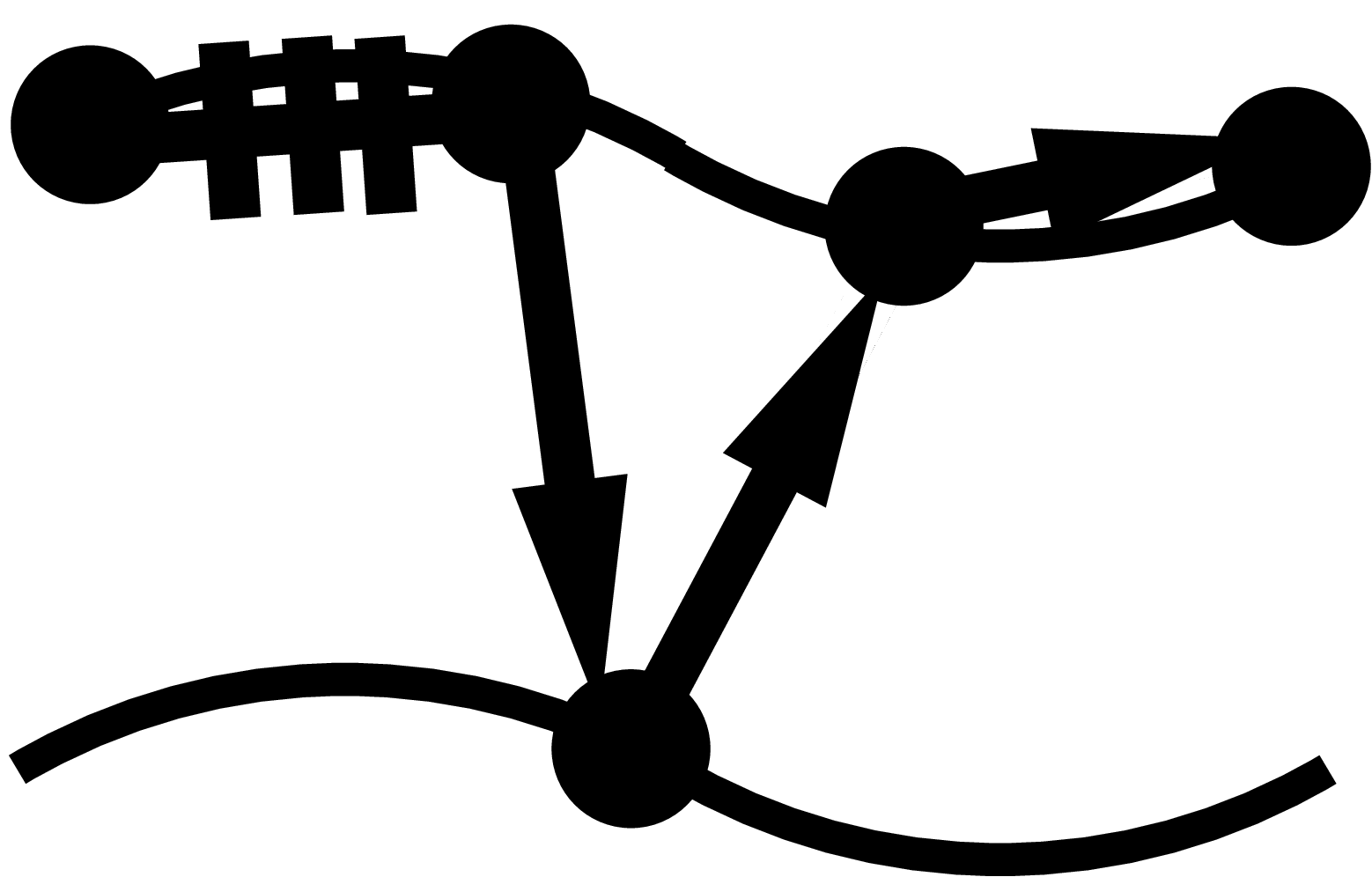}+\fig{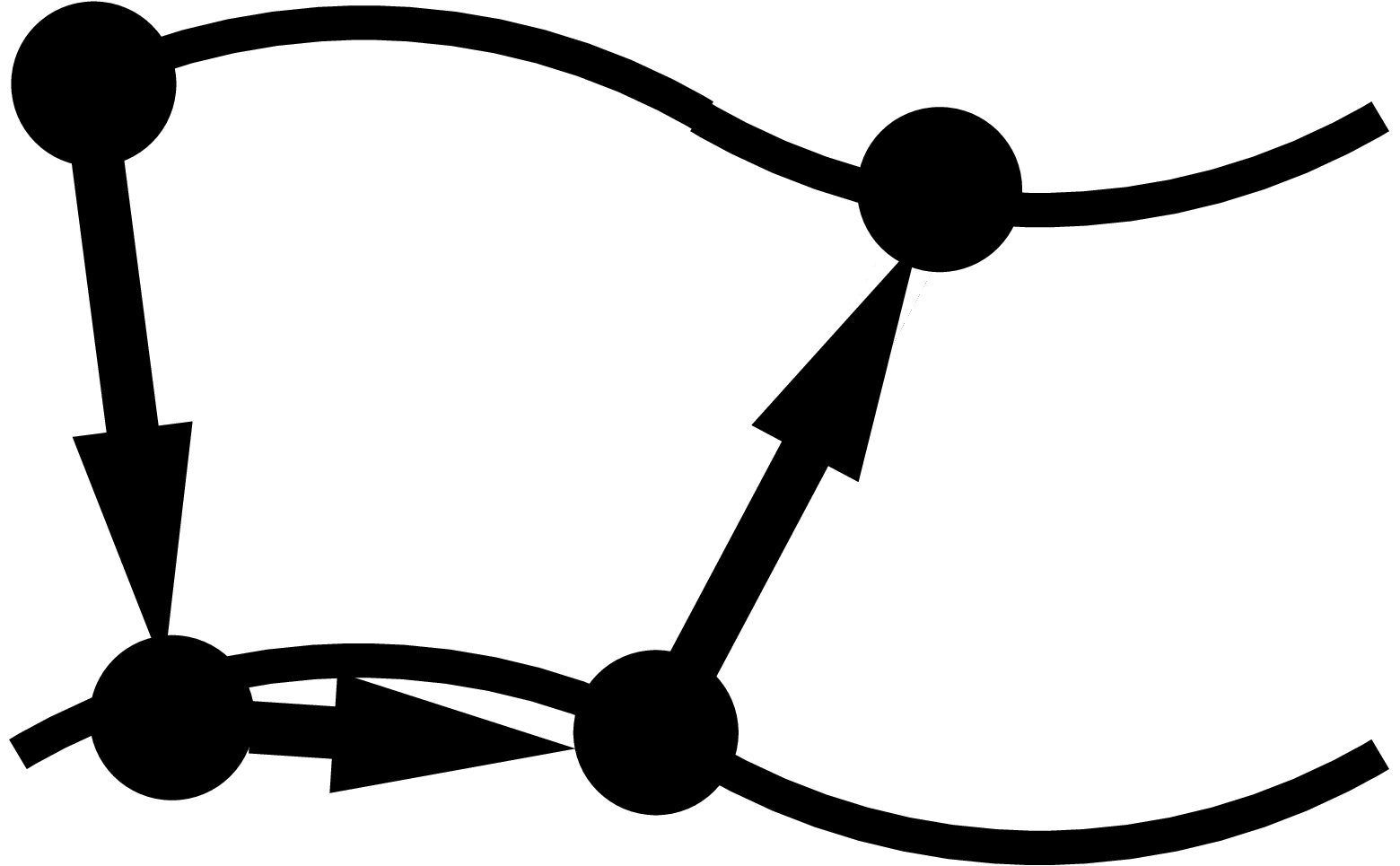}\nonumber\\
&&+\fig{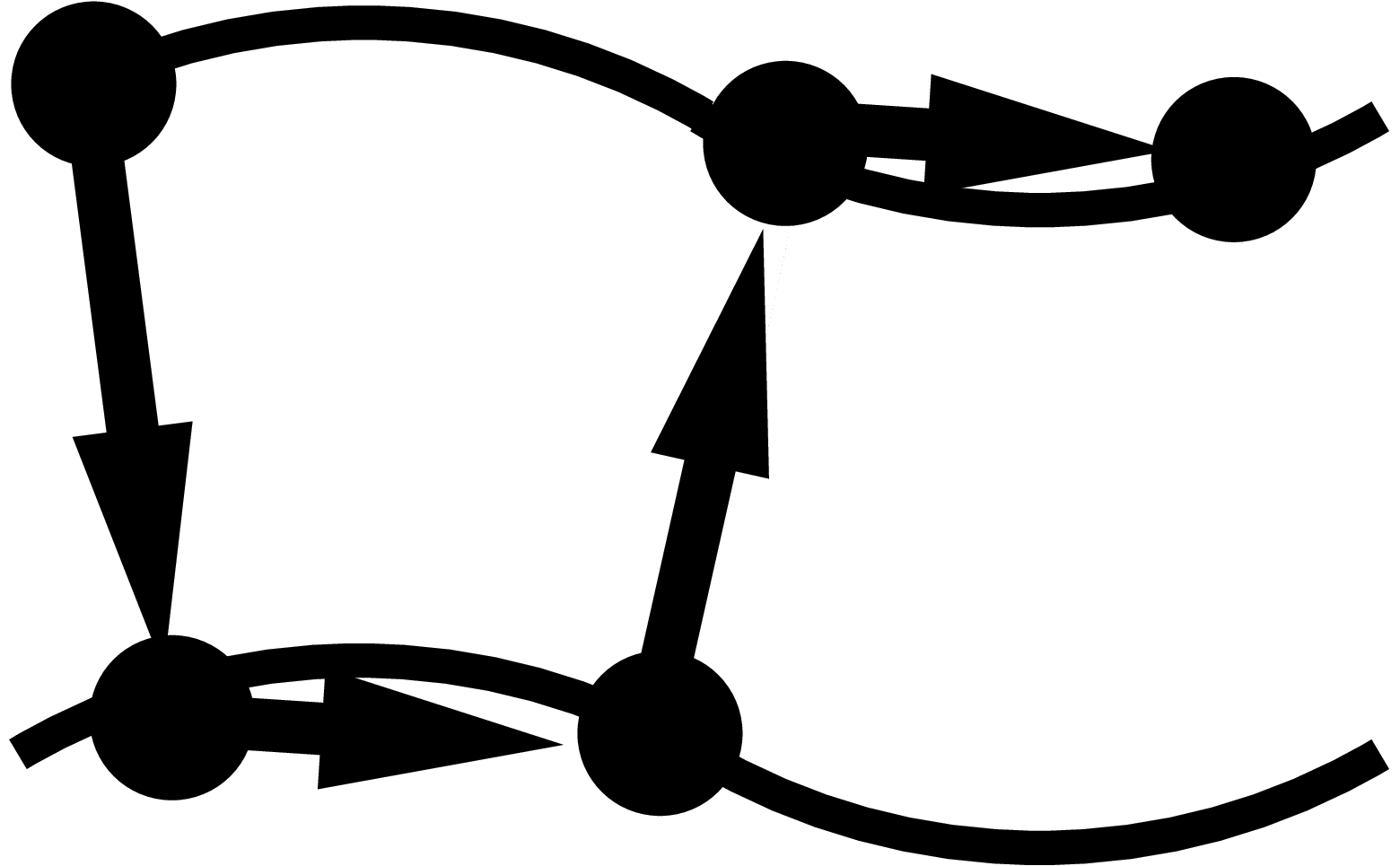}+\cdots,
\label{omega12-2}
\end{eqnarray}
and
\begin{eqnarray}
\Omega_2^1=&&
-\fig{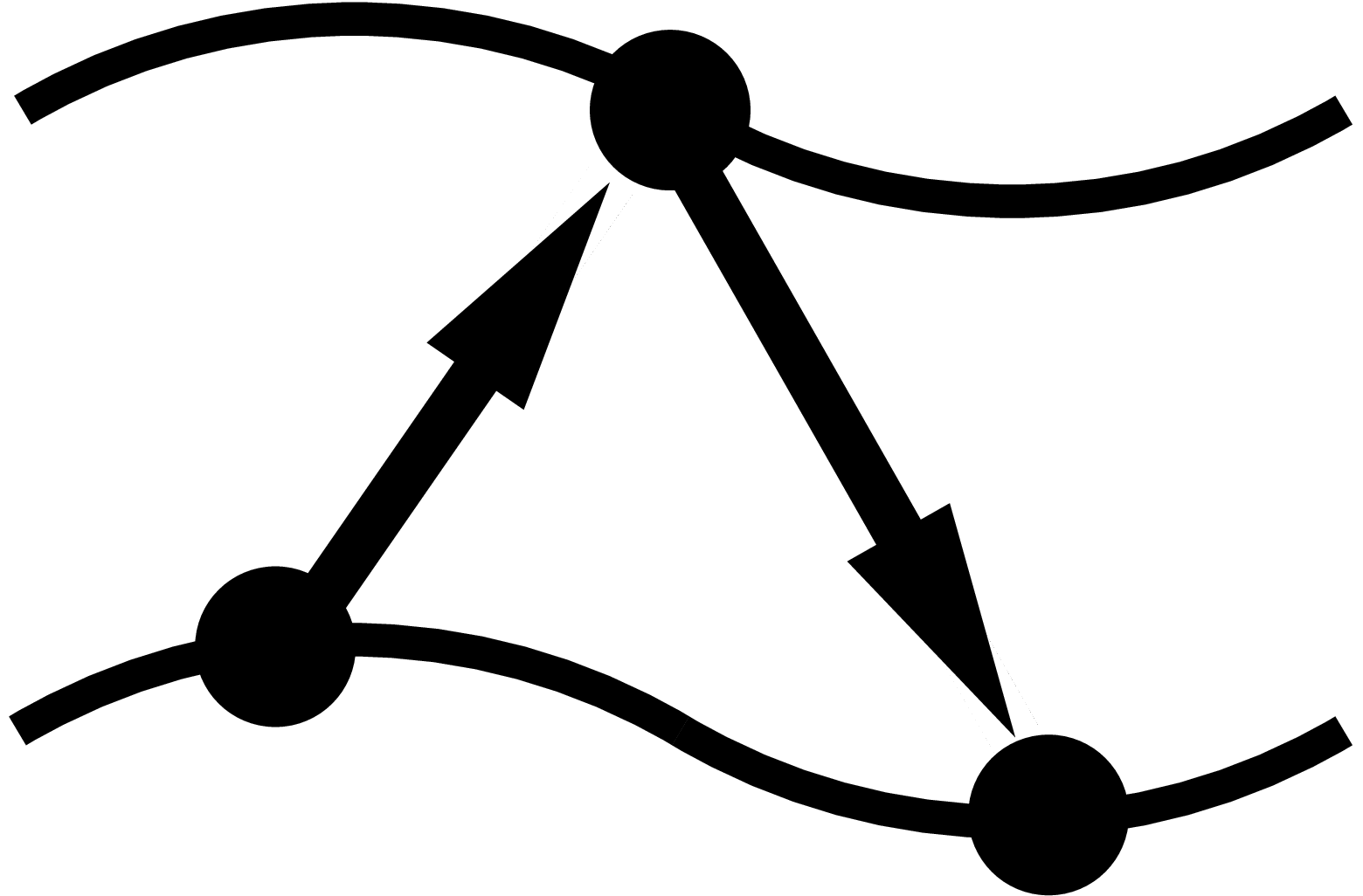}+\fig{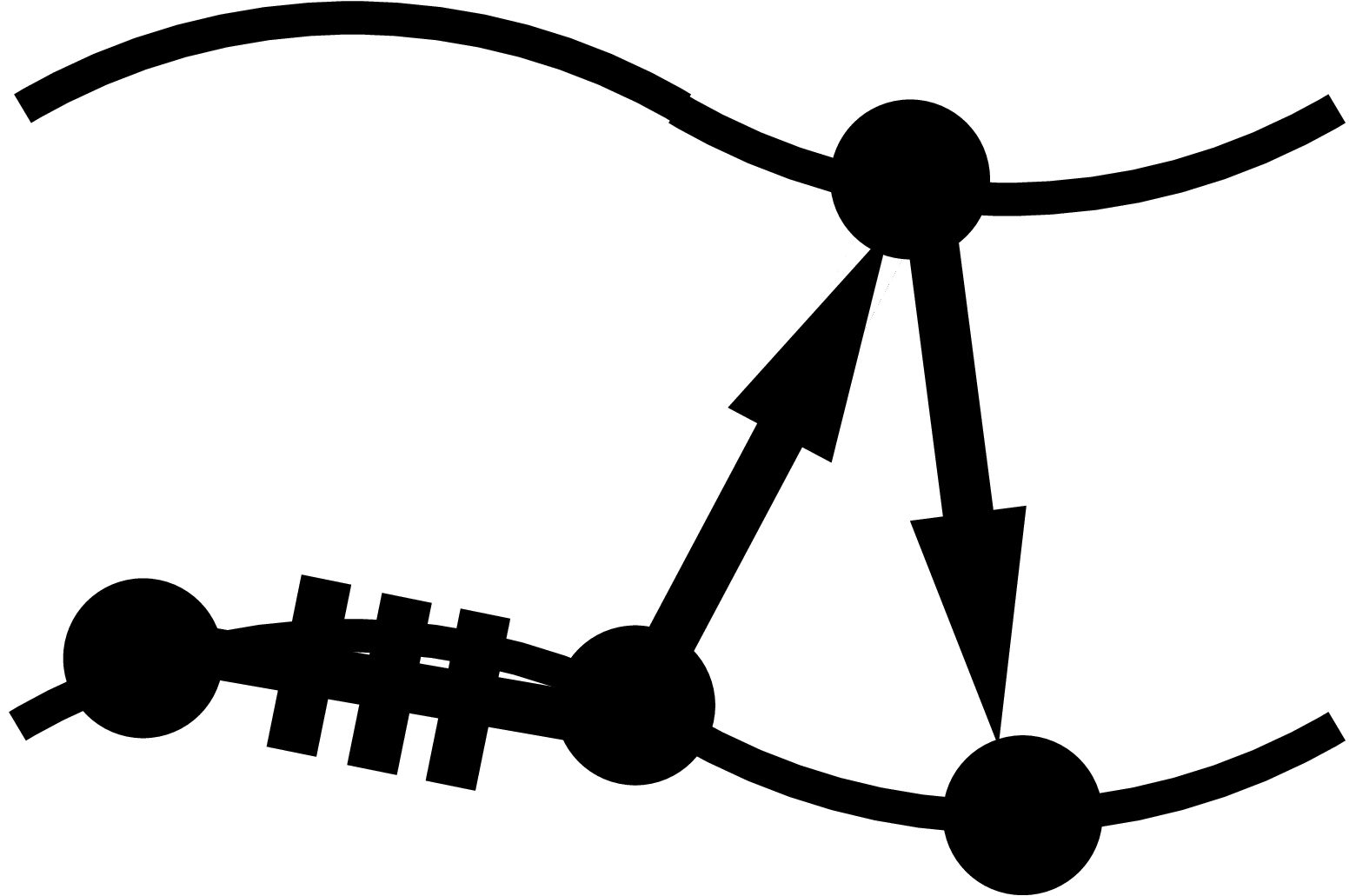}\nonumber\\
&&-\fig{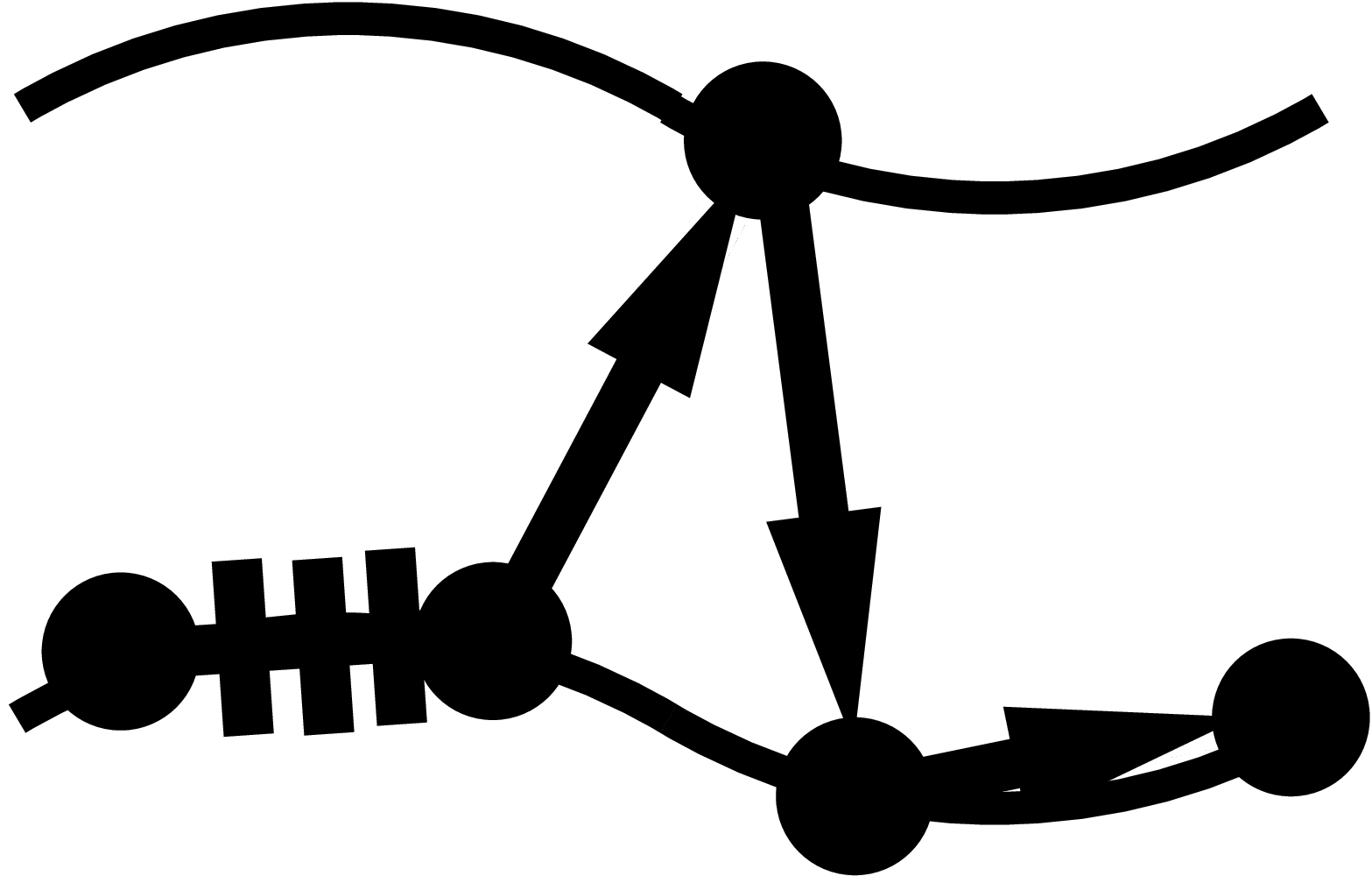}-\fig{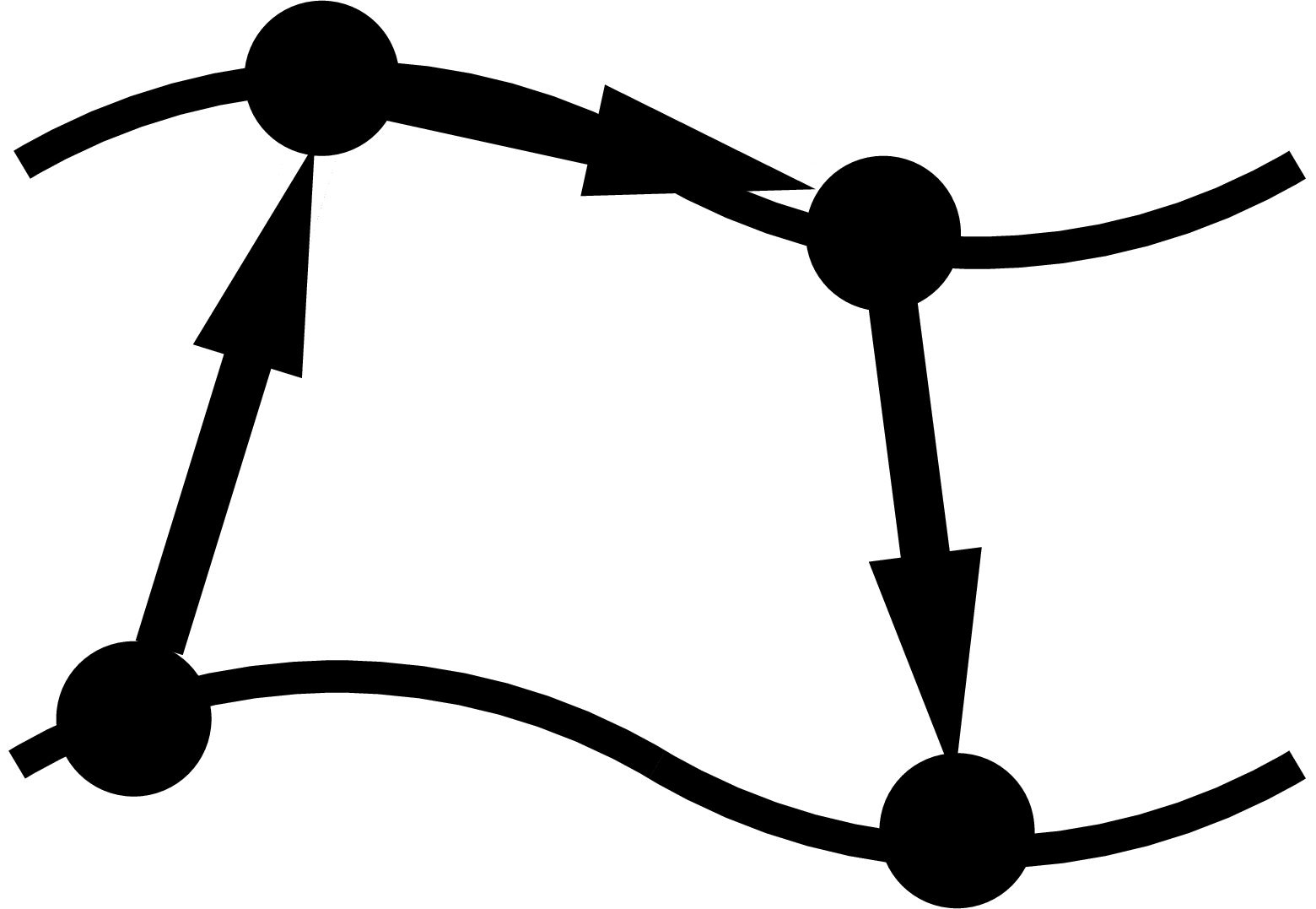}\nonumber\\
&&+\fig{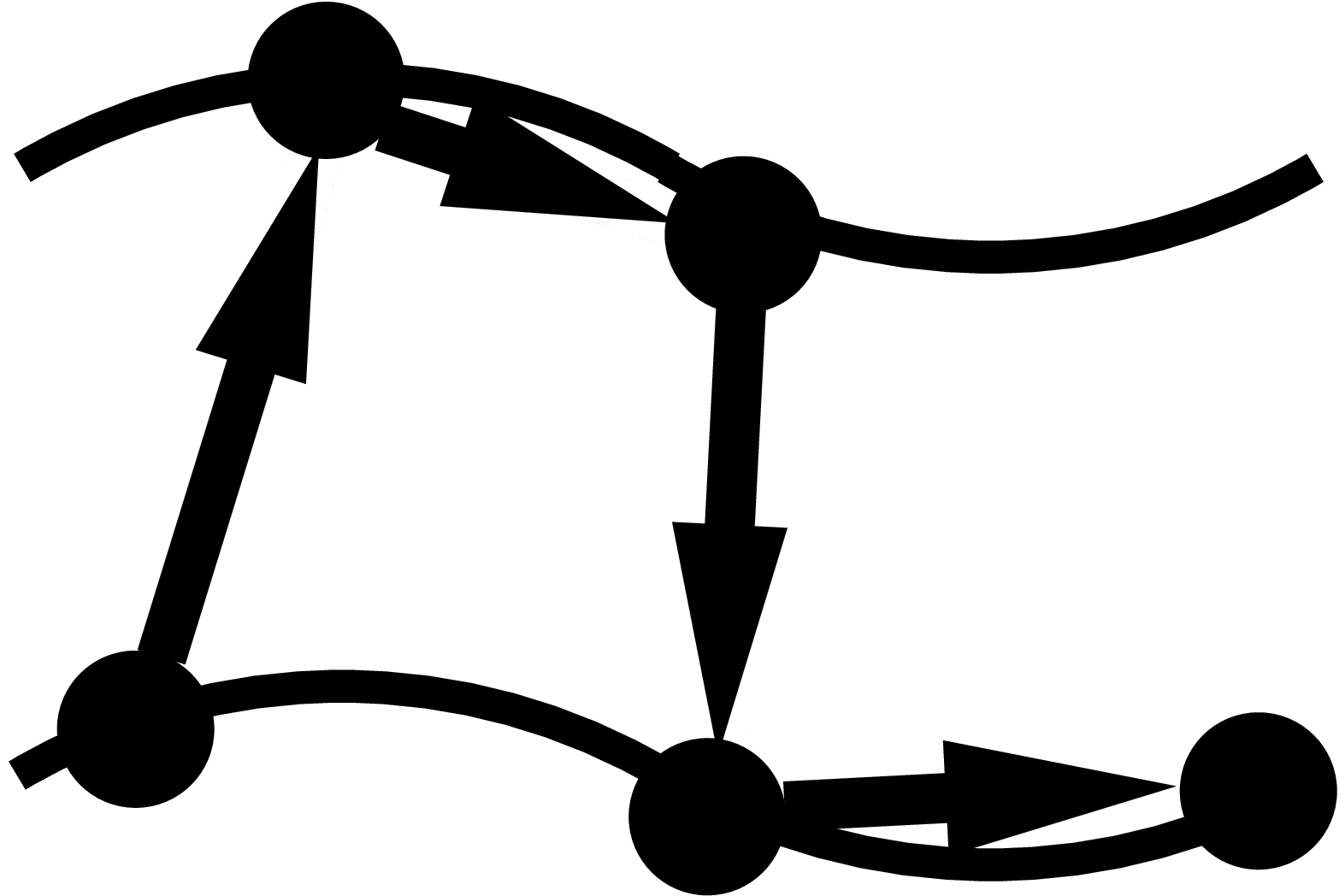}+\cdots,
\label{omega21-2}
\end{eqnarray}
and so on.
Each diagram has segments on each interface connected via $\partial K/\kappa-$type 
bonds that link the wall and the gas-liquid interface.
The leftmost segment can only contain $U-$type bonds (independent of the surface on which it lies). Otherwise,
they only have $\partial K/\kappa$ bonds. The coefficient associated with 
each diagram is now $(-1)^{l+m+o}$, with $l$ 
being the total number of $U-$bonds in the leftmost segment, $m$
the total number of $\partial_n K/\kappa-$type bonds on the substrate (not on the liquid-gas interface), and
$o$ the number of $\partial K/\kappa$ bonds between the wall and the gas-liquid interface that emerge 
from the substrate and point to the gas-liquid interface.

Similarly, the order-parameter profile has an alternative diagrammatic expansion
\begin{eqnarray}
\delta m_\Xi&&=-m_0\Bigg[\fig{diagram59}-\fig{diagram60}+\cdots\nonumber\\
&&-\fig{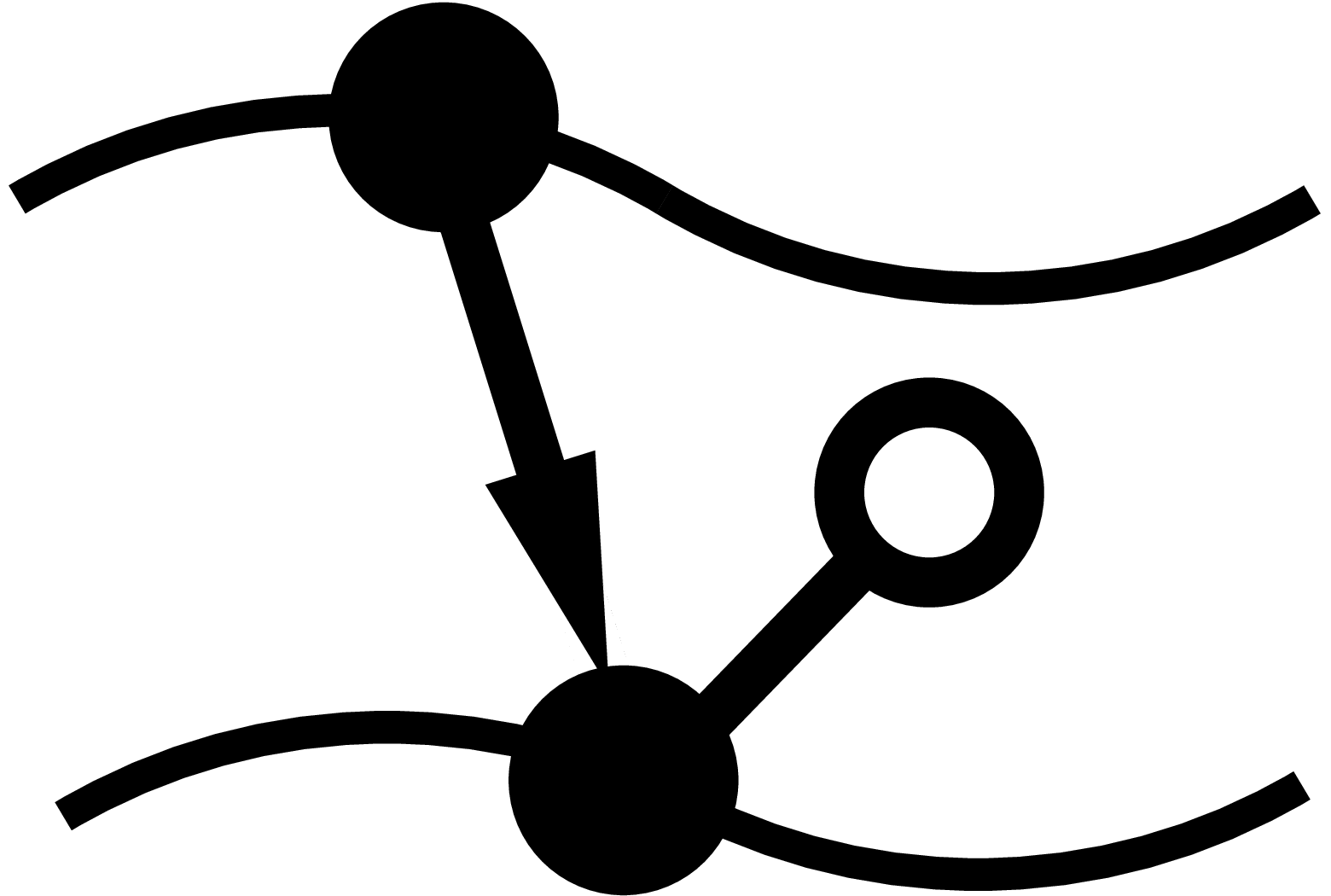}+\fig{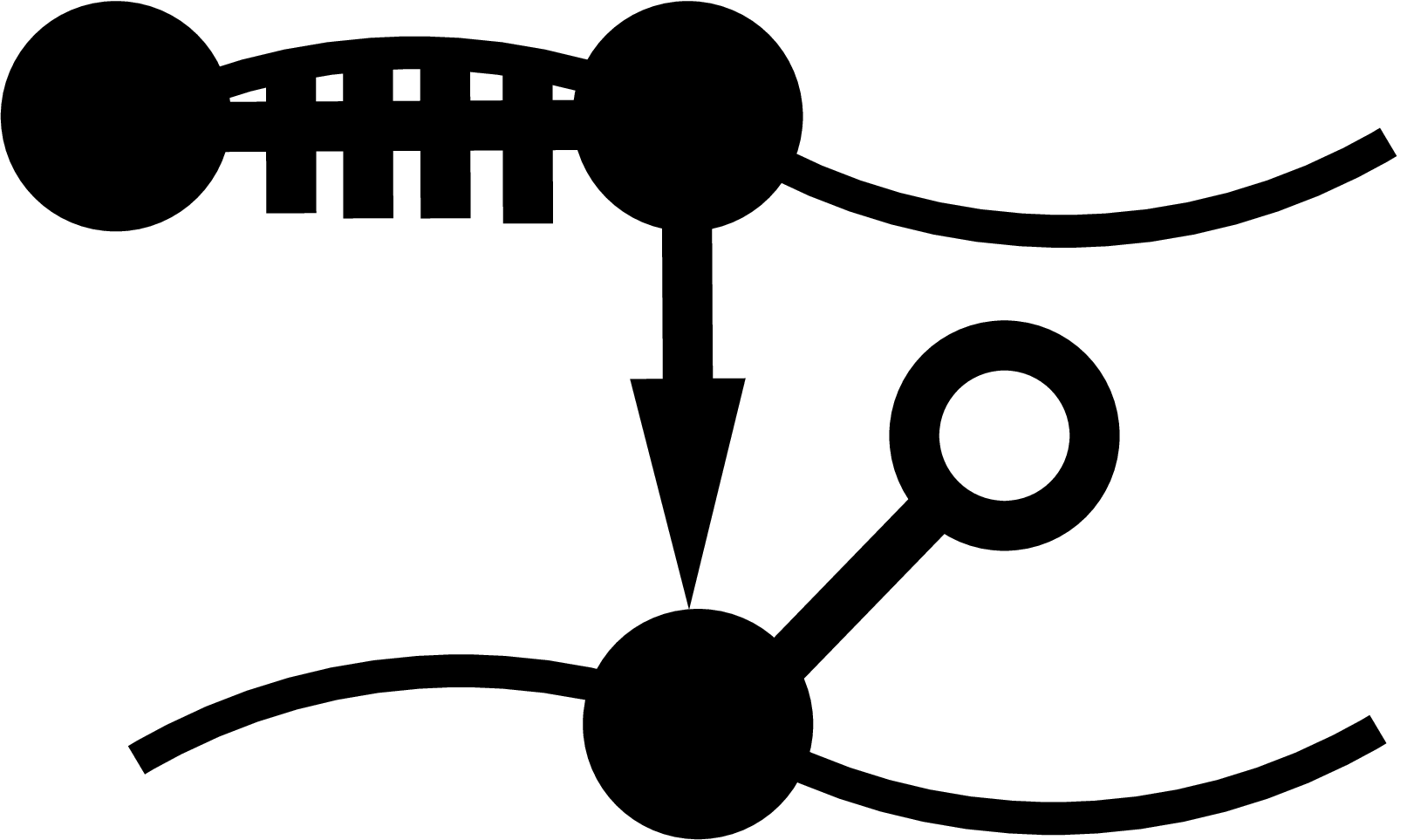}+\cdots\nonumber\\
&&+\fig{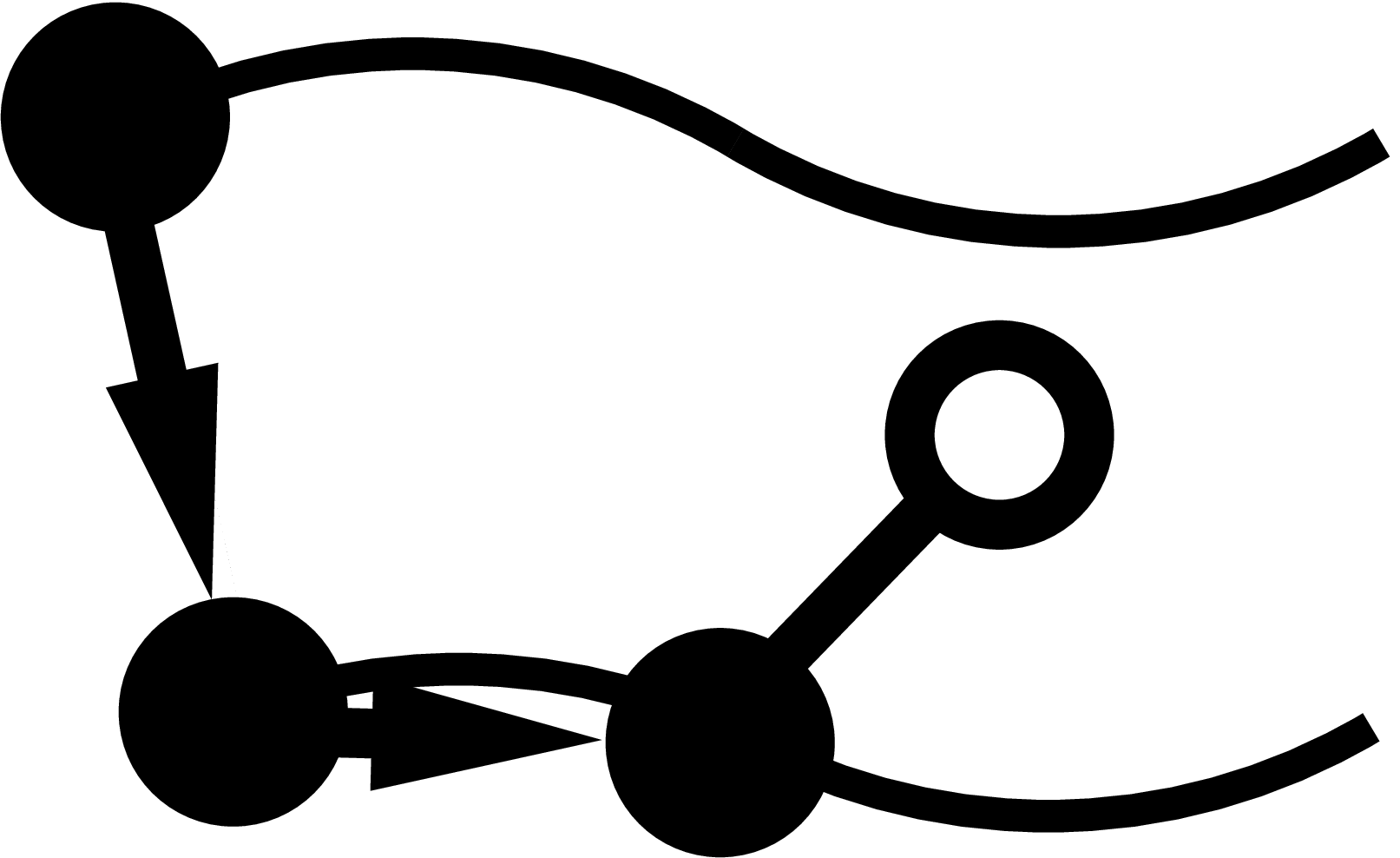}+\cdots\Bigg]\nonumber\\
&&+\delta m_1 \Bigg[\fig{diagram64}-\fig{diagram65}
+\cdots\nonumber\\
&&+\fig{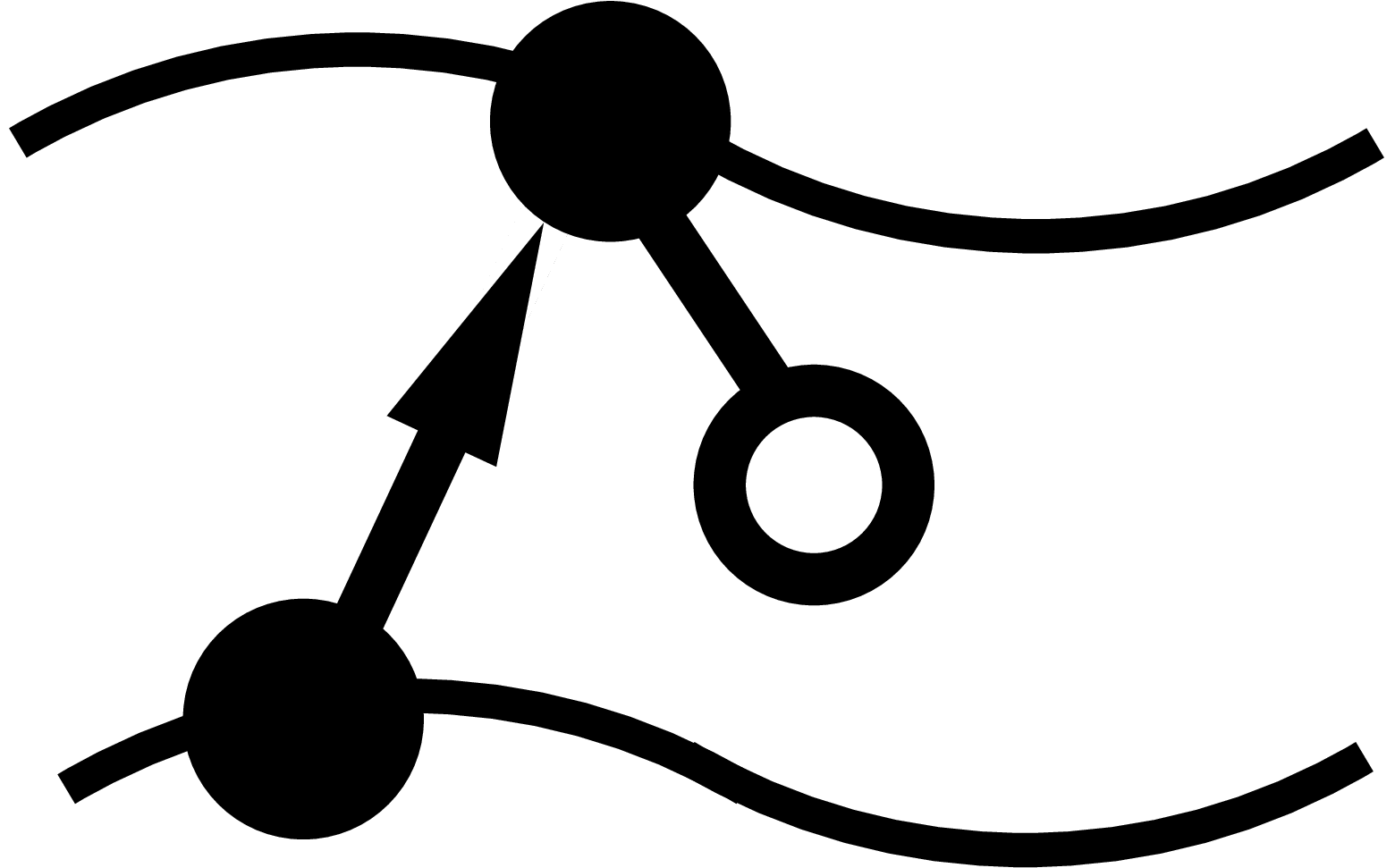}-\fig{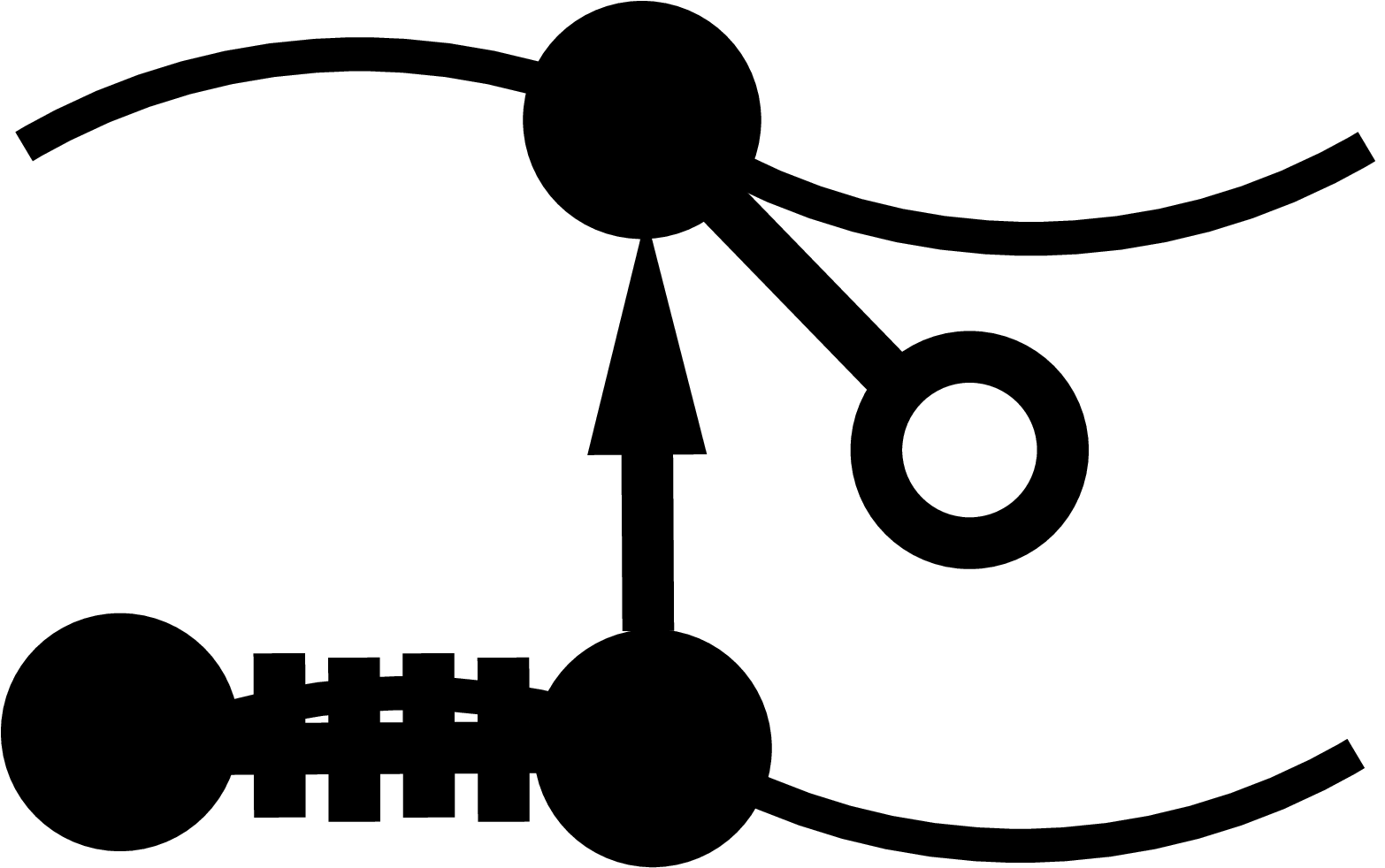}+\cdots \nonumber\\
&&+\fig{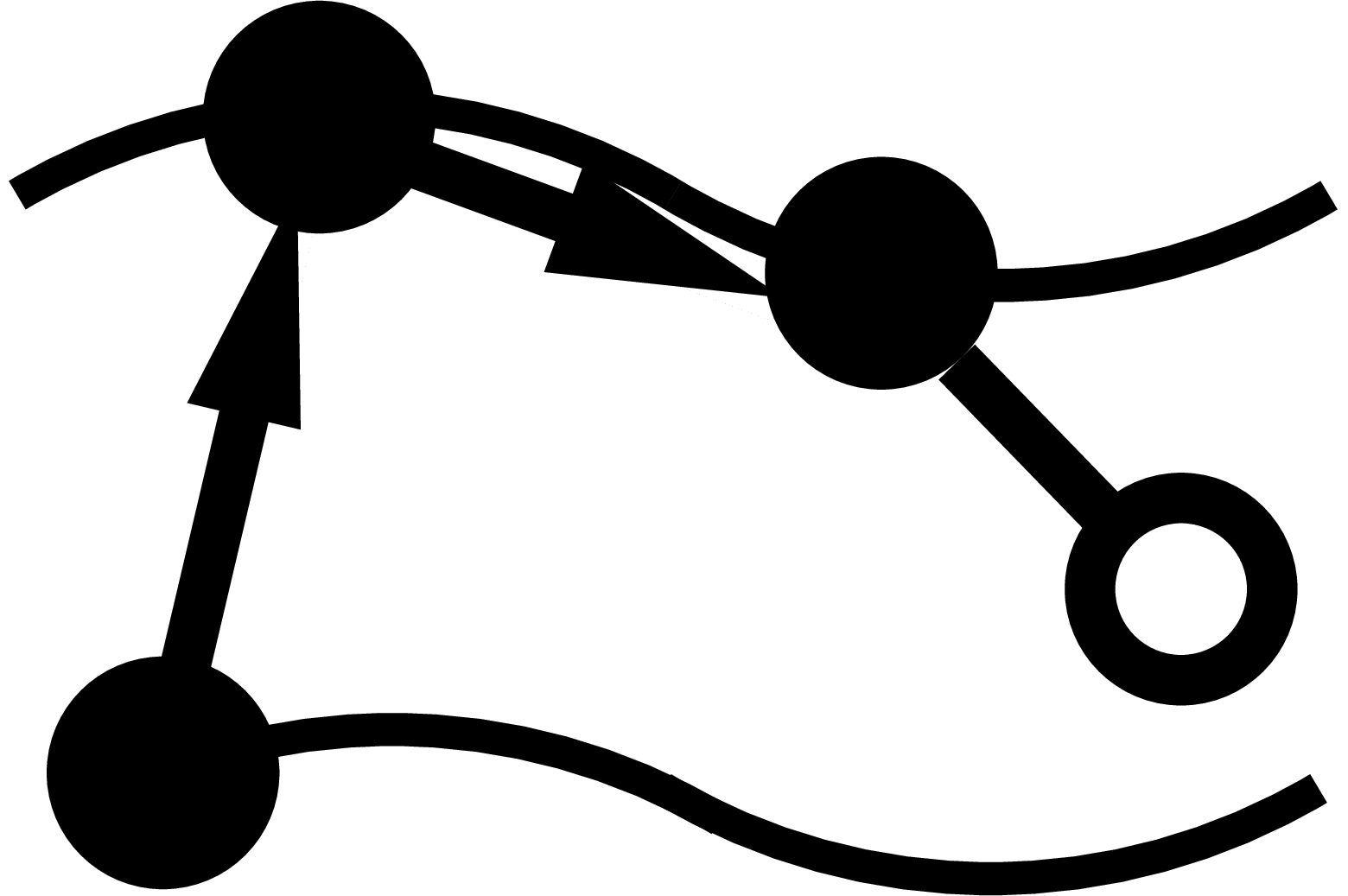}+\cdots\Bigg]
\label{deltamdiagram5}
\end{eqnarray} 
The diagrams start with a (left) segment either on the substrate or on the liquid-gas interface, which 
only can have $U-$type bonds. After that, there are $\partial_n K/\kappa-$type bonds connecting the wall
and the gas-liquid interface, followed
by segments on the corresponding interface that can only have $\partial_n K/\kappa-$bonds. Finally, there is
a $K-$type bond connecting one interface to the position $\mathbf{r}$. 
Now the sign in front of each diagram is $(-1)^{l+m+o'}$, with $l$ being the number of $U-$type 
bonds on the leftmost 
segment, $m$ the number of $\partial_n K/\kappa-$type bonds on the substrate, and $o'$ the number of
$\partial_n K/\kappa$ bonds that emerge from the liquid-gas interface and point to the substrate.

\subsection{Flat substrates}
In the previous sections we pointed out that the decorated diagrams constitute the different features of our formulation of the nonlocal model. In order to better appreciate these aspects, we consider the binding potential for the case of a flat substrate. The exact binding potential admits a curvature expansion, but even at leading order it differs slightly from the binding potential functional of the original formulation.
We start by considering the diagrams contained in $\Omega_{1}^{1}$. From the results of the previous sections we have that
\begin{eqnarray}\nonumber
\label{reduction_01}
\Omega_{1}^{1} & = & -\fig{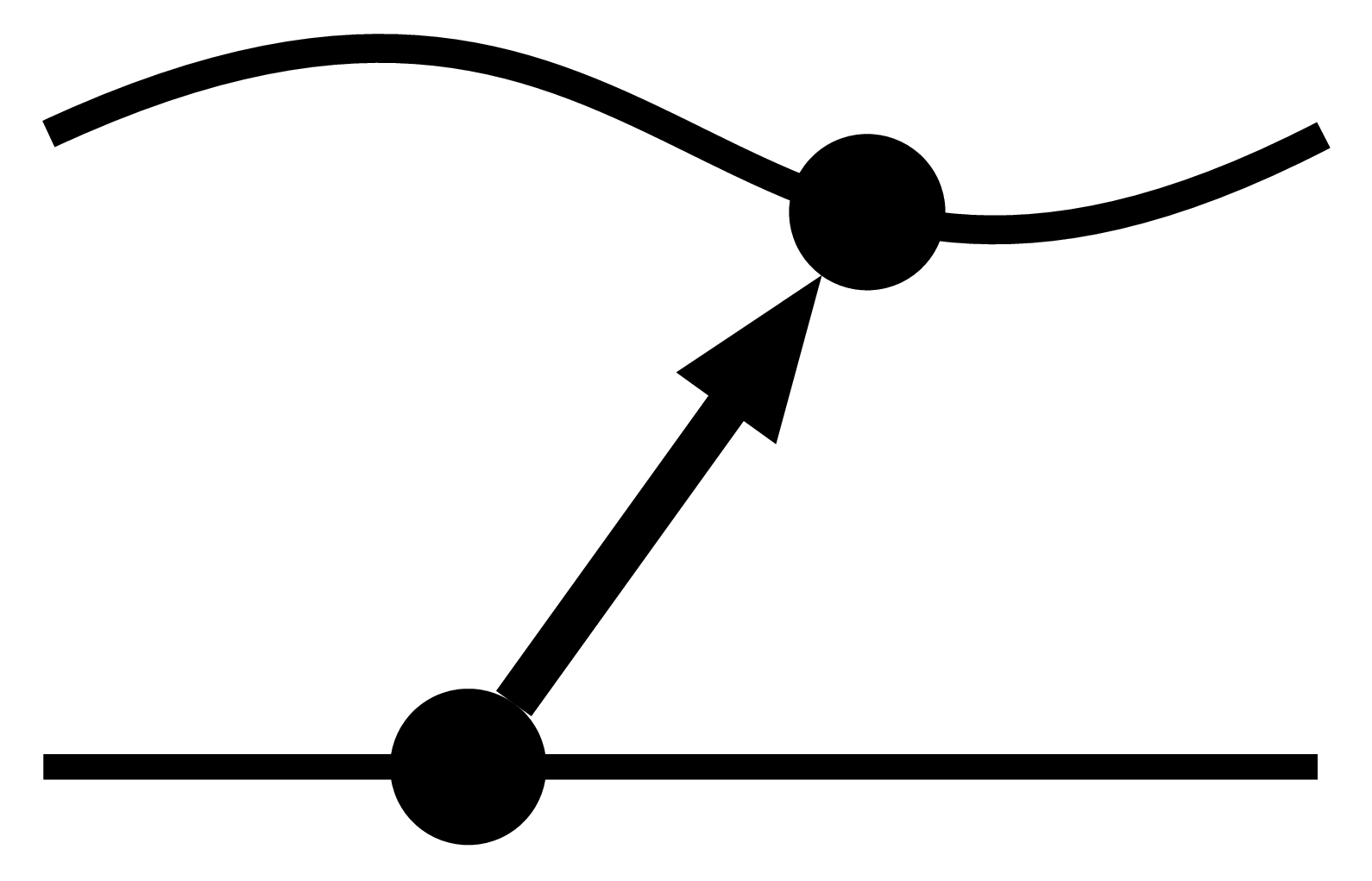} + \fig{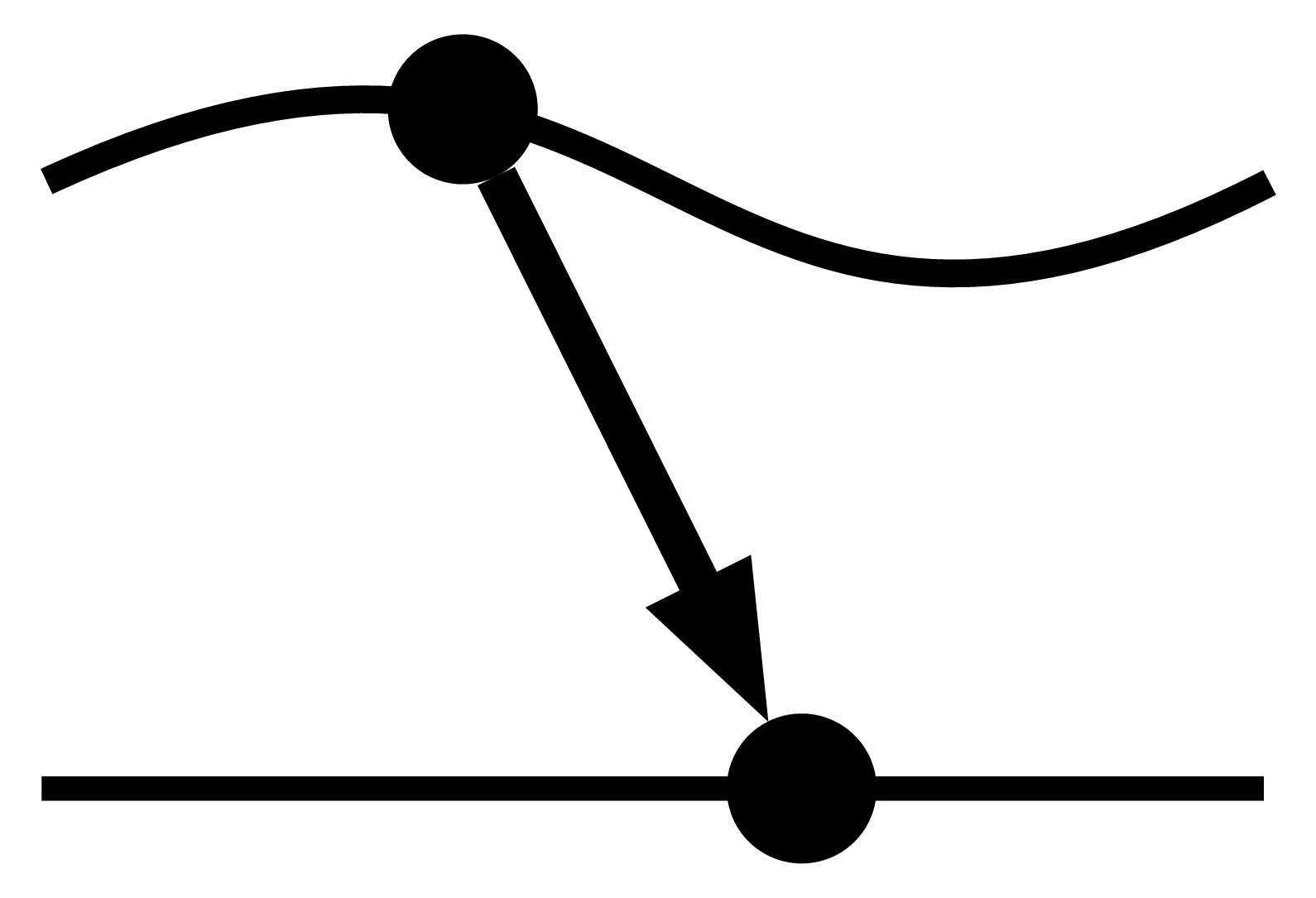} + \fig{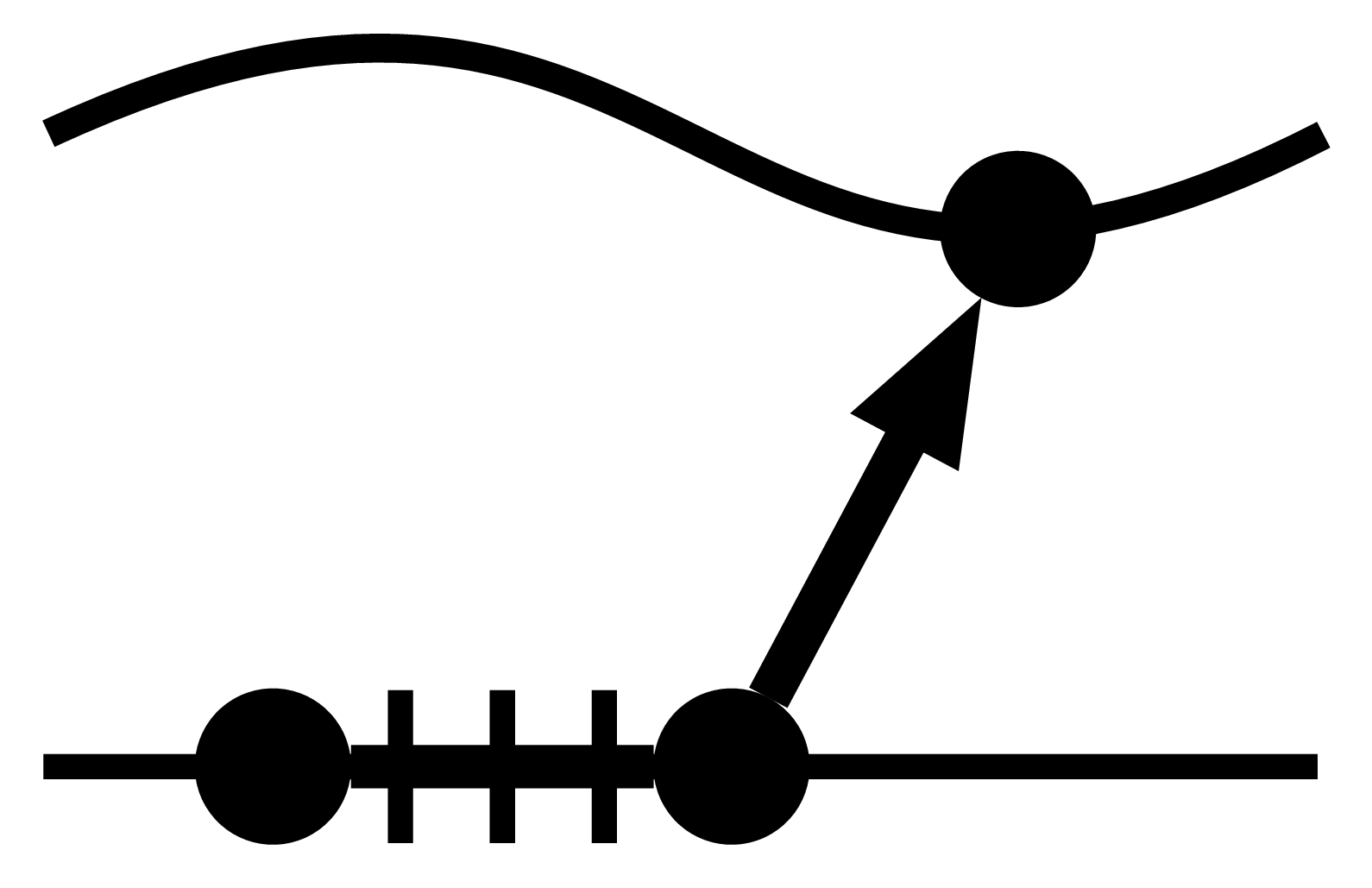}  \\ \nonumber
&& -\fig{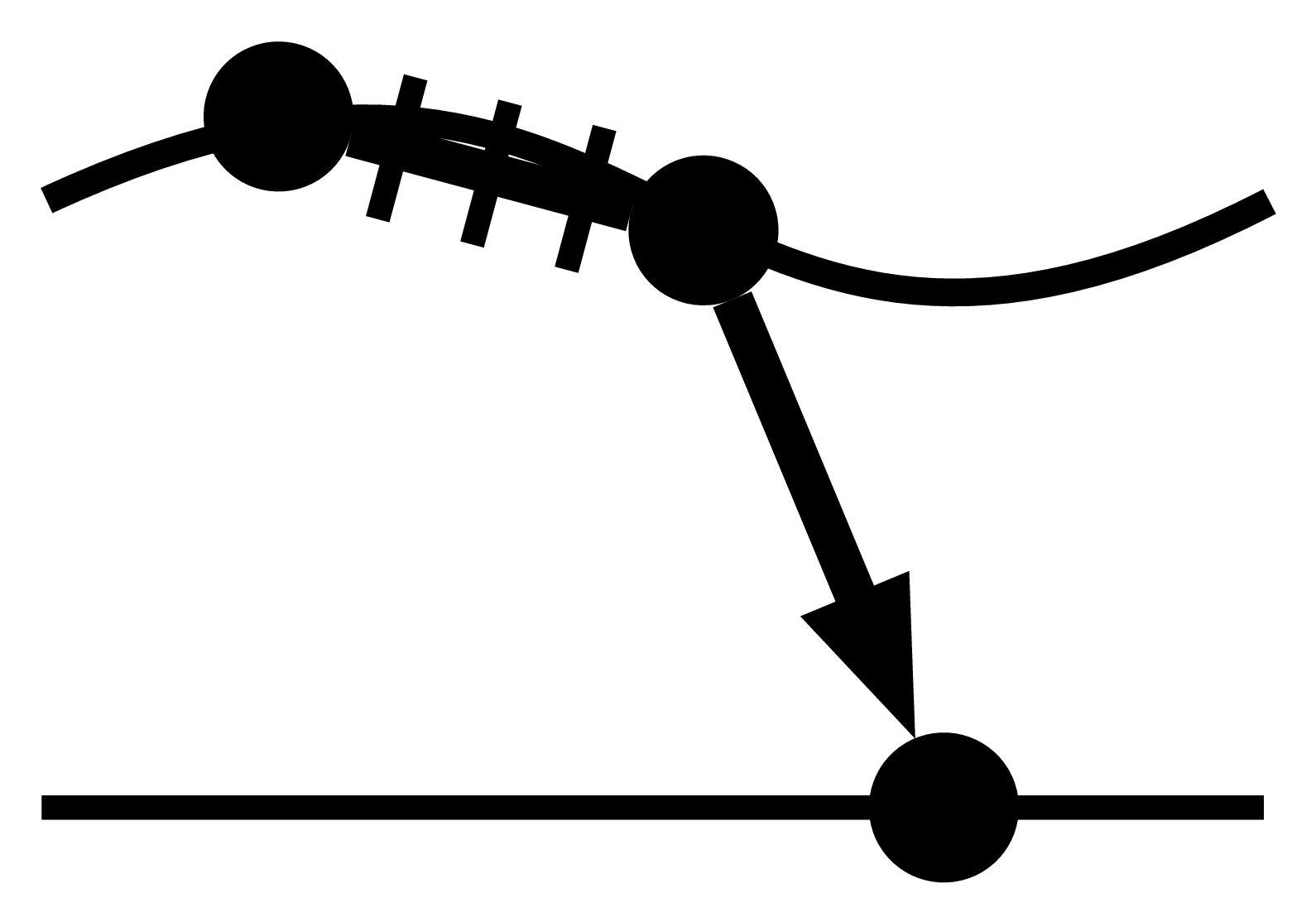} + \fig{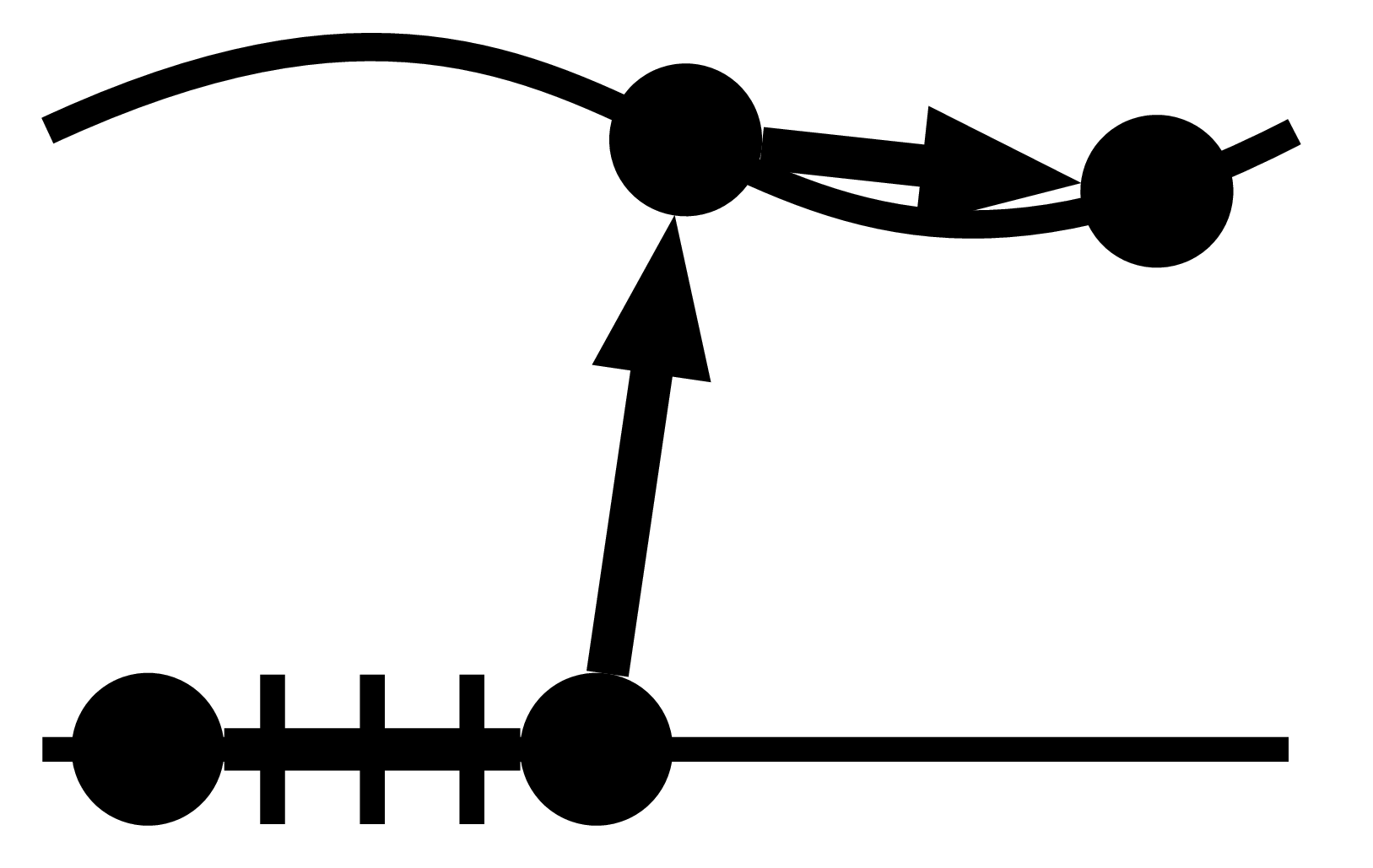} + \fig{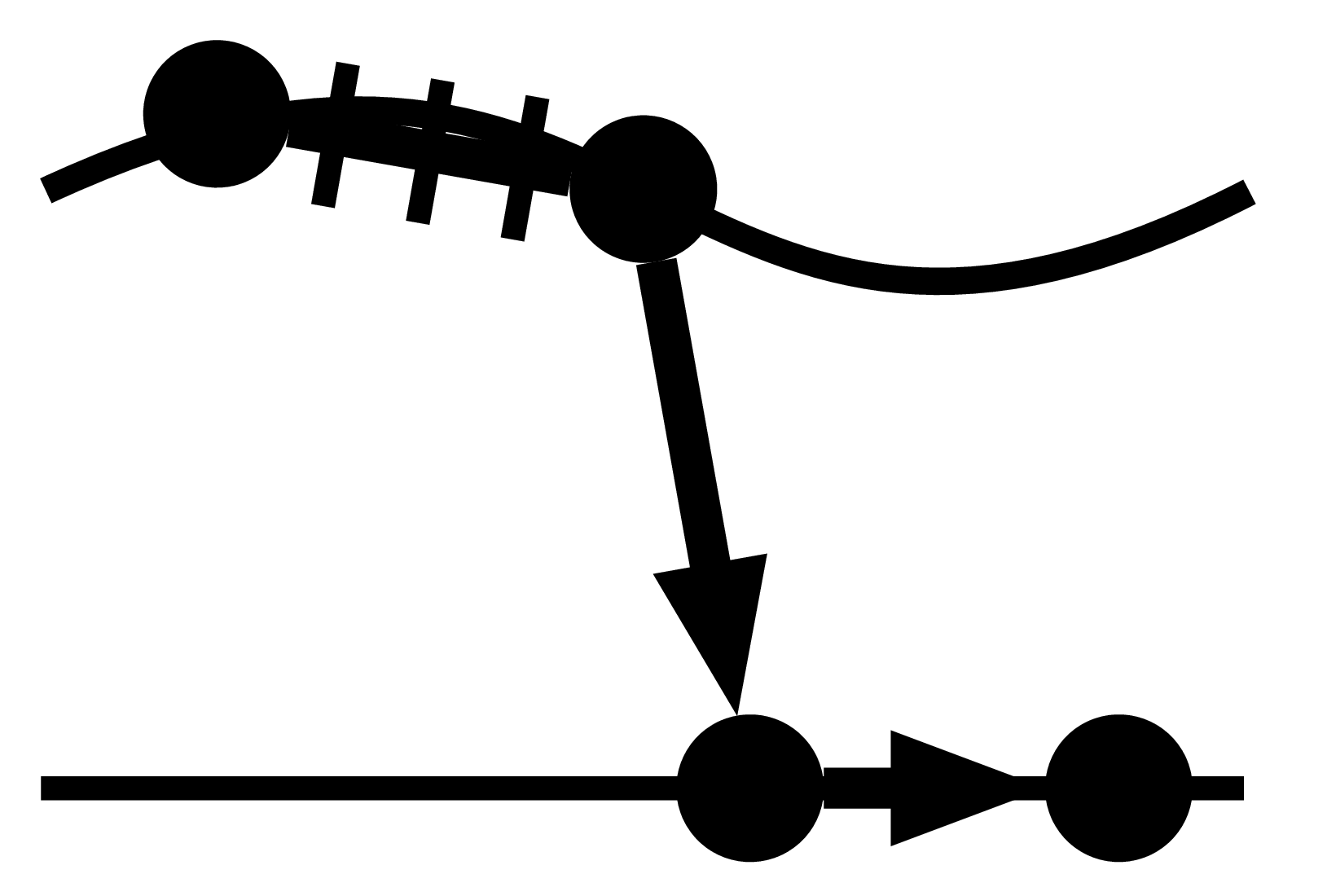} +\cdots , \\
\end{eqnarray}
where the flatness of the substrate has enormously simplified the diagrammatic structure. This 
simplification is due to the vanishing of a large class of wetting diagrams, and can be summarized by the 
following reduction lemmas.\\
\emph{Lemma 1}. We have
\begin{equation}
\label{Lemma1}
\fig{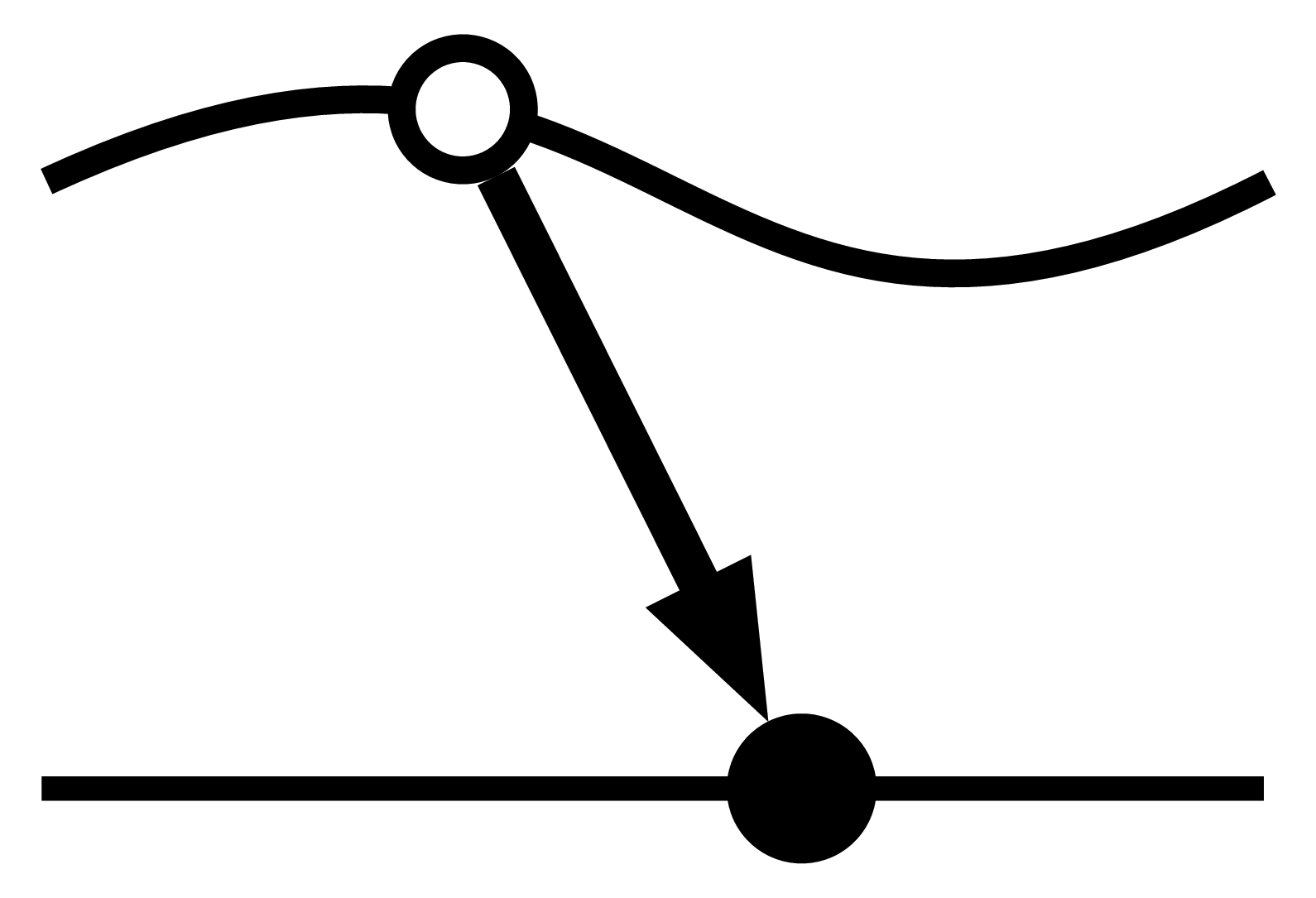} = \fig{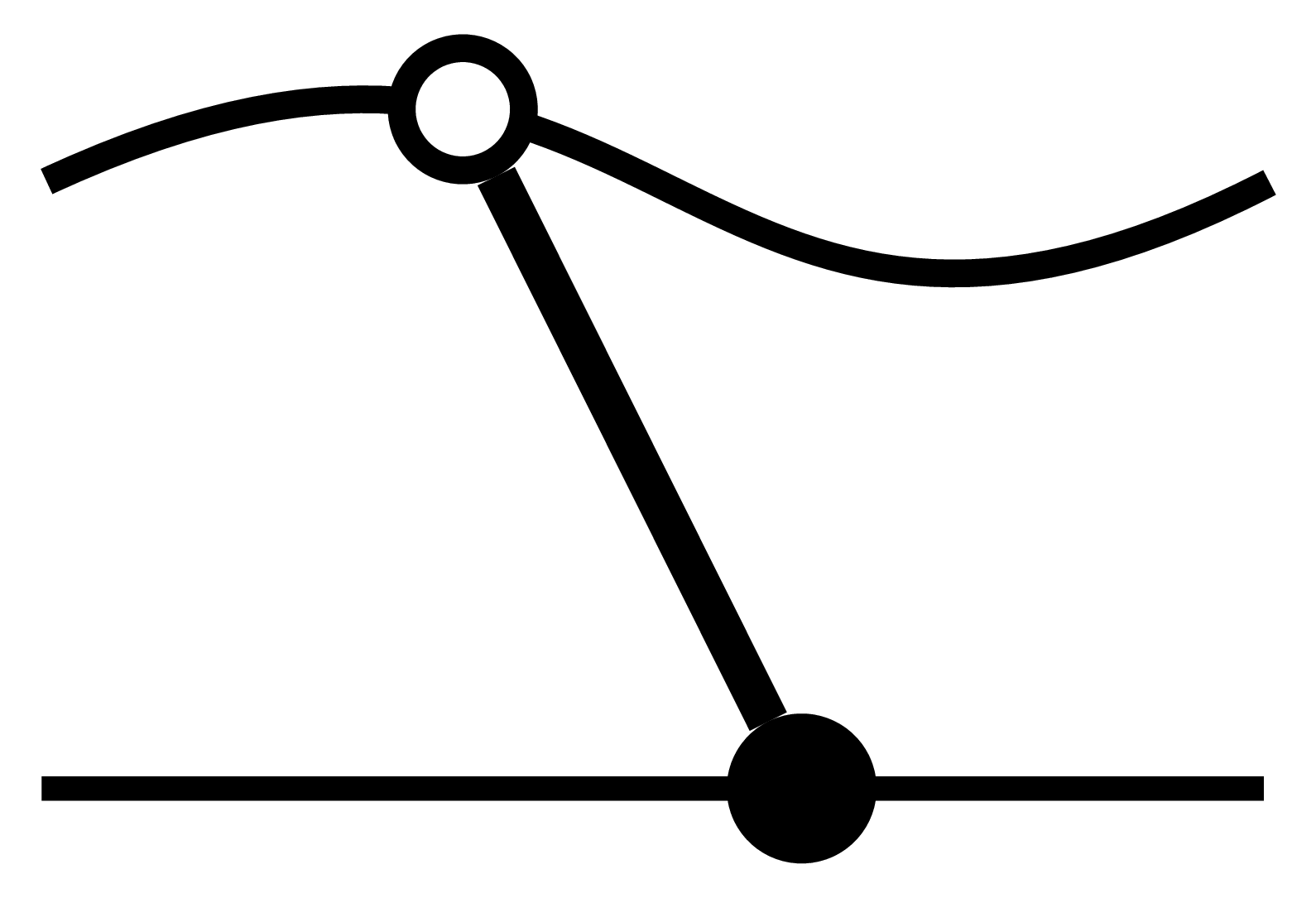}.
\end{equation}
\emph{Lemma 2}. We have
\label{Lemma2}
\begin{equation}
\fig{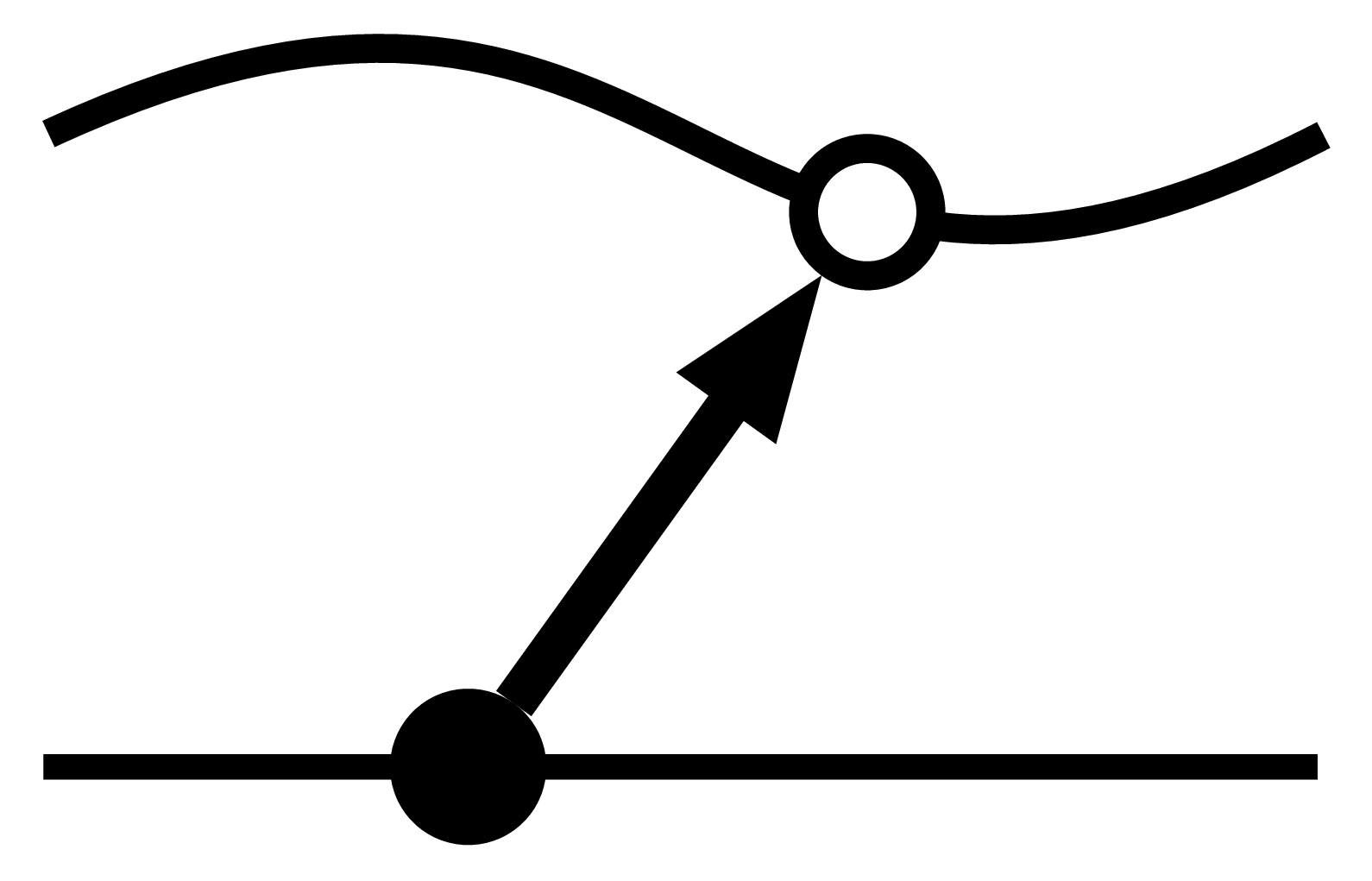} = -\cos\alpha(\textbf{s}) \fig{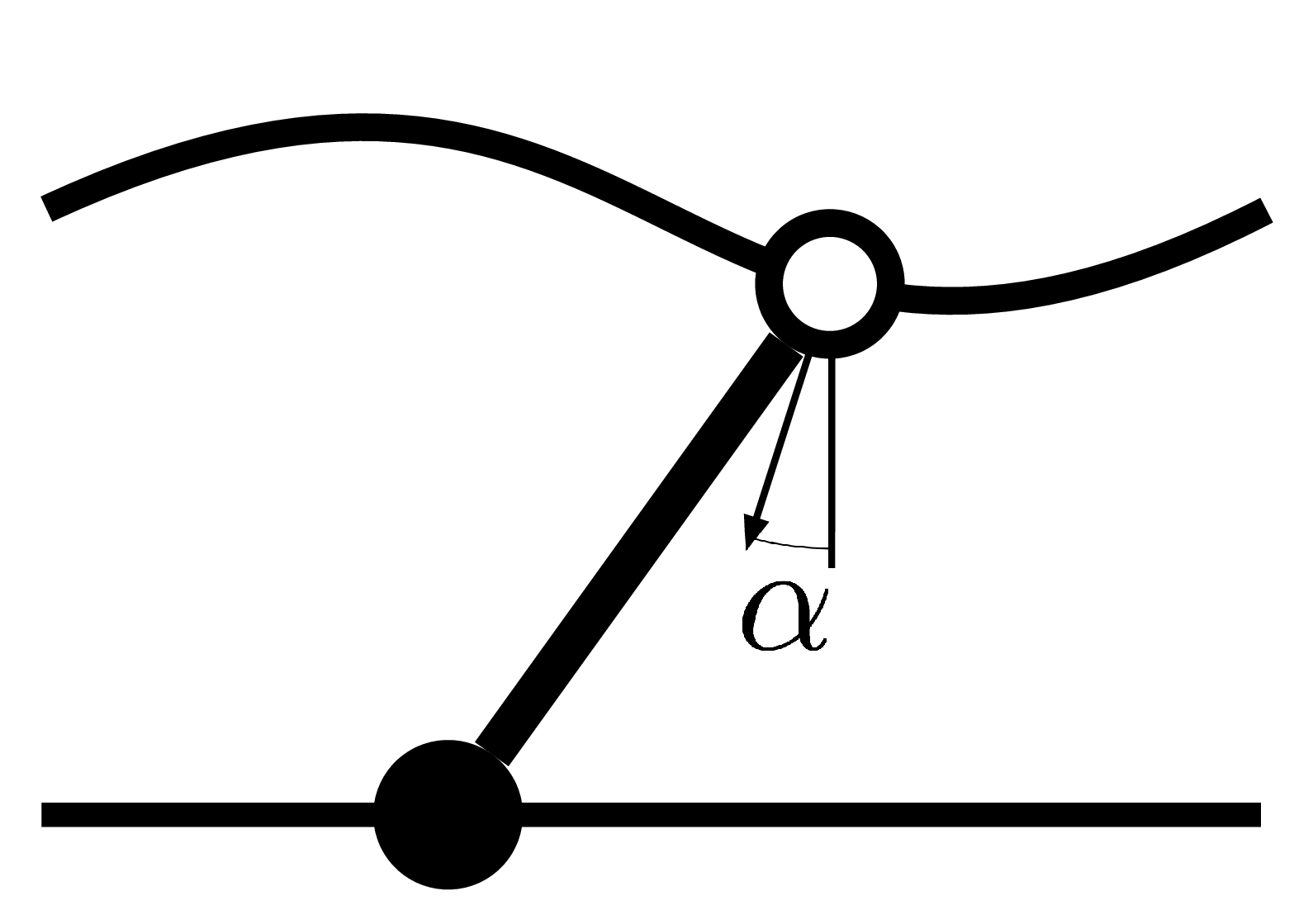},
\end{equation}
where $\alpha(\textbf{s})$ is the angle formed by the normal vector and the vertical direction.\\
\emph{Lemma 3}. We have
\label{Lemma3}
\begin{eqnarray}
\largefig{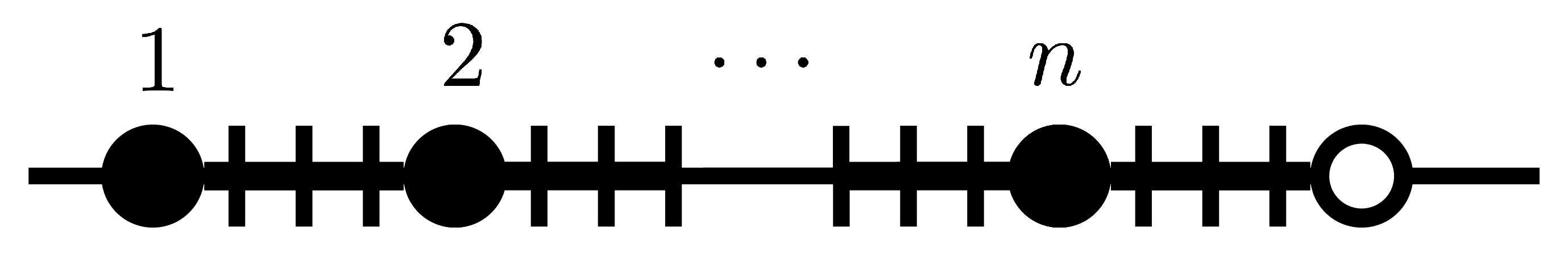} & = & 0 , \\
\largefig{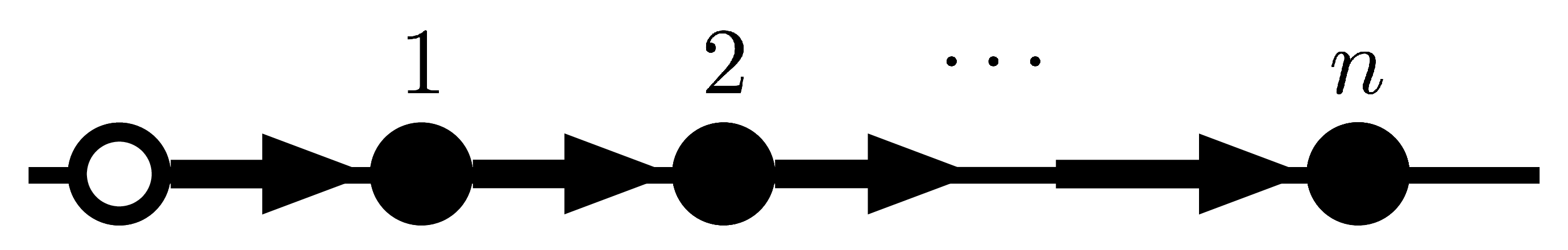} & = & 0 .
\end{eqnarray}
In addition to these rules, we recall that the decorated diagrams contribute higher-order corrections in the curvature expansion. In particular, a wetting diagram with a chain of $n$, $U$-type bonds on the fluid interface belongs to ${O}\left(R^{-2n}\right)$, while instead a chain of $m$ arrow diagrams along the substrate belongs to ${O}\left(R^{-m}\right)$, where $R$ is a typical radius of curvature. Therefore, at the leading order of a curvature expansion only the first two addends survive in Eq.~(\ref{reduction_01}). Then, due to Lemmas
1 and 2, we can further simplify the leading term and we are left with
\begin{equation}
\label{reduction_02}
\Omega_{1}^{1} \approx \left(1+\langle\cos\alpha\rangle_1\right) \fig{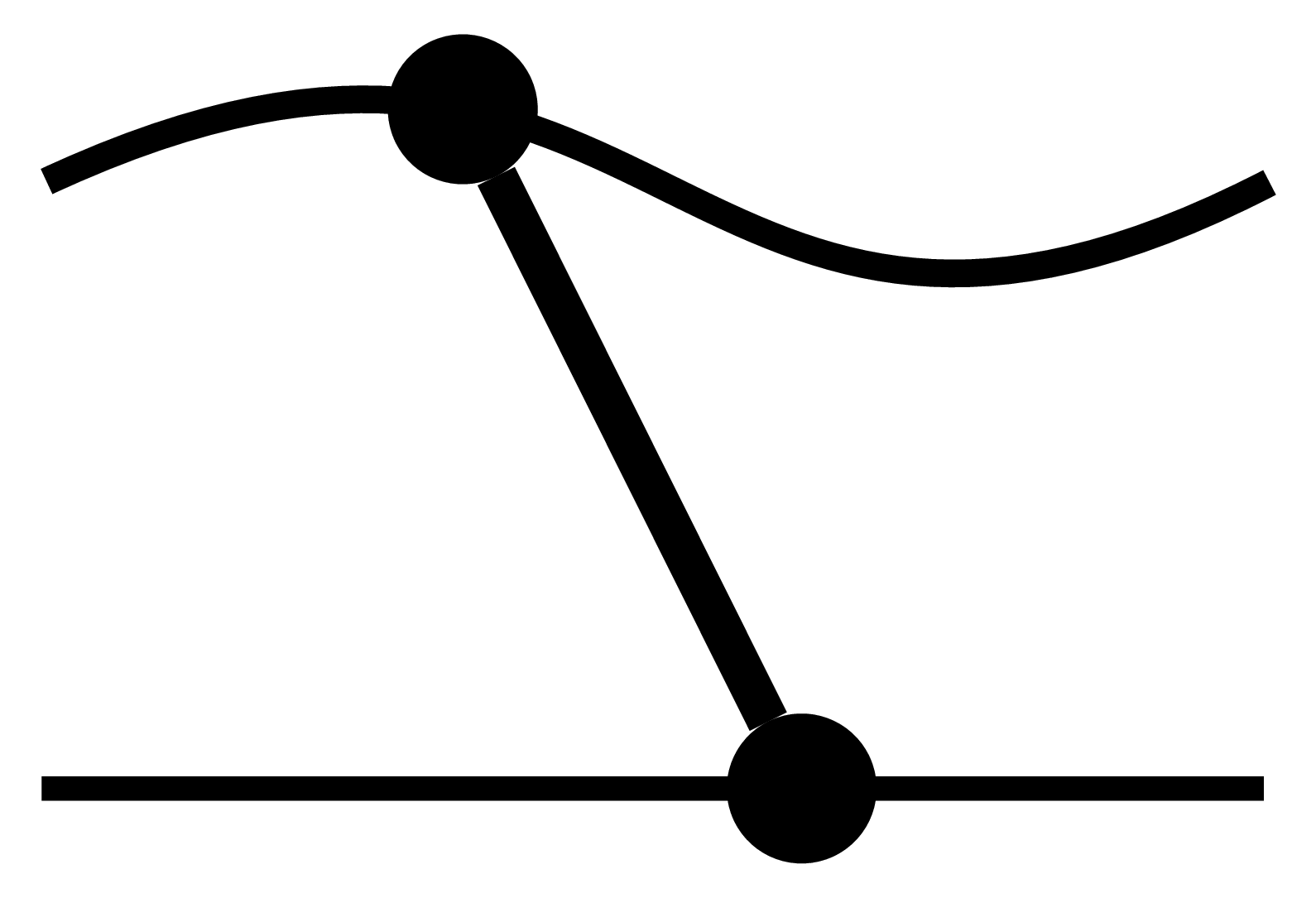}, 
\end{equation}
where
\begin{equation}
\langle\cos\alpha\rangle_n\equiv\frac{\int_\ell d\mathbf{s}
\cos \alpha(\mathbf{s})e^{-n\kappa \ell(\mathbf{s})}}{\int_\ell 
d\mathbf{s} e^{-n\kappa \ell(\mathbf{s})}},
\label{averagecosalpha}
\end{equation}
with $\ell(\mathbf{s})$ being the vertical distance from $\mathbf{s}$ to the 
substrate.
Already at leading order we can appreciate the different features of this exact formulation. Indeed, in the original formulation the expansion (\ref{reduction_01}) starts with the same diagram entering in (\ref{reduction_02}) but with a factor $2$ in front. The factor $1+\cos\alpha(\mathbf{s})$ strongly depends on the local orientation of the interface with respect to the planar wall and it is clear that the two formulations coincide only for parallel interfaces. However, for 
interfacial configurations that have a minimum height $\tilde\ell$ with 
respect to the substrate, the weighted average (\ref{averagecosalpha}) is 
near unity. More precisely, a saddle-point calculation shows that
$\langle \cos \alpha \rangle_1\sim 1-
(\tilde H/\kappa)$, where $\tilde H$ is the mean curvature at the 
interfacial position nearest the substrate.

It then is straightforward to prove, using the above lemmas, that the next-to-leading diagrams appearing in (\ref{reduction_02}) are of the form
\begin{equation}\nonumber
\largefig{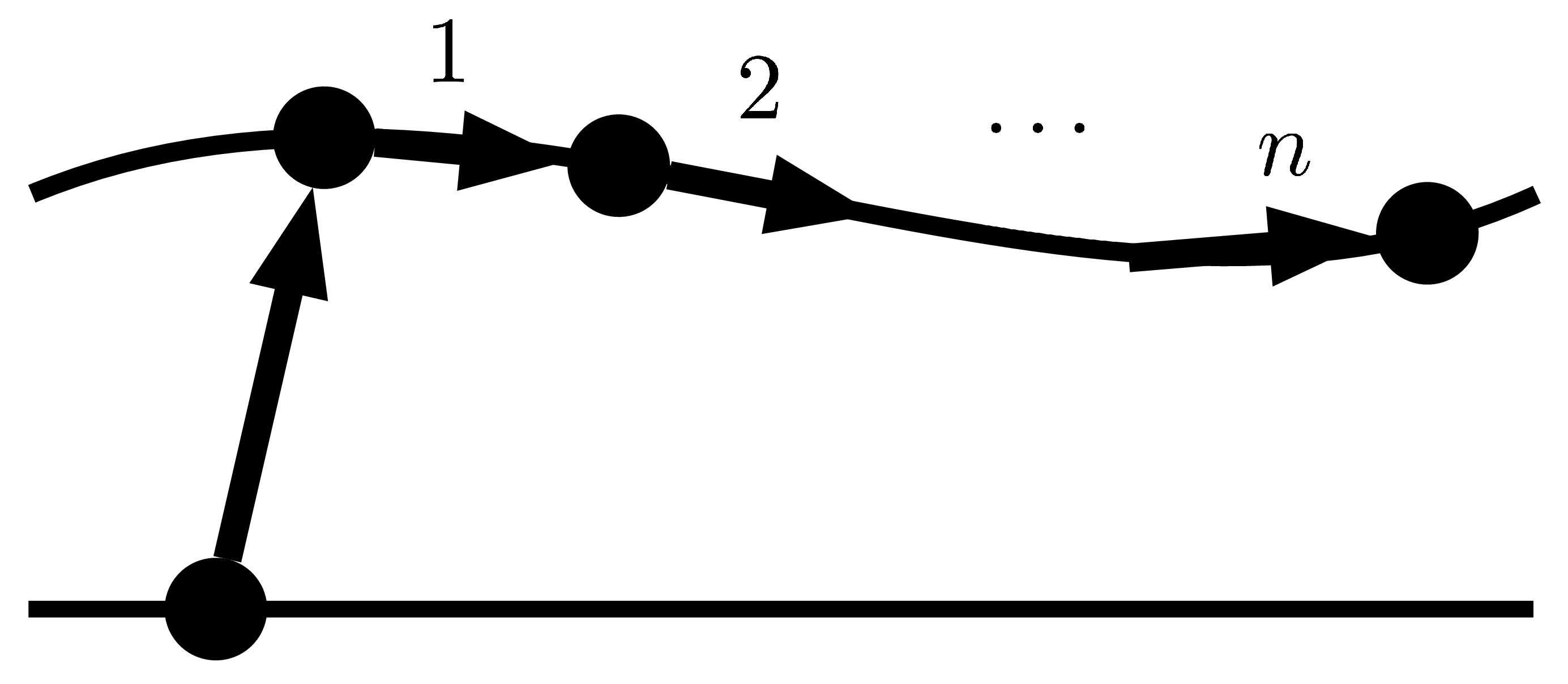} ,\qquad \largefig{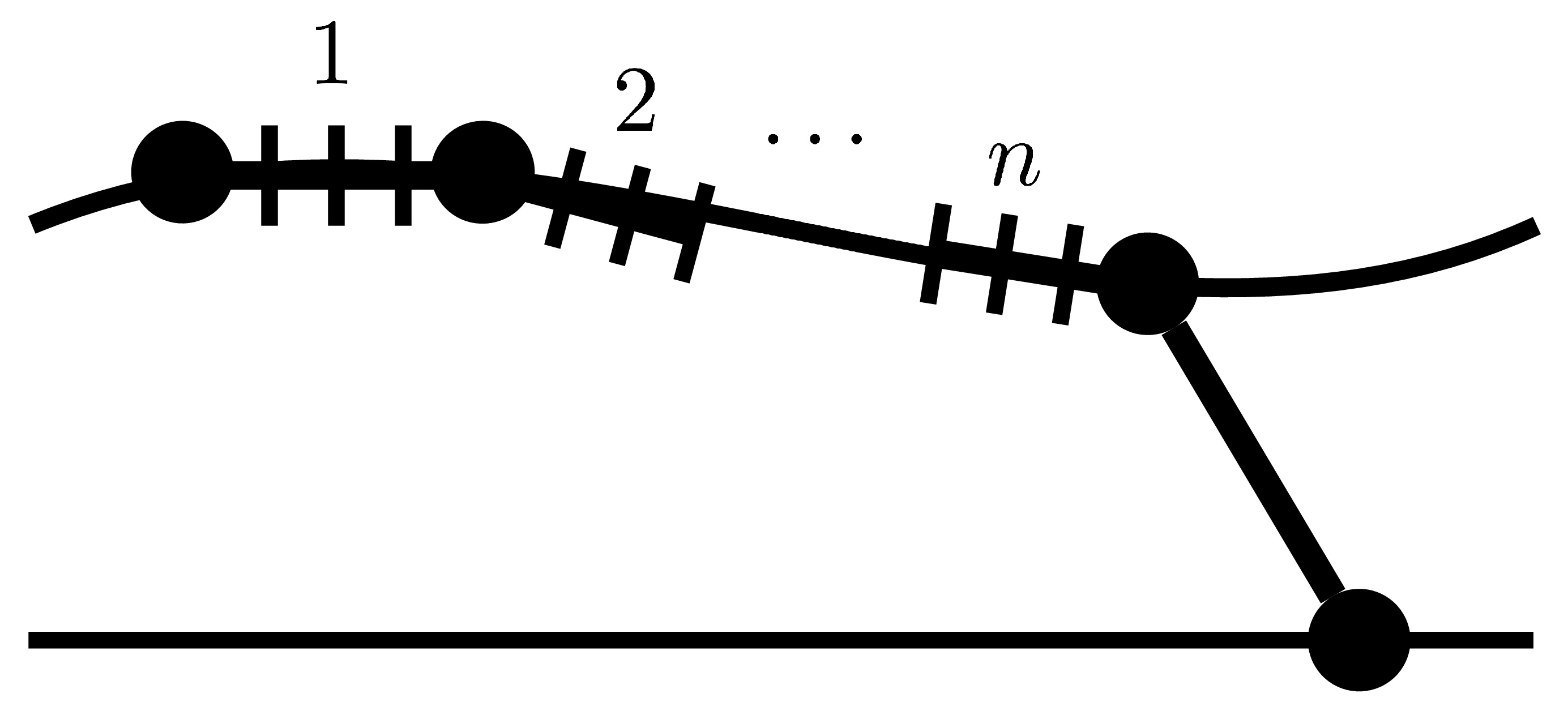} ,
\end{equation}
with a prefactor $-1$ and $(-1)^{n}$, respectively, at ${O}\left(R^{-n}\right)$ and ${O}\left(R^{-2n}\right)$, respectively.
These considerations apply also for the remaining classes of diagrams; in particular, for $\Omega_{2}^{1}$ we have
\begin{equation}
\label{reduction_03}
\Omega_{2}^{1} = - \fig{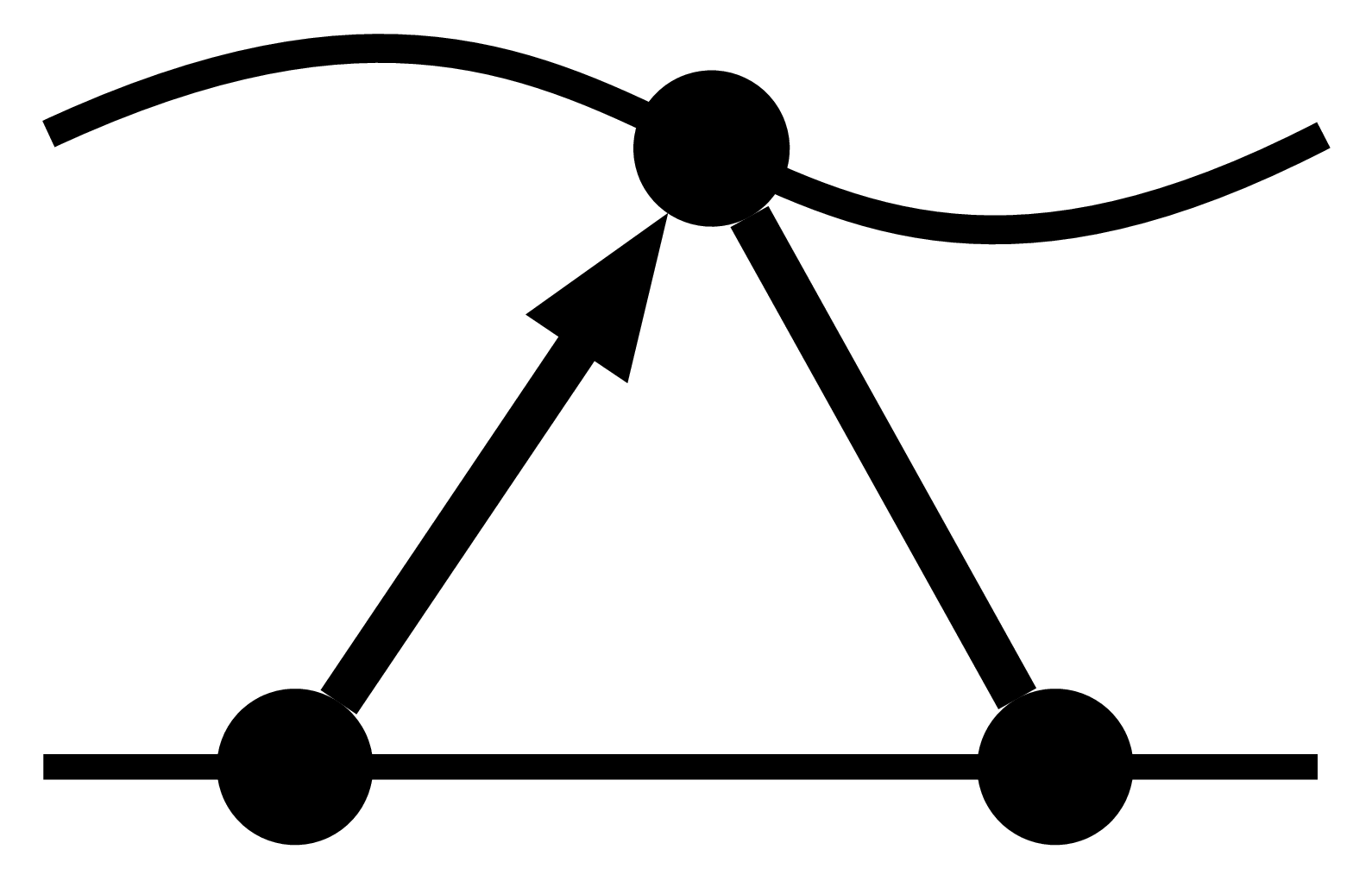} - \fig{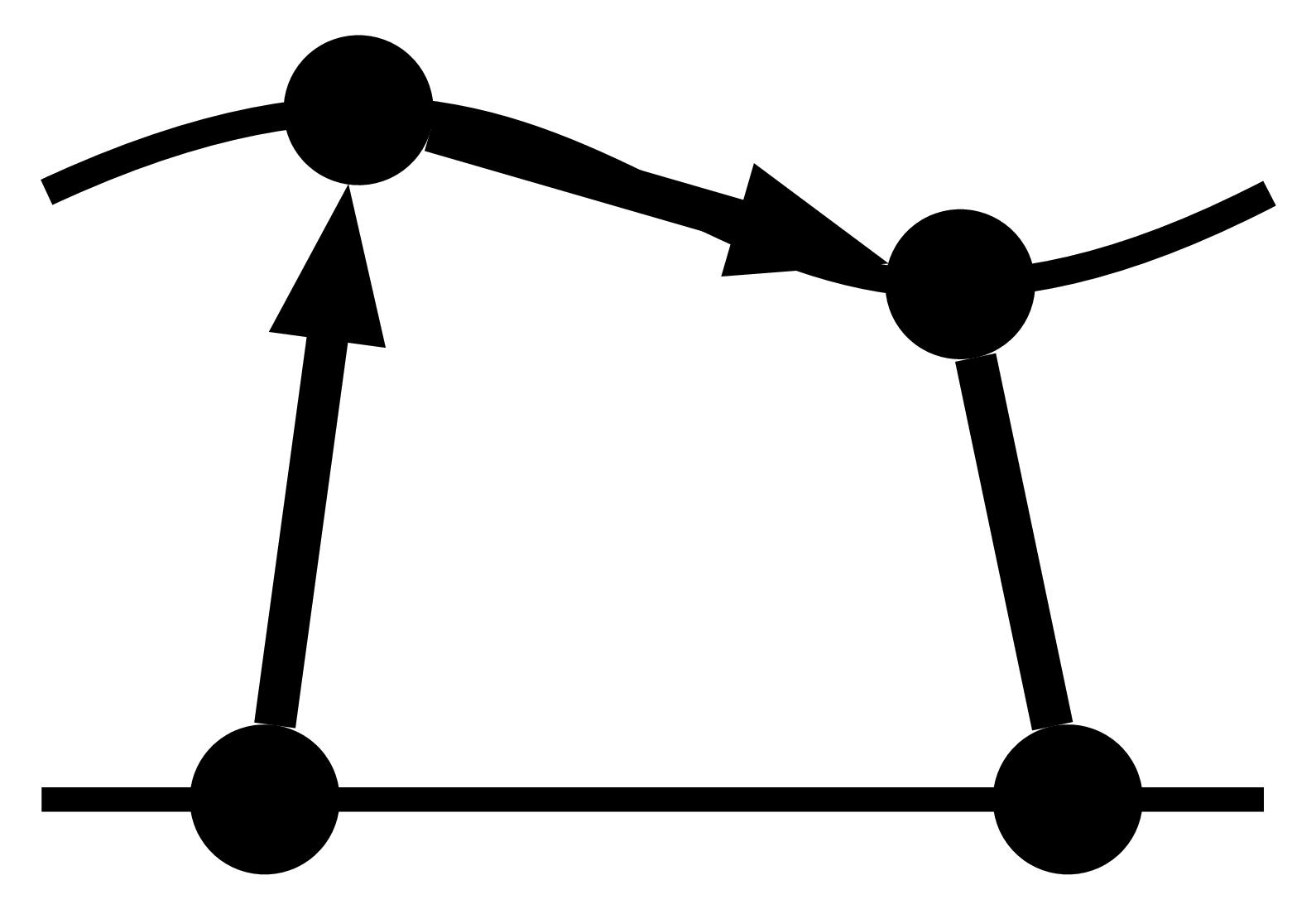} - \fig{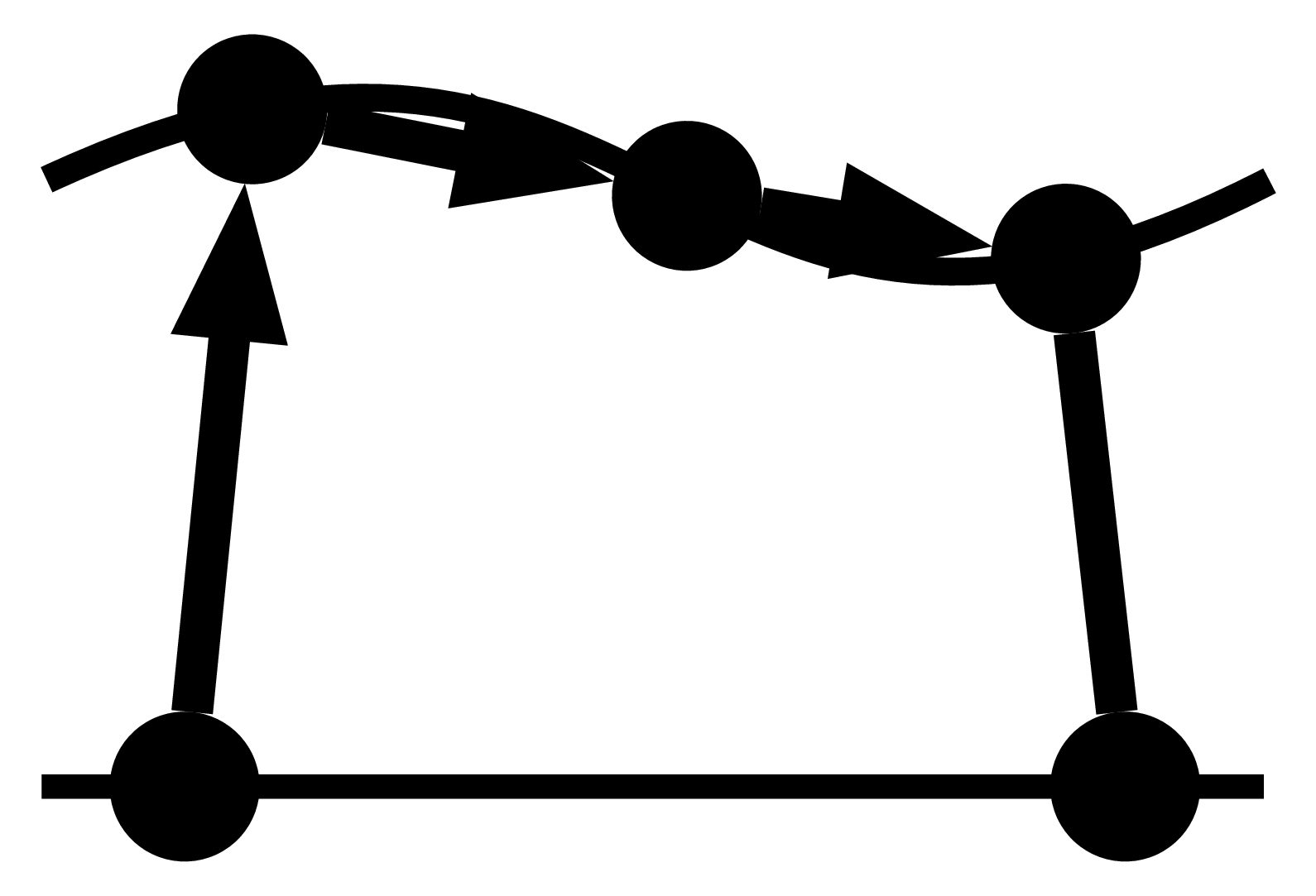} - \fig{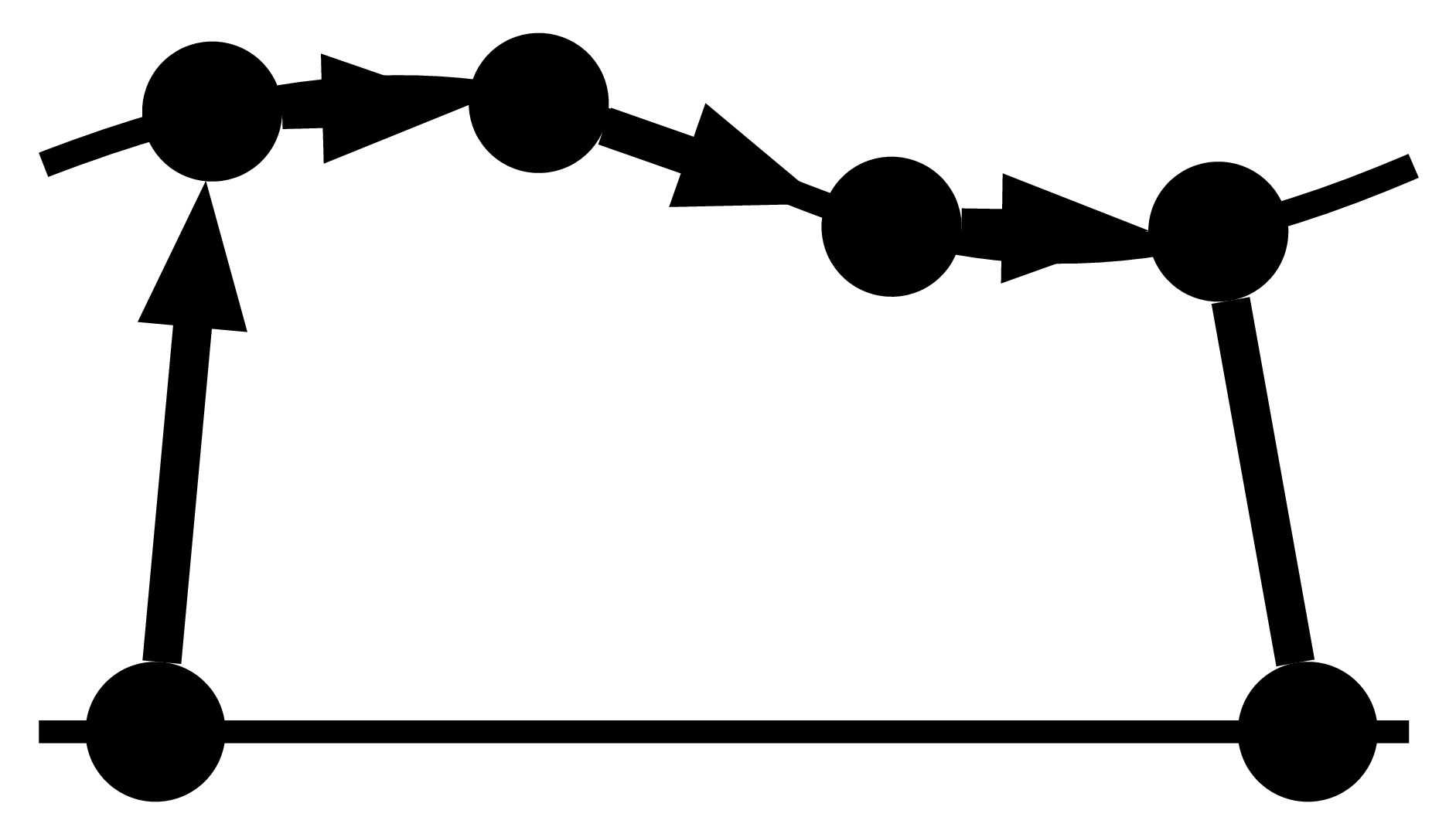} - \cdots,
\end{equation}
where the $n$th diagram belongs to ${O}\left(R^{1-n}\right)$. 
Again, by using the previous lemmas, 
the leading term of $\Omega_{2}^{1}$ can be written as
\begin{equation}
-\fig{diagram117}=\langle \cos \alpha\rangle_2 \fig{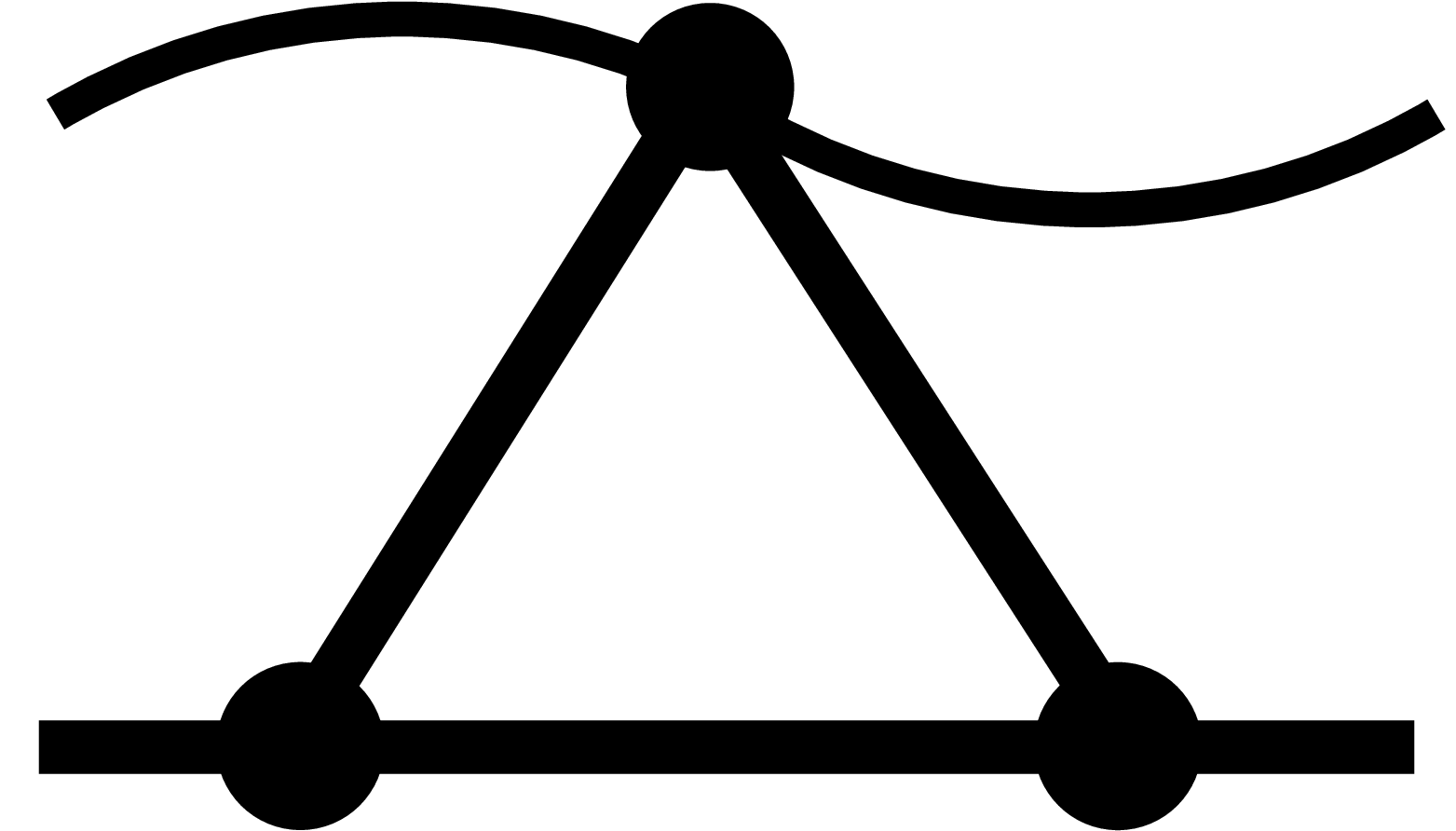},
\end{equation}
where $\langle \cos\alpha \rangle_2\sim 1-\tilde H/2\kappa$ by a 
saddle-point calculation.

The effect of the reduction is less effective for the class $\Omega_{1}^{2}$, for which the segments are located on the fluid interface. However, again
a saddle-point calculation shows that, up to ${O}(R^{-1})$ terms,
\begin{equation}
-\fig{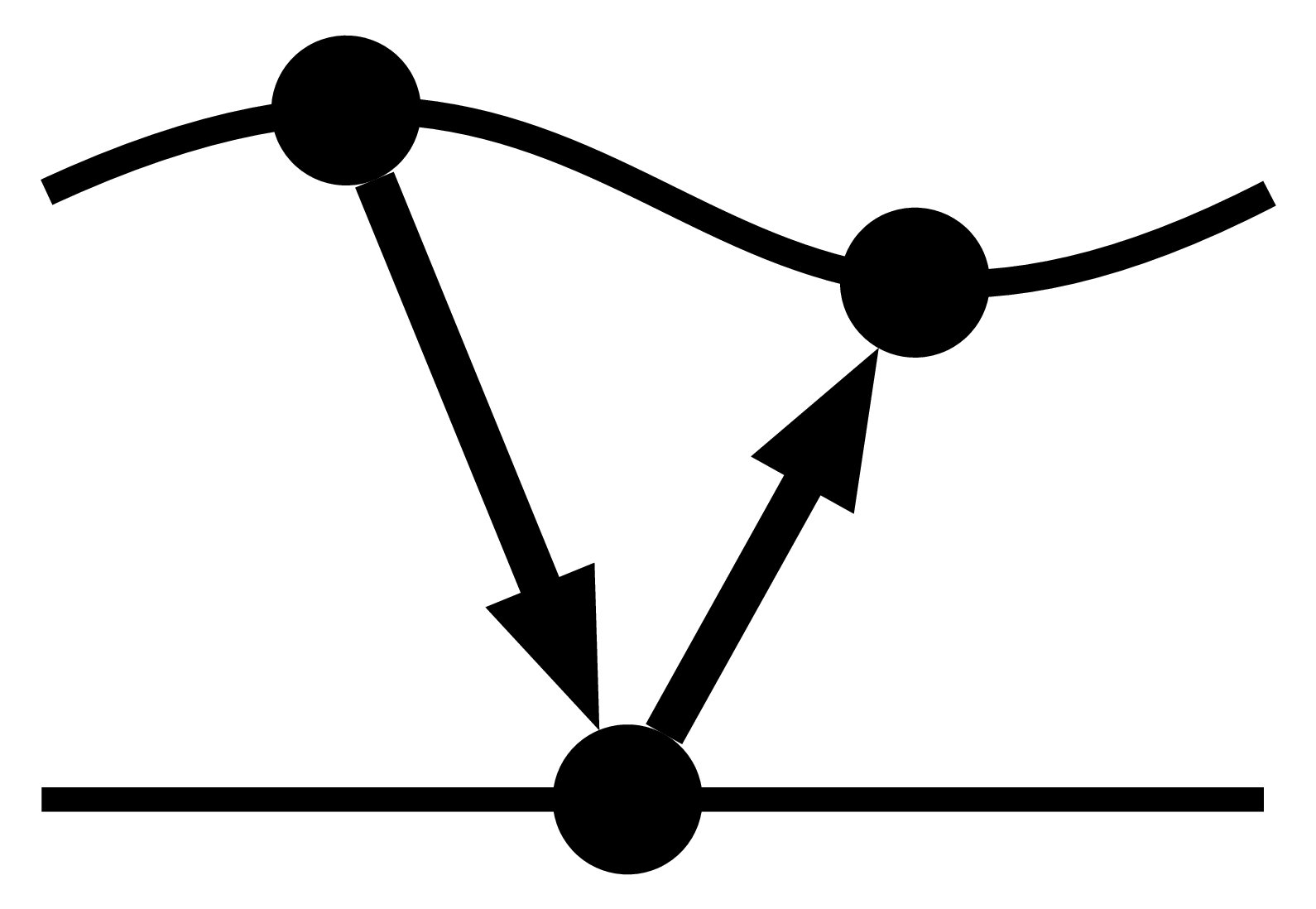}\approx \fig{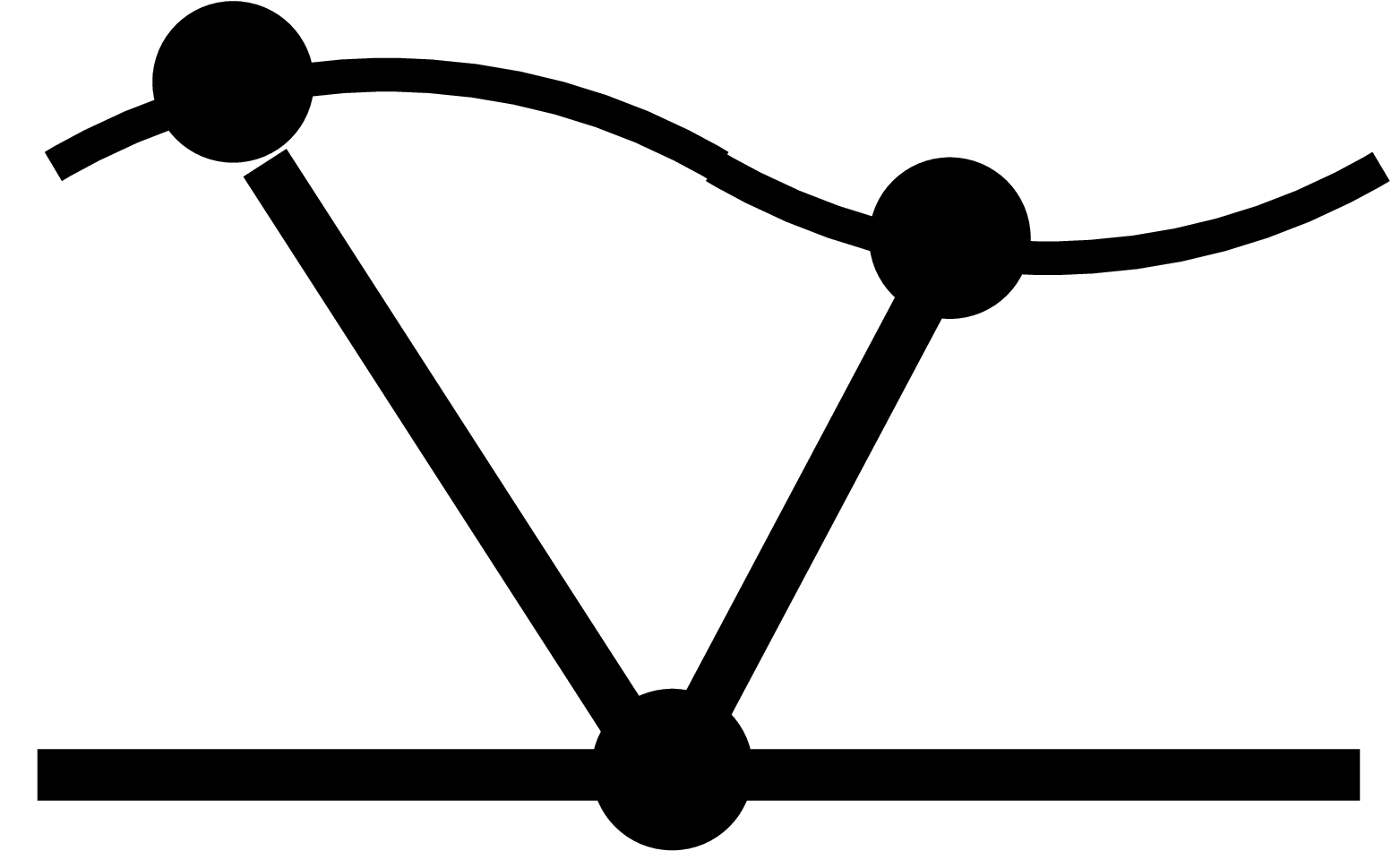},
\end{equation}
recovering the original formulation of the nonlocal model. However, we
should stress that this is a highly nonlocal formulation in the sense that the total binding potential
between the wall and the gas-liquid interface is obtained. However, if we would like    
to characterize the influence of the substrate locally on a portion of the 
liquid-gas interface, we have to resort to the nonlocal model presented in this paper. 
In particular, the functionals $\Omega_{1}^{1}$ and $\Omega_{2}^{1}$ are
local, so their contribution to the binding potential arises from
\begin{eqnarray}
-\fig{diagram102} + \fig{diagram114} + {O}\left(R^{-1}\right) , \\
- \fig{diagram117} + {O}\left(R^{-1}\right), 
\end{eqnarray}
where now the integration on the liquid-gas interface is restricted to the 
portion of the gas-liquid interface in which the binding potential is evaluated.A different feature, absent in the original formulation, 
emerges due to a coupling between the interface position and its orientation.
However, as in the original formulation, the $\Omega_1^2$ functional is highly
nonlocal and has the representation
\begin{equation}
\fig{diagram122}=-\int_\ell d\mathbf{s}_1 d\mathbf{s}_2 \textrm{e}^{-\kappa
[\ell(\mathbf{s}_1)+\ell(\mathbf{s}_2)]}\bar{S}(x_{12},\bar{\ell}),
\end{equation}
where $x_{12}$ is the projection of $\mathbf{s}_2-\mathbf{s}_1$ 
onto the substrate plane and $\bar{S}$ is the effective two-body interaction
between the interfacial area elements located around $\mathbf{s}_1$ and
$\mathbf{s}_2$. As the corresponding interaction $S\equiv S(x_{12},\bar{\ell})$ in the original nonlocal
model \cite{parry1,parry4,parry5}, $\bar{S}$ depends on the   
interfacial heights via $\bar{\ell}\equiv [\ell(\mathbf{s}_1)+\ell(\mathbf{s}_2)]/2$,
which can be analyzed by 
using the same renormalization-group (RG) flow equations derived in Refs.~\cite{parry1,parry4,parry5}. 
More specifically,
\begin{equation}
\bar{S}(x_{12},\bar\ell)=\frac{e^{2\kappa \bar\ell}}{\kappa}\mathbf{n}(\mathbf{s}_2)\mathbf{\cdot}\boldsymbol \nabla_2
\mathcal{K}\left[\sqrt{x_{12}^2+(2\bar\ell)^2}\right],
\end{equation}
where $\boldsymbol\nabla_2$ is the 3D gradient acting on the liquid-gas interfacial position
$\mathbf{r}_2=(\mathbf{s}_2,\ell(\mathbf{s}_2))$. 
For large $\ell$, a 
saddle-point calculation shows that
\begin{equation}
\bar{S}(x_{12},\bar\ell)\approx -\kappa\cos\alpha_2/2\pi\bar{\ell}
\textrm{e}^{-\frac{\kappa x_{12}^2}{2\bar{\ell}}}=
-\cos \alpha_2 S(x_{12},\bar{\ell}),
\end{equation}
where $\alpha_2$ is the angle formed by the normal vector and the vertical 
direction at the liquid-gas interface position $\mathbf{r}_2$. Thus, as $S$, the two-body
interaction $\bar{S}$ has a Gaussian form, with the nonlocal length
$\xi_{NL}=\sqrt{\langle l\rangle/\kappa}$ precisely as identified in the original formulation 
\cite{parry4,parry5}. However, our improved
formulation 
introduces as a different feature the coupling of the two-body interaction to the 
surface orientation through the factor $\cos \alpha_2$. 
A detailed comparison of the RG flows of this effective two-body repulsion within the original and present, 
exact, formulations in the context of critical wetting is beyond the scope of the present paper.

\subsection{Summary and remarks}
In this section we have applied the boundary integral diagrammatic method to determine the binding potential functional and order-parameter configuration when a wetting layer intrudes between the bulk phase and the wall. Our results are decorated versions of those appearing in the original formulation and, in particular, contain  
$U-$type kernels
on the fluid interface, while on the substrate they show $U-$type and 
$\kappa^{-1}(\partial_{n}K)-$type bonds. The effect of the diagrams can be readily understood for small curvature, where the pertinent multiply embedded  convolutions can be resummed, leading to renormalized diagrams involving the orientation of the surface. In this way, the full nonlocality is replaced by a weaker version, which can be used to build a more readily usable effective binding potential functional.

As expected, our formulation reveals curvature corrections to the original formulation that are reliable when the substrate and fluid interfacial configurations are parallel or concentric, as in the case of spherical and cylindrical symmetry. Strictly speaking, when the interfaces are nonparallel the present improved formulation must be used; the analysis of filling transitions for fluids adsorbed in wedge geometries is a natural place for investigating this.

When the surface order parameter at the wall is fixed (i.e., Dirichlet boundary conditions) and the system is at the location of the critical wetting transition ($m_1=m_0$), as pertinent to the critical isotherm, the only diagram of relevance remaining is $\Omega_1^2$. This term is strongly nonlocal and has a structure very similar to that appearing in the original nonlocal formulation. Once  again, this highlights the influence of an effective two-body Gaussian interfacial interaction controlled by a nonlocal length $\xi_{NL}=\sqrt{\langle l\rangle/\kappa}$ that is missing entirely from the original local effective Hamiltonian treatments of the critical
wetting phase transition.

\section{Conclusions}

In this paper we have presented a rigorous derivation of the nonlocal effective interfacial Hamiltonian model for interfaces and wetting in systems possessing short-ranged forces. The present derivation, which is based on a boundary integral formulation, improves on the one given originally, because the boundary conditions at the interface and wall are now handled exactly rather than approximately. The first point to emphasize is that this systematic analysis can indeed be done at all, at least using a simple DP potential and the crossing criterion definition of the interface position (to which we will return later). 

This analysis can also be expressed diagrammatically; a glossary of the elementary diagrams from which all other diagrams follow is given in Appendix~C, together with their algebraic expressions. As with the original formulation, each diagram containing a line that spans the liquid wetting layer, thus connecting the liquid-gas interface and wall, can be thought of as an interaction between these surfaces mediated by a bulklike correlation. Among other things, this rigorous formulation allows us to consider, in a systematic fashion, the nature of the curvature corrections appearing in the appropriate free energy. More specifically, we applied the boundary integral method to three situations with the following conclusions.

\begin{itemize}
\item[(i)]
{\emph{The wall-single-phase interface}}. First we considered a nonplanar wall-liquid interface, where a wetting layer is absent. We showed that the leading-order curvature corrections to the surface tension term involve the local mean and Gaussian curvatures, in the spirit of the Helfrich free energy, with bending and saddle-splay rigidity coefficients, respectively, the values of which are identified. However, the curvature expansion does not truncate at this, or indeed any, order and the free energy does not conform to the morphological thermodynamics hypothesis \cite{konig}.
\item[(ii)]
{\emph{The free liquid-gas interface}}. Extending this analysis to the free (but constrained) liquid-gas interface, we showed that the curvature corrections can be expressed more precisely as an interfacial self-interaction, the form of which is identical to that proposed in Ref.~\cite{PR} using less rigorous methods. Indeed, the order-parameter profiles are also identical, lending strong support to the approximate methods used previously to discuss nonlocality.
\item[(iii)]
{\emph{ The binding potential functional}}. For the case in which a wetting layer is present we have derived a generalized diagrammatic representation of the binding potential and order parameter, which contains decorated versions of those diagrams appearing in the original formulation. These generate, in addition to curvature corrections, a coupling between the interfacial orientation and position, which is missing entirely in the original theory. Indeed, strictly speaking, even for small curvatures the diagrams do not converge to those of the original formulation, {\emph{unless}} the interfacial configurations are nearly parallel to the substrate. However, when our formulation of the nonlocal model is applied to a flat substrate, we find features that are very similar to the original version of the nonlocal model. In particular, the contributions $\Omega_1^1$ and $\Omega_2^1$ to the binding potential functional are local, while the $\Omega_1^2$ contribution remains {\emph{nonlocal}} and can be expressed as a two-body Gaussian interfacial self-interaction, mediated by the substrate, having a lateral range given by the same nonlocal lengthscale $\xi_{NL}=\sqrt{\overline{\ell}/\kappa}$. Thus, the criticism of what is missing in local interfacial Hamiltonian descriptions of critical wetting in Refs.~\cite{BHL,FH}, including the size of the critical regime and also the paradoxical prediction of possible fluctuation-induced first-order transitions \cite{Jin2,Jin4}, remains unchanged (see Refs.~\cite{parry4,parry5}). Nevertheless, it would be interesting to include the coupling of orientation and position into renormalization group and simulation studies of the nonlocal model.
\end{itemize}

Having formulated the problem exactly for the DP potential, it is  possible to make extensions to more general potentials perturbatively, by using a Feynman-Hellmann theorem similar to the approximate analysis of Ref.~\cite{parry3}. For the wall-liquid and free liquid-gas interface, this would generate further curvature corrections to the free energy, although this will not alter the diagrammatic structure only altering the values of the coefficients. However, in applications to the wetting film, the binding potential will now contain decorated versions of the $\chi$ diagram identified in Ref.~\cite{parry3}. To identify the curvature corrections to this term, further resummation of the diagrammatic series is required, similar to the decorated diagrams in the DP model discussed here. Generalizations to heterogeneous walls are also technically possible using the boundary integral approach.

Our rigorous and rather technical derivation of the nonlocal model is still subject to a number of criticisms. For example, we have assumed that the surface field $h_1$ and enhancement $g$ are not altered by the surface curvature of a structured wall, which is very probably an oversimplification.  In addition, of course, the continuum LGW model~(\ref{HLGW}) does not in any way account for volume exclusion and local layering present when a high-density fluid is adsorbed at a wall. There are also alternative definitions of \emph{the} interfacial position. For example, Fisher and Jin discuss integral criteria, and show that these alter the coefficients appearing in the binding potential function (see Refs.~\cite{Jin1,Jin3}). Hopefully, the diagrammatic structure of the binding potential functional is not altered  when using a different definition of the interfacial position, although we should expect that the values of all coefficients and curvature corrections are altered.

We should also mention that, of course, as soon as long-ranged forces
are present all results here change dramatically \cite{mecke}.
For example, exponential terms are replaced by 
algebraic terms in the binding potential. Additionally, for
Lennard-Jones forces the curvature expansion of the 
interfacial free energy fails completely, due to nonanalytic
logarithmic corrections.

However, there are deeper issues concerning the connection between mesoscopic and microscopic descriptions, which highlight some
of the fundamental problems still open in the theory of
interfacial phenomena discussed here. For example, within the crossing criterion,
for any potential $\phi(m)$,  there is no escape from having a
negative bending rigidity $\kappa_B$ (the positive saddle-splay
rigidity plays no part since the principal radius of curvature 
along the wedge is infinite). However,
the very meaning of having a negative bending coefficient has been questioned
by Chac\'on and Tarazona \cite{tarazona13}, who have argued that
the continuum LGW Hamiltonian is already too coarse
grained to enable a direct determination of the rigidity
from a constrained minimization of the model. At a 
microscopic level, they argue that there must be a molecular
top to the capillary-wave spectrum, which leads to a 
positive rigidity. While density-functional models may be
consistent with this feature when we look closely at the
structure of the equilibrium density-density correlation
function, a constrained minimization of any model 
functional will not suffice. Alternatively, they propose that
the constrained minimization is replaced by a weighted
convolution, which smears the interface location over a
region comparable with the bulk correlation length. This
means, of course, that the interface position no
longer has a strict crossing-criterion interpretation. In
fact, it has been shown that the crossing criterion does
not distinguish correctly bulk from interfacial 
contributions present in the mean-field correlation function and
therefore cannot be used naively to determine any wave-vector-dependent corrections to the surface tension \cite{parry8}.
These concerns must also be married with the observation that the mean-field identification \cite{Jin1,Jin3} is, strictly
speaking, only valid in the limit of low temperatures (i.e.,
$T\to 0$) where a saddle-point evaluation of the partial
trace suffices. Finite-temperature corrections to the 
interfacial free energy, interfacial Hamiltonian, and binding
potential must be present at some order. Indeed, these
corrections are already allowed for implicitly when, in
the application of the interfacial Hamiltonian, the mean-field 
value of the surface tension is replaced by its true
thermodynamic value. These ideas, which are still under
development, of course mean that the determination of
the binding potential functional for wetting layers and
the values, and indeed the signs, of the coefficients of all
curvature correction terms, are much more difficult to
determine.

\acknowledgments

J.M.R.-E. is grateful for financial support from the Spanish Ministerio de Econom\'{\i}a y Competitividad through
Grants No.~FIS2017-87117-P and No.~FIS2012-32455, Junta de Andaluc\'{\i}a through Grant No. P09-FQM-4938, all co-funded
by the EU FEDER, and the Portuguese Foundation for Science and Technology under Contract No. EXCL/FIS-NAN/0083/2012.
A.S. acknowledges hospitality of the University of Seville. A.O.P. acknowledges support from the EPSRC UK 
Grant No. EP/L020564/1 
``Multiscale Analysis of Complex Interfacial Phenomena.'' 
P.M.G. gratefully acknowledges support from NSF Grant No. DMR12-1207026.
The work of P.M.G. was performed in part at the Aspen Center for Physics,
which is supported by U.S. National Science Foundation Grant No. PHY-1607611.

\appendix

\section{Perturbative solutions of the boundary integral equations}
\label{appendix_A}
In this appendix 
we illustrate how to solve the integral equations which emerge in our
analysis of the curvature expansion. We start with the
evaluation of the first terms in the curvature expansion of the normal derivative $q$ for a single phase
in contact with a substrate $\psi$. After substitution of the curvature expansions of the terms which appear in
(\ref{inteq1}) we find a recursive chain of equations for the $q_{n}$'s up to ${O}(R^{-3})$ 
\begin{eqnarray} 
&&\int_{\mathbb{R}^{2}} d \mathbf{r}_{\bot} \, q_{0} \left[\mathcal{K}(r_{\bot}) - \frac{\kappa}{g}\delta
(\mathbf{r}_{\bot})\right] =  \kappa\left(\frac{h_1}{g}+m_b\right) \label{exph0}\\
\nonumber\\
&&\int_{\mathbb{R}^{2}} d \mathbf{r}_{\bot} \, q_{1} \left[\mathcal{K}(r_{\bot}) - \frac{\kappa}{g}\delta
(\mathbf{r}_{\bot})\right]  \nonumber\\
 &&=-\frac{1}{g}\int_{\mathbb{R}^{2}} d \mathbf{r}_{\bot} \mathcal{W}(r_{\bot}) \Delta\psi(\mathbf{r}_{\bot})
\left(q_0 +h_1+gm_b\right), \label{exph1}
\end{eqnarray}
and
\begin{eqnarray}
&&\int_{\mathbb{R}^{2}} d \mathbf{r}_{\bot} \, q_{2} \left[\mathcal{K}(r_{\bot}) - \frac{\kappa}{g}\delta
(\mathbf{r}_{\bot})\right]  \nonumber\\
&& =\frac{1}{2}\int_{\mathbb{R}^{2}} d \mathbf{r}_{\bot}\biggl[ \mathcal{W}(r_{\bot}) 
\Delta\psi(\mathbf{r}_{\bot})\left(q_0 \Delta\psi(\mathbf{r}_{\bot}) - \frac{2q_1}{g}\right)
\nonumber\\&&-q_0 \left(\mathcal{K}(r_{\bot}) \left(\boldsymbol\nabla_{\bot}\Delta\psi(\mathbf{r}_{\bot})
\right)^{2}+\frac{2}{g} \mathcal{W}(r_{\bot}) \chi (\mathbf{r}_{\bot})\right) \biggr]\label{exph2} 
\end{eqnarray}
where $\mathcal{K}(x)=\kappa \exp(-\kappa x)/2\pi x$, $\mathcal{W}(x)
=(1+\kappa x)\mathcal{K}(x)/x^2$ 
and we have extended the integral to $\mathbb{R}^{2}$, ignoring exponentially decaying terms on $\kappa R$. 
Note that self-consistency means that only the leading terms of $\Delta \psi$ and $\chi\equiv {\mathbf{r}_\bot}{\mathbf{\cdot}}
\boldsymbol\nabla_{\bot}\Delta\psi-2\Delta \psi$, 
which scale as $R^{-1}$ and $R^{-2}$, respectively, should be used. 

For a flat interface $q_{0}(\mathbf{s})$ is translationally invariant; therefore, it can be factorized from the integral, but since $\int_{\mathbb{R}^{2}} d \mathbf{r}_{\bot} \, \mathcal{K}(r_{\bot}) =1$ we have that $q_{0}$ is
given by Eq.~(\ref{qflat}). 
However, it will be useful to develop a further technique to solve the Eqs. 
(\ref{exph0})-(\ref{exph2}). We define a parallel Fourier transform, in which only the fluctuating modes parallel to the interface are considered. The kernel reads
\begin{equation}
K(\mathbf{s}-\mathbf{s}^{\prime}) = \int_{\mathbb{R}^{2}} \frac{d^{2} \mathbf{q}}{(2\pi)^{2}} \, \textrm{e}^{i \mathbf{q} \mathbf{\cdot} (\mathbf{s}-\mathbf{s}^{\prime})} \, \widetilde{K}(\mathbf{q}) ,
\end{equation}
and with a simple complex integration, we get the inverse Fourier transform
\begin{equation}
\label{kerneltransform}
\widetilde{K}(\mathbf{q}) = \int_{\mathbb{R}^{2}} d \mathbf{s} \, \textrm{e}^{-i \mathbf{q} \mathbf{\cdot} (\mathbf{s}-\mathbf{s}^{\prime})} \, K(\mathbf{s}-\mathbf{s}^{\prime}) = \frac{\kappa}{\sqrt{\kappa^{2}+q^{2}}} .
\end{equation}
With these definitions the convolution equation for $q_{0}$ becomes an algebraic equation for the Fourier modes
\begin{equation}
\widetilde{q_{0}}(\mathbf{q}) = (2\pi)^{2} \kappa \left(\frac{h_1}{g}+m_b\right) 
\frac{\delta(\mathbf{q})}{\widetilde{K}(-\mathbf{q})-\frac{\kappa}{g}} ,
\end{equation}
and transforming back to real space we find $q_{0}=-\kappa (h_1+gm_{b})/(\kappa-g)$. 
Let us consider now the equations for the ${O}(R^{-1})$ and ${O}(R^{-2})$. 
If the leading contributions to $\Delta \psi$ and $\chi$ in powers of $(\kappa R)^{-1}$ are used 
in Eqs. (\ref{exph1}) and (\ref{exph2}), the integrations over 
$\mathbf{r}_{\bot}$ on their right-hand sides can be performed in polar coordinates. 
After a few simple calculations we find
\begin{eqnarray} \nonumber
&&\int_{\psi} d \mathbf{r}_{\bot} \, q_{1} \left[\mathcal{K}(r_{\bot})-\frac{\kappa}{g}\delta(\mathbf{r}_{\bot})\right] 
\nonumber\\ &&  =\left(\frac{h_1+gm_{b}}{\kappa-g}\right)
\left(\frac{k_{1}+k_{2}}{2} \right) = \frac{h_1+gm_{b}}{\kappa-g} H
\end{eqnarray}
and
\begin{eqnarray}
&&\int_{\psi} d \mathbf{r}_{\bot} \, q_{2} \left[\mathcal{K}(r_{\bot})-\frac{\kappa}{g}\delta(\mathbf{r}_{\bot})\right] 
  =\nonumber\\
&&  \kappa\frac{h_1+g m_{b}}{8(\kappa-g)} \left( \frac{k_{1} - k_{2}}{\kappa} \right)^{2} 
\nonumber\\
&&-\frac{1}{g}\int_{\mathbb{R}^{2}} d \mathbf{r}_{\bot} q_1 \mathcal{W}(r_{\bot}) \Delta\psi(\mathbf{r}_{\bot})
=\nonumber\\
&&  \frac{h_1+g m_{b}}{2\kappa(\kappa-g)} \biggl[ H^{2} -K_{G} \biggr] 
\nonumber\\
&&-\frac{1}{g}\int_{\mathbb{R}^{2}} d \mathbf{r}_{\bot} q_1 \mathcal{W}(r_{\bot}) \Delta\psi(\mathbf{r}_{\bot}),
\label{eqq2}
\end{eqnarray}
Note that the results of the integrations are expressed in terms of the mean and Gaussian curvatures of the 
interface, both evaluated at the origin. 
If we denote by $\mathcal{R}_{1,2}(\mathbf{s})$ the right-hand sides of these equations, their formal solution reads
\begin{equation}
q_{1,2}(\mathbf{s}) = \int_{\mathbb{R}^{2}}\frac{d^{2} \mathbf{q}}{(2\pi)^{2}} \, \textrm{e}^{i \mathbf{q} \mathbf{\cdot} \mathbf{s}} \frac{\widetilde{\mathcal{R}}_{1,2}(\mathbf{q})}{\frac{1}{\sqrt{1+\frac{q^2}{\kappa^2}}}-\frac{\kappa}{g}} .
\end{equation}
For our substrate $\kappa R\gg 1$, so the integral is dominated by the slow sector of Fourier modes. Hence 
it is reasonable to expand the square root in powers of the small parameter $q/\kappa$, thus
\begin{eqnarray} \nonumber
\label{q2gradient}
q_{1,2}(\mathbf{s})  &\approx&  \int_{\mathbb{R}^{2}}\frac{d^{2} \mathbf{q}}{(2\pi)^{2}} \, \frac{\textrm{e}^{i \mathbf{q} 
\mathbf{\cdot} \mathbf{s}}}{1-\frac{\kappa}{g}} \Biggl( 1 + \frac{q^{2}}{2\kappa^{2}(1-\frac{\kappa}{g})} 
\Biggr) \, \widetilde{\mathcal{R}}_{1,2}(\mathbf{q}) \\ \nonumber
& = &  -\frac{g}{\kappa-g}\Biggl(\mathcal{R}_{1,2}(\mathbf{s})+\frac{g}{2\kappa^{2}}\frac{\nabla^{2}_\bot\mathcal{R}_{1,2}(\mathbf{s})}{\kappa-g}\nonumber\\&+&{O}\left(\nabla^{4}_\bot\mathcal{R}_{1,2}(\mathbf{s})\right)\Biggr) .
\end{eqnarray}
The same result is obtained if we make a Taylor expansion of $q_{1,2}$ around the origin and substitute in Eqs. 
(\ref{exph1}) and (\ref{exph2}). However, we note that $\nabla_\bot^2 \mathcal{R}_{1,2} \sim \mathcal{R}_{1,2}/R^2$,
so the derivative terms contribute to higher-order curvature terms and thus they can be neglected. The solutions
are then given by the leading contributions of Eq.~(\ref{q2gradient}), which correspond to Eqs. (\ref{q1}) and 
(\ref{q2}).

In a similar way, the perturbative scheme for the computation of the normal derivatives $q$ for a free interface is 
\begin{eqnarray}
\label{pert_q_int_1}
\int_{\mathbb{R}^{2}} d \mathbf{r}_{\bot} \, q_{0}^{\pm} \mathcal{K}(r_{\bot}) & = & -\kappa m_{0} \\
\label{pert_q_int_2}
\int_{\mathbb{R}^{2}} d \mathbf{r}_{\bot} \, q_{1}^{\pm} \mathcal{K}(r_{\bot}) & = & \pm m_{0} \int_{\mathbb{R}^{2}} d \mathbf{r}_{\bot} \, \mathcal{W}(r_{\bot}) \Delta\ell(\mathbf{r}_{\bot}), 
\end{eqnarray}
and
\begin{eqnarray}
\int_{\mathbb{R}^{2}} d \mathbf{r}_{\bot} \, q_{2}^{\pm} \mathcal{K}(r_{\bot}) & = & \frac{1}{2}\int_{\mathbb{R}^{2}} d \mathbf{r}_{\bot} \, q_{0}^{\pm} \biggl\{ \mathcal{W}(r_{\bot}) \Delta\ell(\mathbf{r}_{\bot})^{2}
\label{pert_q_int_3}
\nonumber\\&-&\mathcal{K}(r_{\bot}) \left[\boldsymbol\nabla_{\bot}\Delta\ell(\mathbf{r}_{\bot})\right]^{2} \biggr\}.
\end{eqnarray}
These have the solutions
\begin{eqnarray}
\label{pert_q_sol_1}
q_{0}^{\pm}& =& -\kappa m_{0},\\
\label{pert_q_sol_2}
q_{1}^{\pm} & = & \pm \kappa m_{0} \frac{k_{1}+k_{2}}{2\kappa}  = \pm m_{0} H(\mathbf{s}),\\ \nonumber
\label{pert_q_sol_3}
\end{eqnarray}
and
\begin{eqnarray}
q_{2}^{\pm} & = & \frac{\kappa m_{0}}{8} \left( \frac{k_{1} - k_{2}}{k} \right)^{2} \nonumber\\
& = &\frac{m_{0}}{2\kappa} \biggl[ H(\mathbf{s})^{2} -K_{G}(\mathbf{s}) \biggr] .
\label{exph22}
\end{eqnarray}

Finally, the curvature expansion of $\Psi$ for the single phase in contact with the substrate can be obtained from
Eq.~(\ref{singlelayer2}). After substitution of Eq.~(\ref{boundary2}) and the curvature expansions of $q$, 
the kernel $K$, and the elementary area $d\mathbf{s}$ into Eq.~(\ref{singlelayer2}), we 
obtain the equations
\begin{eqnarray}
\label{psi_1}
&&\int_{\mathbb{R}^{2}} d \mathbf{r}_{\bot} \, \Psi_{0} \mathcal{K}(r_{\bot})  =  -\frac{h_1 + gm_b}{g}-
\frac{q_0}{g}, \\
\label{psi_2}
&&\int_{\mathbb{R}^{2}} d \mathbf{r}_{\bot} \, \Psi_{1} \mathcal{K}(r_{\bot})  =  -\frac{q_1}{g},
\end{eqnarray}
and
\begin{eqnarray} 
&&\int_{\mathbb{R}^{2}} d \mathbf{r}_{\bot} \, \Psi_{2} \mathcal{K}(r_{\bot})  =  -\frac{q_2}{g}
-\frac{1}{2}\int_{\mathbb{R}^{2}} d \mathbf{r}_{\bot} \, \frac{\Psi_0}{g} 
\nonumber
\\&\times&
\biggl\{ \mathcal{W}(r_{\bot}) \Delta\psi(\mathbf{r}_{\bot})^{2}
-\mathcal{K}(r_{\bot}) \left[\boldsymbol\nabla_{\bot}\Delta\psi(\mathbf{r}_{\bot})\right]^{2} \biggr\}.
\label{psi_3}
\end{eqnarray}
where $\Psi_0$, $\Psi_1$, and $\Psi_2$ stand for the first terms in the curvature expansion of $\Psi$.  
The solutions of these integral equations are
\begin{eqnarray}
\label{psi_1_sol}
\frac{\Psi_0}{2\kappa}&=&\frac{h_1+gm_b}{\kappa-g},
\\
\label{psi_2_sol}
\frac{\Psi_1}{2\kappa}&=&\left(\frac{h_1+gm_b}{\kappa-g}\right)\left(\frac{\kappa}{\kappa-g}\right)\frac{H}{\kappa},
\end{eqnarray}
and
\begin{eqnarray} 
\frac{\Psi_1}{2\kappa}&=&\left(\frac{h_1+gm_b}{\kappa-g}\right)\Bigg[\left(\frac{1}{2}\frac{g}{\kappa-g}+\frac{\kappa^2}
{(\kappa-g)^2}\right)\left(\frac{H}{\kappa}\right)^2\nonumber\\
&-&\frac{1}{2}\frac{g}{\kappa-g}\frac{K_G}{2\kappa^2}\Bigg].
\label{psi_3_sol}
\end{eqnarray}

\section{Derivation of the liquid-gas interfacial self-interaction Hamiltonian}
\label{appendix_B}
\begin{figure}[htbp]
\centering
\includegraphics[width=9cm]{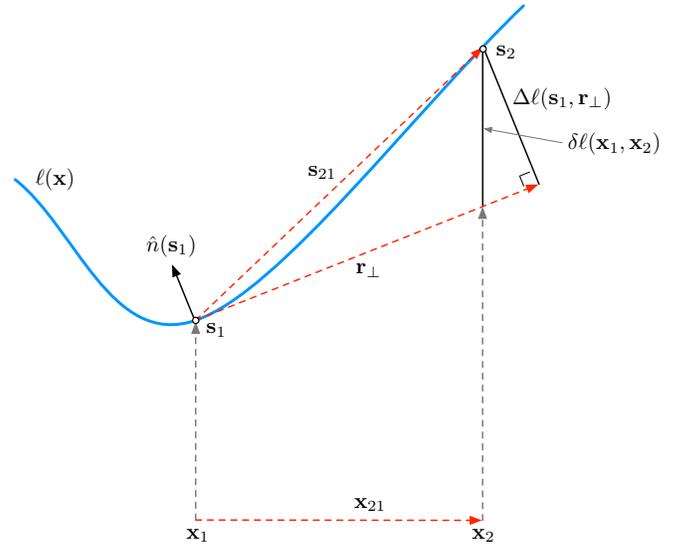}
\caption{Schematic illustration of the coordinates, vectors, and geometry appearing in the curvature expansion for a constrained interfacial configuration. The symbols are described in the text.}
\label{construction}
\end{figure}
In this appendix we derive (\ref{ham4}). 
Consider two points on the interface with $\mathbf{s}_{1}$ as the origin and $\mathbf{s}_{2}$ as in Fig. 
\ref{construction}. We supposed that the surface $\ell$ can be approximated, locally, as a paraboloid.
Taking into account the right-hand side of Eq.~(\ref{exph22}), the interfacial free-energy functional ~ 
(\ref{ham1}) can be written as 
\begin{eqnarray}
\label{ham2}
H[\ell] &\approx& \sigma \mathcal{A}_{lg} - \frac{\sigma}{2} \int_{\ell} d \mathbf{s}_1 \, \int_{\mathbb{R}^{2}} d \mathbf{r}_{\bot} \, \mathcal{K}(r_{\bot}) \left[\boldsymbol\nabla_{\bot}\Delta \ell(\mathbf{s},\mathbf{r}_{\bot})\right]^{2} \nonumber\\&+& \frac{\sigma}{2} \int d \mathbf{s}_{\ell} \, \int_{\mathbb{R}^{2}} d \mathbf{r}_{\bot} \mathcal{W}(r_{\bot}) \Delta \ell(\mathbf{s},\mathbf{r}_{\bot})^{2} ,
\end{eqnarray}
where $\mathbf{r}_{\bot}$ is the projection of $\mathbf{s}_2-\mathbf{s}_1$ on the tangent plane 
$\pi_{\mathbf{s}_1}$
to the interface at $\mathbf{s}_1$ and $\Delta\ell$ is the vertical displacement from $\pi_{\mathbf{s}_1}$,
\begin{equation}
\label{orthogonaldisplacement}
\Delta \ell(\mathbf{s}_1,\mathbf{r}_{\bot}) = \mathbf{n}(\mathbf{s}_1) \mathbf{\cdot} (\mathbf{s}_2 - \mathbf{s}_1) .
\end{equation}

The last step is to convert the surface integrations in integrals over the reference plane. 
In order to do that we need the mapping between the charts $\{\mathbf{x}_{1},\mathbf{x}_{2}\}$ and 
$\{\mathbf{s},\mathbf{r}_{\bot}\}$. The expressions of the mapping can be obtained from 
$\mathbf{s}_{2} - \mathbf{s}_{1} = \mathbf{n}(\mathbf{s}_{1}) \Delta \ell(\mathbf{s}_{1},\mathbf{r}_{\bot}) + 
\mathbf{r}_{\bot}$, supplemented by (\ref{orthogonaldisplacement}) and 
\begin{equation}
\mathbf{n}(\mathbf{s}_{1})=\frac{1}{\sqrt{1+[\boldsymbol\nabla \ell(\mathbf{x}_{1})]^{2}}}\left(-\boldsymbol\nabla \ell (\mathbf{x}_{1}),1\right) ,
\end{equation}
where $\boldsymbol\nabla$ represents the 2D gradient on the reference plane coordinates
\begin{equation}
J = \biggl | \frac{\partial\left(\mathbf{s}_{1},\mathbf{r}_{\bot}\right)}{\partial\left(\mathbf{x}_{1},\mathbf{x}_{2}\right)} \biggr| .
\end{equation}
We can show that the mapping Jacobian $J=1$ if quadratic terms on the gradients are neglected.
In this limit $|\mathbf{r}_{\bot}|\approx |\mathbf{x}_{21}|$, the orthogonal 
displacement (\ref{orthogonaldisplacement}) can be replaced with the vertical displacement
\begin{equation}
\Delta\ell(\mathbf{s}_{1},\mathbf{r}_{\bot}) \simeq \delta\ell(\mathbf{x}_{1},\mathbf{x}_{2}) \equiv \ell(\mathbf{x}_{2})-\ell(\mathbf{x}_{1})-\mathbf{x}_{21} \mathbf{\cdot}\boldsymbol \nabla\ell(\mathbf{x}_{1}) ,
\end{equation}
and taking the 2D gradient, $\boldsymbol\nabla_{\bot} \Delta\ell \simeq \boldsymbol\nabla\ell(\mathbf{x}_{1})-
\boldsymbol\nabla\ell(\mathbf{x}_{1})$. Finally,
\begin{equation}
\mathcal{A}_{lg} \simeq \mathcal{A}_{\pi} + \frac{1}{2} \int d \mathbf{x} \, \left[\boldsymbol\nabla\ell(\mathbf{x})\right]^{2} ,
\end{equation}
where $\mathcal{A}_{\pi}$ is the area of the surface obtained from the projection of the surface $\ell$ onto the reference plane. The Hamiltonian (\ref{ham2}) becomes
\begin{eqnarray} \nonumber
\label{ham3}
&&H[\ell]  \approx  \sigma \mathcal{A}_{\pi} + \frac{\sigma}{2} \int d \mathbf{x} \, \left[\boldsymbol\nabla\ell(\mathbf{x})\right]^{2} \nonumber\\& - & \frac{\sigma}{2} \int d \mathbf{x}_{1} d \mathbf{x}_{2} \, \mathcal{K}(x_{12}) \bigl[ \boldsymbol\nabla\ell(\mathbf{x}_{1}) - \boldsymbol\nabla\ell(\mathbf{x}_{2} ) \bigr]^{2} \\
& + & \frac{\sigma}{2} \int d \mathbf{x}_{1} d \mathbf{x}_{2} \, \mathcal{W}(x_{12}) \bigl[ \ell(\mathbf{x}_{2})-\ell(\mathbf{x}_{1})-\mathbf{x}_{12}\mathbf{\cdot}\boldsymbol\nabla\ell(\mathbf{ x}_{1}) \bigr]^{2} .
\nonumber
\end{eqnarray}
The expressions in (\ref{ham3}) can be further simplified. First we compute the squares, isolating the term 
proportional to the difference in vertical displacement. New terms will be created and for them we use the 
identities:
\begin{eqnarray} 
&&\frac{1}{2}\int d \mathbf{x}_{1} d \mathbf{x}_{2} \, \mathcal{K}(x_{12}) \bigl[\boldsymbol \nabla\ell(\mathbf{x}_{1}) - \boldsymbol\nabla\ell(\mathbf{x}_{2} ) \bigr]^{2}\\  \nonumber
&=&  \int d \mathbf{x}_{1} d \mathbf{x}_{2} \, \mathcal{K}(x_{12}) \bigl\{ \left[\boldsymbol\nabla\ell(\mathbf{x}_{1})\right]^{2} - \boldsymbol\nabla\ell(\mathbf{x}_{1})\mathbf{\cdot}\boldsymbol\nabla\ell(\mathbf{x}_{2}] \bigr\} \\ \nonumber
& = & \int d \mathbf{x} \, \left[\boldsymbol\nabla\ell(\mathbf{x})\right]^{2} - \int d \mathbf{x}_{1} d \mathbf{x}_{2} \, \mathcal{K}(x_{12}) \, \boldsymbol\nabla\ell(\mathbf{x}_{1})\mathbf{\cdot}\boldsymbol\nabla\ell(\mathbf{x}_{2}) 
\end{eqnarray}
and
\begin{eqnarray} 
&&\int d \mathbf{x}_{1} d \mathbf{x}_{2} \, \mathcal{W}(x_{12}) \bigl[ \mathbf{x}_{12}\mathbf{\cdot}\boldsymbol{\nabla}\ell(\mathbf{ x}_{1}) \bigr]^{2} \\ \nonumber 
& = & \int d \mathbf{x}_{1} d \mathbf{x}_{12} \, \mathcal{W}(x_{12}) \Bigl\{ x_{21}^{2} \left[ \partial_{x_{1}}\ell(\mathbf{x}_{1}) \right]^{2} + y_{21}^{2} \left[ \partial_{y_{1}}\ell(\mathbf{x}_{1}) \right]^{2} \Bigr\} \\ \nonumber
& = & \frac{1}{2} \int d \mathbf{x}_{12} \, \mathbf{x}_{12}^{2} \, \mathcal{W}(x_{12}) \int d \mathbf{x}_{1} \left[\boldsymbol\nabla\ell(\mathbf{x}_{1})\right]^{2} \\
& = & \int d \mathbf{x} \left[\boldsymbol\nabla\ell(\mathbf{x})\right]^{2} .
\end{eqnarray}
There is also a term of the form
\begin{eqnarray}
&&\int d \mathbf{x}_{1} d \mathbf{x}_{2} \biggl\{\mathcal{K}(x_{12}) \boldsymbol\nabla\ell(\mathbf{x}_{1}) \mathbf{\cdot}\boldsymbol \nabla\ell(\mathbf{x}_{2}) \nonumber \\ &&- \mathcal{W}(x_{12}) \bigl[\ell(\mathbf{x}_{2})-\ell(\mathbf{x}_{1})\bigr] \bigl[\mathbf{x}_{21} \mathbf{\cdot}\boldsymbol \nabla\ell(\mathbf{x}_{1})\bigr]\biggr\} .
\end{eqnarray}
Grouping the integral over $\mathbf{x}_{1}$ we have
\begin{eqnarray}
&&\int d \mathbf{x}_{1} \boldsymbol\nabla\ell(\mathbf{x}_{1}) \mathbf{\cdot} \int d \mathbf{x}_{2} \biggl\{\mathcal{K}(x_{12}) \boldsymbol\nabla\ell(\mathbf{x}_{2}) \nonumber\\ &&- \mathcal{W}(x_{12}) \bigl[\ell(\mathbf{x}_{2})-\ell(\mathbf{x}_{1})\bigr] \mathbf{x}_{21} \biggr\} ,
\end{eqnarray}
but since $- \mathcal{W}(x_{12}) \mathbf{x}_{21} = \boldsymbol\nabla_{\mathbf{x}_{2}}\mathcal{K}(x_{12})$, the second integrand can be written as a gradient of a scalar function
\begin{equation}
\int d \mathbf{x}_{1} \boldsymbol\nabla\ell(\mathbf{x}_{1}) \mathbf{\cdot} \int d \mathbf{x}_{2} \boldsymbol\nabla_{\mathbf{x}_{2}}\biggl\{\bigl[\ell(\mathbf{x}_{2})-\ell(\mathbf{x}_{1})\bigr]\mathcal{K}(x_{12})\biggr\}
\end{equation}
and so it reduces to a boundary contribution, which we can neglect. Collecting all the remaining terms, we are 
left with (\ref{ham4}).

\section{Wetting diagrams}
\label{appendix_C}
In this appendix we collect the definitions for the various wetting diagrams used in the main text. The diagrams
\begin{eqnarray}
\fig{diagram1} & = & \int \textrm{d}\mathbf{s}, \\
\fig{diagram2} & = & \int \textrm{d}\mathbf{s} \, H(\mathbf{s}) /\kappa, \\
\fig{diagram3} & = & \int \textrm{d}\mathbf{s} \, H^{2}(\mathbf{s}) /\kappa^{2}, \\
\fig{diagram4} & = & \int \textrm{d}\mathbf{s} \, K_{G}(\mathbf{s}) /\kappa^{2}
\end{eqnarray}
involve only local interfacial properties. The circle represents the area element, while $H$ denotes the local mean curvature and $K_{G}(\mathbf{s})$  the Gaussian curvature. Open symbols such as the one appearing in (\ref{dashedk}) and (\ref{dashedk2}) stand for the evaluation of the corresponding weight functions at a specified point $\mathbf{s}$ on the surface.

The Ornstein-Zernike kernel of (\ref{OZ2}) is represented by a thick black line with two open circles at the extrema. For instance, if $\mathbf{s}$ belongs to the surface $\ell$ and $\mathbf{r}$ to the upper region we will write
\begin{equation}
\fig{diagram21} = \int_{\ell}\textrm{d}\mathbf{s} \, K(\mathbf{s},\mathbf{r}) \, ,
\end{equation}
and similarly
\begin{equation}
\fig{diagram24} = \kappa^{-2} \int_{\ell}\textrm{d}\mathbf{s} \, K_{G}(\mathbf{s}) \, K(\mathbf{s},\mathbf{r}) \, .
\end{equation}
Then we have the dashed and arrow diagrams
\begin{eqnarray}
\fig{diagram5} & = & \fig{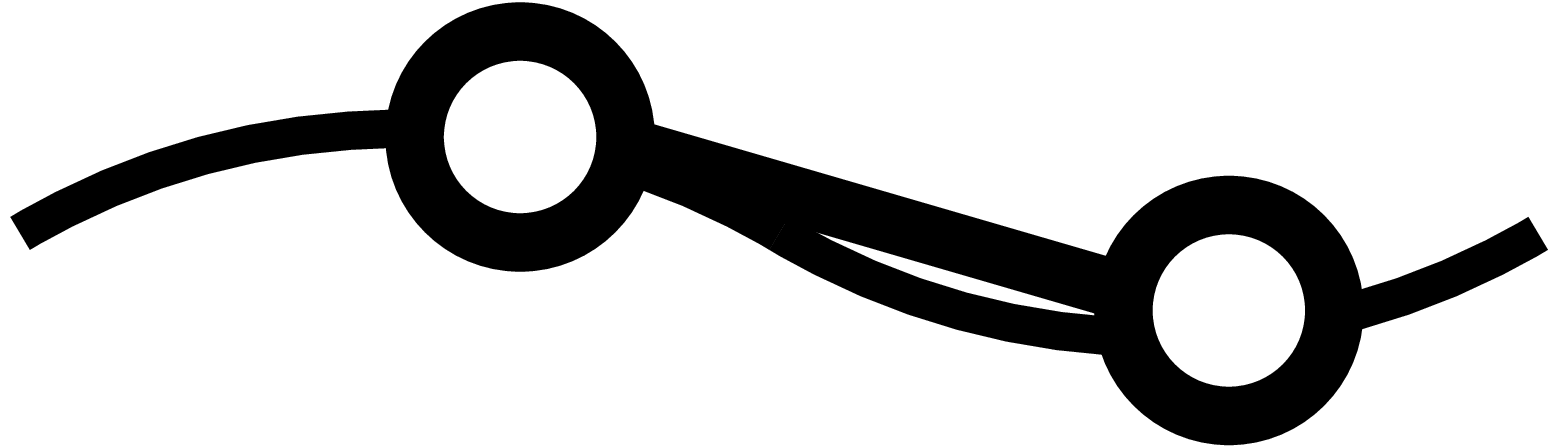}-\delta(\mathbf{s}-\mathbf{s}') = U(\mathbf{s},\mathbf{s}') \, , \\
\fig{diagram6} & = & \frac{1}{\kappa}\partial_{n} K(\mathbf{s},\mathbf{s}') \, ,
\end{eqnarray}
where in the latter diagram the arrow points to the position where the normal derivative is taken, as in the example
\begin{equation}
\int_\ell \textrm{d}\mathbf{s}_1 \textrm{d}\mathbf{s}_2 U(\mathbf{s},\mathbf{s}_1) \frac{1}{\kappa}\partial_{n_1} K(\mathbf{s}_1,\mathbf{s}_2) \, =\fig{diagram7} \, .
\end{equation}
The arrow diagram can also span between two interfaces, for example
\begin{equation}
\fig{diagram80} = \int_{\psi}\textrm{d}\mathbf{s}_{\psi}\int_{\ell}\textrm{d}\mathbf{s}_{\ell} \, \frac{1}{\kappa}\partial_{n_\ell} K(\mathbf{s}_\ell,\mathbf{s}_\psi) \, .
\end{equation}
The algebraic expressions for all other diagrams can be reconstructed in terms of these elementary building blocks.

\section{Binding potential for planar, spherical and cylindrical interfacial configurations. \label{appendixe}}

In this Appendix we will review the known form for planar, spherical and cylindrical interfacial configurations and 
how they are reproduced from the nonlocal representations of the binding potential we have discussed in Sec.~V.

\subsection{Planar interfaces}

We first consider the simplest case of a planar wall ($\psi=0$) and a planar interface of constant thickness, 
$\ell(\mathbf{x})=\ell$. In this case, $\mathcal{F}_{wl}[\psi]=\sigma_{wl} \mathcal{A}_{wl}$ and $H[\ell]=\sigma \mathcal{A}_{lg}$,
where $\mathcal{A}_{lg}=\mathcal{A}_{wl}=\mathcal{A}$ is the interfacial area and $\sigma_{wl}=(\kappa/2)(h_1+gm_b)^2/g(g-\kappa)$ and $\sigma=\kappa m_0^2$ are the surface
tensions defined for the planar wall-liquid and liquid-gas interfaces, respectively. On the other hand, the binding 
potential is \cite{laura,parry2,parry3}
\begin{eqnarray}
&&\frac{W[\ell,\psi]}{\mathcal{A}}=2\kappa m_0 \left(\frac{h_1+gm_0}{\kappa-g}\right) \frac{e^{-\kappa \ell}}{1-
\frac{g+\kappa}{g-\kappa}e^{-2\kappa \ell}}
\nonumber\\ &&+ \kappa \left(\frac{h_1+gm_0}{\kappa-g}\right)^2 \frac{e^{-2\kappa\ell}}{1-
\frac{g+\kappa}{g-\kappa}e^{-2\kappa \ell}}\nonumber\\&& + \frac{g+\kappa}{g-\kappa}
\kappa m_0^2 \frac{e^{-2 \kappa \ell}}
{1-\frac{g+\kappa}{g-\kappa}e^{-2\kappa \ell}}.
\label{bindingplanar}
\end{eqnarray}
The basic diagrams to obtain the decorated version of the original nonlocal model are
\begin{equation}
\fig{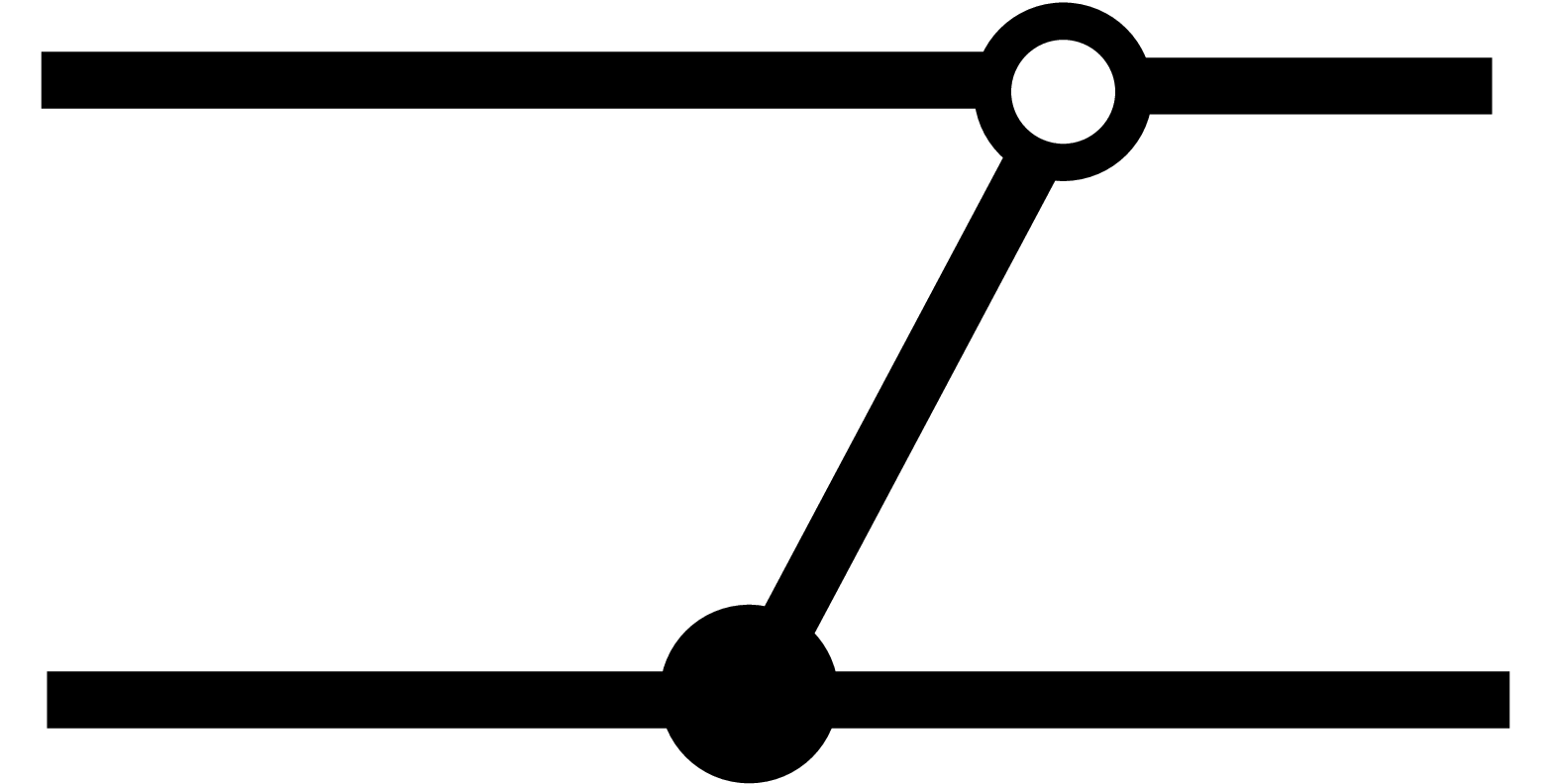}=\fig{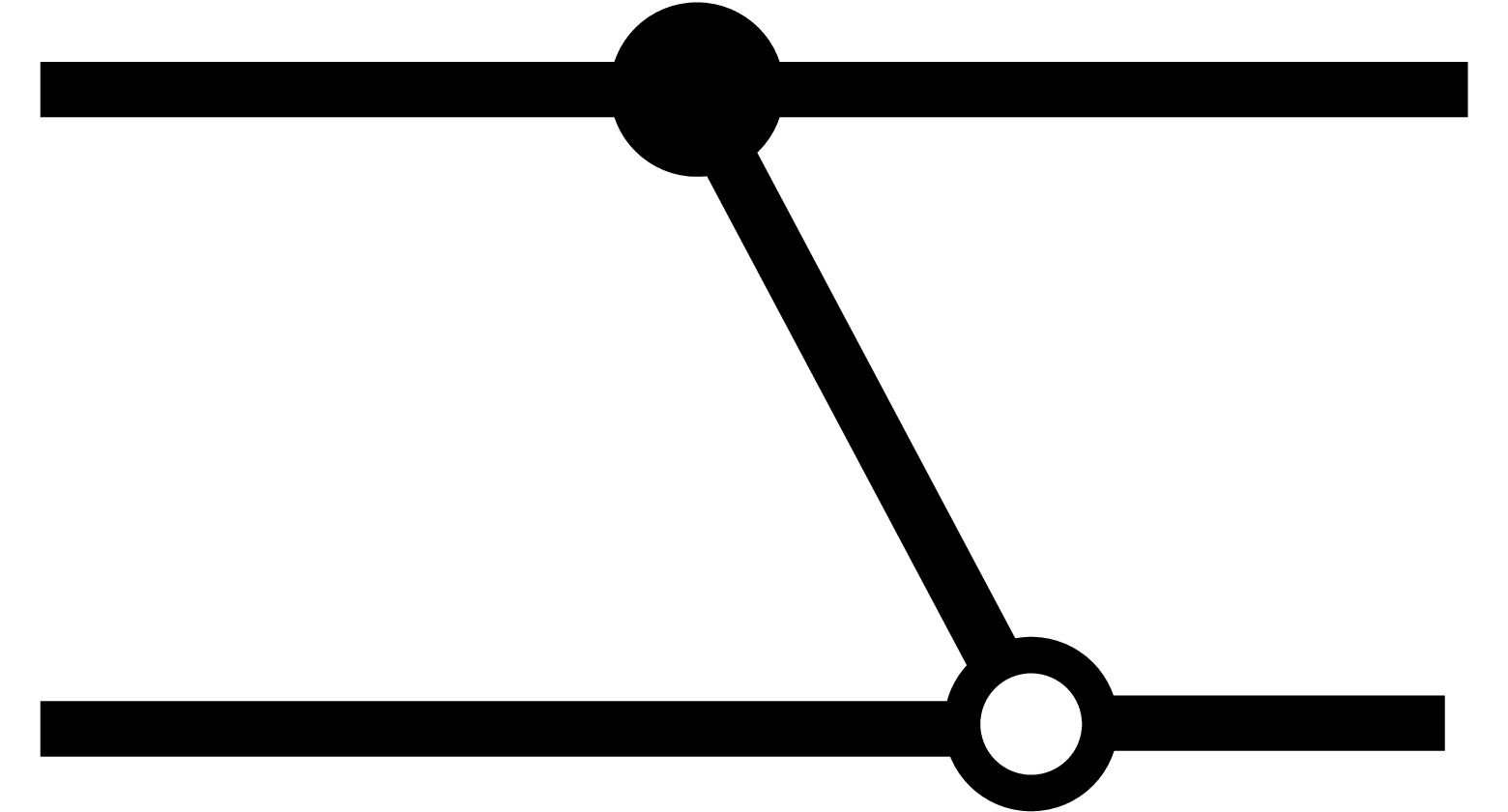}=e^{-\kappa \ell}
\end{equation}
and 
\begin{equation}
\fig{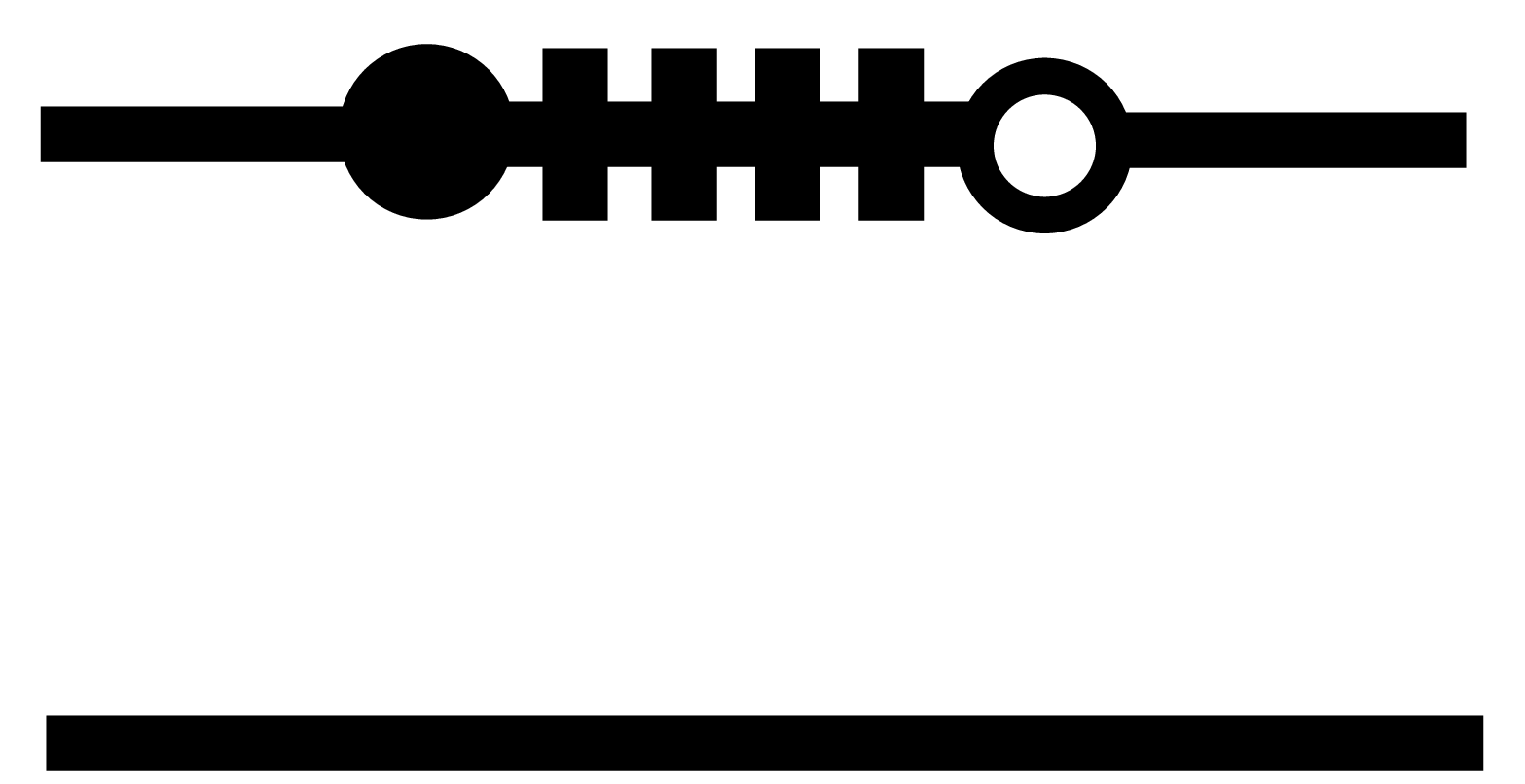}=\fig{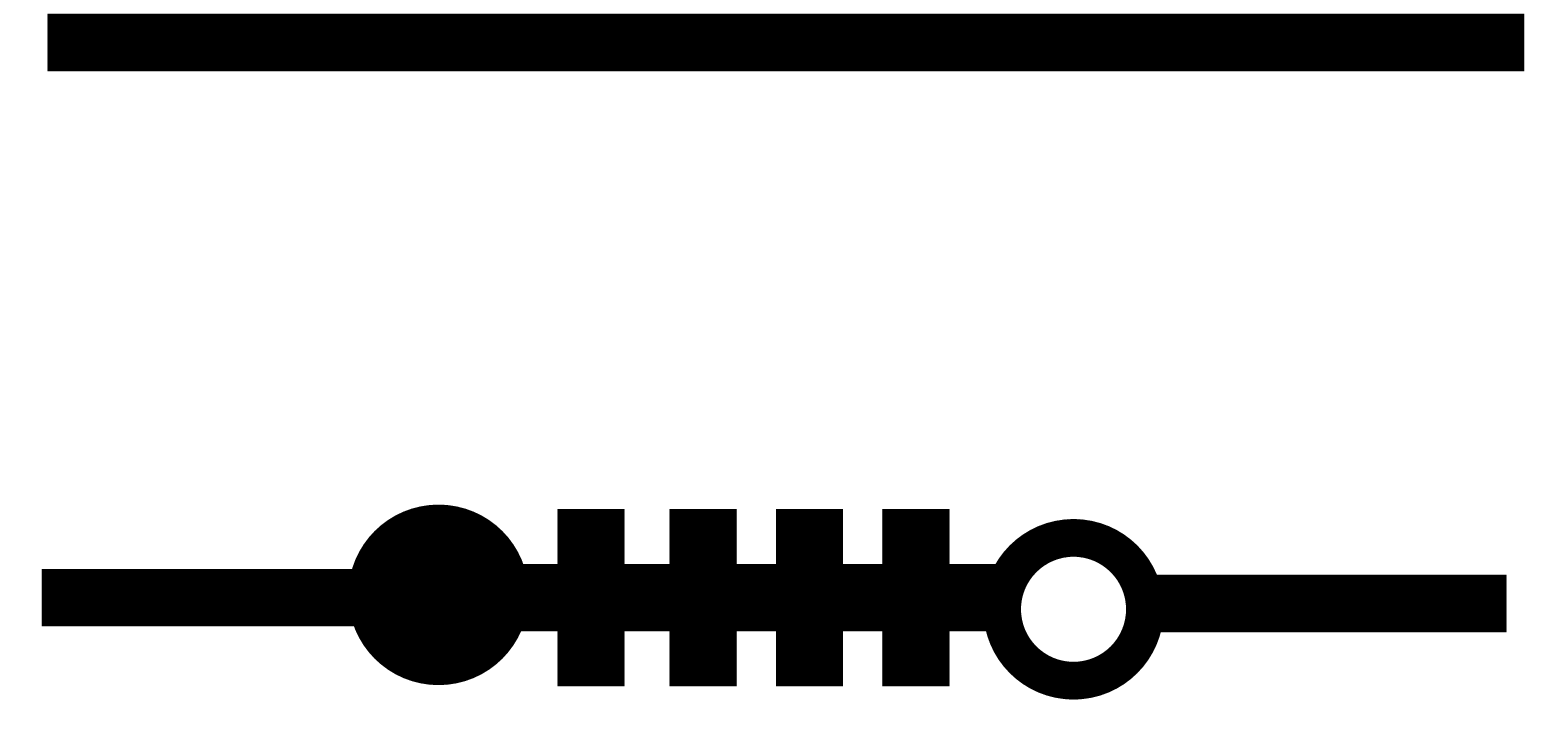}=\fig{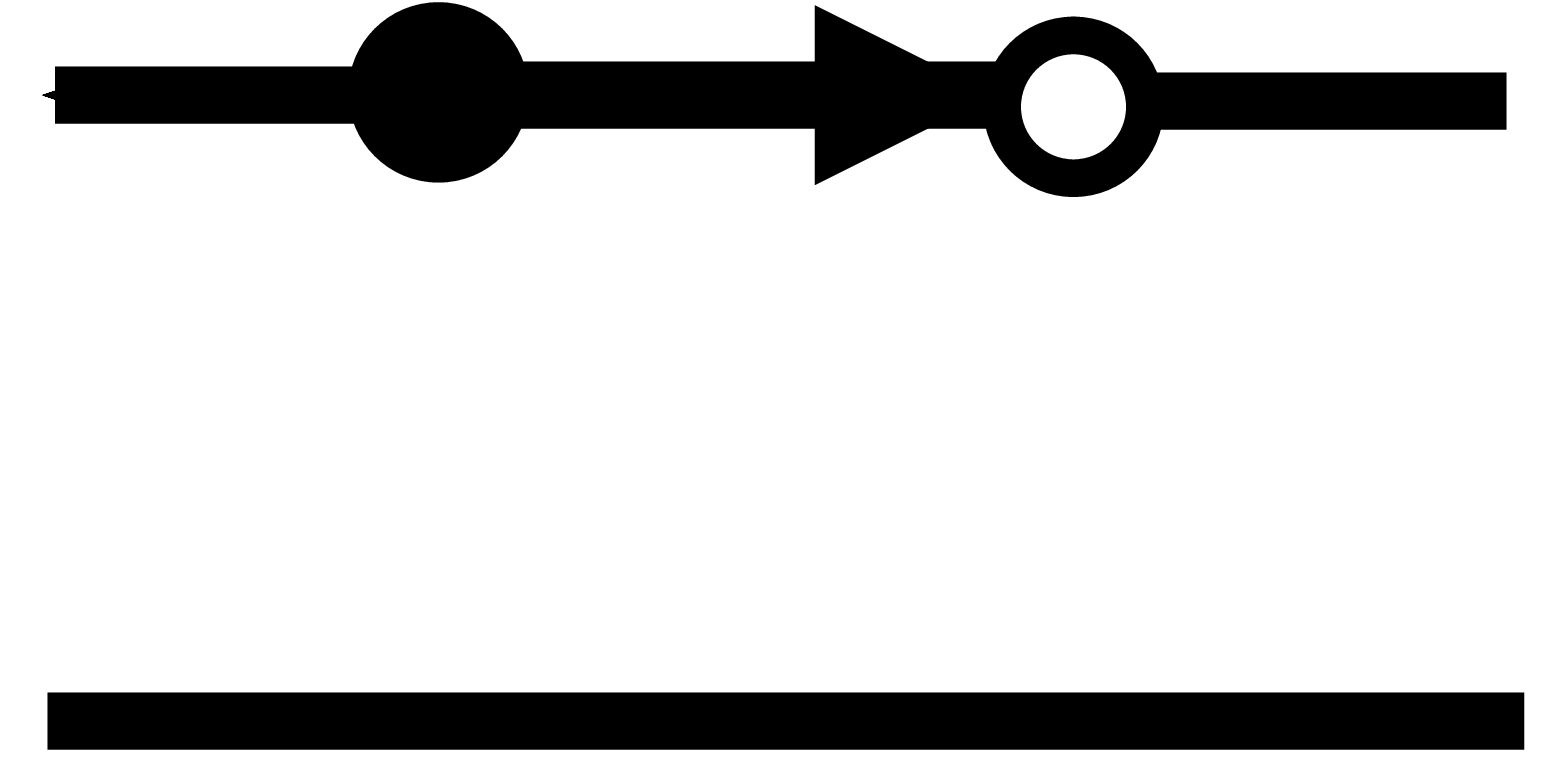}=\fig{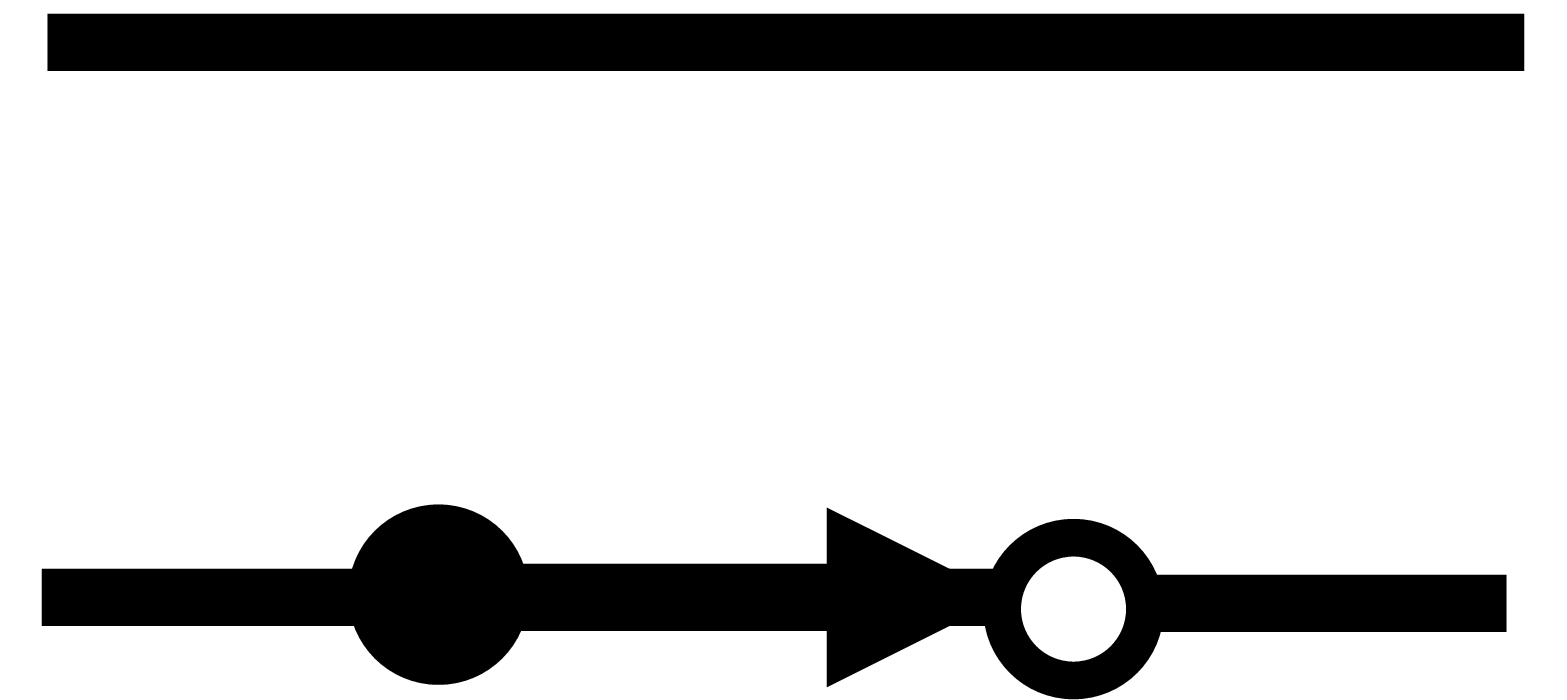}=0.
\label{flatukbonds}
\end{equation}
We note that these diagrams do not depend on the position associated with the open circle, so any diagram can be split
into the contribution of its bonds. For example,
\begin{equation}
\fig{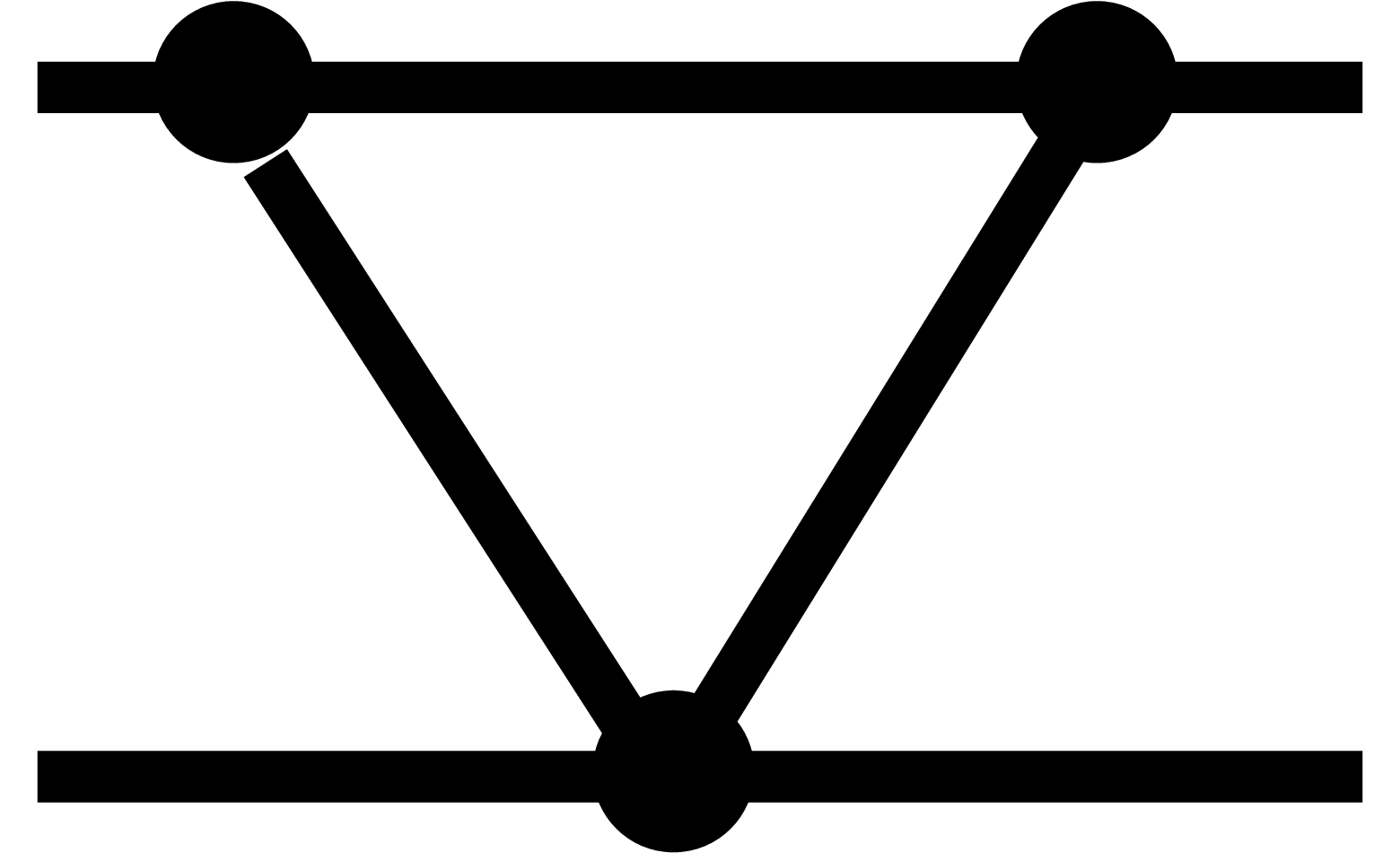}=\left(\fig{diagram126}\right)\times\left(\fig{diagram125}\right)\times \left(\fig{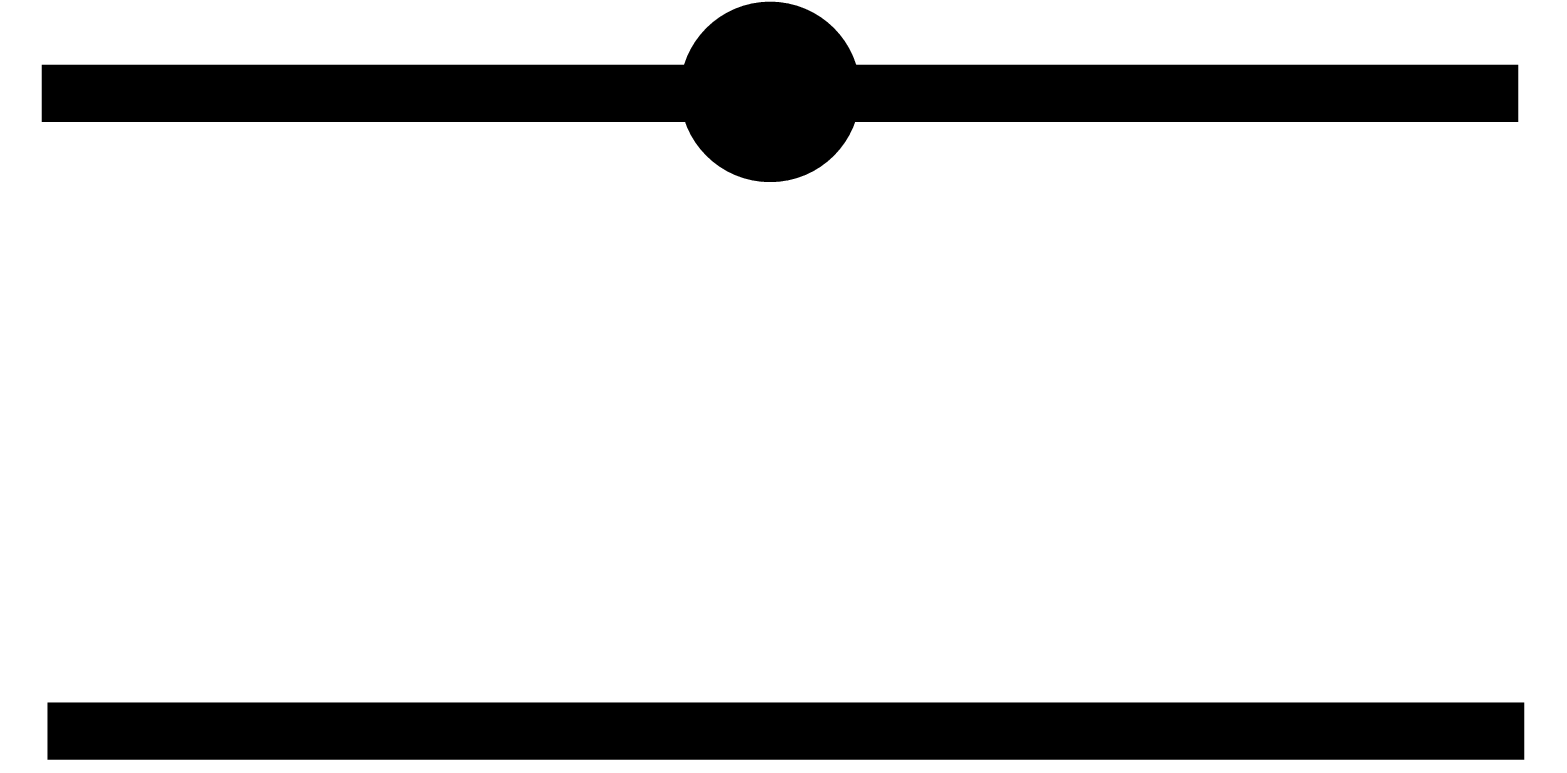}\right),
\end{equation}
with 
\begin{equation}
\fig{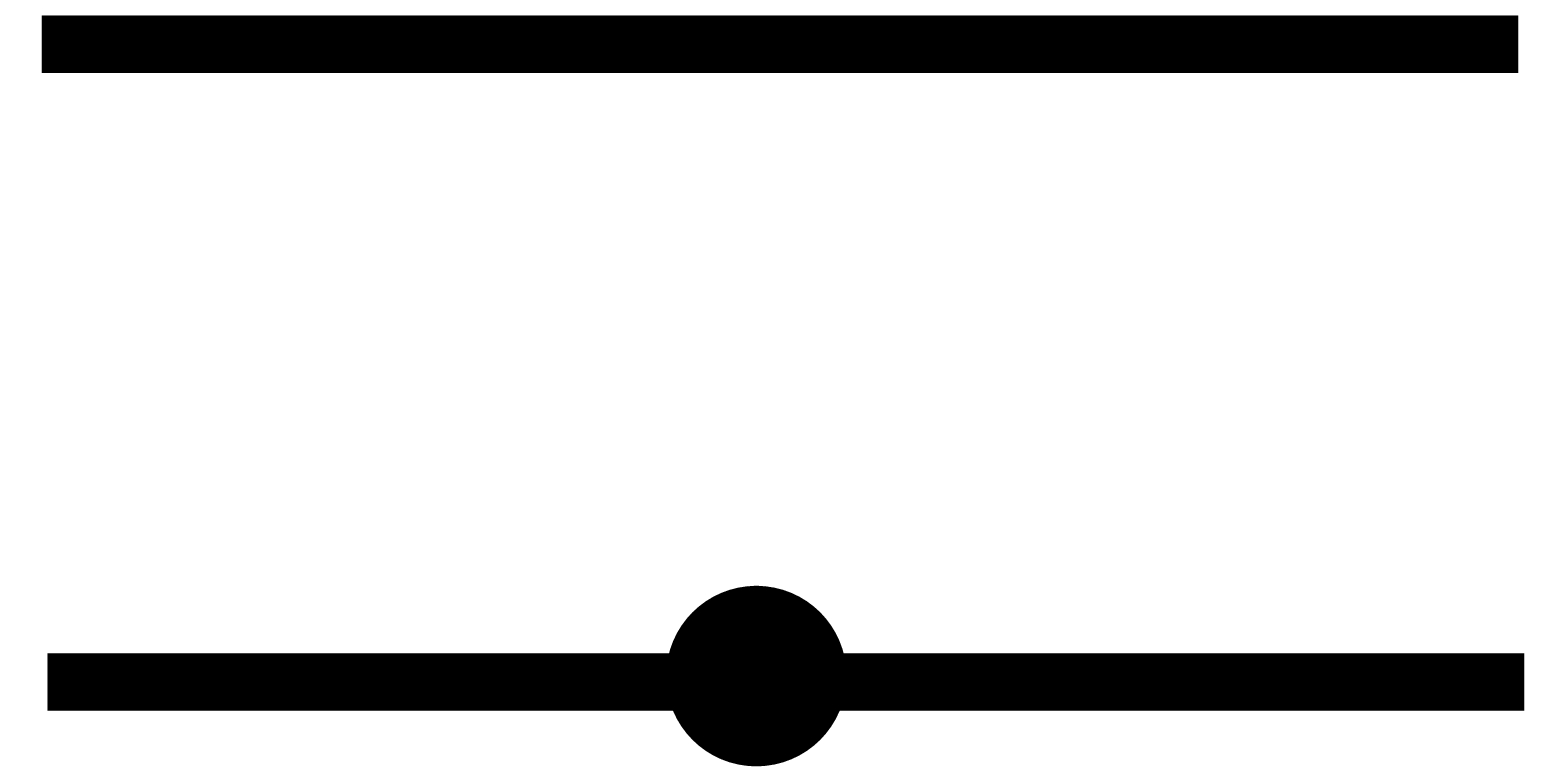}=\fig{diagram132}=\mathcal{A}.
\end{equation}
Due to the expression (\ref{flatukbonds}), the nonvanishing diagrams are those of the original nonlocal model. In 
particular, Eqs. (\ref{omega11}), (\ref{omega12}) and (\ref{omega21}) reduce,
respectively, to  
\begin{eqnarray}
\Omega_1^1 &=& \frac{2g}{g-\kappa} \fig{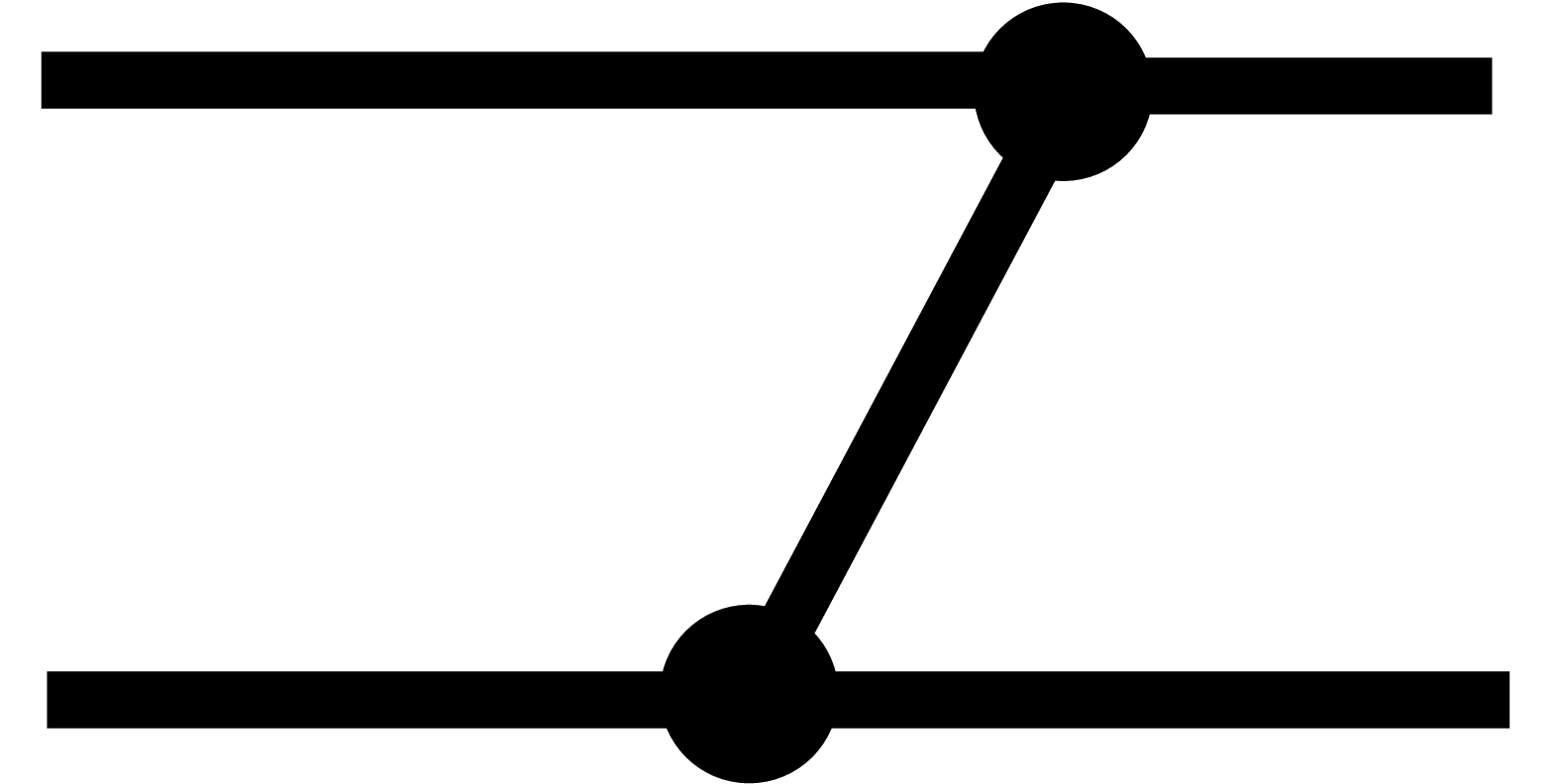} = \frac{2g}{g-\kappa} \mathcal{A} e^{-\kappa \ell},\\
\Omega_1^2 &=& \frac{g+\kappa}{g-\kappa} \fig{diagram131} = \frac{g+\kappa}{g-\kappa} \mathcal{A} e^{-2\kappa \ell},\\
\Omega_2^1 &=& \left(\frac{g}{g-\kappa}\right)^2 \fig{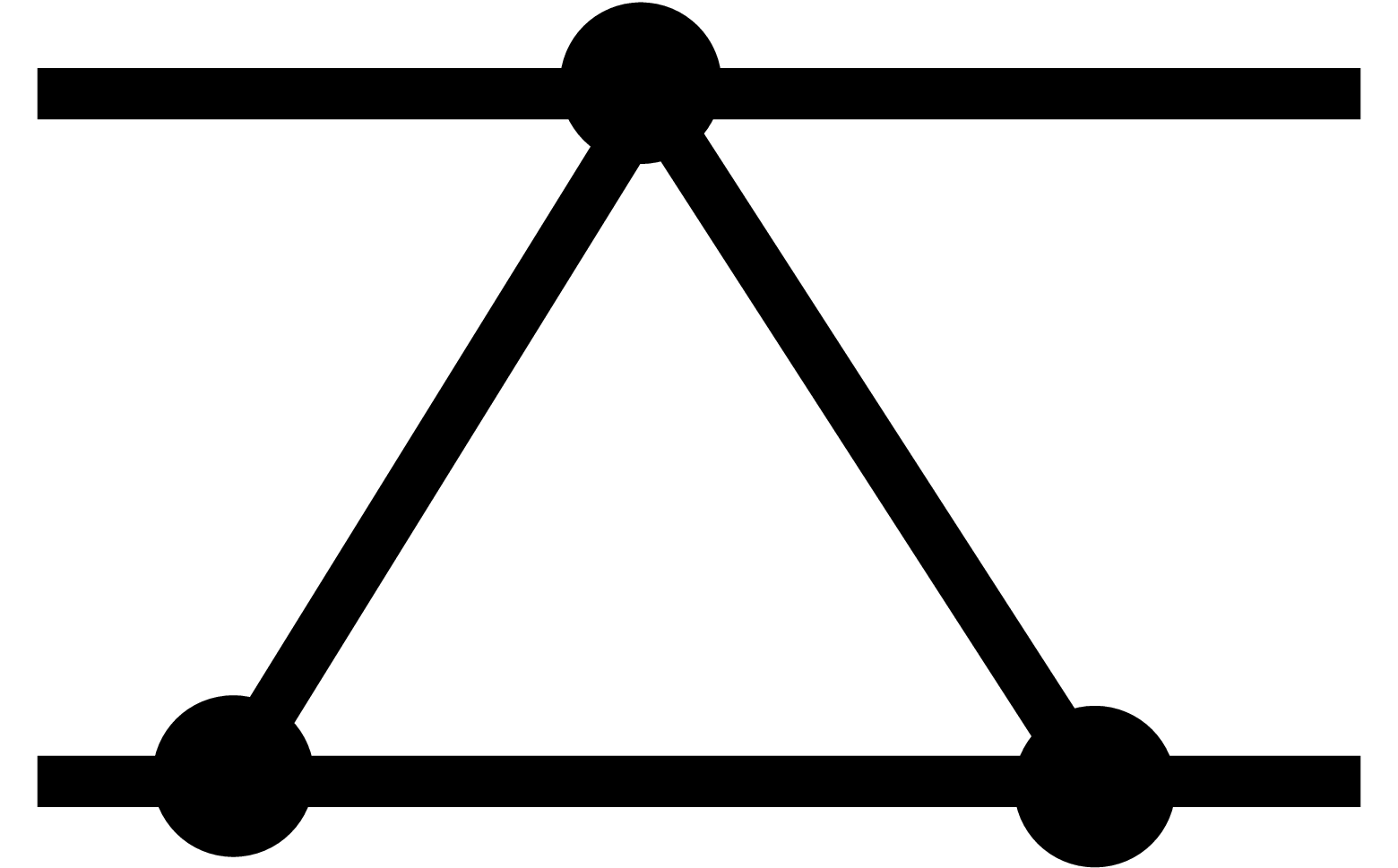} = \left(\frac{g}{g-\kappa}\right)^2  
\mathcal{A} e^{-2\kappa \ell},
\end{eqnarray}
which are consistent with the expressions in Refs. \cite{laura,parry3}, although our notation differs slightly 
from that used in these references. The higher-order terms in the functional can also be easily
evaluated:
\begin{eqnarray}
\Omega_n^n &=& \Omega_1^1 \left(\frac{g+\kappa}{g-\kappa}e^{-2\kappa \ell}\right)^{n-1},\\
\Omega_n^{n+1} &=& \Omega_1^2 \left(\frac{g+\kappa}{g-\kappa}e^{-2\kappa \ell}\right)^{n-1},\\
\Omega_{n+1}^n &=& \Omega_2^1 \left(\frac{g+\kappa}{g-\kappa}e^{-2\kappa \ell}\right)^{n-1}. 
\end{eqnarray}
If we substitute these expressions into Eq.~(\ref{Hmin6}), we reobtain Eq.~(\ref{bindingplanar}) after a trivial 
resummation. Finally, the expressions obtained in Ref. \cite{parry2} for fixed boundary conditions on the wall
are reobtained by taking the limit $g\to -\infty$ and
$-h_1/g-m_0\to \delta m_1\equiv m_1-m_0$ in our equations, where $m_1$ is the order parameter on the wall. 

Now we turn to the formulation for the nonlocal model we have introduced in this paper for fixed boundary conditions 
on the wall. The basic diagrams for this formulation are
\begin{equation}
\fig{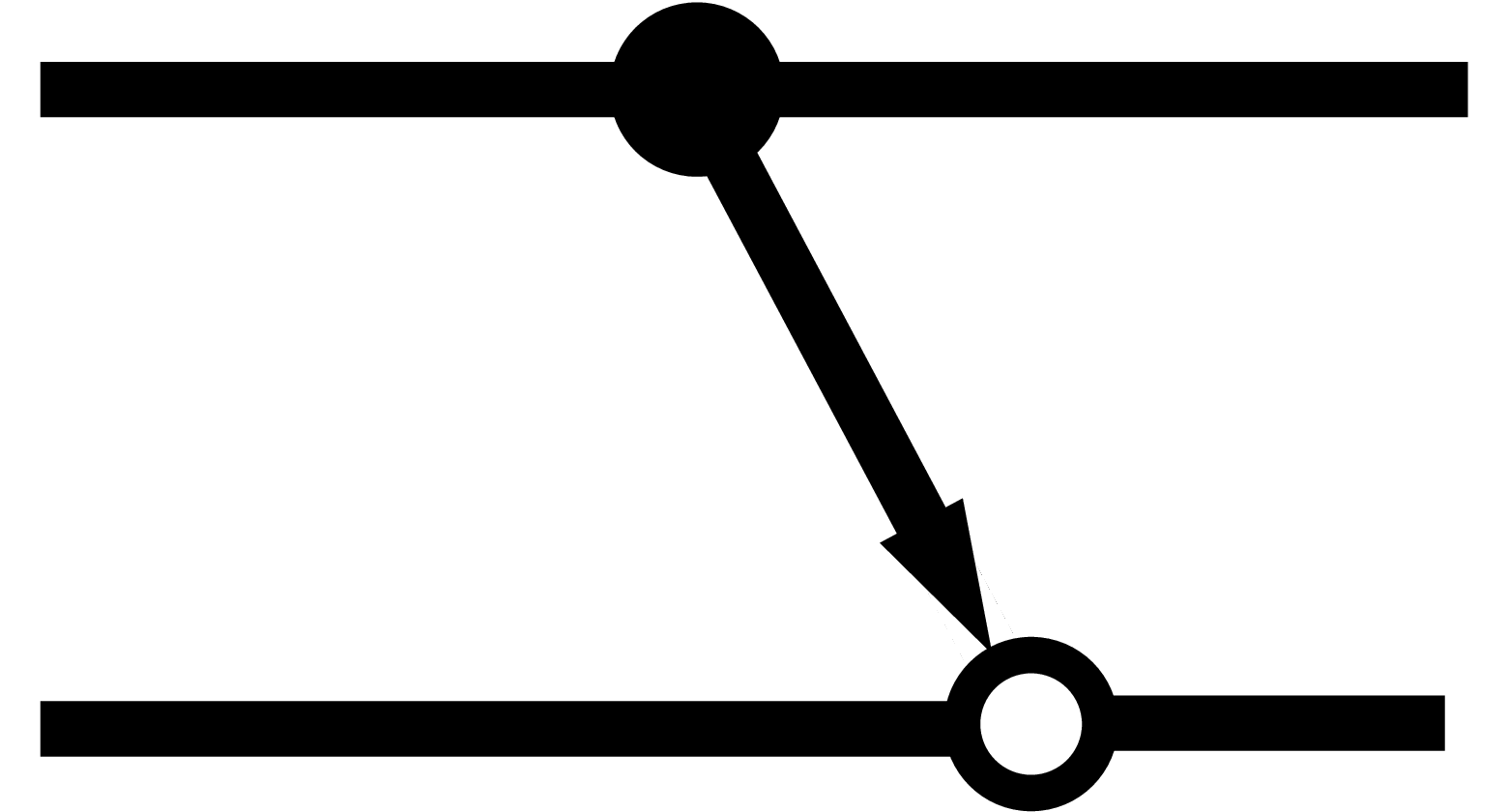}=-\fig{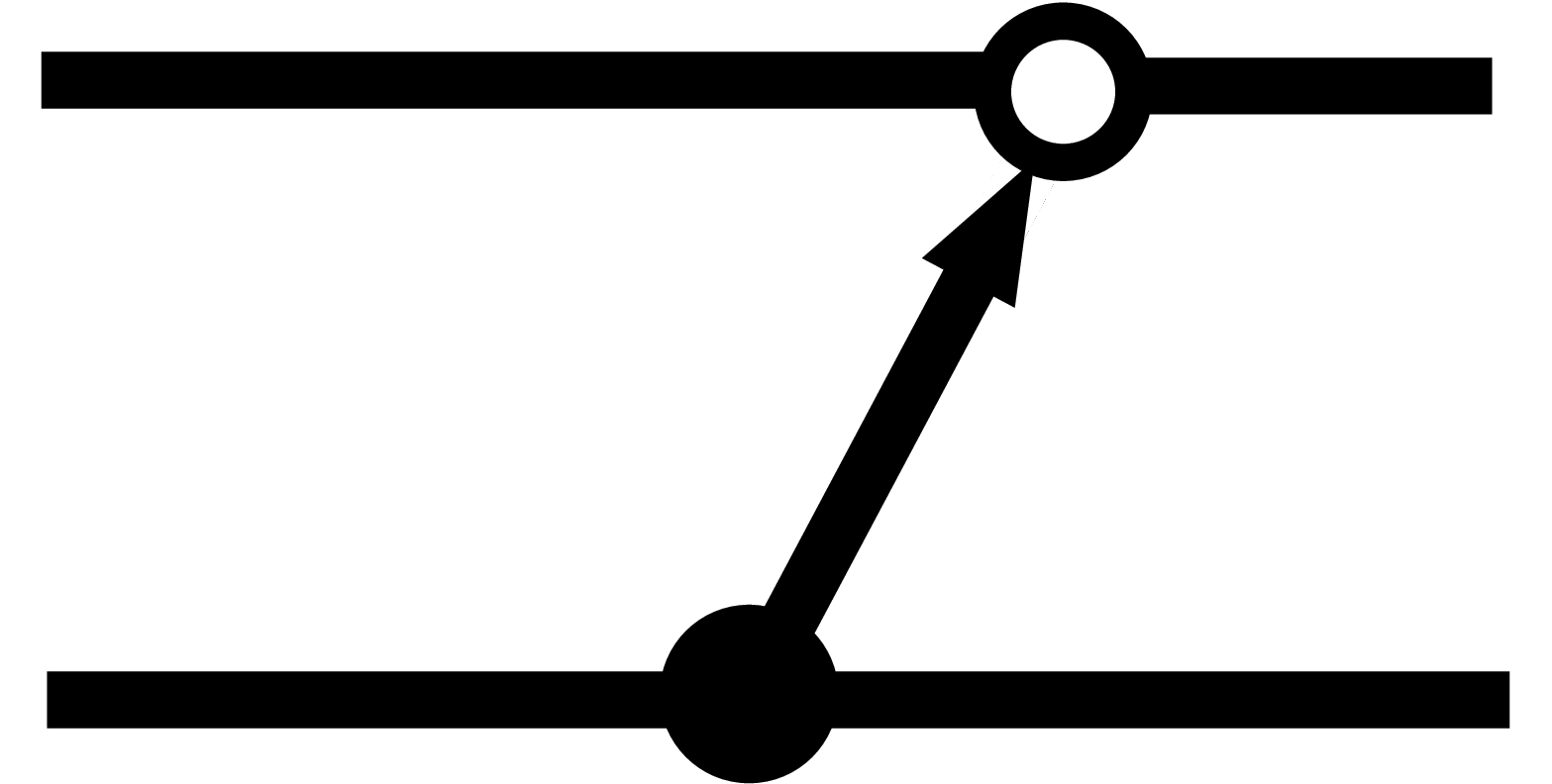}=e^{-\kappa \ell}.
\end{equation}
The expressions for $\Omega_1^1$, $\Omega_1^2$, and $\Omega_2^1$ are obtained from
Eqs. (\ref{omega11-2}), (\ref{omega12-2}) and (\ref{omega21-2}), respectively, 
as
\begin{eqnarray}
\Omega_1^1 &=& \fig{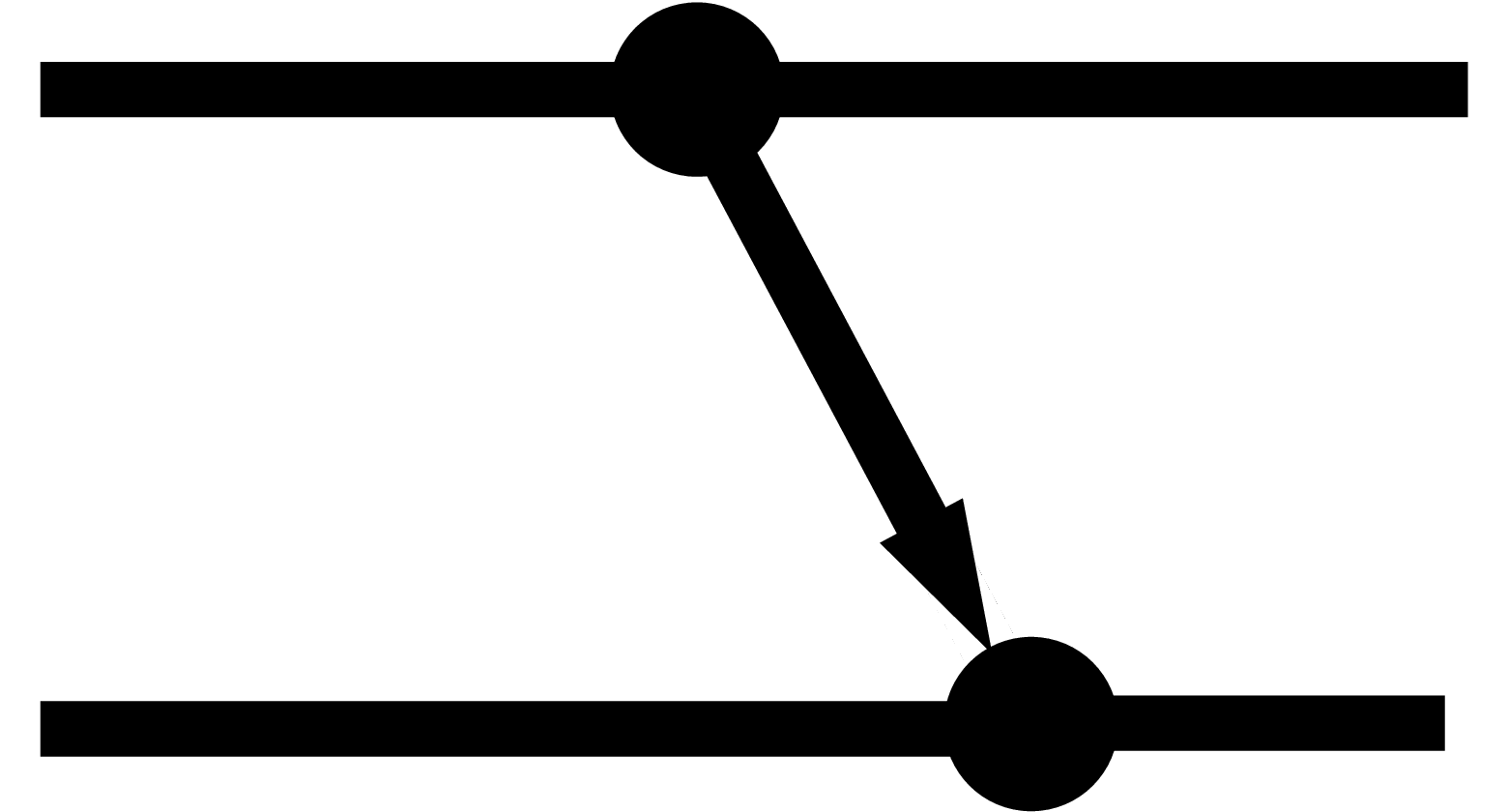}-\fig{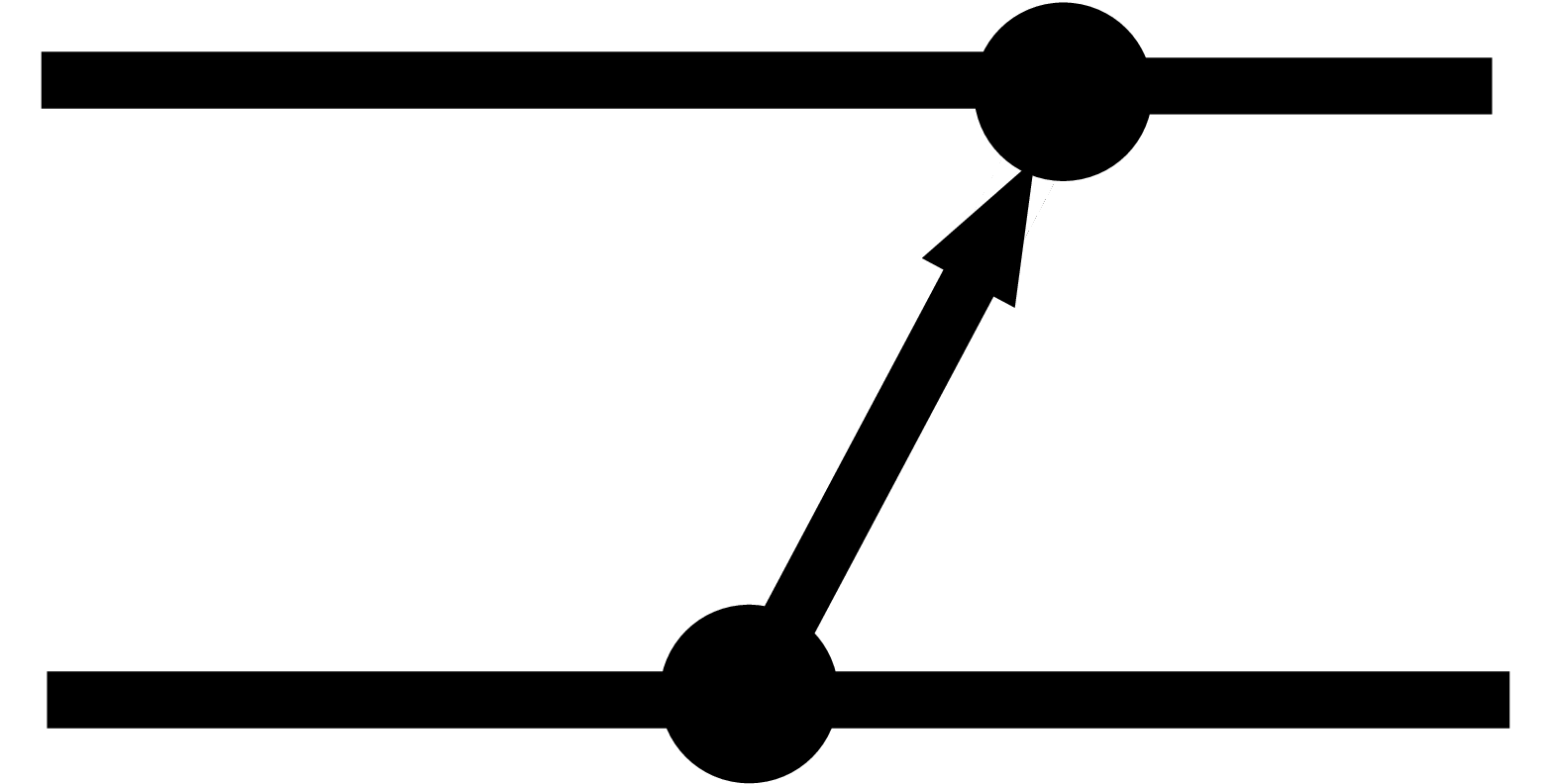} = 2\mathcal{A} e^{-\kappa \ell},\\
\Omega_1^2 &=& - \fig{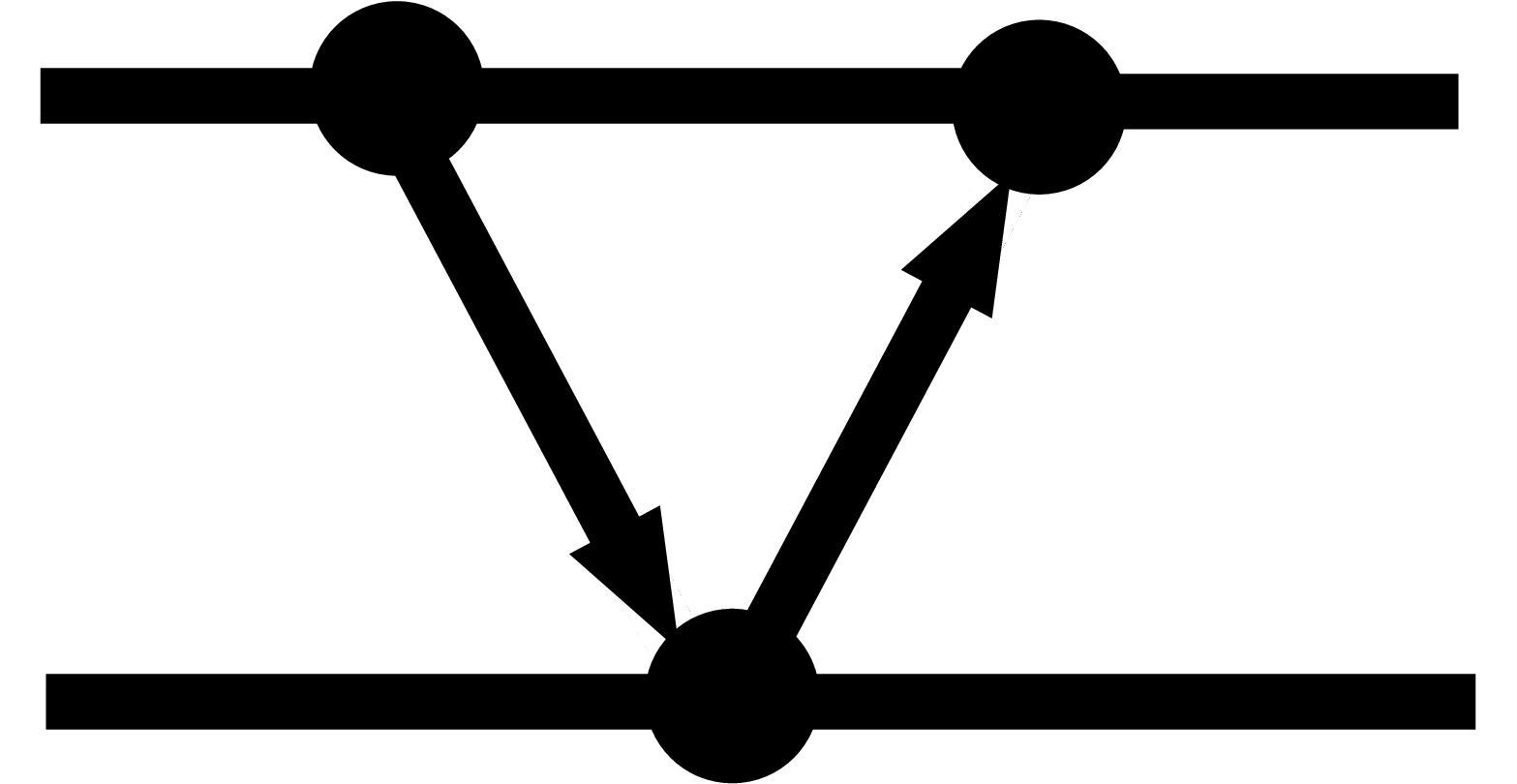} = \mathcal{A} e^{-2\kappa \ell},\\
\Omega_2^1 &=& - \fig{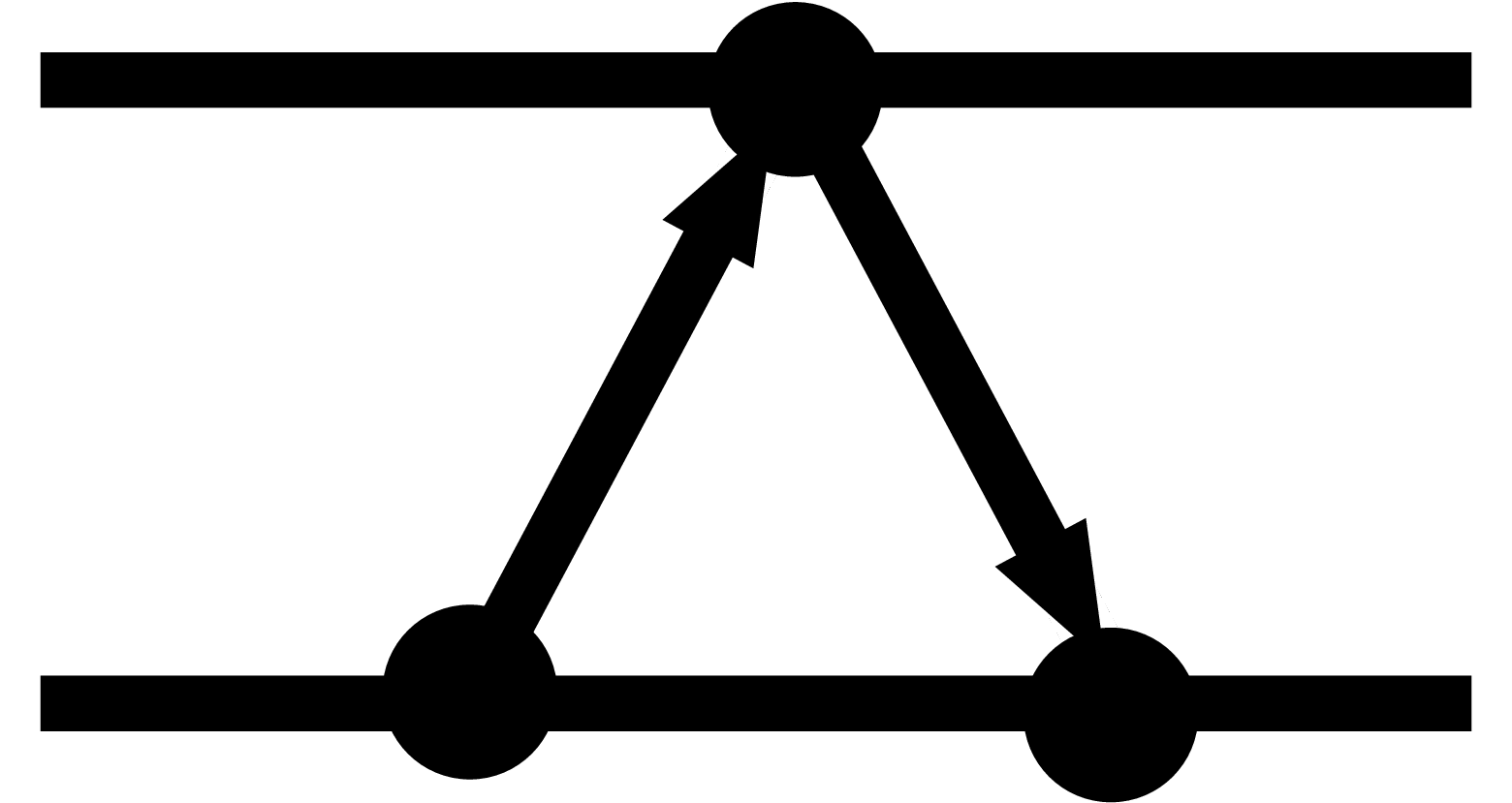} =  \mathcal{A} e^{-2\kappa \ell},
\end{eqnarray}
and for higher-order contributions we get that
\begin{eqnarray}
\Omega_n^n &=& \Omega_1^1 e^{-2(n-1)\kappa \ell},\\
\Omega_n^{n+1} &=& \Omega_1^2 e^{-2(n-1)\kappa \ell},\\
\Omega_{n+1}^n &=& \Omega_2^1 e^{-2(n-1)\kappa \ell}.
\end{eqnarray}
Substitution of these expressions into Eq.~(\ref{Hmin7}) leads to the expression
\begin{eqnarray}
&&\frac{W[\ell,\psi]}{\mathcal{A}}=\sum_{n=1}^\infty \Bigg\{2 \kappa m_0\delta m_1 e^{-(2n-1)\kappa \ell}\nonumber\\
&&  
+\left[\kappa (\delta m_1)^2 + \kappa (m_0)^2\right] e^{-2 n \kappa \ell}
\Bigg\}
\label{Hmin7-2}
\end{eqnarray}
which can be resummed as
\begin{eqnarray}
&&\frac{W[\ell,\psi]}{\mathcal{A}}=\frac{2 \kappa m_0\delta m_1 e^{-\kappa \ell}}{1-e^{-2\kappa \ell}}\nonumber\\
&&  
+\left[\kappa (\delta m_1)^2 + \kappa (m_0)^2\right] \frac{e^{-2\kappa \ell}}{1-e^{-2\kappa \ell}},
\label{bindingplanar2}
\end{eqnarray}
which is Eq.~(\ref{bindingplanar}) in the limit of fixed boundary conditions on the wall.

\subsection{Spherical interfaces}

A similar calculation can be performed for the problem of wetting around a sphere. We suppose that the sphere is of 
radius $R$ and consider an interfacial configuration corresponding to a concentric sphere of radius $R+\ell$.
In this case, $\mathcal{F}_{wl}[\psi]$ is given by \cite{laura}
\begin{equation}
\mathcal{F}_{wl}[\psi]=\sigma_{wl} \mathcal{A}_{wl} \left(\frac{1+\frac{1}{\kappa R}}{1+\frac{1}{(\kappa - g)R}}\right),
\label{fwlsphere}
\end{equation}
where $\sigma_{wl}$ is the surface tension for the planar wall-liquid interface
and the area of the sphere is $ \mathcal{A}_{wl}=4\pi R^2$. Note that the mean curvature and 
the Gaussian curvature on the sphere are $H=-1/R$ and $K_G=1/R^2$, respectively. 
Thus Eq.~(\ref{fwlsphere}) satisfies Eq.~(\ref{deltafwlexpansion}) for large $R$.
It is instructive to reobtain Eq.~(\ref{fwlsphere}) from the diagrammatic expansion Eq.~(\ref{F_chain_1}).
The relevant diagrams for this calculation are
\begin{eqnarray}
\fig{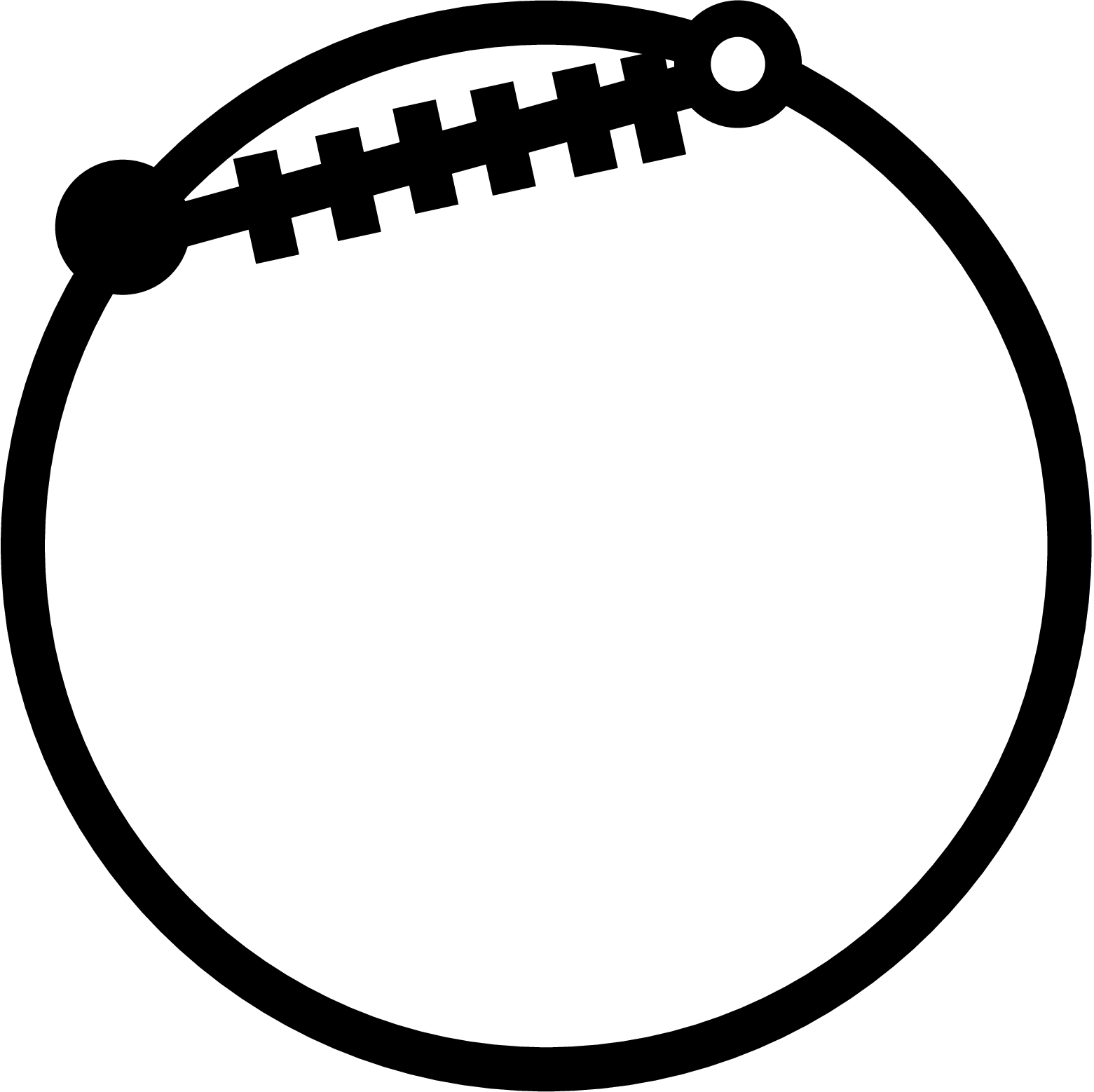}&=&-e^{-2\kappa R},\\
\fig{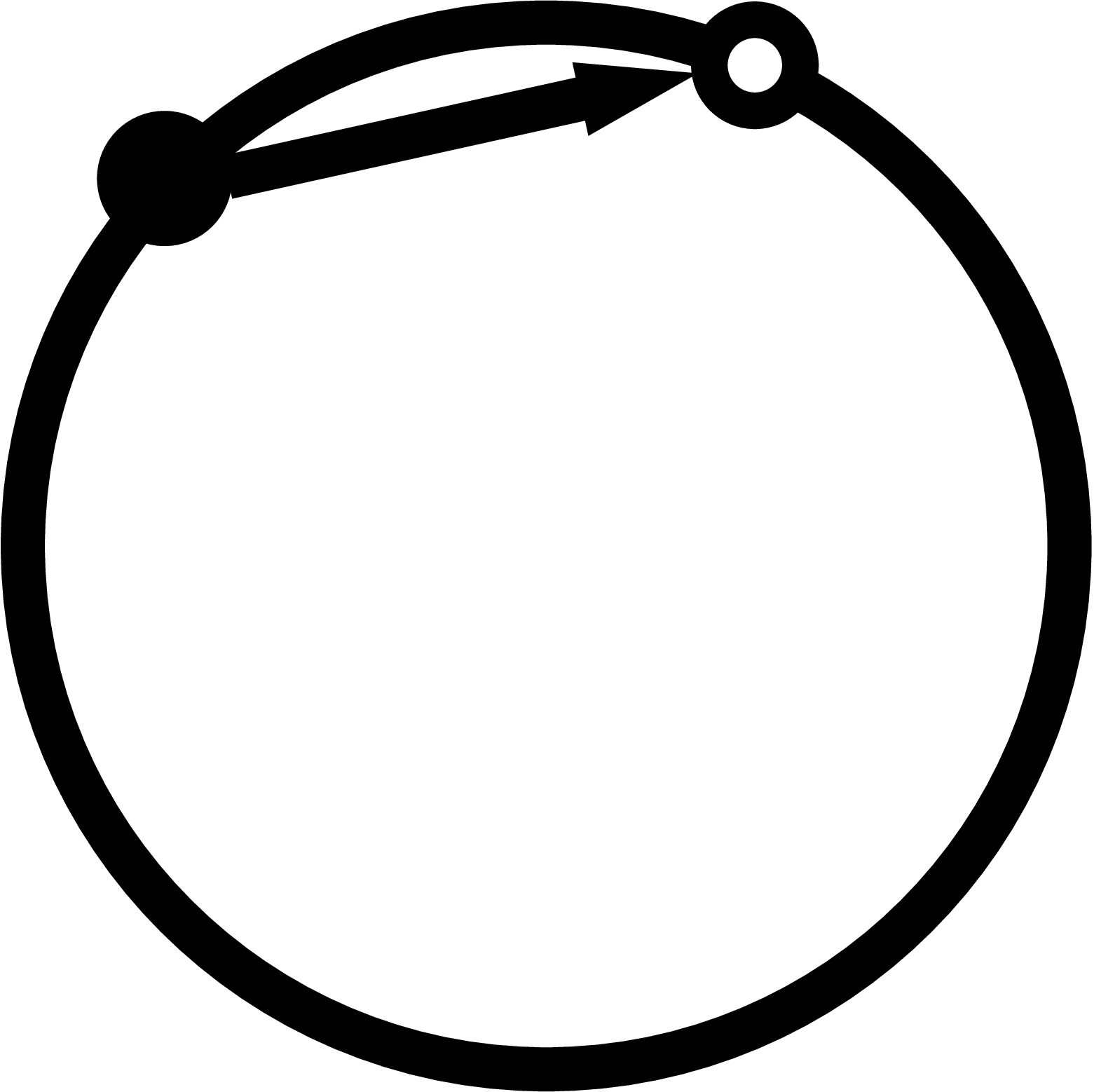}&=&-\frac{1}{\kappa R}\left[1-(1+\kappa R)e^{-2\kappa R}\right],
\\
\fig{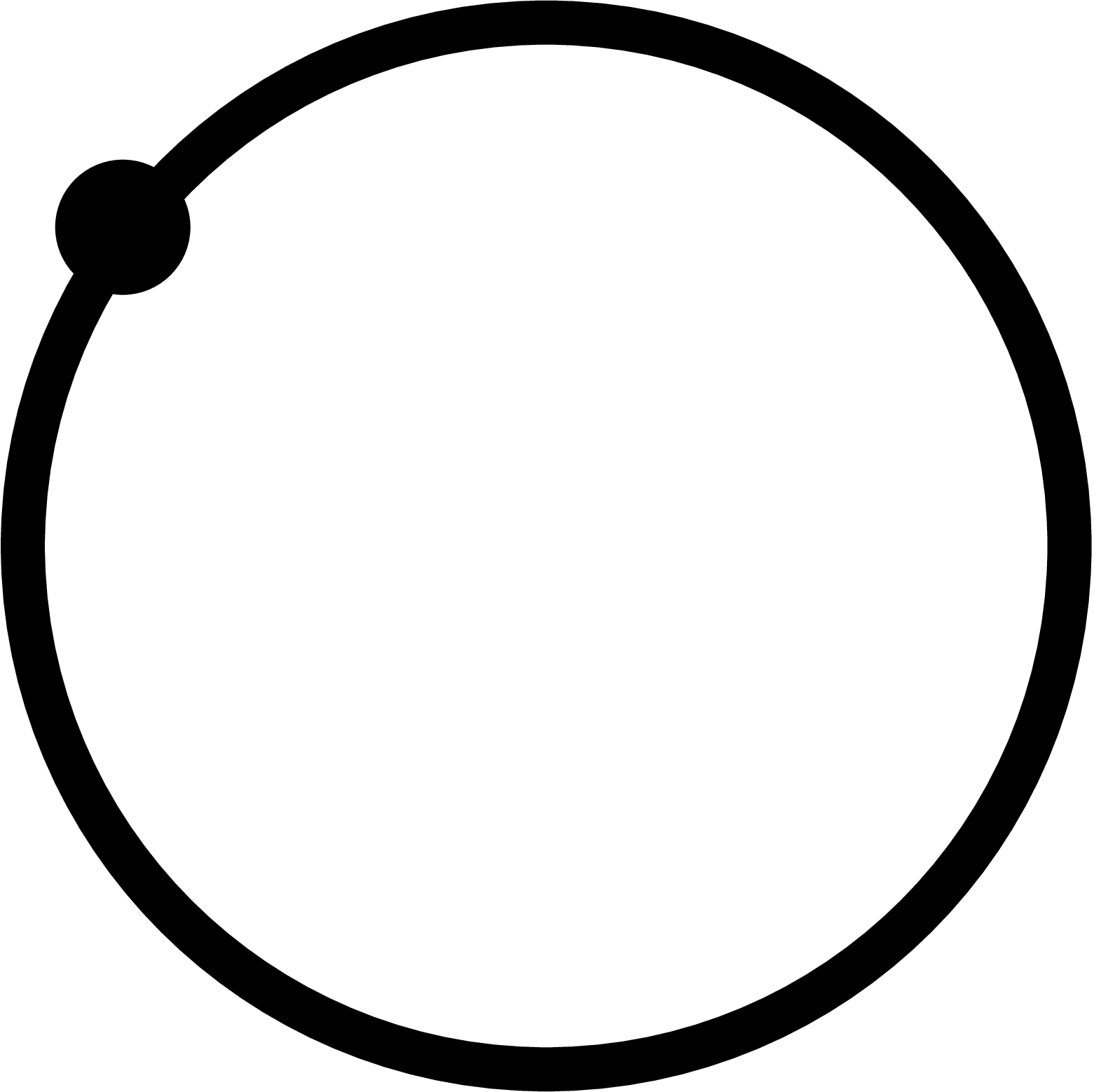}&=&4\pi R^2,
\end{eqnarray}
which are independent of the position of the open circle, as in the planar case. So, again, each diagram
is just the product of its bonds. In order to sum the contributions of the diagrams, we note that each diagram is a
chainlike sequence of $U$ and $\partial K/\kappa$ bonds, with coefficients given by Eq.~(\ref{factor}). We first
sum the diagrams without $\partial K/\kappa$ bonds. Their total contribution $S_0$ to $\mathcal{F}_{wl}$ is
\begin{eqnarray}
S_0&=&\sigma_{wl}\mathcal{A}_{wl}\sum_{n=0}^\infty \left[\frac{g}{\kappa - g}\left(-e^{-2\kappa R}\right)\right]^n
\nonumber\\
&=&\sigma_{wl}\mathcal{A}_{wl}\frac{1}{1+\frac{g}{\kappa - g}e^{-2\kappa R}}.
\end{eqnarray}
Now we consider the diagrams with only one $\partial K/\kappa$ bond. Their total contribution $S_1$ to $\mathcal{F}_{wl}$
can be written as
\begin{eqnarray}
S_1&=&\sigma_{wl}\mathcal{A}_{wl} \left(\sum_{n_1=0}^\infty \left[\frac{g}{\kappa - g}
\left[-e^{-2\kappa R}\right)\right]^{n_1}\right]\nonumber\\
&\times& \left(\frac{\kappa}{\kappa-g}\right)\left[-\frac{1}{\kappa R}\left(1-(1+\kappa R)e^{-2\kappa R}\right)\right]
\nonumber\\
&\times& \left(\frac{g}{\kappa}+\sum_{n_2=1}^\infty \left[\frac{g}{\kappa - g}
\left(-e^{-2\kappa R}\right)\right]^{n_2}\right),
\end{eqnarray}
which can be written as
\begin{eqnarray}
S_1&=& \sigma_{wl}\mathcal{A}_{wl} \frac{g}{\kappa}\frac{1-e^{-2\kappa R}}{1+\frac{g}{\kappa - g}e^{-2\kappa R}}
\nonumber\\
&\times& \frac{-\frac{1}{(\kappa-g)R}\left[
1-(1+\kappa R)e^{-2\kappa R}\right]}{1+\frac{g}{\kappa - g}e^{-2\kappa R}}.
\end{eqnarray}
For diagrams with $m>1$, $\partial K/\kappa$ bonds, their contribution $S_n$ to $\mathcal{F}_{wl}$ can be obtained
similarly as
\begin{eqnarray}
S_m&=& \sigma_{wl}\mathcal{A}_{wl} \frac{g}{\kappa}\frac{1-e^{-2\kappa R}}{1+\frac{g}{\kappa - g}e^{-2\kappa R}}
\nonumber\\
&\times& \left(\frac{-\frac{1}{(\kappa-g)R}\left(
1-(1+\kappa R)e^{-2\kappa R}\right)}{1+\frac{g}{\kappa - g}e^{-2\kappa R}}\right)^m .
\end{eqnarray}
So, $\mathcal{F}_{wl}[\psi]$ has the expression
\begin{eqnarray}
\mathcal{F}_{wl}[\psi]&=&\sum_{m=0}^{\infty} S_m = \frac{\sigma_{wl}\mathcal{A}_{wl}}{1+\frac{g}{\kappa-g}e^{-2\kappa R}}
\nonumber\\
&\times&
\Bigg[1+\frac{g}{\kappa}\left(1-e^{-2\kappa R}\right)
\nonumber\\
&\times& \frac{\frac{-1}{(\kappa-g)R}\left(\frac{1-(1+\kappa R)e^{-2\kappa R}}{1+\frac{g}{\kappa-g}e^{-2\kappa R}}\right)}
{1+\frac{1}{(\kappa-g)R}\left(\frac{1-(1+\kappa R)e^{-2\kappa R}}{1+\frac{g}{\kappa-g}e^{-2\kappa R}}\right)}\Bigg],
\end{eqnarray}
which after some algebra reduces to Eq.~(\ref{fwlsphere}). 

Similarly, $H[\ell]$ has the expression
\begin{equation}
H[\ell]=\frac{\sigma\mathcal{A}_{lg}}{1-e^{-2\kappa (R+\ell)}},
\label{hsphere}
\end{equation}
where $\sigma$ is the 
surface tension for the planar liquid-gas interface,
with area $ \mathcal{A}_{lg}=4\pi (R+\ell)^2$. The diagrammatic expansion ~(\ref{F_chain}) can be evaluated 
explicitly \cite{PR}, where now the basic diagrams are
\begin{eqnarray}
\fig{diagram142}&=&-e^{-2\kappa (R+\ell)},\\
\fig{diagram144}&=&4\pi (R+\ell)^2,
\end{eqnarray}
leading, after resummation, to Eq.~(\ref{hsphere}).

We turn to the evaluation of $W[\ell,\psi]$, which has the expression
\begin{eqnarray}
&&W[\ell,\psi]=2\kappa m_0 \left(-\frac{h_1+gm_0}{g-\kappa-\frac{1}{R}}\right)
\frac{\sqrt{\mathcal{A}_{wl}\mathcal{A}_{lg}} e^{-\kappa \ell}}{1-
\frac{g+\kappa-\frac{1}{R}}{g-\kappa-\frac{1}{R}}e^{-2\kappa \ell}}
\nonumber\\ 
&& + \frac{g+\kappa-\frac{1}{R}}{g-\kappa-\frac{1}{R}}
\left(\frac{\kappa m_0^2}{1-e^{-2\kappa (R+\ell)}} \right)
\frac{\mathcal{A}_{lg}e^{-2 \kappa \ell}}
{1-\frac{g+\kappa-\frac{1}{R}}{g-\kappa-\frac{1}{R}}e^{-2\kappa \ell}}
\nonumber\\
&&+ \kappa \left(\frac{h_1+gm_0}{g-\kappa-\frac{1}{R}}
\right)^2 \frac{\mathcal{A}_{wl}e^{-2\kappa\ell}}{1-
\frac{g+\kappa-\frac{1}{R}}{g-\kappa-\frac{1}{R}}e^{-2\kappa \ell}} .
\label{bindingspherical}
\end{eqnarray}
This expression reduces to Eq.~(\ref{bindingplanar}) for $R\to \infty$, and
it is consistent with those reported in Refs. \cite{laura,parry2} if the
exponential term $\exp[-2\kappa(R+\ell)]$ in Eq.~(\ref{bindingspherical}) 
is neglected.

As in the planar case, we will reproduce this result within the nonlocal 
model. We will first consider the decorated version of the original 
nonlocal model. In this formalism, 
the basic diagrams for this model are 
\begin{eqnarray}
\fig{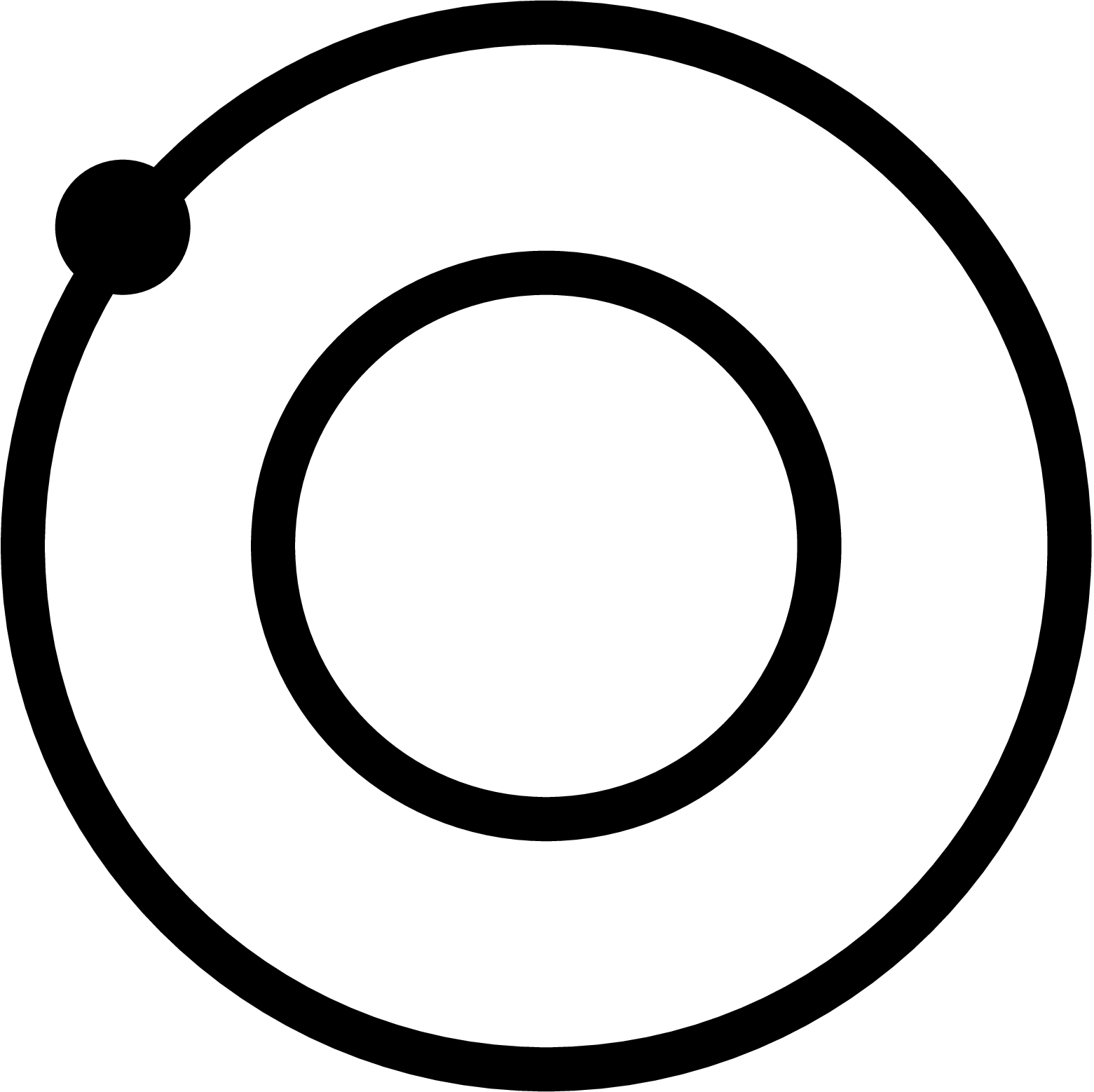}&=&4\pi (R+\ell)^2,\\
\fig{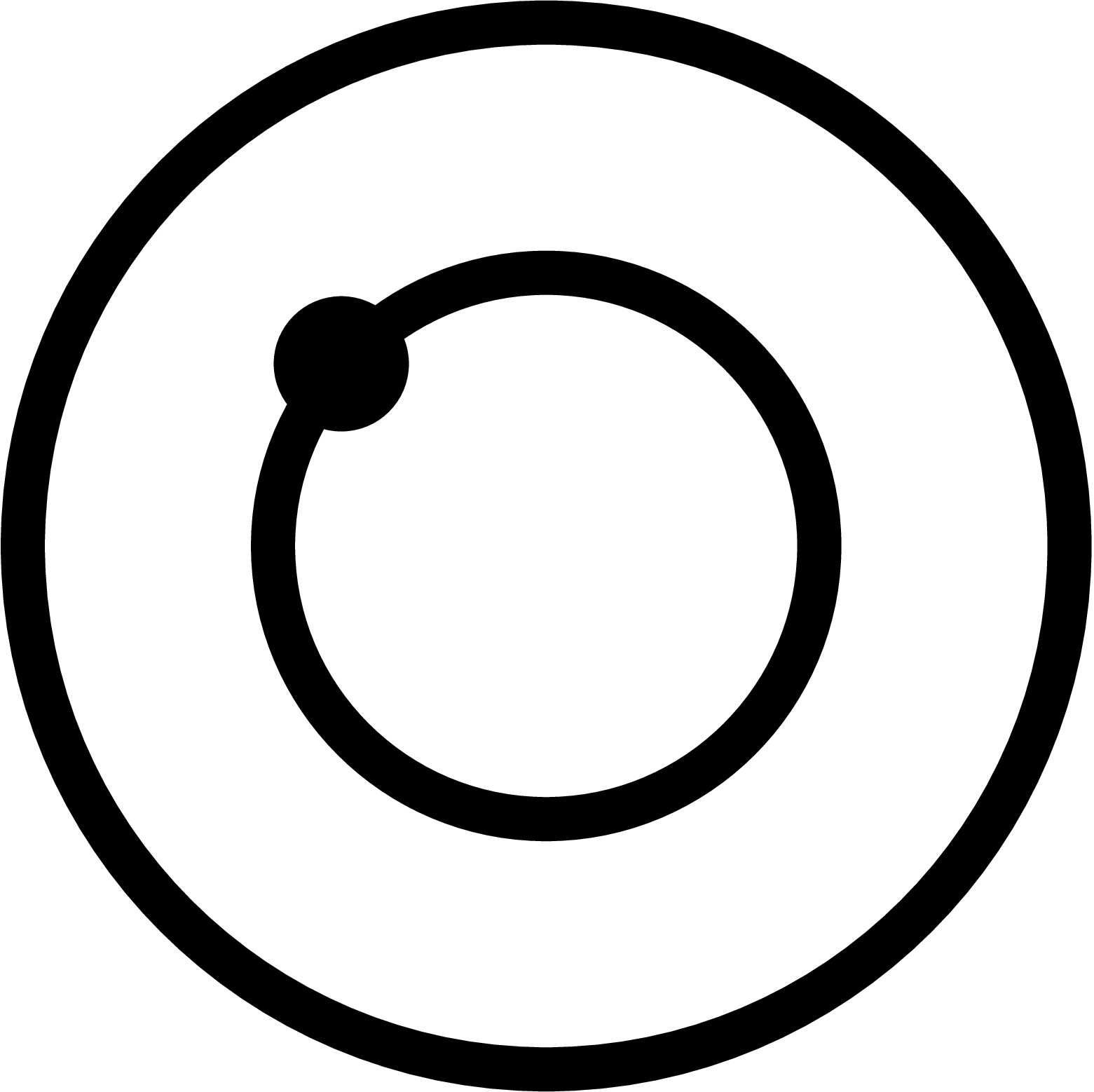}&=&4\pi R^2,\\
\fig{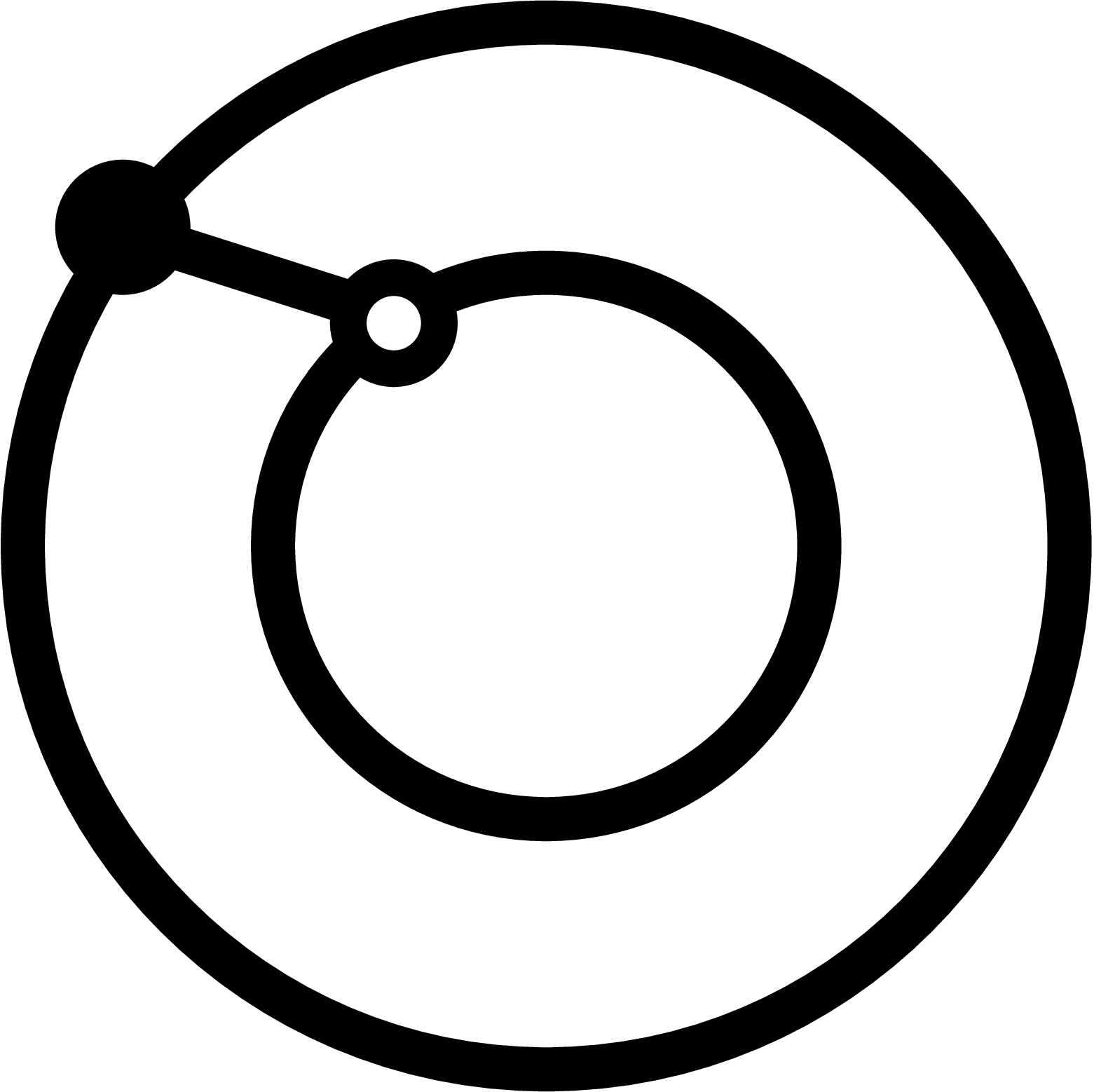}&=&\left(1+\frac{\ell}{R}\right)\left(1-e^{-2\kappa R}\right)e^{-\kappa\ell},\\
\fig{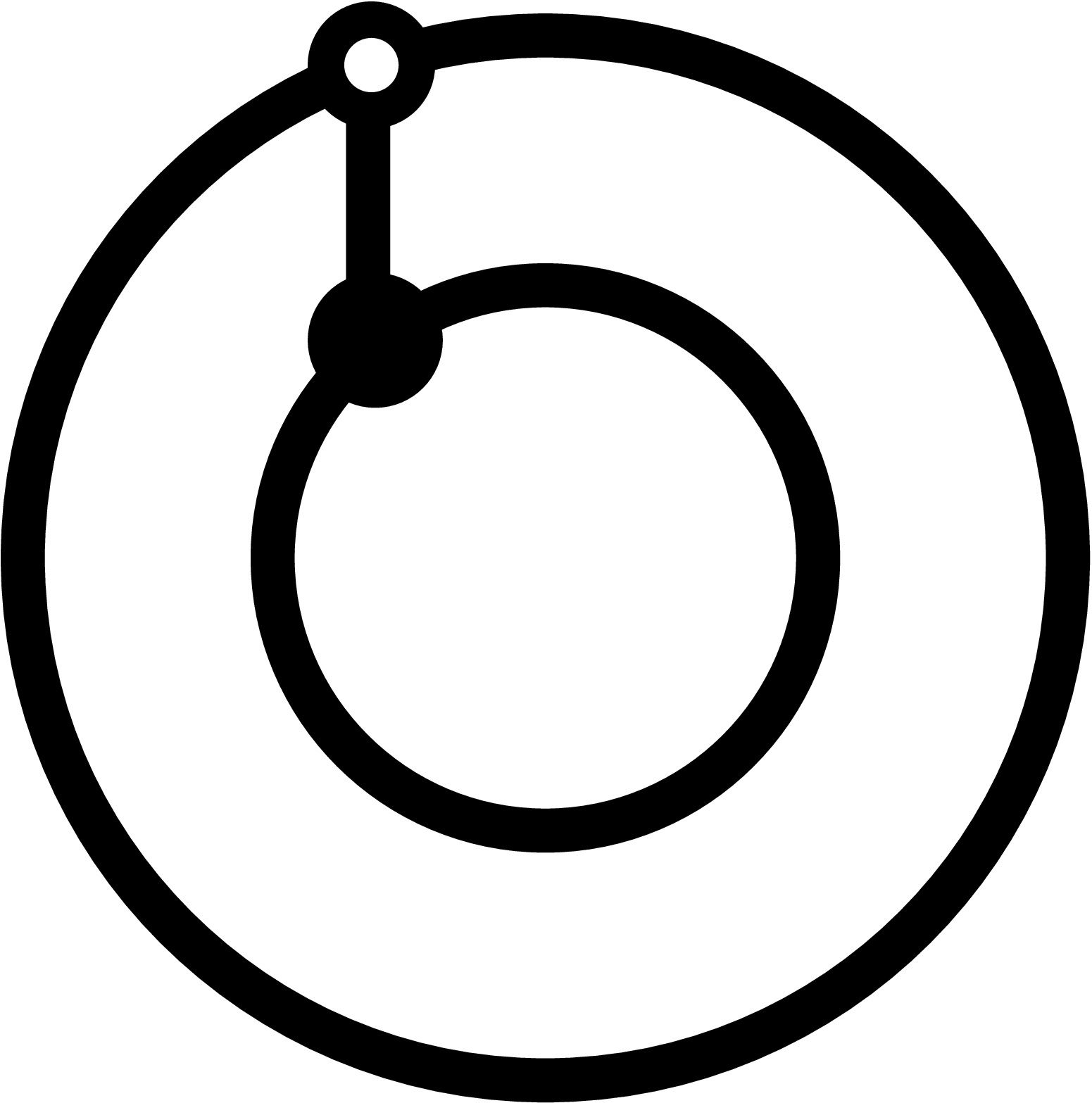}&=&\frac{1-e^{-2\kappa R}}{1+\frac{\ell}{R}}e^{-\kappa\ell}\\
\fig{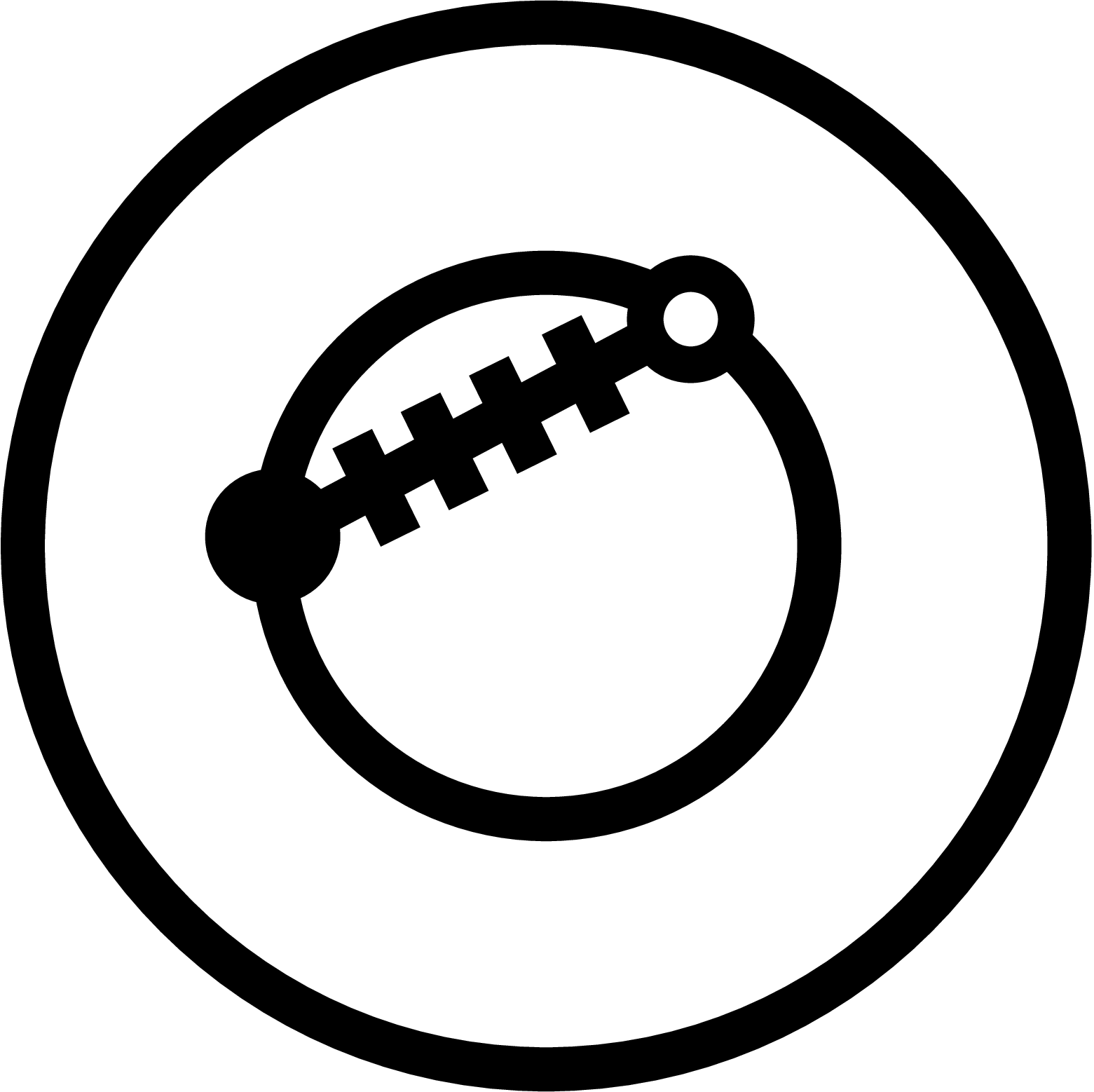}&=&-e^{-2\kappa R},\\
\fig{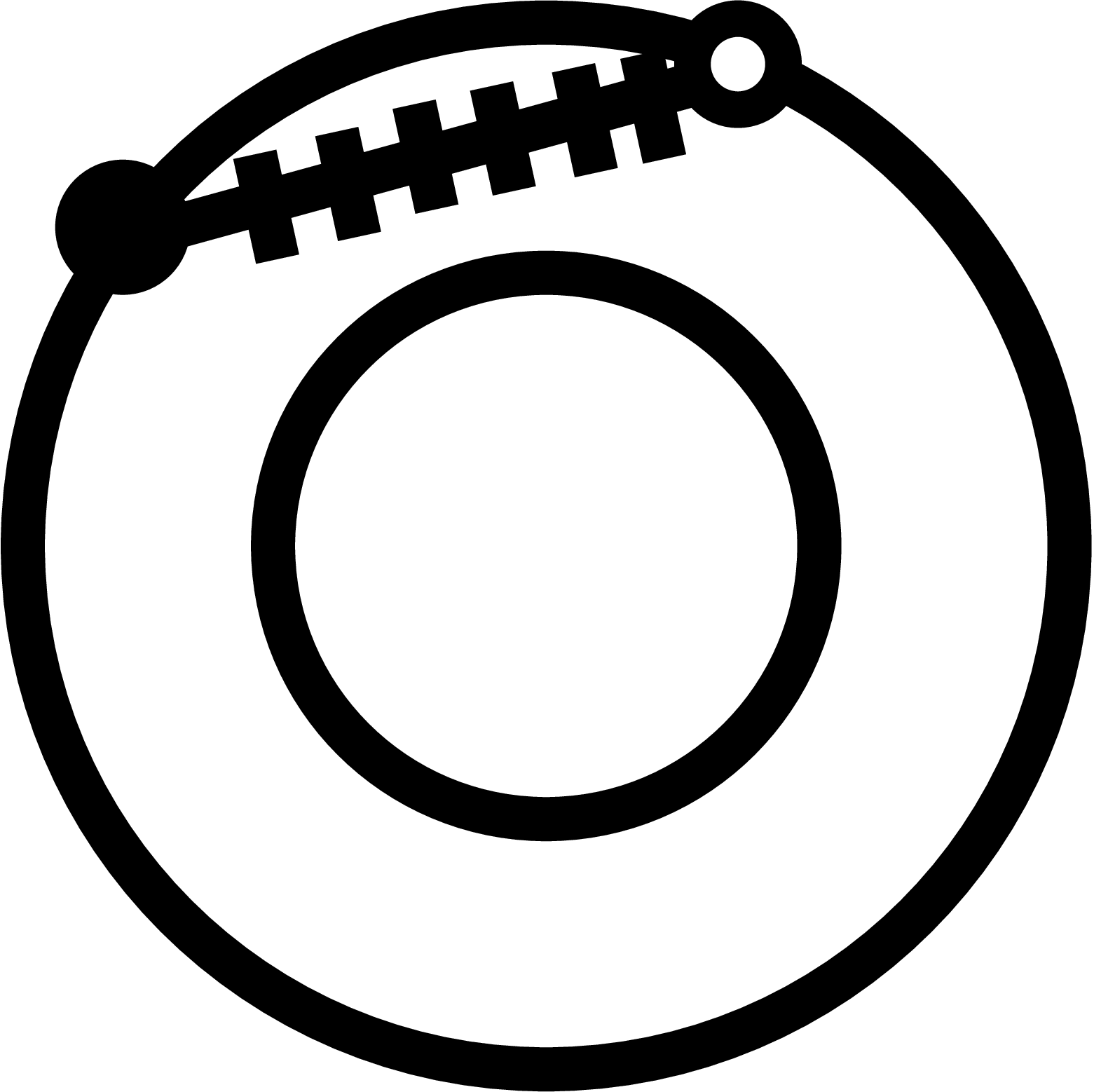}&=&-e^{-2\kappa (R+\ell)},\\
\fig{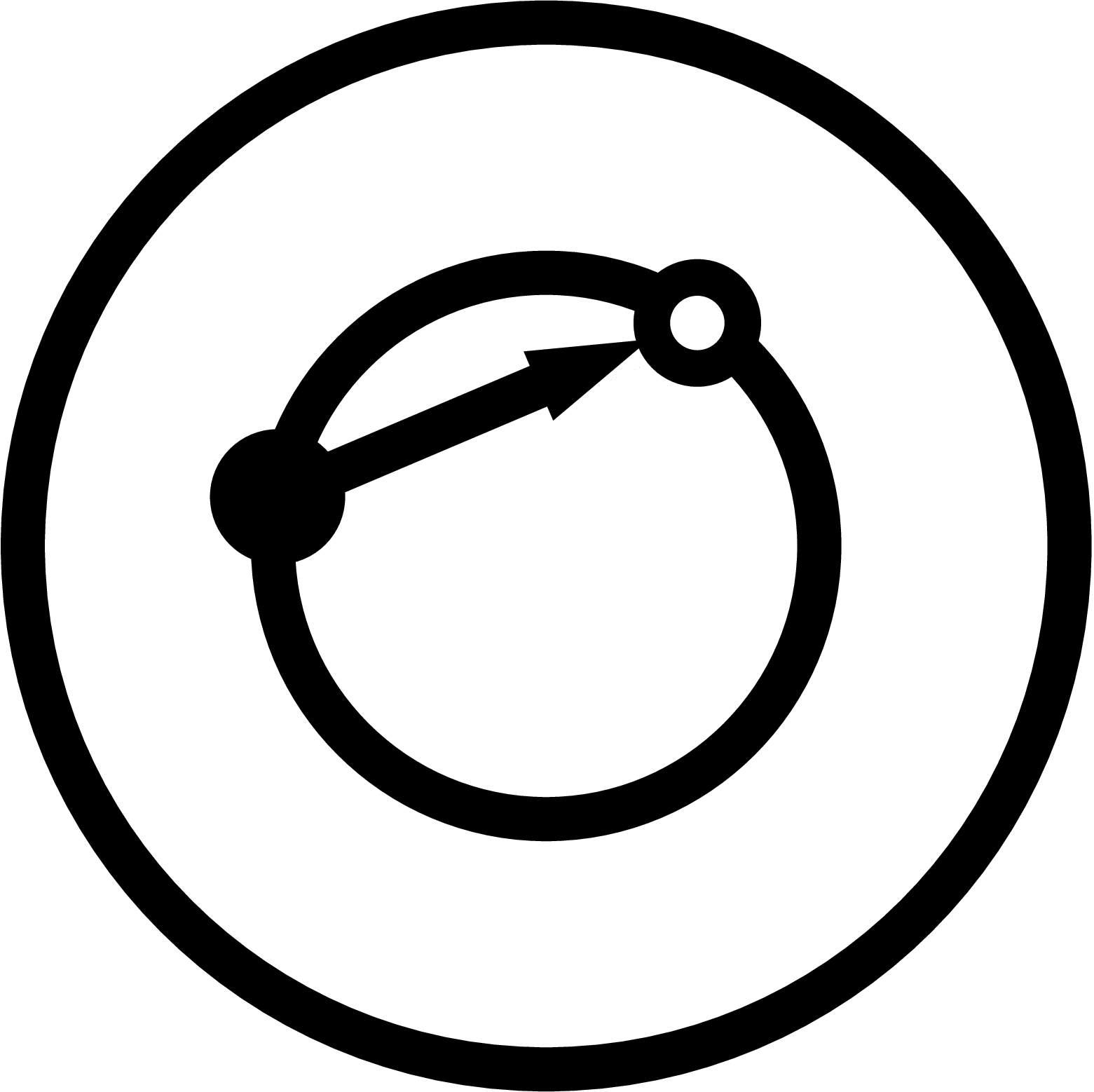}&=&-\frac{1}{\kappa R}\left[1-(1+\kappa R)e^{-2\kappa R}\right].
\end{eqnarray}
Note that, as for the planar case, the diagrams do not depend on the position
associated with the open circle, so a general diagram can be obtained as a
product of its bond contributions. In order to resum all the contributions 
to $\Omega_n^n$, $\Omega_{n+1}^n$, and $\Omega_{n}^{n+1}$, we have to sum
the contributions of all possible segments either on the wall or on the 
liquid-gas interface in a similar manner as we did for the evaluation of 
$\mathcal{F}_{wl}$. The resummation of all the contributions of the segments
of consecutive ($U$) bonds on the liquid-gas interface is
\begin{equation}
I_\ell=\frac{1}{1-e^{-2\kappa (R+\ell)}}.
\label{il}
\end{equation}
On the other hand, the analogous expression for a segment of consecutive bonds
on the wall depends on its position in the diagram. If the segment is on an 
extreme of the full diagram, its contribution is
\begin{equation}
I_\psi^1=\frac{g}{g-\kappa-\frac{1}{R}} \frac{1}{1-e^{-2\kappa R}}.
\label{ipsi1}
\end{equation}
Otherwise, the contribution of the segment is 
\begin{equation}
I_\psi^2=\left(\frac{1}{1-e^{-2\kappa R}}\right)\left(1+\frac{2\kappa}
{g-k-\frac{1}{R}}\frac{1}{1-e^{-2\kappa R}}\right).
\end{equation}   
With these results, Eqs. (\ref{omega11}), (\ref{omega12}) and (\ref{omega21}) reduce, respectively, to  
\begin{eqnarray}
&&\Omega_1^1 = 8\pi R(R+\ell)\left(\frac{g}{g-k-\frac{1}{R}}\right)
\frac{e^{-\kappa \ell}}{1-e^{-2\kappa(R+\ell)}},\label{omega11sphere}\\
&&\Omega_1^2 = 4\pi (R+\ell)^2 \left(\frac{1-e^{-2\kappa R}}{1-e^{-2\kappa(R+\ell)}}\right)\nonumber\\
&&\times\left(1+\frac{2\kappa}
{g-k-\frac{1}{R}}\frac{1}{1-e^{-2\kappa R}}\right)\frac{e^{-2 \kappa \ell}}{1-e^{-2\kappa(R+\ell)}},\label{omega12sphere}\\
&&\Omega_2^1 = 4 \pi R^2 \left(\frac{g}{g-k-\frac{1}{R}}\right)^2  
\frac{e^{-2\kappa \ell}}{1-e^{-2\kappa(R+\ell)}}.
\label{omega21sphere}
\end{eqnarray}
For higher-order contributions
\begin{eqnarray}
&&\Omega_n^n = \Omega_1^1 \Bigg[\left(\frac{1-e^{-2\kappa R}}{1-e^{-2\kappa(R+\ell)}}\right)
,\label{omegannsphere}\nonumber\\&&\times\left(1+\frac{2\kappa}
{g-k-\frac{1}{R}}\frac{1}{1-e^{-2\kappa R}}\right)e^{-2\kappa \ell}\Bigg]^{n-1}\\
&&\Omega_n^{n+1} = \Omega_1^2 \Bigg[\left(\frac{1-e^{-2\kappa R}}{1-e^{-2\kappa(R+\ell)}}\right)
,\label{omegann1sphere}\nonumber\\&&\times\left(1+\frac{2\kappa}
{g-k-\frac{1}{R}}\frac{1}{1-e^{-2\kappa R}}\right)e^{-2\kappa \ell}\Bigg]^{n-1}\\ 
&&\Omega_{n+1}^n = \Omega_2^1 \Bigg[\left(\frac{1-e^{-2\kappa R}}{1-e^{-2\kappa(R+\ell)}}\right)
\nonumber\\&&\times\left(1+\frac{2\kappa}
{g-k-\frac{1}{R}}\frac{1}{1-e^{-2\kappa R}}\right)e^{-2\kappa \ell}\Bigg]^{n-1}
.\label{omegan1nsphere}
\end{eqnarray}
As in the planar case, the resummation of the series ~(\ref{Hmin7}) leads to Eq.~(\ref{bindingspherical}).

In order to check our formulation for the nonlocal model for fixed boundary conditions 
on the wall, we will make use of the diagrams
\begin{eqnarray}
&&\fig{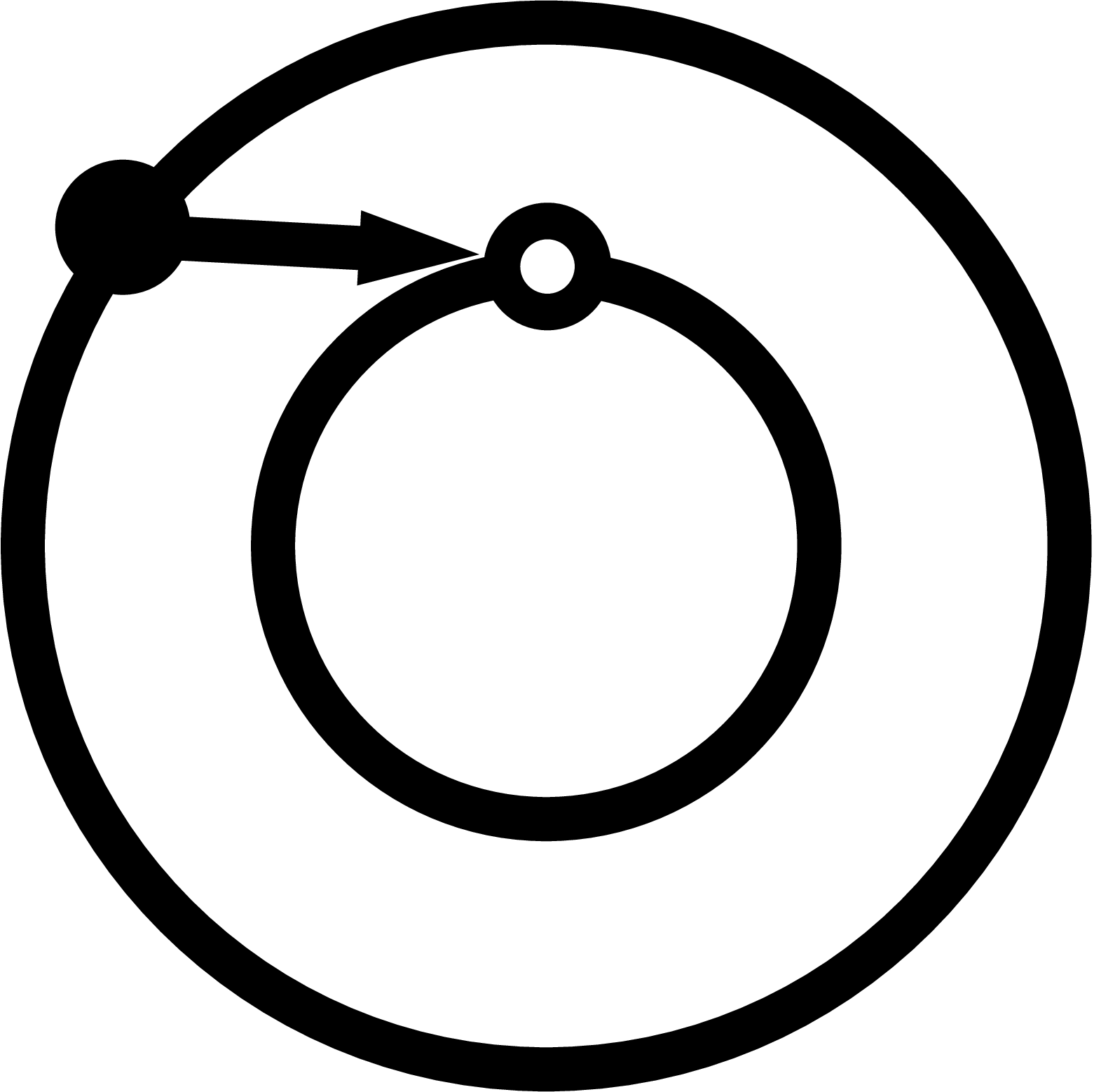}=\left(1+\frac{\ell}{R}\right)e^{-\kappa \ell}\nonumber\\
&&\times \left(2 -\left(1 + \frac{1}{\kappa R}\right)\left(1-e^{-2\kappa R}\right)\right), \\
&&\fig{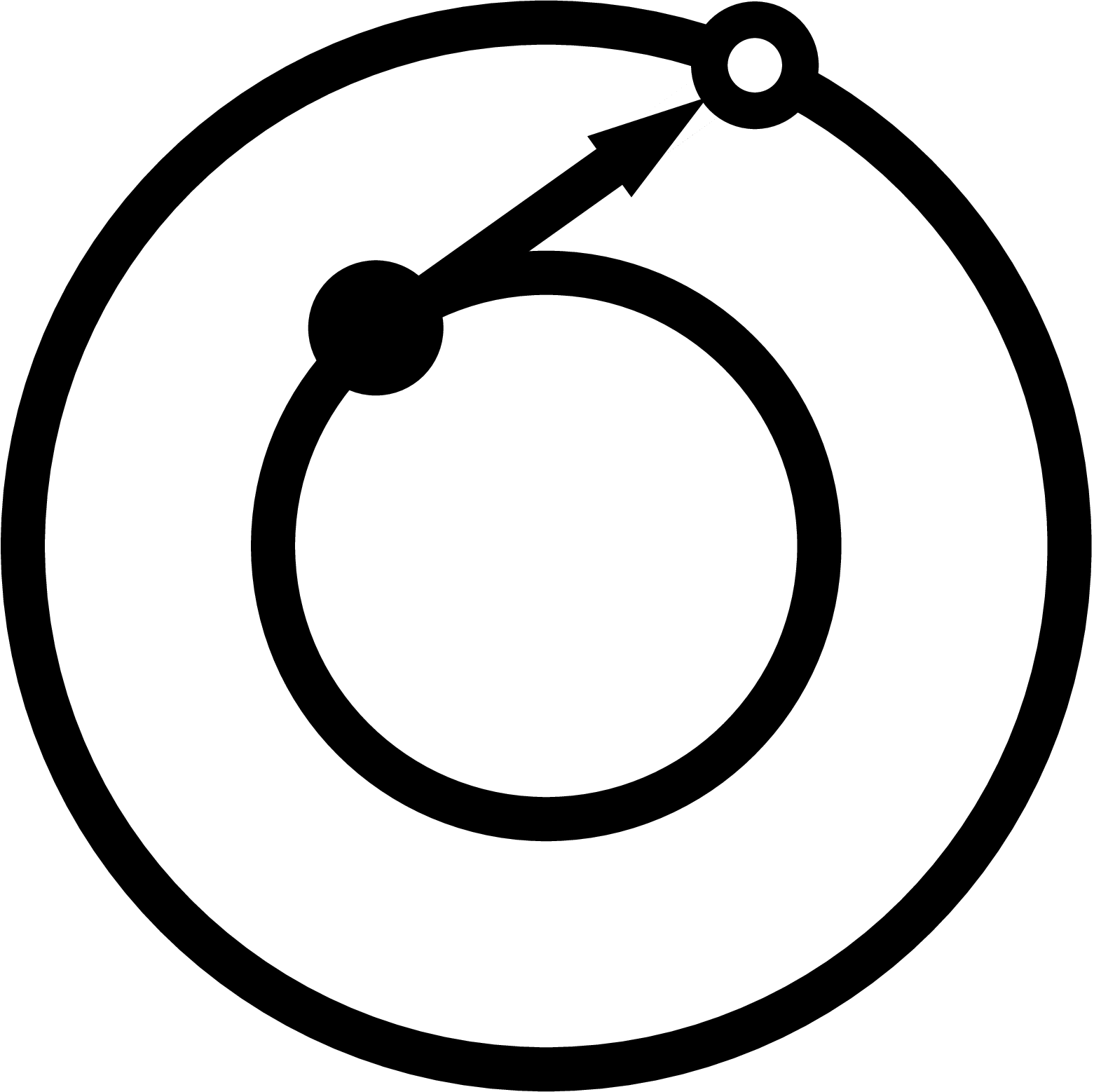}=-\frac{e^{-\kappa \ell}}{1+\frac{\ell}{R}}\nonumber\\
&&\times \left(\left(1 + \frac{1}{\kappa(R+\ell)}\right)\left(1-e^{-2\kappa R}\right)\right),
\\
&&\fig{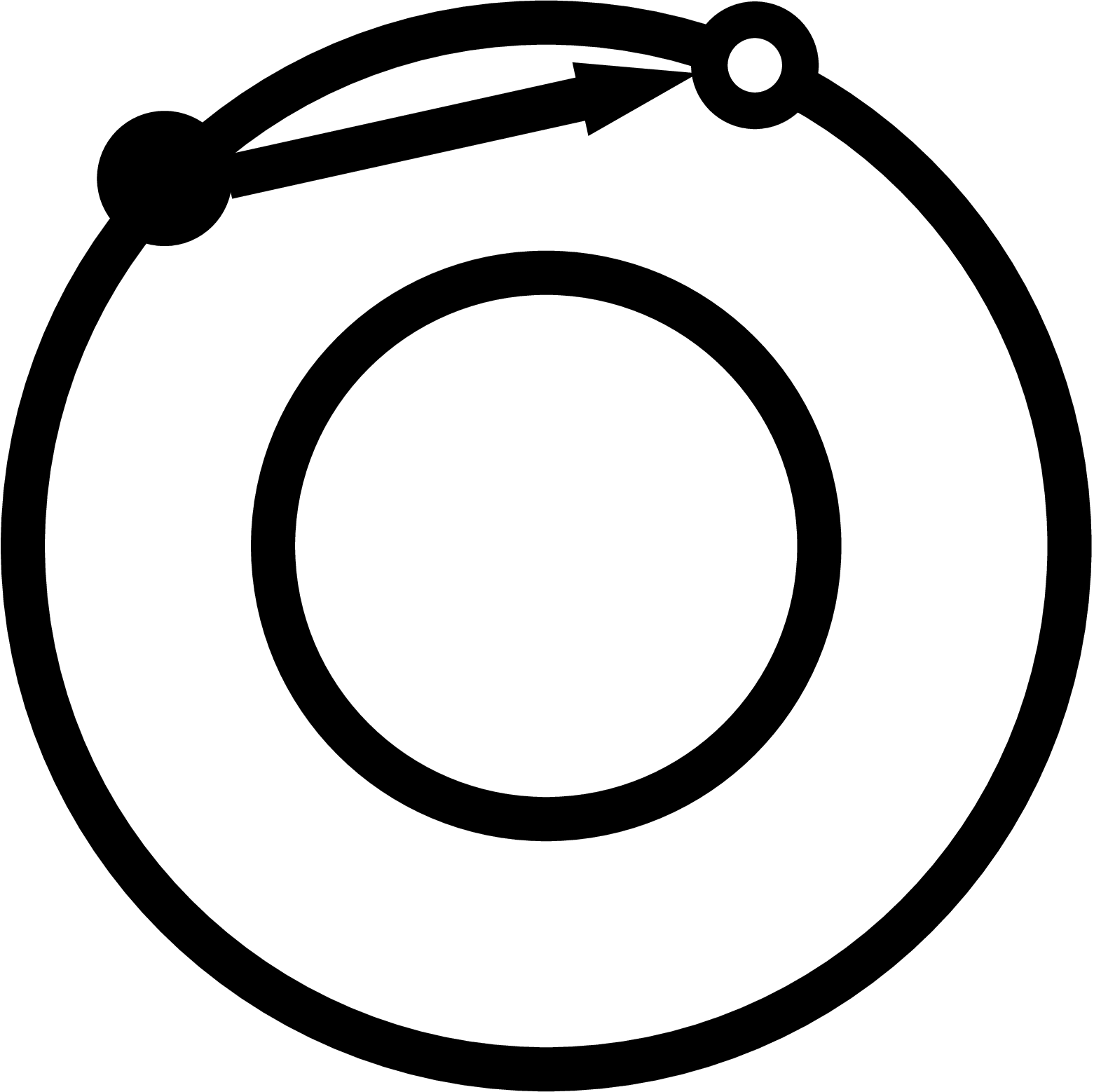} =
-\frac{1}{\kappa (R+\ell)}\nonumber\\
&&\times\left\{1-[1+\kappa (R+\ell)]e^{-2\kappa (R+\ell)}\right\}.
\end{eqnarray} 
The total contribution of the segments of the diagrams on the liquid-gas interface now depends on its positions. The leftmost
one is composed of $U$ bonds and it has a contribution $I_\ell$ given by Eq.~(\ref{il}). Otherwise, the segments 
are composed of $\partial K/\kappa$ bonds, with a contribution 
\begin{equation}
I^\prime_\ell= \frac{1}{\left(1 + \frac{1}{\kappa(R+\ell)}\right)\left(1-e^{-2\kappa (R+\ell)}\right)}.
\end{equation}
Similarly, the segments of the diagrams on the wall on the left extreme of the diagram contribute as 
\begin{equation}
(I^{\prime})^1_\psi= \frac{1}{1-e^{-2\kappa R}},
\end{equation}
which is the limit of Eq.~(\ref{ipsi1}) when $g\to -\infty$, 
and otherwise as
\begin{equation}
(I^{\prime})^2_\psi= \frac{1}{2-\left(1 + \frac{1}{\kappa R}\right)\left(1-e^{-2\kappa R}\right)}.
\end{equation}
Thus, the expressions from Eqs. (\ref{omega11-2}), (\ref{omega12-2}), and (\ref{omega21-2})
for $\Omega_1^1$, $\Omega_1^2$, and $\Omega_2^1$ can be resummed as 
\begin{eqnarray}
\Omega_1^1 &=& 8\pi R(R+\ell)
\frac{e^{-\kappa \ell}}{1-e^{-2\kappa(R+\ell)}},\\
\Omega_1^2 &=& 4\pi (R+\ell)^2 \left(\frac{1-e^{-2\kappa R}}{(1-e^{-2\kappa(R+\ell)})^2}\right)
e^{-2 \kappa \ell},\\
\Omega_2^1 &=& 4 \pi R^2   
\frac{e^{-2\kappa \ell}}{1-e^{-2\kappa(R+\ell)}},
\end{eqnarray}
and for higher-order contributions we get that
\begin{eqnarray}
&&\Omega_n^n = \Omega_1^1 
\left[\left(\frac{1-e^{-2\kappa R}}{1-e^{-2\kappa(R+\ell)}}\right)
e^{-2\kappa \ell}\right]^{n-1},\\
&&\Omega_n^{n+1} = \Omega_1^2 
\left[\left(\frac{1-e^{-2\kappa R}}{1-e^{-2\kappa(R+\ell)}}\right)
e^{-2\kappa \ell}\right]^{n-1},\\
&&\Omega_{n+1}^n = \Omega_2^1 
\left[\left(\frac{1-e^{-2\kappa R}}{1-e^{-2\kappa(R+\ell)}}\right)
e^{-2\kappa \ell}\right]^{n-1},
\end{eqnarray}
which coincide with the expressions Eqs. (\ref{omega11sphere})-(\ref{omegan1nsphere}) in the limit $g\to -\infty$.

\subsection{Cylindrical interfaces}

Finally, we will consider the problem of wetting around a cylinder of radius $R$ and length $L$ (large enough
to neglect border effects), where the 
liquid-gas interfacial configuration is a concentric cylinder of radius $R+\ell$.
In this case, $\mathcal{F}_{wl}[\psi]$ is given by 
\begin{equation}
\mathcal{F}_{wl}[\psi]=\sigma_{wl} \mathcal{A}_{wl} \frac{\frac{K_1(\kappa R)}{K_0(\kappa R)}}{1-\frac{\kappa}{\kappa-g}
\left(1-\frac{K_1(\kappa R)}{K_0(\kappa R)}\right)} ,
\label{fwlcylinder}
\end{equation}
where $\sigma_{wl}$ is the surface tension for the planar wall-liquid interface,
$\mathcal{A}_{wl}=2\pi R L$ is the area of the cylinder,  
and $K_0$ and $K_1$ are the modified Bessel functions of the second kind and order 0 and 1, respectively. 
For large $\kappa R$, this expression can be approximated as
\begin{eqnarray}
&&\mathcal{F}_{wl}[\psi]=\sigma_{wl} \mathcal{A}_{wl}\Bigg[1-\left(\frac{g}{\kappa-g}\right)\frac{1}{2\kappa R}\nonumber\\
&&
+ \left(\frac{g}{\kappa-g}\right)\left(1+\frac{2\kappa}{\kappa-g}\right)\frac{1}{(2\kappa R)^2}+{O}(R^{-3})\Bigg] ,
\end{eqnarray}
which satisfies Eq.~(\ref{deltafwlexpansion}) since 
the mean curvature and
the Gaussian curvature on the cylinder are $H=-1/2R$ and $K_G=0$, respectively.
Equation~(\ref{fwlcylinder}) can be obtained in a similar way as in the spherical case 
from the diagrammatic expansion ~(\ref{F_chain_1}), where 
the relevant diagrams are 
\begin{eqnarray}
\fig{diagram142}&=&2\kappa R I_0(\kappa R) K_0(\kappa R)-1,\\
\fig{diagram143}&=&2 \kappa R I_1(\kappa R)K_0(\kappa R)
\nonumber\\
&=&1-2\kappa R I_0(\kappa R)K_1(\kappa R),
\\
\fig{diagram144}&=&2\pi R L,
\end{eqnarray}
where $I_0$ and $I_1$ are the modified Bessel function of the first kind and 
order 0 and 1, respectively.
Note that as in the planar and spherical cases, they are independent of the position of the open circle. Thus, as in these
previous cases, the contribution of each diagram is the product of its bonds. 

Similarly, $H[\ell]$ has the expression
\begin{equation}
H[\ell]=\frac{\sigma\mathcal{A}_{lg}}{2\kappa R K_0[\kappa (R+\ell)]I_0[\kappa (R+\ell)]},
\label{hcylinder}
\end{equation}
where $\sigma$ is the 
surface tension for the planar liquid-gas interface,
with area $ \mathcal{A}_{lg}=2\pi (R+\ell)L$. 
For large $\kappa R$, Eq.~(\ref{hcylinder}) yields
\begin{equation}
H[\ell]\approx \sigma\mathcal{A}_{lg}\left(1-\frac{1}{8(\kappa R)^2}\right),
\end{equation}
in agreement with Eq.~(\ref{Helfrich}). 
The diagrammatic expansion Eq.~(\ref{F_chain}) can be also evaluated 
explicitly in this case, where now the basic diagrams are
\begin{eqnarray}
\fig{diagram142}&=&2\kappa (R+\ell) I_0[\kappa (R+\ell)]K_0[\kappa (R+\ell)]-1,\nonumber\\
\\
\fig{diagram144}&=&2\pi (R+\ell)L.
\end{eqnarray}
After resummation of Eq.~(\ref{F_chain}), we recover Eq.~(\ref{hcylinder}).

Finally, the binding potential $W[\ell,\psi]$ has the expression
\begin{eqnarray}
&&W[\ell,\psi]=
\left(\frac{\pi L}{\kappa\Delta}\right)
\Bigg[2\kappa m_0 \left(-\frac{h_1+gm_0}{g}\right)
\nonumber\\ 
&&+ \kappa \left(\frac{h_1+gm_0}{g}
\right)^2 
\left(\frac{1}{1-\frac{\kappa}{g}\frac{K_1(\kappa R)}{K_0(\kappa R)}}\right)
\left(\frac{K_0[\kappa (R+\ell)]}
{K_0(\kappa R)}\right)
\nonumber\\
&& + 
\kappa m_0^2 \frac{I_0(\kappa R)}{I_0[\kappa (R+\ell)]}\left(1+\frac{\kappa}{g}
\frac{I_1(\kappa R)}{I_0(\kappa R)}\right)
\Bigg]
\label{bindingcylindrical}
\end{eqnarray}
where $\Delta$ is defined as
\begin{eqnarray} 
\Delta&=&I_0[\kappa (R+\ell)]K_0(\kappa R)\left(1-\frac{\kappa}{g}
\frac{K_1(\kappa R)}{K_0(\kappa R)}\right)\nonumber\\
&-&K_0[\kappa (R+\ell)]I_0(\kappa R)
\left(1+\frac{\kappa}{g}
\frac{I_1(\kappa R)}{I_0(\kappa R)}\right).
\end{eqnarray}
The modified Bessel functions can be approximated asymptotically for 
large values of their arguments as
\begin{eqnarray}
&&K_0(x)\sim K_1(x) \sim \sqrt{\frac{\pi}{2x}}e^{-x}\left(1+{O}
\left(\frac{1}{x}\right)\right),\nonumber\\
&&I_0(x)\sim I_1(x)\sim 
\sqrt{\frac{1}{2\pi x}}e^x\left(1+{O}\left(\frac{1}{x}\right)\right).
\end{eqnarray}
Thus, Eq.~(\ref{bindingcylindrical}) reduces, for large $\kappa R$, to
\begin{eqnarray}
&&W[\ell,\psi]=2\kappa m_0 \left(-\frac{h_1+gm_0}{g-\kappa}\right)
\frac{\sqrt{\mathcal{A}_{wl}\mathcal{A}_{lg}} 
e^{-\kappa\ell}}{1-\frac{g+\kappa}{g-\kappa}
e^{-2\kappa \ell}}
\nonumber\\ 
&& + \frac{g+\kappa}{g-\kappa}
\kappa m_0^2
\frac{\mathcal{A}_{lg}
e^{-2\kappa\ell}}{1-\frac{g+\kappa}{g-\kappa}
e^{-2\kappa \ell}}
\nonumber\\
&&+ \kappa \left(\frac{h_1+gm_0}{g-\kappa}
\right)^2 \frac{\mathcal{A}_{wl}
e^{-2\kappa\ell}}{1-\frac{g+\kappa}{g-\kappa}
e^{-2\kappa \ell}}
\label{bindingcylindrical2}
\end{eqnarray}
up to corrections of order $(\kappa R)^{-1}$ and $[\kappa(R+\ell)]^{-1}$. 
This expression 
is consistent with the expression reported in Ref. \cite{laura} and
it reduces to Eq.~(\ref{bindingplanar}) for $R\to \infty$.

This result can be also obtained within the nonlocal 
model. We will first consider the decorated version of the original 
nonlocal model. In this formalism, 
the basic diagrams model are 
\begin{eqnarray}
\fig{diagram145}&=&2\pi (R+\ell) L,\\
\fig{diagram146}&=&2\pi R L,\\
\fig{diagram147}&=&2\kappa (R+\ell) I_0(\kappa R)K_0[\kappa (R+\ell)],\\
\fig{diagram148}&=&2\kappa R I_0(\kappa R)K_0[\kappa (R+\ell)],\\
\fig{diagram149}&=&2\kappa R I_0(\kappa R)K_0(\kappa R),\\
\fig{diagram150}&=&2\kappa (R+\ell) I_0[\kappa(R+\ell)]K_0[\kappa (R+\ell)], \\
\fig{diagram151}&=&2\kappa R I_1(\kappa R) K_0(\kappa R)-1.
\end{eqnarray}
Again these diagrams do not depend on the position
associated with the open circle. We proceed as in the spherical case to obtain
the expressions of 
$\Omega_n^n$, $\Omega_{n+1}^n$, and $\Omega_{n}^{n+1}$. 
The resummation of all the contributions of the segments
of consecutive ($U$) bonds on the liquid-gas interface is
\begin{equation}
I_\ell=\frac{1}{2\kappa (R+\ell) I_0[\kappa (R+\ell)] K_0[\kappa (R+\ell)]}.
\label{ilcyl}
\end{equation}
On the other hand, 
segments on the wall contribute as 
\begin{equation}
I_\psi^1=\left(\frac{1}{1-\frac{\kappa}{g}\frac{K_1(\kappa R)}{K_0(\kappa R)}} 
\right)
\frac{1}{2\kappa R I_0(\kappa R)K_0(\kappa R)}
\label{ipsi1cyl}
\end{equation}
if the segment is at any of the extremes of the diagram and otherwise
\begin{equation}
I_\psi^2=
\left(\frac{1+\frac{\kappa}{g}\frac{I_1(\kappa R)}{I_0(\kappa R)}}
{1-\frac{\kappa}{g}\frac{K_1(\kappa R)}{K_0(\kappa R)}} 
\right)
\frac{1}{2\kappa R I_0(\kappa R)K_0(\kappa R)}.
\label{ipsi2cyl}
\end{equation}   
With these results, Eqs. (\ref{omega11}), (\ref{omega12}), and (\ref{omega21}) reduce, respectively, to  
\begin{eqnarray}
&&\Omega_1^1 = \frac{2\pi L}{\kappa K_0(\kappa R)I_0[\kappa (R+\ell)]}\left(\frac{1}{1-\frac{\kappa}{g}\frac{K_1(\kappa R)}{K_0(\kappa R)}}\right)
,\label{omega11cylinder}\\
&&\Omega_1^2 = \frac{\pi L}{\kappa K_0(\kappa R)I_0[\kappa (R+\ell)]}
\left(\frac{I_0(\kappa R)}{I_0[\kappa(R+\ell)]}
\right)\nonumber\\&&\times\left(\frac{1+\frac{\kappa}{g}\frac{I_1(\kappa R)}{I_0(\kappa R)}}{1-\frac{\kappa}{g}\frac{K_1(\kappa R)}{K_0(\kappa R)}}\right)
,\label{omega12cylinder}\\
&&\Omega_2^1 = \frac{\pi L}{\kappa K_0(\kappa R)I_0[\kappa (R+\ell)]}
\left(\frac{K_0[\kappa (R+\ell)]}{K_0(\kappa R)}
\right)\nonumber\\&&\times\left(\frac{1}{1-\frac{\kappa}{g}\frac{K_1(\kappa R)}{K_0(\kappa R)}}\right)^2
.\label{omega21cylinder}
\end{eqnarray}
For higher-order contributions
\begin{eqnarray}
&&\Omega_n^n = \Omega_1^1 
\Bigg[\left(\frac{I_0(\kappa R)K_0[\kappa (R+\ell)]}{I_0[\kappa(R+\ell)]
K_0(\kappa R)}\right)
\nonumber\\
&&\times\left(\frac{1+\frac{\kappa}{g}\frac{I_1(\kappa R)}
{I_0(\kappa R)}}{1-\frac{\kappa}{g}\frac{K_1(\kappa R)}{K_0(\kappa R)}}\right)
\Bigg]^{n-1},\label{omeganncylinder}\\
&&\Omega_n^{n+1} = \Omega_1^2 
\Bigg[\left(\frac{I_0(\kappa R)K_0[\kappa (R+\ell)]}{I_0[\kappa(R+\ell)]
K_0(\kappa R)}\right)
\nonumber\\
&&\times\left(\frac{1+\frac{\kappa}{g}\frac{I_1(\kappa R)}
{I_0(\kappa R)}}{1-\frac{\kappa}{g}\frac{K_1(\kappa R)}{K_0(\kappa R)}}\right)
\Bigg]^{n-1}, \label{omegann1cylinder}\\
&&\Omega_{n+1}^n = \Omega_2^1 
\Bigg[\left(\frac{I_0(\kappa R)K_0[\kappa (R+\ell)]}{I_0[\kappa(R+\ell)]
K_0(\kappa R)}\right)
\nonumber\\
&&\times\left(\frac{1+\frac{\kappa}{g}\frac{I_1(\kappa R)}
{I_0(\kappa R)}}{1-\frac{\kappa}{g}\frac{K_1(\kappa R)}{K_0(\kappa R)}}\right)
\Bigg]^{n-1}, 
\label{omegan1ncylinder}
\end{eqnarray}
which lead to Eq.~(\ref{bindingcylindrical}) after resummation
of the series (\ref{Hmin7}).

For our formulation of the nonlocal model for fixed boundary conditions 
on the wall, we consider the diagrams
\begin{eqnarray}
&&\fig{diagram152}=2\kappa (R+\ell)I_1(\kappa R)K_0[\kappa (R+\ell)],
\\
&&\fig{diagram153}=-2\kappa RI_0(\kappa R)K_1[\kappa (R+\ell)],
\\
&&\fig{diagram154} = 
2\kappa (R+\ell) I_1[\kappa (R+\ell)] K_0[\kappa (R+\ell)]
-1.
\nonumber\\
\end{eqnarray} 
The contributions of segments on the liquid-gas interface are given by
Eq.~(\ref{ilcyl}) if the segment is on the left extreme and otherwise by
\begin{equation}
I^\prime_\ell= 
\frac{1}{2\kappa (R+\ell) I_0[\kappa(R+\ell)] K_1[\kappa(R+\ell)]}.
\end{equation}
Similarly, the segments of the diagrams on the wall on the left extreme of the diagram contribute as 
\begin{equation}
(I^{\prime})^1_\psi= 
\frac{1}{2\kappa R I_0(\kappa R) K_0(\kappa R)}
\end{equation}
and otherwise as
\begin{equation}
(I^{\prime})^2_\psi= \frac{1}{2\kappa R I_1(\kappa R)K_0[\kappa (R+\ell)]}.
\end{equation}
The expressions from Eqs. (\ref{omega11-2}), (\ref{omega12-2}), and (\ref{omega21-2})
for $\Omega_1^1$, $\Omega_1^2$, and $\Omega_2^1$ are 
\begin{eqnarray}
&&\Omega_1^1 = \frac{2\pi L}{\kappa K_0(\kappa R)I_0[\kappa (R+\ell)]},
\label{omega11cylinder-2}\\
&&\Omega_1^2 = \frac{\pi L}{\kappa K_0(\kappa R)I_0[\kappa (R+\ell)]}
\left(\frac{I_0(\kappa R)}{I_0[\kappa(R+\ell)]}
\right),
\label{omega12cylinder-2}\\
&&\Omega_2^1 = \frac{\pi L}{\kappa K_0(\kappa R)I_0[\kappa (R+\ell)]}
\left(\frac{K_0[\kappa (R+\ell)]}{K_0(\kappa R)}
\right),
\label{omega21cylinder-2}
\end{eqnarray}
and for higher-order contributions
\begin{eqnarray}
&&\Omega_n^n = \Omega_1^1 
\left(\frac{I_0(\kappa R)K_0(\kappa (R+\ell))}{I_0[\kappa(R+\ell)]
K_0(\kappa R)}\right)
^{n-1},\label{omeganncylinder-2}\\
&&\Omega_n^{n+1} = \Omega_1^2 
\left(\frac{I_0(\kappa R)K_0(\kappa (R+\ell))}{I_0[\kappa(R+\ell)]
K_0(\kappa R)}\right)
^{n-1}, \label{omegann1cylinder-2}\\
&&\Omega_{n+1}^n = \Omega_2^1 
\left(\frac{I_0(\kappa R)K_0[\kappa (R+\ell)]}{I_0[\kappa(R+\ell)]
K_0(\kappa R)}\right)
^{n-1}, 
\label{omegan1ncylinder-2}
\end{eqnarray}
which correspond to Eqs. (\ref{omega11cylinder})-(\ref{omegan1ncylinder}) in the limit $g\to -\infty$.

\end{document}